\documentclass[11pt,oneside,english]{article}
\usepackage{bm}
\usepackage{booktabs}
\usepackage{geometry}
\usepackage{authblk}
\usepackage{caption}
\usepackage{amsfonts,amssymb,amsmath}
\numberwithin{equation}{section}
\usepackage{mathrsfs}
\usepackage[utf8]{inputenc}
\usepackage{mathtools}
\usepackage[hidelinks]{hyperref}
\usepackage{hyperref}
\usepackage{xcolor}
\hypersetup{
    colorlinks,
    linkcolor={red!70!black},
    citecolor={blue!90},
    urlcolor={blue!80!black}
}
\usepackage{tensor}
\usepackage{tikz}
\usetikzlibrary{arrows,decorations.markings}
\tikzset{
  big arrow/.style={
    decoration={markings,mark=at position 1 with {\arrow[scale=2,#1]{>}}},
    postaction={decorate},
    shorten >=0.4pt},
  big arrow/.default=black}
\usepackage[nameinlink,capitalize]{cleveref}
\DeclarePairedDelimiter\ceil{\lceil}{\rceil}
\DeclarePairedDelimiter\floor{\lfloor}{\rfloor}

\makeatletter
\renewenvironment{abstract}{%
    \if@twocolumn
      \section*{\abstractname}%
    \else %% <- here I've removed \small
      \begin{center}%
        {\bfseries \normalsize\abstractname\vspace{\z@}}%  %% <- here I've added \Large
      \end{center}%
      \quotation
    \fi}
    {\if@twocolumn\else\endquotation\fi}
\makeatother

\crefname{figure}{Figure}{Figures}
\labelcrefformat{subequation}{#2(#1)#3}
\labelcrefrangeformat{subequation}{#3(#1)#4 to #5(#2)#6}

\let\originalleft\left
\let\originalright\right
\renewcommand*{\left}{\mathopen{}\mathclose\bgroup\originalleft}
\renewcommand*{\right}{\aftergroup\egroup\originalright}

    \DeclareMathOperator{\Adj}{Adj}
    \DeclareMathOperator{\SO}{SO}
    \DeclareMathOperator{\Sp}{Sp}
    \DeclareMathOperator{\SU}{SU}
    \DeclareMathOperator{\U}{U}
    \DeclareMathOperator{\gE}{E}

    \newcommand*{\au}{\ensuremath\mathfrak{u}}

    \newcommand*{\cN}{\mathcal{N}}
    \newcommand*{\cO}{\mathcal{O}}

    \newcommand*{\bP}{\mathbb{P}}
    \newcommand*{\F}{\mathbb{F}}
    \newcommand*{\Z}{\mathbb{Z}}
    \newcommand*{\Q}{\mathbb{Q}}
    \newcommand*{\C}{\mathbb{C}}
    \newcommand*{\R}{\mathbb{R}}
    \newcommand*{\sfG}{\mathsf{G}}
    \newcommand*{\sfr}{\mathsf{r}}
    \newcommand*{\sfR}{\mathsf{R}}
    \newcommand*{\scB}{\mathscr{B}}
    \newcommand*{\scI}{\mathscr{I}}
    \newcommand*{\scL}{\mathscr{L}}
    \newcommand*{\scO}{\mathscr{O}}
    \newcommand*{\scV}{\mathscr{V}}

    \newcommand*{\SM}{\ensuremath(\SU(3) \times \SU(2) \times \U(1)) / \Z_6}

    \DeclareMathOperator{\diag}{diag}
    \DeclareMathOperator{\GL}{GL}
    \DeclareMathOperator{\Id}{Id}
    \DeclareMathOperator{\PD}{PD}
    \DeclareMathOperator{\rk}{rk}
    \DeclareMathOperator{\sign}{sign}
    \DeclareMathOperator{\Span}{span}
    \DeclareMathOperator{\trace}{tr}

    \newcommand*{\transpose}{\ensuremath\mathrm{t}}

    \newcommand*{\cF}{\mathcal{F}}
    \newcommand*{\ag}{\mathfrak{g}}

    \newcommand*{\mult}[1]{\ensuremath n_{#1}}

    \newcommand*{\chIndex}[1]{\ensuremath\chi_{#1}}
    \newcommand*{\httv}{\ensuremath H^{2,2}_\text{vert}} %cohomology
    \newcommand*{\hvtt}{\ensuremath H^\text{vert}_{2,2}} %homology
    \newcommand*{\diff}{\mathop{}\!\mathrm{d}}
    \newcommand*{\bigdiff}{\mathop{}\!\mathrm{D}}
\newcommand{\remove}[1]{}

\title{
\Huge  Chiral matter multiplicities
 and resolution-independent structure
in 4D F-theory models}
\author[$\dag$]{\Large Patrick Jefferson} 
\author[$\dag$]{\Large Washington Taylor} 
\author[$\dag\dag$]{\Large Andrew P. Turner}
\affil[$\dag$]{\normalsize \emph{Center for Theoretical Physics, Department of Physics, Massachusetts Institute of Technology, 77 Massachusetts Avenue, Cambridge, MA 02139, USA}}
\affil[$\dag\dag$]{\normalsize \emph{Department of Physics and Astronomy, University of Pennsylvania, Philadelphia, PA 19104, USA}}

\date{\texttt{pjeffers@mit.edu}~~ \texttt{wati@mit.edu} ~~\texttt{turnerap@sas.upenn.edu}}
\begin{document}
\maketitle
\begin{tikzpicture}[remember picture,overlay]
   \node[anchor=north east,inner sep=0pt] at (current page.north east)
              {$\begin{array}{ccc}&&\\ \\ \text{MIT-CTP-5293}&&\end{array}$};
\end{tikzpicture}
\thispagestyle{empty}
\begin{abstract}
\noindent Motivated by questions related to the landscape of flux compactifications, we combine new and existing techniques into a
systematic, streamlined approach for computing vertical fluxes and chiral matter multiplicities in 4D F-theory models.  A central feature of our
approach is the conjecturally resolution-independent intersection pairing of the vertical part
of the integer middle cohomology of smooth elliptic Calabi-Yau fourfolds, relevant for computing chiral indices and related aspects of 4D F-theory flux vacua.
We illustrate our approach by analyzing vertical flux backgrounds for F-theory
models with simple, simply-laced gauge groups and generic matter content, as well as models with $\U(1)$ gauge factors.
%$\SU(N)$, $\SO(4 k + 2)$, $\gE_6$, and $(\SU(2)
%\times \U(1))/\Z_2$ and
We explicitly analyze resolutions of these F-theory models in which
the elliptic fiber is realized as a cubic in $\bP^2$ over an
arbitrary (e.g., not necessarily toric) smooth base, and confirm the
independence of the intersection pairing of the vertical part of the middle
cohomology for the resolutions we study. In
each model, we find that vertical flux
backgrounds can produce nonzero multiplicities for a spanning set of anomaly-free
chiral matter field combinations, suggesting that F-theory geometry
 imposes no additional linear constraints on allowed matter
 representations
 beyond those implied by 4D
anomaly cancellation.
%We find the number of anomaly-free families of chiral matter
%multiplets is related to the rank of the reduced intersection pairing
%matrix in models free of non-minimal codimension-three singularities,
%with the curious exception of the $\SO(11)$ model. We use the
%methods developed in this paper to analyze Minimal Supersymmetric
%Standard Model--like spectra in the universal $\SM$ model of
%1912.10991 in a forthcoming publication.
\end{abstract}
\flushbottom
\newpage
\tableofcontents
\addtocontents{toc}{\protect\thispagestyle{empty}}
\setcounter{page}{1}
%%%%%%%%%%%%%%%%%%%%%%%%%%%%%%%%%%%%%%%%%%%%%%%%%%%%%%%%%%%%%%%%%%%%%%%%%%%%%%
%%%%%%%%%%%%%%%%%%%%%%%%%%%%%%%%%%%%%%%%%%%%%%%%%%%%%%%%%%%%%%%%%%%%%%%%%%%%%%
%%%%%%%%%%%%%%%%%%%%%%%%%%%%%%%%%%%%%%%%%%%%%%%%%%%%%%%%%%%%%%%%%%%%%%%%%%%%%%
\section{Introduction}
\label{sec:intro}

F-theory \cite{VafaF-theory, MorrisonVafaI, MorrisonVafaII} provides a
powerful geometric framework for describing a large class of
supersymmetric string theory vacua.  In particular, F-theory can be
used to describe a vast number of 4D $\cN = 1$ supergravity
theories with gauge symmetries.

Because F-theory provides a uniquely broad nonperturbative perspective
on the set of supersymmetric string vacuum solutions, there are two
rather fundamental questions about this set that can be explored
fruitfully within this particular branch of string
theory.  First is the question of the
extent to which F-theory, or string theory more generally, can provide
a UV description of any low-energy field theory that has
no known
obstruction to coupling to quantum gravity; this question has been usefully
framed as the problem of delineating the \emph{swampland}
\cite{VafaSwamp, OoguriVafaSwamp} of apparently consistent
low-energy
effective theories of gravity not realized in string theory.  Second
is the question of how the gauge group, chiral matter content, and
other physical features of the observed Standard Model of particle
physics can be realized in string theory, and the extent to which this
physics is typical or requires extensive fine tuning.

There has been a great deal of work on each of these questions in the
context of F-theory over the last two decades
(for some recent reviews see, e.g.,
\cite{Cvetic:2022fnv,Marchesano:2022qbx}). However, neither question has been answered
definitively.
%, and progress on these questions appears to be hindered by a combination of unresolved conceptual issues and
%limitations of available computational techniques for analyzing
%flux vacua.

In this paper, we
investigate some aspects of F-theory flux backgrounds that are relevant for both
of these questions. As part of our investigation, we bring together a variety of methods (some known and some new) to
frame a systematic approach for characterizing chiral matter in broad classes of 4D F-theory models.%We use these methods to
%analyze the chiral matter content of a universal class of
%geometries with a tuned gauge group matching that of the Standard
%Model.  We also investigate more generally the extent to which

Many of the known methods we employ in our approach have been explored
in different threads of the literature, as there has been extensive research on understanding how chiral matter arises from
fluxes in 4D F-theory models. Chiral matter in F-theory GUT models
was described locally in \cite{Donagi:2008ca, BeasleyHeckmanVafaI,
  BeasleyHeckmanVafaII}, and a more systematic description in terms of
fluxes and global geometry was developed in
\cite{Marsano:2010ix,Grimm:2011tb,Braun_2012,Marsano_2011,KRAUSE20121,Grimm:2011fx}, among others.  Much of this work is reviewed in
\cite{WeigandTASI};  many of these papers compute the multiplicities of chiral matter by identifying
geometric ``matter surfaces'' (i.e. specific holomorphic four-cycles in the elliptic Calabi-Yau fourfold) through which fluxes can be integrated to obtain the chiral indices, whereas by contrast \cite{Grimm:2011fx} and some
related works  \cite{Grimm:2011sk,Cvetic:2012xn}
indirectly compute the chiral indices by identifying fluxes through various holomorphic cycles
with one-loop Chern--Simons couplings in 3D (which can be interpreted as linear combinations of the chiral indices). We follow the latter approach for explicit
computations in this paper, though the resulting insights may shed
light on some subtle aspects of the geometry of matter
surfaces.

Our approach for studying 4D F-theory vacua offers computational and
conceptual simplifications relevant for the two questions posed above.  The computational simplification offered by our approach
is that it combines the results of the previous work cited above with the
techniques of \cite{Esole:2017kyr} (used for computing intersection
numbers) into a streamlined algorithm for analyzing chiral matter and
vertical fluxes, which allows us to easily survey large families of
F-theory flux vacua. We demonstrate the utility of our
approach by analyzing numerous examples, some not previously studied
in the literature, of flux vacua in models with fixed gauge group
$\sfG$ over arbitrary smooth threefold base.  Conceptually, our
approach is simpler in that while previous work on chiral matter in 4D
F-theory models has %been based
relied in an essential way upon specific choices of resolution of
the singularities in the Weierstrass model defining the F-theory
compactification, in this paper we take steps towards analyzing the chiral
multiplicities, as well as the linear constraints they satisfy, in
terms of (conjecturally) resolution-independent geometric structure intrinsic to the global elliptic Calabi-Yau fourfold.

The main resolution-independent structure that we make use of here is
related to the intersection pairing on a particular subgroup of
the middle cohomology $H^4(X,\Z)$ of a smooth elliptic Calabi--Yau
(CY) fourfold $X$ resolving the singular Weierstrass
model. Specifically, we study the \emph{nondegenerate} intersection
pairing $M_\text{red}$ acting on the (``vertical'') cohomology
subgroup $\httv(X,\Z) \subset H^4(X,\Z)$ generated by products of divisors in $X$. The
intersection pairing $M_\text{red}$ can be obtained by assembling the
quadruple intersection numbers of $X$ into a matrix $M$ and
removing its nullspace. While the quadruple intersection numbers of
divisors are not generally independent of the choice of resolution $X$ of
the singular Weierstrass model, we find evidence that for all models we study
$M_\text{red}$ (and hence implicitly $M$ as well) is independent of
the choice of $X$, up to an
integral change of basis.  Since $M_\text{red}$ encodes fluxes relevant for
computing chiral matter multiplicities, we highlight the importance of
$M_\text{red}$ as the primary geometric object of interest for
analyzing chiral matter and
vertical flux backgrounds in a manifestly
resolution-independent manner.
The apparent resolution-independence of $M_\text{red}$ and $M$
suggests that
this intersection structure is in some sense an intrinsic mathematical
feature of the singular elliptic CY fourfold that defines a 4D F-theory vacuum and may
have a direct interpretation in this geometric language as well as in type
IIB string theory, without any need for explicit resolution, although
to our knowledge this statement has not been proven in the
mathematical literature.
 There is perhaps a useful analogy to be made here:
Just as the resolution-independent Dynkin diagram associated with a Kodaira singularity type
encodes the resolution-invariant physics of the nonabelian gauge
algebra of an F-theory compactification, this (conjecturally) resolution-independent
part of the intersection structure encodes the resolution-invariant physics connecting
vertical fluxes and chiral matter.\footnote{Note that other resolution-independent structures encoded in the intersection numbers of CY resolutions have been identified in the context of F-theory and M-theory compactifications. For example, the combined fiber diagrams (CFDs) of \cite{Apruzzi:2019kgb} appearing in non-flat resolutions of singular elliptic CY threefolds were shown to be manifestly flop-invariant. Furthermore,
%the Smith normal form of
the intersection pairing between  divisors and certain curve classes
in smooth CY threefolds was shown to have invariant Smith normal form
in \cite{Morrison:2020ool}. We thank S. Schafer-Nameki for bringing
these references to our attention.}

%More generally, we investigate chiral matter in a variety of other
%gauge groups, some not previously studied in the literature.
%The fact that our approach is a convenient
The set of tools that this analysis provides for exploring the
landscape of 4D F-theory flux vacua positions us to clarify aspects of
the first question raised at the beginning of the paper. While 4D anomaly cancellation is satisfied by all
F-theory constructions that have been studied\footnote{For example,
analyses of such conditions were carried out in
\cite{LinWeigandG4, Bies_2017, Corvilain:2017luj,Cheng:2021zjh}.} and is expected to hold in all
%implying that the subset of
4D $\cN=1$ supergravity theories that can be constructed in
F-theory,
% are consistent quantum gravity theories,
it is unknown whether or not all anomaly-free families of chiral
matter can be realized in F-theory. Interestingly, it turns out that
in all %such
 cases we study this is indeed true, 
at least in the sense that for the most generic matter representations
associated with a given gauge group,
fluxes are available that produce
 combinations of massless chiral matter fields that span the linear
 space of anomaly-free matter representations.
%hinting that (at least
%at the level of the number of independent linear combinations of chiral matter allowed for
%the most generic matter representations associated with a given gauge
%group) every 
%anomaly-free configuration of 4D chiral multiplets may have a UV
%completion in F-theory.
%Our approach offers further insight into this question, as 
More specifically,
we 
find in all cases that we study that the number of independent
vertical flux backgrounds in $H_{2,2}^{\text{vert}}(X,\Z)$ that lift
to consistent F-theory flux backgrounds with unbroken gauge
group---equivalently, the rank of $M_\text{red}$ minus the number of
constraints required to preserve 4D local Lorentz and full gauge
symmetry---is the same, and is in particular greater than or equal to
the number of allowed independent families of anomaly-free chiral
matter.
%It is therefore tempting to
%conjecture
Part of this result is to be expected: since the physics of any F-theory model is presumed
resolution-invariant, given the relationship between chiral multiplicities and vertical flux backgrounds, it should follow that the
number of independent vertical flux backgrounds corresponding to independent families of chiral matter multiplets is also a resolution-invariant property of the theory.
%; for models free of
%$(4, 6)$ points, 
%the results of our analysis support this expectation.
%these quantities are in fact equal to one another. 
%What is perhaps unexpected is that
 Our results further suggest that this number is at least as large as the total
number of linearly-independent families allowed by 4D anomaly
cancellation.
%\footnote{We leave for the future a more detailed
%  exploration of
%%In a future publication \cite{46}, we explore
%  models containing $(4, 6)$ loci, in which the rank of $M_\text{red}$
%  is increased and may include extra flux backgrounds describing
%  additional, strongly coupled chiral degrees of freedom localized at
%  the $(4, 6)$ points.} 
%Resolution-independence of $H_{2,2}^{\text{vert}}(X,\Z)$ thus implies
%that the possible set of independent vertical flux backgrounds in
%$H_{2, 2}^\text{vert}(X,\Z)$ \wati{?}  corresponding to independent
%families of chiral matter should be characterizable in a
%resolution-invariant fashion that is closely related to the gauge
%theory description of the low-energy effective theory.  Furthermore,
Since resolution-independence of the lattice pairing $M_{\text{red}}$
also implies that $M$ is resolution-independent, it may be possible to
characterize part of the nullspace of $M$ in a canonical manner that
is related to the 4D anomaly cancellation conditions.  Since the
nullspace of $M$ restricted to the subspace of 4D symmetry-preserving
fluxes can be identified with the set of linear constraints (of which
the 4D anomaly cancellation conditions must necessarily be a subset),
this potentially points to a more systematic method for exploring
possible swampland-like conditions obstructing the F-theory
realization of certain families of chiral matter multiplets, or
showing that no such additional linear conditions can exist, as we
essentially conjecture here.

Regarding the second question raised at the beginning of this paper, one of the initial motivations was to analyze
chiral matter in the family of $\SM$ models found in
\cite{Raghuram:2019efb}. This model has three
independent families of generic chiral matter fields that satisfy 4D
anomaly cancellation, one of which corresponds to the
matter content of the Minimal Supersymmetric Standard Model (MSSM).
This seems to be the broadest class of F-theory models that have a
tuned Standard Model--like gauge group\footnote{That is, a gauge group
  that is directly tuned in the Weierstrass model, as opposed to one
  that arises from breaking a larger GUT group or that is imposed as a
  generic feature of the F-theory base geometry.}, and which naturally
includes Standard Model--like matter.  One subclass of these models
arises naturally through a toric fiber (``$F_{11}$'') construction
\cite{KleversEtAlToric}, and has only the single family of chiral
matter fields associated with the MSSM; chiral matter in some Standard
Model--like $F_{11}$ constructions was recently intensively
investigated in \cite{CveticEtAlQuadrillion}. The approach developed
here gives us a means to check whether F-theory models of the more
general tuned $\SM$ type naturally contain chiral matter in the other
two allowed families, or whether these are forbidden by string
geometry for some reason and hence belong to the swampland. We find
that indeed all three of the allowed chiral matter types are allowed;
we briefly summarize these results here and report further on the
details of this analysis in a forthcoming publication \cite{Jefferson:2022yya}.

%\subsection{Outline}

The structure of this paper is as follows: in \cref{sec:overview}, we
give an overview of the main ideas, technical contributions, and
results of the paper; in
% \cref{sec:4dflux} and
%we explain how to (indirectly) compute the
%sublattice $ \Lambda_C \subset \Lambda_S$ of flux backgrounds
%preserving 4D local Lorentz and gauge symmetry, giving in particular
%an explicit expression for the restriction $M_C$ of the intersection
%pairing $M$ to $\Lambda_C$; in
\cref{sec:4dflux,sec:constraints-homology}, we
%explore a complementary approach to this analysis, whereby we compute
%the lattice $H_{2,2}^{\text{vert}}(X,\Z) = \Lambda_S/ \sim$
%defined by removing the nullspace of $M$ spanned by homologically
%trivial cycles in $\Lambda_S$;
explore two complementary approaches to analyzing the set of
cohomologically-distinct fluxes
that preserve 4D local Lorentz and gauge symmetry, where the two approaches differ by the order in which the symmetry constraints and equivalence
relations in (co)homology are imposed;
\cref{3Dcompare} reviews   the strategy we use %, following
%\cite{Grimm:2011sk,Cvetic:2012xn},
for
determining the precise relationship between chiral indices
$\chi_{\sfr}$ and vertical fluxes, which exploits their relationship
to one-loop Chern--Simons couplings appearing in the 3D low-energy
effective action describing the F-theory Coulomb branch; in
\cref{sec:exampleADE}, we use our approach to study various F-theory models with
simple gauge group $\sfG_\text{na}$; in \cref{sec:abelianmodel}, we
discuss the generalization of our analysis to models with gauge group
$\sfG = (\sfG_\text{na} \times \U(1))/\Gamma$, using the
$(\SU(2) \times \U(1)) / \Z_2$ and $\SM$ models to illustrate various
aspects of the analysis; finally, \cref{sec:conclusions} contains concluding
remarks and future directions. A number of technical results related
to e.g., anomaly cancellation, intersection theory, and resolutions of singular Weierstrass models, are
collected in the appendices.

%In \cref{sec:4dflux}, we give a more detailed
%description of the relevant parts of the structure of the intersection
%form on middle cohomology and describe the general framework that we
%use to compute chiral matter indices in 4D F-theory flux
%compactifications.  In \cref{sec:example}, we apply this methodology
%to F-theory compactifications whose gauge group is a simple ADE group,
%and in \cref{F6model,sec:example-SM} we apply the methodology to theories
%with respective gauge groups $(\SU(2) \times \U(1))/\Z_2$ and $\SM$.  \S \ref{sec:conclusions} contains some
%concluding remarks, and a number of appendices go into technical
%details on group theory, resolutions, and other aspects of the
%analysis.

%%%%%%%%%%%%%%%%%%%%%%%%%%%%%%%%%%%%%%%%%%%%%%%%%%%%%%%%%%%%%%%%%%%%%%%%%%%%%%
%%%%%%%%%%%%%%%%%%%%%%%%%%%%%%%%%%%%%%%%%%%%%%%%%%%%%%%%%%%%%%%%%%%%%%%%%%%%%%
%%%%%%%%%%%%%%%%%%%%%%%%%%%%%%%%%%%%%%%%%%%%%%%%%%%%%%%%%%%%%%%%%%%%%%%%%%%%%%
\section{Overview}
\label{sec:overview}

In this section, we give an overview of the steps needed to
systematically describe a class of 4D F-theory models with a given
gauge group and ultimately to compute the chiral matter content from vertical fluxes using our approach.  In
particular, we try to make clear how various techniques in the
existing literature are integrated into our approach, and where this paper
makes novel contributions.
The current state of knowledge for
many parts of this analysis
is reviewed in more detail in a transcription of Weigand's excellent TASI lectures
\cite{WeigandTASI}.

We are interested in finding a general formulation of the
chiral matter multiplicities for a variety of F-theory constructions
with different gauge groups, in a way that can be expressed succinctly in terms
of the geometry of the base of the F-theory compactification and a
choice of fluxes.  In particular, for a given choice of gauge group
and generic\footnote{See \cref{sec:Weierstrass} for a precise definition of the notion of ``generic matter'' in F-theory compactifications.} matter representations, we are interested in identifying closed
form expressions for the chiral matter multiplicities in a
base-independent\footnote{By ``base-independent'', we mean in a manner
  that does not rely on a specific choice of base. Clearly, the choice
  of base can change the physics of the F-theory vacuum.} (and
resolution-independent) fashion. 
Expressions of this type have been found previously in
the literature using related but distinct combinations of techniques
for various gauge groups, such as $\SU(5)$ \cite{Marsano_2011}, $\text{E}_6$ \cite{Kuntzler:2012bu}, $\U(1) \times \U(1)$,
$(\SU(5) \times \U(1)\times\U(1)) / \Z_5$ \cite{Cveti__2014}, and
$(\SU(3) \times \SU(2) \times \U(1)^2) / \Z_6$
\cite{LinWeigandG4} (see also \cite{Marsano:2009gv}).

At a very heuristic level, the analysis can be described as follows:
for any specific choice of gauge group, it should be possible to
identify a multi-parameter family of Weierstrass models that describes
F-theory models over an arbitrary base with that gauge group and
generic matter.  A resolution $X$ of any of the corresponding CY
fourfolds
gives rise to a well-defined set of intersection numbers, which can be
organized into a matrix $M$ containing the intersection pairing $M_{\text{red}}$ acting on the set of homology
classes $H_{2, 2}^\text{vert}(X,\Z)$.
The intersection pairing is relevant for computing fluxes through certain homology classes dubbed ``matter surfaces''
that encode the multiplicities of chiral matter fields.
The chiral matter multiplicities that are fixed by the choice of gauge
flux and the intersection numbers can also be related directly to the
3D physics arising from a circle reduction of the F-theory
model.
While the choice of
resolution and its associated intersection numbers are not unique,
it should be possible in general to describe
the multiplicities of chiral matter fields in the 4D limit
 in a resolution-independent fashion that depends
only on the intersection structure of the compactification base and a
choice of fluxes in an appropriate basis.
One of the key ingredients in this paper is the identification of a conjecturally
resolution-independent piece of the intersection structure of $X$, namely $M_{\text{red}}$, that
is relevant for understanding the chiral multiplicities.

We now describe each of the
steps in this procedure in a bit more detail, framing the analysis of
the remainder of the paper.

\subsection{Selection of the base}

The first step in choosing an F-theory compactification is the choice
of complex threefold base $B$.
From the IIB string theory point of view, the 10D IIB
theory is compactified on $B$, which we take here to be a
compact K\"ahler threefold.  Note that $B$ need not be a CY
manifold, i.e., the canonical class $K$ of $B$ need not be trivial,
though $-K$ must be an effective class.  The F-theory model
\cite{VafaF-theory, MorrisonVafaI, MorrisonVafaII} is described by a
Weierstrass model
\begin{equation}
y^2 = x^3 + f x + g\,,
\end{equation}
defining an elliptic CY fourfold $X_0$
with base $B$, where $f, g$ are sections of the line bundles $\scO(-4K),\scO(-6K)$, respectively.
In general, the CY fourfold $X_0$ has singularities associated
with loci in the base where the elliptic fiber degenerates.
Degenerations over codimension-one loci in the base are associated
with the gauge group of the F-theory model and degenerations over
codimension-two loci are associated with matter.
F-theory is frequently analyzed as a limit of M-theory on a smooth
resolution $X$ of $X_0$, but the physics should in principle be
independent of resolution as discussed further in \cref{sec:resolution}.

The number of possible bases $B$ is quite large.  The primary constraint is that $B$ cannot contain a divisor $\Sigma$ (codimension-one
algebraic surface) that has a normal bundle that is so negative that
$(f, g)$ need to vanish to orders $(4, 6)$ on $\Sigma$. When such a
divisor exists, the singularity structure of the total space of the
elliptic fibration goes beyond the classification of Kodaira
\cite{Kodaira} and N\'{e}ron \cite{Neron}; there is no smooth CY
resolution and the resulting geometry lies at infinite distance in the
moduli space of compactifications.\footnote{Note that there can be higher-codimension $(4, 6)$
  singularities without a crepant (CY) resolution (see,
  e.g., \cite{Klemm:1996ts}); these geometries, however, lie at finite
  distance in moduli space and seem physically relevant as F-theory
  compactifications.}  A large
range of elliptic CY fourfolds have been studied in the
literature, see, e.g., \cite{Klemm:1996ts,Kreuzer:1997zg,Gray:2013mja}. Restricting to the simple case of toric
$B$, the number of possible bases has been shown by explicit
construction to be at least $10^{755}$ \cite{HalversonLongSungAlg} and
is estimated through Monte Carlo analysis to be of order closer to
$10^{3000}$ \cite{TaylorWangLandscape}.
Many of these bases have codimension-two loci where $(f, g)$ vanish to
orders $(4, 6)$.  These codimension-two loci are generally associated with nonperturbative massless excitations
in the low-energy 4D theory, see,
e.g., \cite{Candelas:2000nc,Achmed-Zade:2018idx,Hayashi:2009ge}; in 6D,
such excitations are generally associated with a superconformal sector
in the theory \cite{SeibergSCFT, HeckmanMorrisonVafa}, and while there
are some parallel aspects of 4D F-theory models
\cite{Apruzzi:2018oge} the structure of
these sectors in four space-time dimensions is less well
understood.

Much of the detailed analysis of chiral matter in 4D F-theory models
has been done in the context of toric geometry.  One advantage of
toric bases is that there are many powerful and simple tools for
computing resolutions, intersection numbers, and other relevant
features of toric varieties that extend to many elliptic CY
fourfolds over toric bases that can be described as hypersurfaces in
toric varieties. At least
in the case of elliptic CY threefolds with relatively large
Hodge numbers over complex surface bases, toric constructions seem to
give a good representative sample of the set of possibilities
\cite{TaylorWangNon-toric}, although for 4D F-theory models with
chiral matter, some features such as GUT breaking are not easily seen
in purely toric contexts (see, e.g., \cite{Marsano:2009wr,
  Braun:2014xka}).  Toric geometry has been used with great efficacy in many
examples in the literature, e.g.,
\cite{KleversEtAlToric,Knapp:2011wk,Braun:2011ux,Buchmuller:2017wpe}.  By
contrast, our analysis employs resolution techniques developed to study
Weierstrass models defined over an arbitrary (toric or non-toric) base
for certain gauge group and matter structures---see,
e.g., \cite{Esole:2011sm,Esole:2014hya}.
%As we demonstrate, this is
%particularly relevant to models with gauge group $(\SU(3) \times
%\SU(2) \times \U(1))/\Z_6$, which do not admit a simple known toric
%realization, even over toric bases.
Resolutions of general classes of
elliptic fourfolds including non-toric constructions have also been
considered in, e.g., \cite{Cveti__2014, LinWeigandG4}, using somewhat
different approaches.

\subsection{Non-degeneracy of the intersection pairing on the base}
\label{sec:nondegeneracy}

For any threefold base $B$, there is a triple intersection form
$D_\alpha \cdot D_\beta \cdot D_\gamma$
on the space $ H_{2, 2} (B,\Z) \cong H^{1, 1} (B,\Z)$ of divisors on $B$.  One feature of a
general F-theory threefold base that we will use in various places is
the observation that for any such smooth $B$, the triple
intersection form is nondegenerate, in the sense that for any
divisor $A = A^\alpha D_\alpha$,
there exists some  $D',  D''$ for which
$A \cdot D' \cdot D'' \ne 0$, so that there
exists a curve $C$ whose class is of the form $C = D' \cdot D''$ with $C \cdot A \ne
0$.
For a toric base, this follows from the standard result that the ring
in intersection theory generated by the divisors (i.e., the Chow
ring) generates the full linear space of homology classes $H_{i, i}
(B,\Z)$, combined with Poincar\'{e} duality, which states that the
space of curves $H_{1, 1} (B,\Z)$ is dual to the space of divisors
under the intersection product.  More generally, the stated
result follows from the hard Lefschetz theorem (see, e.g., \cite{Griffiths:433962}), which asserts that
$J:H^{1,1} (B,\Q) \to H^{2, 2}(B,\Q)$ is an isomorphism
over $\Q$ for any  compact K\"ahler manifold $B$, where $J$ is a
K\"ahler class (equivalently, a
cohomology class Poincar\'e dual to the pullback of the hyperplane section in a projective realization of
$B$ when $B$ is a smooth complex projective variety.) This nondegeneracy plays a useful role in our analysis of the
structure of fluxes and the intersection numbers of CY
fourfolds that can be realized as elliptic fibrations over $B$.

\subsection{Weierstrass model: gauge group and matter content}
\label{sec:Weierstrass}

A central feature of a 4D F-theory model is the gauge group $\sfG$ realized in
the effective 4D theory constructed by compactifying F-theory on a
Weierstrass model defined over a given threefold base $B$.  In general, $\sfG$ is encoded in the
Kodaira type of the singularities in the elliptic fibration over various
divisors in $B$.

The gauge group $\sfG$ can arise either because it is forced from the geometry
of $B$ or through explicit tuning of the Weierstrass model.  In
the first case, geometrically ``non-Higgsable'' gauge group factors
can arise when certain divisors in $B$ have normal bundles that
are sufficiently negative that $(f, g)$ are forced to vanish to orders
at least $(1, 2)$ over those divisors \cite{MorrisonTaylorClusters, MorrisonTaylor4DClusters}.
Virtually all of the large number of threefold bases that support
elliptic CY fourfolds have multiple non-Higgsable gauge group
factors \cite{TaylorWangMC, HalversonLongSungAlg,
    TaylorWangLandscape}.  The gauge group can also be tuned by choosing
a Weierstrass model where $f, g$, and the discriminant $\Delta =
4 f^3 +27 g^2$ vanish to the appropriate orders over a given divisor in $B$
necessary to guarantee a desired nonabelian gauge factor.  $\U(1)$ gauge factors
can also be non-Higgsable \cite{MartiniTaylorSemitoric,
  morrison2016nonhiggsable, WangU1s} or tuned, and are subtler, as
they rely on the global structure of the Mordell-Weil group of
rational sections.

The allowed matter content in a given theory depends on the more
detailed structure of singularities in the elliptic fibration over
codimension-two loci in $B$.  There is a natural distinction in
F-theory between ``generic'' matter content for given $\sfG$,
associated with the simplest codimension-two singularity types, and
more exotic matter representations that can be realized through more
complicated singularities.  This notion of genericity can be made precise
in 6D, where generic matter content is associated with the
branch of moduli space of the largest dimension for a fixed $\sfG$ and
anomaly coefficients that are not ``too large'' \cite{TaylorTurnerGeneric}.  For given $\sfG$, in general we expect that there is a
universal construction of a multi-parameter family of Weierstrass
models that realize the full geometric moduli space of elliptic
CY varieties over an arbitrary base that realize $\sfG$ and have generic matter content for that gauge
group.
Such ``universal'' $\sfG$ models were studied in
\cite{Raghuram:2019efb}\footnote{In that paper these universal
  Weierstrass model constructions were referred to as ``generic'';
  here we change terminology to ``universal'' to avoid confusion with
  other uses of the term generic.}, where the universal $\SM$
Weierstrass model with generic matter was constructed, and a
moduli-counting argument was introduced to check that a universal
$\sfG$ model is fully parameterized; other universal $\sfG$
models with generic matter representations include the Tate-tuned
models with various nonabelian gauge factors (see,
e.g., \cite{BershadskyEtAlSingularities}), and the Morrison--Park
universal $\U(1)$ model \cite{MorrisonParkU1}.
In general, the parameters of the universal Weierstrass construction
for a given $\sfG$
include discrete parameters associated with the divisor classes
supporting the gauge factors and continuous parameters associated with
complex structure moduli of the associated $X_0$. These discrete parameters, along with the canonical class of the base, form what for us will be the characteristic data of the F-theory model.
While the definition
of ``generic'' matter representations is most clear in 6D theories, the same representations are naturally generic
for 4D F-theory constructions in terms of the dimension of the
geometric moduli space and the complexity of the singularities;
universal F-theory models with fixed $\sfG$ and these generic
matter representations such as the Tate-tuned and Morrison--Park
models take the same parameterized form in 6D and in 4D theories.

In this paper, we work with various universal $\sfG$ models
with generic chiral matter content in 4D, meaning
that we consider multi-parameter Weierstrass models with $\sfG = \SU(N),
\SO(4k+2), \text{E}_6,(\SU(2) \times \U(1))/\Z_2$ over
arbitrary threefold bases $B$ that need not be toric. Note that even over
toric $B$, only some universal $\sfG$ models have known toric
constructions with fibers that can be constructed torically as
elliptic curves within toric 2D fibers. For example, some $\SU(N)$
models can be constructed in this way torically, and a subset
of the universal $\SM$ models can be so constructed torically, but not all.

In this paper we focus on the degrees of freedom of 4D F-theory models
encoded in the Weierstrass model through the axiodilaton of type IIB
theory and the fluxes that come from the 3-form field $C_3$ in the
M-theory picture.  There can be additional degrees of freedom such as
``T-branes'' \cite{Cecotti:2010bp} encoded in the world-volume
dynamics of the 7-branes of the IIB theory; in the analysis here we do
not consider the matter or other structures that these degrees of
freedom may produce in the effective 4D $\mathcal N=1$ supergravity theory.

\subsection{Resolution and intersection numbers}
\label{sec:resolution}

As described above, we are interested in general families of Weierstrass models with particular
structures of codimension-one and codimension-two singularities.
%, such as are realized by universal $\sfG$ models described through
%multi-parameter Weierstrass models.
Given a Weierstrass model in such
a family, the standard
approach taken for understanding F-theory models is to resolve the
singular Weierstrass geometry into a smooth elliptic
CY manifold and analyze the theory as a limit of M-theory,
see, e.g., \cite{Grimm:2010ks,WeigandTASI}.
While this approach gives the best understood way of analyzing the
physics of the resulting 4D F-theory model,  the physics
should be independent of the specific resolution; indeed, from the
nonperturbative type IIB point of view, the physics should be
well-defined directly in the context of the singular Weierstrass model.
Note further that there can be terminal
singularities at higher (i.e., $\geq 2$) codimension that do not admit a CY
resolution at all; in many cases, these can be present without any apparent
significant effects on the resulting physical model
\cite{arras2016terminal,Grassi:2018rva}.
One of the broad motivations for the methods we explore in this
paper is to find ways of characterizing the resolution-independent
aspects of the physics of 4D chiral matter.
Given a Weierstrass
model defined by a singular elliptic CY fourfold,
there are, in general, multiple distinct resolutions $X$ that preserve
the CY structure of the singular fourfold. For most of the analysis here we do not
concern ourselves with terminal singularities or higher-codimension
$(4, 6)$ loci where there is no ``flat'' resolution $X$
respecting the elliptic structure (although, see \cref{sec:46} for
some further comments on codimension-three $(4, 6)$ loci).\footnote{\label{partial} In
  this paper, when we refer to a CY fourfold $X$ as a ``resolution''
  of a singular Weierstrass model $X_0$
obtained by a sequence of blowups, we  mean that $X$ is at least a partial resolution that is smooth through codimension-three loci in $B$ (for elliptic fibrations whose geometry does not force tunings leading to additional singular fibers beyond those suggested by the Weierstrass model), but may nonetheless contain singularities over special codimension-three loci (i.e., points) in $B$. These codimension-three singularities do not affect the results of our analysis, hence we ignore them and permit this abuse of terminology. Note that when $B$ is restricted to be a twofold $B^{(2)}$, these codimension-three singularities are absent; in the models we consider here there are also no codimension-two terminal singularities, hence the resulting CY spaces $X$ are in general genuine resolutions over $B^{(2)}$. A comprehensive analysis of the network of genuine CY fourfold resolutions (i.e., through codimension-three in $B$) using the physics of the low-energy effective 3D $\cN = 2$ description of the F-theory Coulomb branch is presented in \cite{Hayashi:2014kca} (see also \cite{Lawrie:2012gg}); in \cite{Hayashi:2014kca}, particular attention is given to the geometry of the singular elliptic fibers over codimension-two and codimension-three loci.} Since simple and manifestly
resolution-independent methods are currently lacking for a complete
analysis of physics like chiral matter, we use
specific resolutions for explicit calculations and try to extract and
identify the
resolution-independent parts of the results. 

One of the key features of a resolved CY fourfold $X$ is the
set of quadruple intersection numbers of divisors $\hat{D}_I$.\footnote{We use hats to denote divisors in the fourfold $X$, as
  opposed to divisors in the base $B$; a glossary of notation commonly used throughout the paper is
  given in \cref{sec:notation}.}
Expanding an arbitrary set of divisors $\hat C, \hat D, \hat E, \hat F$ in an
appropriate basis $\hat D_I$, we may write
\begin{equation}
\hat C \cdot \hat D \cdot \hat E \cdot \hat F = C^I D^J E^K F^L(\hat D_I \cdot \hat D_J \cdot \hat D_K \cdot \hat D_L)\,.
\end{equation}
These intersection numbers can also describe
aspects of the dual pairing associated with Poincar\'{e} duality
between divisors
(codimension-one algebraic surfaces) and curves
(codimension three) in the fourfold, e.g., when a curve is realized as
an intersection of three divisors.
As we discuss in further detail below, the quadruple intersection
numbers of $X$ are not in general resolution-independent, although some are resolution-independent.
A natural basis for the divisors in a fourfold with an elliptic
fibration structure is suggested through the Shioda--Tate--Wazir
\cite{Wazir} formula
\begin{equation}
h^{1, 1}(X) = 1 + h^{1, 1} (B) + \rk\sfG\,,
\label{eq:Wazir}
\end{equation}
where the 1 comes from the zero section of the elliptic fibration, the
second term comes from pullbacks of divisors in the base to the total
space of the elliptic fibration, and the last term contains Cartan
divisors of nonabelian gauge factors and additional sections for the
free abelian part of $\sfG$.  In view of this decomposition, and
following standard notation in the literature (e.g., \cite{Grimm:2011sk}), we
use the following conventions for indices $I$:
\begin{itemize}
\item $I=0$ denotes the zero section
\item $I=a$ denotes a generating section associated to a non-Cartan $\U(1)$ gauge factor
\item $I=\alpha$ denotes a divisor $\hat D_\alpha = \pi^*(D_\alpha)$
    realized as a pullback of a divisor in the base
\item $I=i_s$ denotes a Cartan divisor of a nonabelian factor $\sfG_s \subset \sfG$.
\end{itemize}
To make contact with the low-energy gauge theoretic description of the Coulomb branch physics in the M-theory duality frame, we sometimes convert to the ``physical'' basis $\hat D_{\bar I} = \sigma_{\bar I}^I \hat D_I$ (see \labelcref{physbasis} for the definition of $\sigma_{\bar I}^I$), where
	\begin{itemize}
		\item $\bar I = \bar 0$ denotes the $\U(1)_{\text{KK}}$ divisor
		\item $\bar I = i=(\bar a,i_s)$ collectively denotes all other $\U(1)$ divisors.
	\end{itemize} 
Many of the intersection numbers are independent of resolution and are
known for a general smooth elliptically-fibered CY.  For example
\cite{Grimm:2011sk}, when $\hat D_0$ is a holomorphic section and
there are no abelian factors, one can use the fact that the quadruple intersection numbers of
$X$ can be pushed down to the base $B$\footnote{Strictly speaking, the intersection products $\hat D_I \cdot \hat D_J \cdot \hat D_K \cdot \hat D_L$ live in the Chow ring of the variety $X$ (the Chow ring encodes the intersection structure in a smooth
  algebraic variety; see \cref{fluxesint} for further discussion). However, since the pushforward is
  computed with respect to the canonical projection $\pi\colon X
  \to B$, the resulting intersection product $\pi_*(\hat D_I \cdot \hat D_J \cdot \hat D_K \cdot \hat D_L)$ lives in the Chow ring of
 $B$. For simplicity of notation we are often sloppy and omit explicit pushforward maps such as $\pi_*$ when the appropriate Chow ring is otherwise clear from the context.} to write
\begin{equation}
\begin{aligned}
\pi_*(\hat D_0^4 ) =  K^3\,, ~~~~\pi_*(\hat D_0 \cdot \hat D_\alpha \cdot \hat D_\beta \cdot \hat D_\gamma) =
D_\alpha \cdot D_\beta \cdot D_\gamma\,.
\end{aligned}
\end{equation}
%where the intersection products on the right hand side of the above
%equations are understood to be evaluated in the Chow ring of the base.
The above intersection numbers are clearly independent of the geometric properties of the elliptic fibration, and only depend on the intersection numbers of the base. Other intersection numbers, particularly those with three or four indices of
type $i_s$ or $a$, depend not only on the matter content of the theory
but also on the particular resolution; see \cref{intersection} for a
more comprehensive discussion
of the structure of these intersection numbers for rather general classes of elliptic fibrations.

%The results of this paper make direct use of the fact that for a set of gauge groups $\sfG$
%of
%interest, and a specific class of resolutions, a generalization of the methods of \cite{Esole:2017kyr} can be used to compute
%general formulae for all the intersection numbers of the CY
%fourfold associated with the resolution of the Weierstrass model for
%universal $\sfG$ models over a completely general base.

One issue that arises in certain situations, for instance when $\sfG = (\sfG_\text{na}\times\U(1)^k)/\Gamma$, is
that there may not be a holomorphic section of the Weierstrass model;
this occurs, in particular, when the section associated with the
identity element of the Mordell--Weil group intersects with one or more
sections associated with generators of $\U(1)$ factors over the
discriminant locus. In some of these cases, the procedure of
resolving the singular CY geometry and analyzing various physical
properties in the dual M-theory frame on $X$ is more easily
accomplished in models in which the elliptic fiber is realized as a
general cubic in $\bP^2$ rather than the usual Weierstrass model
(e.g., the general cubic is used to define the resolutions studied in
\cite{KleversEtAlToric}.)

In this paper, we adapt the mathematical techniques of
\cite{Esole:2017kyr} to compute intersection numbers of resolutions of
models in which the elliptic fiber is presented as a general cubic in
$\bP^2$. These mathematical techniques enable us to evaluate the
quadruple intersection numbers in terms of linear combinations of the
triple intersection numbers of an arbitrary base $B$, much in the same manner
as described above:
    \begin{equation}
    \label{firstpush}
        \pi_*(\hat D_I \cdot \hat D_J \cdot \hat D_K \cdot \hat D_L)=W_{IJKL}= W^{\alpha \beta \gamma}_{IJKL}  D_\alpha \cdot D_\beta \cdot D_\gamma\,.
    \end{equation}
With the aid of a symbolic computing tool, the action of the above map
$\pi_*$ can easily be used to compute intersection numbers (and other
characteristic numbers\footnote{For example, the same methods have
  been used to compute the generating function of the Euler
  characteristics of smooth (up to codimension two) elliptic $n$-folds
  resolving singular Weierstrass models with gauge symmetry $\sfG$; see
  \cite{Esole:2017kyr} for further details.}) of $X$ in terms of
rational expressions involving divisor classes of the
ambient fivefold in which $X$ is realized as a hypersurface.\footnote{It should be
  possible to straightforwardly adapt these techniques to models in
  which the elliptic fiber is realized as a complete intersection in
  $\bP^n$.} The techniques used to evaluate the pushforward map
$\pi_*$ are described in detail in \cref{pushapp}.
Note that the
physical results obtained from a given resolutions $X$ should, and do in the
cases we analyze, match with the expected physics from any other
resolution, including the structure of the tensors $W_{IJKL}^{\alpha \beta \gamma}$ (see \cref{intersection}) appearing on the right hand side of \cref{firstpush}.

\subsection{Fluxes, consistency conditions, and linear algebra}
\label{sec:fluxes-algebra}

In order to obtain a chiral matter spectrum in 4D, it is necessary to
turn on a nontrivial flux background, which in the M-theory duality
frame corresponds to a nontrivial profile for the 4-form field
strength $\diff C_3$ whose key properties we now summarize.

\subsubsection{Flux conditions}

It was argued in \cite{Witten:1996md} that the cohomology class $G =\diff
C_3 \in H^4(X,\R)$ satisfies a shifted quantization
condition\footnote{It turns out that $c_2$ belongs to
  $H^{2,2}_{\text{vert}}(X,\Z)$ as defined in
  \cref{eq:verticalcohomology}. %Since in this paper we solely consider
  % vertical flux backgrounds $\mathrm{d}C_3 \in
  % H^{2,2}_{\text{vert}}(X,\Z)$, we implicitly work with the
  % shifted flux background $G = \mathrm{d}C_3 - c_2/2$ in cases when
  % $c_2$ is not an even class to maintain integer quantization
  % \labelcref{eq:c2quant}. The need for the shifted quantization was
  % explained in \cite{Witten:1996md} where it was shown, for example,
  % that the Hirzebruch--Riemann--Roch theorem in combination with Wu's
  % formula implies any flux background obeying the quantization
  % condition \labelcref{eq:c2quant} is sufficient to produce an
  % integer-valued M2 brane tadpole \labelcref{D3tadpole}, whereas
  % without the shift, the M2 brane tadpole is not in general
  % integer-valued. When $c_2$ is even, we simply absorb the shift into
  % the definition of $\mathrm{d}C_3$.
  }
%\footnote{We assume that appropriate factors of $2\pi$ have been absorbed into the definition of $G$.}
    \begin{equation}
        \label{eq:c2quant}
        G_\Z = G - \frac{c_2(X)}{2} \in H^{4}(X,\Z)\,.
    \end{equation}
To preserve supersymmetry, $G$ must satisfy
    \begin{equation}
        G \in H^{2,2}(X, \R) \cap H^4(X, \Z/2)
    \end{equation}
along with the primitivity condition
    \begin{equation}
        J \wedge G = 0\,,
    \end{equation}
where $J$ is the K\"ahler form of $X$ \cite{Becker:1996gj,Gukov:1999ya}.  Finally, there is a tadpole condition
\cite{Sethi:1996es,Dasgupta:1996yh,Dasgupta:1999ss} requiring that the net number of M2-branes (which are dual to D3-branes in the F-theory frame) is non-negative to ensure a stable vacuum,
    \begin{equation}
        \label{D3tadpole}
        N_\text{M2} = \frac{\chi}{24} - \frac{1}{2} \int_X  G \wedge G \in \Z_{\ge 0}\,,
    \end{equation}
where $\chi = \int_X c_4$ is the Euler characteristic of
$X$; the integrality of $N_{\text{M2}}$ follows from \cref{eq:c2quant}, as explained in \cite{Witten:1996md}.
%\footnote{For a CY fourfold $X$, the Hirzebruch--Riemann--Roch theorem in combination with Wu's formula $x^2 = c_2 \wedge x \, \text{mod 2}$ implies that for any flux obeying the quantization condition $x = G-c_2/2 \in H_4(X,\Z)$, $N_\text{M2}$ as defined in \labelcref{D3tadpole} is integer-valued; see \cite{Witten:1996md}.}

Additional conditions must be imposed to ensure that $G$ dualizes to a
suitable F-theory flux background. To preserve Poincar\'{e} symmetry, we
require that the following fluxes vanish:
\cite{Dasgupta:1999ss},
    \begin{equation}
        \label{eq:Poincare}
        \int_{S_{0\alpha}} G =0\,, \quad \int_{S_{\alpha \beta} } G = 0\,.
    \end{equation}
Furthermore, to ensure that the flux background does not break the gauge symmetry $\sfG$ in the 4D limit, it is necessary to impose the conditions (see, e.g., \cite{Donagi:2008ca})
    \begin{equation}
        \label{eq:gauge}
        \int_{S_{i \alpha} } G = 0\,,
    \end{equation}
where we emphasize that $i$ collectively indexes all divisors dual to gauge $\U(1)$s on the F-theory Coulomb branch.
Note that since $G$ may be a half-integral
cohomology class, in principle it seems there could be circumstances under which
no flux satisfies these conditions; in all cases we have considered
here,
however, the integrals \cref{eq:Poincare} and \cref{eq:gauge} take
integer values and the constraints can be satisfied, and this is
likely always true for the Poincar\'{e} symmetry
constraints---in particular, the results of
\cite{Collinucci:2010gz} show that the fluxes appearing in \cref{eq:Poincare}
are always integer-valued for any $G$ on a smooth elliptic fourfold.
We describe these conditions explicitly in some families of models with gauge groups $\SU(2), \SU(5)$ in \cref{sec:su2}
and \cref{sec:su5}.

\subsubsection{Vertical fluxes and intersection pairing}
\label{sec:vertical-pairing}

In addition to the usual Hodge decomposition, the cohomology group $H^4(X)$ admits the finer orthogonal decomposition \cite{Greene:1993vm,Braun:2014xka}
    \begin{equation}
    \label{ortho}
        H^{4}(X,\C) = \httv(X,\C)  \oplus H^{2,2}_\text{rem}(X,\C) \oplus H^{4}_\text{hor}(X,\C)\,. 
    \end{equation}
As in much of the previous literature, for the most part
in this paper we focus on
integral ``vertical'' fluxes, i.e., flux backgrounds $G$
belonging to the subgroup $H^{2,2}_{\text{vert}}(X, \R) \cap
H^4(X,\Z)$ spanned by wedge products of cohomology classes in
$H^{1,1}(X)$, where $H^{1,1}(X)$ has a basis $\PD(\hat D_I)$ of
harmonic $(1,1)$-forms on $X$ dual to divisors $\hat D_I$.\footnote{Note that when $c_2 (X)$ is not even, $G$ is a
  half-integer class; we neglect this refinement in our notation in
  various places, essentially restricting to the simplified cases
  where $c_2 (x)$ is even,
  except when it is directly relevant to the discussion.}  More precisely, for the
purposes of this paper, we focus on fluxes belonging to the sublattice $H_{\text{vert}}^{2,2}(X,\Z) \subset H^{2,2}(X, \R) \cap
H^4(X,\Z)$, which we define as follows:
    \begin{equation}
        \label{eq:verticalcohomology}
        \httv(X,\Z) := \Span_{\Z}(H^{1,1}(X,\Z) \wedge H^{1,1}(X,\Z))\,,
    \end{equation}
which is to say that a ``vertical'' class for us means a class belonging to
the integer linear span of forms $\PD(\hat D_I) \wedge \PD(\hat
D_J)$. Note that it is in principle possible for there to exist integral
vertical cohomology classes that do not lie in
$H^{2,2}_{\text{vert}}(X,\Z)$ as given by this definition
and therefore it is possible that
our definition excludes some consistent vertical flux backgrounds that
could be included by permitting non-integer coefficients. 

As reviewed in \cite{WeigandTASI}, vertical fluxes play a primary role
in determining the chiral matter content of a 4D F-theory
compactification, and for the most part we
 ignore components in
 $H^{4}_\text{hor}(X,\C) \oplus H^{2,2}_\text{rem}(X,\C)$
for flux backgrounds.\footnote{The possibility that Poincar\'{e}
duality and the inclusion of fluxes in
$H^{4}_\text{hor}(X,\C)\oplus H^{2,2}_\text{rem}(X,\C)$ may give a broader class of possible
matter multiplicities is explored in \cref{sec:quantization-1},
and more specifically in the case of the $\SU(5)$ model with generic
matter in \cref{sec:su5}.} Denoting by $\Lambda_S$ the $[h^{1, 1}(X) (h^{1, 1}(X) + 1)/2]$-dimensional integral lattice
spanned by the
% basis of  homology cycles
surfaces
 $S_{IJ} = \hat D_I \cap \hat
D_J$ (here treated as formally independent objects for each $IJ$ pair),
we can
conveniently encode
%their
the
flux integrals over vertical surfaces via the intersection pairing
matrix
\begin{equation}
\label{eq:m-o}
\begin{aligned}
    M &\colon \Lambda_S \times \Lambda_S \rightarrow \Z\,, \\
    M_{(IJ) (KL)} &=
    S_{IJ} \cdot S_{KL} =
    \int_X \PD(S_{IJ}) \wedge \PD(S_{KL})\,.
\end{aligned}
\end{equation}
In the second line above, $\cdot$ indicates the intersection pairing
on homology. As we explain in more detail in \cref{fluxesint}, $M$ can
%equivalently
thus
be viewed as an integral bilinear form
on
 vectors $\phi = \phi^{IJ} S_{IJ} \in \Lambda_S$, giving
 $\Lambda_S$ the structure of an integral lattice.\footnote{While sometimes
 physicists refer to any subgroup of $\R^n$ that is isomorphic to
 $\Z^n$ as a lattice without reference to
 any associated bilinear form, throughout this paper we reserve the
 term lattice for a free abelian group of finite rank with a symmetric
 bilinear form.} Following \cite{Grimm:2011fx}, we define the fluxes
\begin{equation}
\Theta_{IJ} = \int_{S_{IJ}} G =  \int_X G \wedge \PD(S_{IJ})
= M_{(IJ) (KL)} \phi^{KL}\,,
\label{eq:constraints-theta-m}
\end{equation}
where $\phi \in \Lambda_S$ represents the components of the Poincar\'{e} dual of the
flux background $G$ expanded in a collection of  classes
$S_{IJ}$. Throughout the paper, we
refer to $\phi$ as a ``flux background''. 
(Note that when $c_2 (X)$ is not even, the possible values of
 $\phi$  are shifted appropriately
by a half-integer lattice element $\PD(c_2 (X)/2) \in
\Lambda_S/2$.)
In terms of the above notation, the symmetry constraints \labelcref{eq:Poincare,eq:gauge} can be expressed as
    \begin{equation}
        \label{eq:general-constraints}
        \Theta_{I\alpha} = 0\,,
    \end{equation}
where we note that for vertical fluxes the above conditions are both
necessary and sufficient to preserve 4D gauge symmetry and local
Lorentz symmetry.
%We later make extensive use of the fact that the
%coefficients of
The fluxes $\Theta_{IJ}$ can be written as linear functions of the flux backgrounds
%, viewed as polynomials in
 $\phi^{IJ}$, with coefficients given by
the pushforwards of the intersection
numbers of divisors of $X$.
In explicit computations we can for certain resolutions of $\mathsf G$ models (defined over arbitrary $B$) formally solve the equations
\cref{eq:general-constraints} for a subset of the $\phi$s, so that
the nonzero $\Theta$s that encode the chiral matter multiplicities
are again linear functions of the remaining $\phi$s with coefficients
that are polynomial functions of the intersection numbers.

Imposing the symmetry constraints is equivalent to restricting the
flux background $\phi^{IJ}$ to lie in a sublattice $\Lambda_C \subset
\Lambda_S$. For %appropriately chosen
a given resolution $X$, the sublattice
$\Lambda_C$ can be viewed as the lattice of $\phi^{IJ}$ whose image
under $M$ is the sublattice of $\Theta_{IJ}$ satisfying
\labelcref{eq:general-constraints},
%related by rational linear transformations, to
which in turn encodes the multiplicities of 4D chiral matter, as we
review in more detail in the following subsection. We emphasize that while the
intersection numbers entering the matrix $M$ are generically
resolution-dependent, we expect that
% the set of allowed $\Theta_{IJ}$
%and consequently
 the allowed chiral multiplicities must be
resolution-independent, consistent with the expectation that every set
of M-theory vacua defined by a set of distinct resolutions $X$ lifts to a common set of
F-theory vacua on a singular elliptic CY fourfold $X_0$.

%Additionally there is an orthogonal decomposition \cite{WeigandTASI}
%\begin{equation}
%  H^{2, 2}(X,\C)=
%  H^{2, 2}_\text{hor}(X,\C)\oplus
%  H^{2, 2}_\text{vert}(X,\C)\oplus
%    H^{2, 2}_\text{rem}(X,\C).
%\label{eq:h22-orthogonal}
%\end{equation}

What we have described above is essentially the standard perspective
on analyzing chiral matter in 4D F-theory flux vacua. We now discuss a
complementary perspective that illuminates additional aspects of the
analysis. We begin with the observation that not all the cycles
$S_{IJ}$ are independent in homology \cite{LinWeigandG4,
  Bies_2017}. This implies
that $M$ generically has a nontrivial nullspace, where the elements
of the nullspace represent equivalence relations in homology, and hence
the rank of the matrix $M$ is equal to the dimension $h_{2,
  2}^\text{vert} (X)$ of $H_{2,2}^{\text{vert}}(X,\Z)$. We denote by
$M_\text{red}$ the \emph{nondegenerate} intersection pairing
    \begin{equation}
        M_\text{red}\colon \hvtt(X,\Z) \times \hvtt(X,\Z) \to \Z\,,
    \end{equation}
where we describe $M_\text{red}$ explicitly as a matrix by restricting
the action of the matrix $M$ %to the
%orthocomplement of the
to the quotient of $\Lambda_S$ by the nullspace, which
%by quotienting out the nullspace
projects
$\Lambda_S$
to the
quotient lattice $\hvtt(X,\Z)$.

While the reduced matrix $M_\text{red}$ produces the same results for
the multiplicities of chiral matter as $M$ does in the procedure described above,
$M_\text{red}$ is a simple and useful tool for analyzing various
aspects of fluxes and chiral matter (e.g. the number of independent families of chiral matter combinations can in principle be inferred from the rank of $M_{\text{red}}$, without having to explicitly compute chiral indices.)
Furthermore, we provide evidence suggesting that while the full set of quadruple
intersection numbers are not
in general resolution-independent, $M_\text{red}$ is independent of the choice of
$X$ up to a change of basis. Among other things, this implies
that $M_\text{red}$ makes the resolution-invariance of the chiral
matter multiplicities manifest in terms of a canonical subspace of
homology classes that parametrize the space of vertical fluxes lifting
to consistent F-theory flux backgrounds.
Since the dimension of the null space of $M$ is just $(h^{1, 1}(X) (h^{1, 1}(X) + 1)/2)
-  h_{2,
  2}^\text{vert} (X)$, resolution-independence of $M_\text{red}$ also
implies resolution-independence (up to an integral change of basis) of
$M$; this argument is spelled out more explicitly in \cref{invariance,sec:lattice-reduce}.
Previous work
 \cite{LinWeigandG4} has implemented the quotient by the nullspace taking
$\Lambda_S$ to $H_{2,2}^{\text{vert}}(X)$ in explicit resolutions by using methods
 related to the Stanley-Reisner ideal; here we carry out this quotient directly on the matrix $M$ computed for various $\mathsf G$ models defined over arbitrary
 $B$. To our knowledge, the observation that  $M_\text{red}$
and $M$  are resolution-invariant has not been made previously either in the
 mathematics or physics literature.
% nonetheless,
%much of the important resolution-independent physics can be extracted
%from  $M_\text{red}$ in a uniform fashion even when it varies depending
%upon the details of the resolution.
%\wati{[pass 1]
%  e.g., is it resolution independent for SU(5)? \textcolor{red}{PJ : I claim that the matrices $M_\text{red}$ corresponding to theories with nontrivial flux are not resolution independent. One way to argue this is to note that $\Theta_{IJ}$ can be extracted from the geometric expression for $\text{vol}_{\phi}$ where $\phi$ is the Poincar\'{e} dual to $G$. This is a resolution dependent quantity that changes under flops, and the precise changes under flops can be computed using the expressions for 3D $\cN=2$ CS couplings.}}

%It has been well established in the literature \cite{WeigandTASI} that
%preserving Poincar\'{e} invariance and the full gauge group of the
%theory is only possible when certain components $\Theta_{IJ}$ vanish.

Both the symmetry constraints and the
projection
into nontrivial homology classes
have a simple geometric interpretation, and it is clear that
the composition of these two operations in either order leads to the same sublattice equipped with a
nondegenerate bilinear form.  
Given the original lattice
$\Lambda_S$ with bilinear form $M$, 
the constraints \labelcref{eq:Poincare,eq:gauge}
restrict $\phi$ to the sublattice $\Lambda_C$.
If $\dim(\Lambda_S)= m$ and there are $k$ (non-null) constraints, then $\dim(\Lambda_C) = m-k$.
 Imposing the homological equivalence
 relation $\phi \sim \psi \Leftrightarrow M (\phi-\psi) = 0$ (i.e.,
 quotienting out $\Lambda_C$ by the nullspace $V$ of $M$, which
 satisfies $V\subset \Lambda_C$) gives us the
 lattice of independent vertical flux backgrounds
 $\Lambda_\text{phys} = \Lambda_{C}\mathclose{}/\mathopen{}\sim$, with the nondegenerate
 bilinear form $M_\text{phys}$ that is the restriction of $M_C$ to
 $\Lambda_C \mathclose{}/\mathopen{}\sim$.
%; the image $
% M_\text{phys}\Lambda_\text{phys}$ is the lattice of independent
% fluxes satisfying the symmetry constraints.  
We can describe this procedure explicitly in a given basis for $\Lambda_S$.
In particular, if we define an
 $m \times(m-k)$ matrix $C$ to have columns given by a set of
 generators of the lattice $\Lambda_C\subset\Lambda_S$, then 
$C:\Z^{m-k} \rightarrow\Lambda_S$
describes a lattice embedding of $\Z^{m-k}$ into $\Lambda_C \subset\Lambda_S$, and
$M_C :=C^\transpose M C$
 is the restriction of $M$ to $\Lambda_C$ that results from imposing
 the symmetry constraints, expressed in a natural basis for
 $\Lambda_C$.  The resulting form of $M_C$ plays an important role in
 our analysis, although with the simplest choice of coordinates there
 are some subtleties with integrality conditions that we discuss in
 more detail in \cref{sec:quantization-1}
and \cref{constraintsolutions}.
Alternatively, we can
 first impose the %analogous
 homological equivalence relation on $M$,
 leading to the reduced intersection pairing matrix
 $M_\text{red}$, and then impose the
 symmetry constraints. %The intersection pairing $M_\text{phys}$ on the
% physical fluxes in
%$\Lambda_\text{phys}$ is manifestly nondegenerate since all null
%directions have been removed.
% For certain purposes we find it more convenient to first perform
%the homology projection $\Lambda_S \rightarrow H_{2,2}^{\text{vert}}(X,\Z)$ (equivalently, projecting $M\rightarrow M_\text{red}$) and then to impose the symmetry
%constraints restricting $\phi^{IJ}$ to the sublattice $\Lambda_C$, while in other situations it is
%more convenient to first impose the symmetry constraints and then perform the homology
%projection.
The preferred order in which to perform these two operations depends on the circumstances. Nevertheless, these two operations lead to the same result when both are performed either
over $\Z$ (more generally, over $\R$),
%namely the restriction of the lattice $\Lambda_S$ to the lattice $\Lambda_\text{phys}$ of flux backgrounds preserving 4D Poincar\'{e} and gauge symmetry,
so the analysis can be carried out in either
order---see \cref{fig:commute}.
\Cref{sec:constraints-homology,sec:4dflux} essentially describe different perspectives on
our analysis that arise from performing these two different orders of
operation.

Each of these two approaches has value for understanding the structure
of chiral matter multiplicities; explicit computation of $M_C$ in many
cases provides an efficient method to compute the chiral indices as a
function of the characteristic data, while the structure of
$M_\text{red}$ gives us insight into resolution-independence and the
%discrete sets of allowed chiral matter fields.
discretization structure of allowed chiral matter multiplicities.

To maintain clear control of the
discrete quantization of allowed fluxes $\Theta$, %however, 
some care is needed.
%Since the fluxes $\Theta$ are the image of the map $M$ acting on the sublattice $\Lambda_\text{phys}$ (or equivalently the image of $M_\text{phys}$ acting on $\Lambda_S$), the quantization of the fluxes $\Theta$ is determined by both the determinant of the matrix $M_\text{phys}$ and the integrality conditions imposed on the lattice $\Lambda_S$. We comment in particular on the latter point:
While every integer quantized choice of flux background $\phi \in
\Lambda_S$ must correspond to an integer vertical flux background
$G$ by Poincar\'{e} duality, in some circumstances (i.e., when
non-vertical fluxes are included) there may exist
quantized flux backgrounds $G$ that give rise to more general
fractional choices of $\phi$.  We restrict attention in our analysis
primarily to fluxes corresponding to integrally quantized $\phi \in
\Lambda_S$ (except for the possible half-integer shift from $c_2 (X)/2$), though we discuss in some places the possibility of more
general fluxes, which may in turn lead to a larger set of possible
chiral matter multiplicities.
These issues are discussed further in \cref{sec:quantization-1}.

\begin{figure}
\begin{center}
    \begin{tikzpicture}
        \node(1) at (-2,2) {$\Lambda_{S}$};
        \node(2) at (2,2) {$\Lambda_C$};
        \node(3) at (-2,-2) {$H_{2,2}^{\text{vert}}(X,\Z)$};
        \node(4) at (2,-2) {$\Lambda_\text{phys}$};
        \draw[big arrow] (1) -- node[above,midway]{$$} (2);
        \draw[big arrow] (1) -- node[left,midway]{$$} (3);
        \draw[big arrow] (2) -- node[right,midway]{$$}(4);
        \draw[big arrow] (3) --  node[below,midway]{$$}(4);
    \end{tikzpicture}
\end{center}
    \caption{Our approach to analyzing vertical fluxes and chiral
          matter involves the interplay of two commuting operations on
          the lattice of vertical flux backgrounds $\Lambda_S$
          spanned by the vertical cycles $S_{IJ} = \hat D_I \cap \hat
          D_J$. One operation is the restriction of the $\Lambda_S$ to
          the sublattice $\Lambda_C$ of backgrounds satisfying the
          symmetry constraints \labelcref{eq:Poincare,eq:gauge} necessary to
          preserve 4D local Lorentz and gauge symmetry. The other
          operation is the restriction of $\Lambda_S$ to the
          vertical homology $\hvtt(X,\Z)$ by quotienting $\Lambda_S$ by homologically trivial cycles. Performed in either order, the composition of these two operations
          lead to the same sublattice $\Lambda_\text{phys}$ of
          consistent F-theory flux backgrounds that preserve gauge symmetries in the low-energy effective 4D $\mathcal N=1$ description of the F-theory compactification. We present evidence suggesting that
          $\hvtt(X,\Z)$ equipped with its symmetric bilinear form $M_\text{red}$ is resolution-independent up to an integer change of basis.}
    \label{fig:commute}
\end{figure}

\subsection{Chiral matter multiplicities}
\label{sec:matter-multiplicities}

The by now standard result in the F-theory literature \cite{DonagiWijnholtModelBuilding,Braun_2012,Marsano_2011,KRAUSE20121,Grimm:2011fx}
is that for any complex
representation $\sfr$ of the gauge group $\sfG$, the chiral
index is
\begin{equation}
  \chIndex{\sfr} = n_{\sfr} - n_{\sfr^*} = \int_{S_{\sfr}} G \,,
  \label{eq:chiral-index}
\end{equation}
where the homology class $S_{\sfr} \in H_{4}(X,\Z)$ is a
``matter surface''. For local F-theory matter, it
is expected \cite{Marsano_2011,Borchmann:2013hta,Bies:2017fam} that any cycle belonging to the class $S_{\sfr}$ is topologically the fibration of
an irreducible component $C_w$ of the elliptic fibers (where $w \in \sfr$ is any weight) over an irreducible codimension-two component (i.e., a ``matter curve'', not to be confused with a matter surface) $C_{\sfr} \subset \{\Delta^{(2)} = 0\}$ of the discriminant locus
$\{\Delta = 0\} \subset B$, associated to the local matter transforming in the quaternionic representation $\sfr =\sfr \oplus \sfr^*$. In practice the flux of $G$ through $S_{\sfr}$ is computed by way of Poincar\'e duality, i.e.
	\begin{equation}
	  \int_{S_{\sfr}} G = \int_X G \wedge \text{PD}(S_{\sfr})\,,
	\end{equation}
and hence the analysis of vertical flux backgrounds we describe depends crucially on being able to identify an appropriate cohomology class $\text{PD}(S_{\sfr})$ dual to $S_{\sfr}$ (note that the choice of $\text{PD}(S_{\sfr})$ in general may depend on the choice of resolution $X$). With one exception \cite{Braun_2012}, in all known examples $\text{PD}(S_{\sfr})$ can be characterized as an element of $\httv(X)$ \cite{WeigandTASI} (or equivalently $S_{\sfr} \in H_{2,2}^{\text{vert}}(X)$.) However, the precise
definition is subtle and it is unclear that
$S_{\sfr}$ always has non-trivial components in
$H^{2,2}_{\text{vert}}(X)$; see \cref{sec:puzzle} for a discussion
about this subtlety in the context of certain resolutions of the
$\SU(5)$ model. Our default assumption in this paper is that
$S_{\sfr}$ always contains a non-trivial component in
$H_{2,2}^{\text{vert}}(X,\Z)$. Note that in cases we study where $c_2 (X)$ is not an even class and hence (because of \cref{eq:c2quant}) the flux $G$ is a half-integer class in cohomology, the chiral indices \cref{eq:chiral-index} nonetheless take integer values. This is presumably guaranteed for all physically allowed configurations though we do not know of a complete proof.

As discussed above, the fiber of $S_{\sfr}$ is a curve $C_w$ that an M2-brane wraps leading to 3D
matter characterized by BPS central charges
$C \cdot \hat D_i = w_i$ (here $\hat D_i$ are Cartan divisors associated to $\U(1)$ gauge factors characterizing the low-energy physics and $w_i$ are the Dynkin coefficients of the weight $w$).  Thus by Poincar\'{e} duality, we can
 construct in homology the class $C$ associated
with a particle labeled by any weight of any representation; note
that to utilize Poincar\'{e} duality in this context one must project
out, e.g., the fiber and zero section, as described in
\cite{Morrison:2021wuv}.   We describe an explicit example of a matter
curve and some related quantization subtleties in the simple case of
SU(2) in \cref{sec:su2}.
Unfortunately, however, there is no universal
approach known to explicitly construct $S_{\sfr}$ in homology
simply from topological and representation theoretic considerations, without using
a specific resolution.  The issue is that, as reviewed in
\cite{WeigandTASI}, the image of $S_{\sfr}$ in $\Lambda_C$ does not simply contain
components of the form $S_{i \alpha} =\hat D_i \cap \hat D_\alpha$; indeed, these must be
projected out to preserve gauge invariance.  Rather, the image of $S_{\sfr}$ in $\Lambda_C$ must also include components in the $S_{ij}$ directions (as demonstrated explicitly by \cref{eqn:SC}), and since the intersection properties of the classes $S_{ij}$ in general may depend on the choice of $X$, it follows that the
precise form of $S_{\sfr}$ is not known from first principles in a
resolution-independent fashion.
While the approach described in this paper does not rely on explicit computation of the
matter surfaces, we remark that despite the apparent resolution dependence of $S_{\sfr}$, the resolution independence of $M_\text{red}$
suggests
that there exists a natural description of $H_{2,2}^{\text{vert}}(X,\Z)$ in terms of which
the vertical components of the matter surfaces for any given
anomaly-free combination of representations realized in F-theory can be characterized in a
resolution-independent fashion.

Before addressing the explicit computation of chiral matter indices,
we recall that, as described above, after both imposing the symmetry
conditions and quotienting by the homology relations encoded in the
nullspace of $M$, we are left with %$\rk M_\text{phys}$
a set of
independent flux backgrounds $\phi$, associated with a nontrivial
$(\rk M_\text{phys})$-dimensional sublattice $M_{\text{phys}} \Lambda_{\text{phys}} \subset M \Lambda_{S}$ that for
%appropriately chosen
a given $X$ encodes the
4D chiral matter multiplicities $\chi_{\sfr}$.  Thus, even without explicitly computing $S_{\sfr}$, we can expect in such cases for there to be
 an $m$-dimensional space of $\chi_{\sfr}$ (where $m \leq \rk M_\text{phys}$) that
can be realized by turning on different combinations of
$\phi^{IJ}$.  Since
F-theory constructions are expected to always be consistent at low energies, these
combinations of $\chi_{\sfr}$ should always satisfy 4D anomaly
cancellation.
Therefore, we expect that the rank of $M_\text{phys}$, or equivalently the rank of $M_\text{red}$
minus the number of independent constraints in
(\ref{eq:general-constraints}), places an upper bound on
the number of linearly independent combinations of chiral matter
fields available in the theory.

%In all cases of which we are aware, the full dimensionality of solutions of
%the 4D anomaly cancellation conditions are realized through the rank
%$A$ of the image of $M_\text{red}$ after imposing the Poincar\'{e} and gauge
%invariance conditions.  We show this explicitly in this paper for numerous
%examples, and we
%conjecture that it is true in general.  In particular, since we can
%compute the rank $A$ directly from the $M_\text{red}$ without computing
%matter surfaces explicitly, we can in principle verify that all 4D anomaly-allowed
%chiral matter families can arise through fluxes in F-theory by simply
%checking that $A$ matches the dimensionality of the set of solutions
%to the 4D anomaly equations provided that the fourfold $X$ does not
%contain additional low-energy chiral degrees of freedom.

As is evident from the above discussion, to explicitly compute $\chi_{\sfr}$
one must either identify $S_{\sfr}$, or proceed by
more indirect means.  Here we proceed in the latter fashion and
 follow a strategy similar to that of \cite{Cvetic:2012xn} (see also \cite{Cveti__2014,Cvetic:2015txa}), which exploits
the following relationship between the set of $\Theta_{IJ}$ satisfying the symmetry constraints and linear combinations of $\chi_{\sfr}$ given by 3D one-loop Chern--Simons couplings appearing on the Coulomb branch of the 4D F-theory vacuum compactified on a circle:
    \begin{equation}
    	\Theta_{\bar{I}\bar{J}}  =   -\Theta_{\bar{I}\bar{J}}^{\text{3D}},~~~~ \bar{I} = \bar{0}, i
    \end{equation}
where 
	\begin{equation}
	\Theta_{\bar I \bar J} := \sigma_{\bar I}^I \sigma_{\bar J}^J \Theta_{IJ}\,.
	\end{equation}
(Recall that the index $\bar{0}$ denotes the abelian Kaluza--Klein
gauge field associated to the compact circle and we use the index $i$
to collectively denote all other $\U(1)$ gauge fields; see
\labelcref{physbasis} for the precise definition of $\sigma^I_{\bar
  I}$.) In the above equation the couplings
$\Theta_{\bar{I}\bar{J}}^{\text{3D}}$, which receive contributions
from integrating out all massive fermions on the Coulomb branch, can
be expressed as linear combinations
    \begin{equation}
        \label{eq:thetaconvert}
        \Theta_{ij}^{\text{3D}} = x^{\sfr}_{ij} \chi_{\sfr},~~~~ x^{\sfr}_{ij} \in \Q\,.
    \end{equation}

For every resolution that satisfies our default assumption that each
matter surface has a vertical component,
the above linear system can
be inverted, which allows us to then write an explicit formula
%appearing in the universal $\sfG$ model and a set of
%$A$ independent parameters $\phi^{IJ}$:
    \begin{equation}
        \label{eq:chiconvert}
        \chi_{\sfr} = x^{ij}_{\sfr} \Theta_{ij}^{\text{3D}}\,.
    \end{equation}
%which we expect
We expect the set of allowed chiral multiplicities $\chi_{\sfr}$
that can be realized for integer flux backgrounds $\phi^{ij}$
to be independent of the choice of
resolution $X$, up to a choice of basis for $\phi^{ij}$.

\subsection{Linear constraints and anomaly cancellations}
\label{sec:linear-anomaly}

The anomaly conditions for any 4D theory are linear relations on
$\chi_{\sfr}$ (these conditions are reviewed in
\cref{4Danomalyreview}).  There are also linear relations that automatically
hold on $\Theta_{IJ}$ by virtue of the nullspace of $M$.  Connections
between the anomaly relations and these geometric conditions were
identified in \cite{Bies_2017} (see also \cite{Grimm:2015zea, Corvilain:2017luj}).  Our  finding here is that, in
all cases we consider, the linear relations on $\Theta_{ij}^{\text{3D}}$ imposed by
the nullspace conditions and symmetry constraints are precisely the
same as the anomaly conditions, so that not only does geometry encode
the anomaly conditions, but there are also no
further linear constraints coming from F-theory on the set of allowed
chiral multiplicities, and thus fluxes exist that can turn on all
anomaly-allowed combinations of chiral matter fields in all the cases
we explore.

In general, the linear constraints that hold on the fluxes
$\Theta_{IJ}$ for an F-theory background where the 4D gauge
group is unbroken (and hence equal to the geometric gauge group $\sfG$) are the union
of those that come from the nullspace of $M$ and the constraints
\labelcref{eq:general-constraints}.  It is helpful to consider how this set
of constraints arises in the two approaches characterized in
\cref{fig:commute}.  When the nullspace of $M$ is quotiented
out first, giving $M_\text{red}$, and then the constraints are imposed,
it is clear that the constraints listed above are precisely the constraints on the resulting
$\Theta$s that can arise.  The situation is slightly subtler,
however, when the constraints are imposed first.  In particular, the
signature of the inner product matrix $M$ is not generally
semi-definite, so in principle
there can be vectors of vanishing norm that are not null vectors of
$M$.  If one of the constraints \labelcref{eq:general-constraints} can
be described as $w M \phi$ for a vector $w$ of this
 type, then when the constraints are imposed first the matrix $M_C$
could have additional null vectors beyond those associated with
homological equivalence in $\Lambda_S$; this would
occur when the vector
$w$ also lies in $\Lambda_C$.  The subsequent quotient by
homologically trivial cycles (i.e., null vectors of $M$) does
not remove such null vectors from $M_C$.  Nonetheless, any such vector
would still correspond to a linear combination of the nullspace and symmetry
constraints.\footnote{A simple proof of this statement can be made as
  follows: assume without loss of generality that $M$ has no
  nullspace, and the constraints are of the form $w \Theta =
  w M \phi= 0$ for $w\in W$, and $\Lambda_C$ is the
  orthocomplement $W^\perp$ of the set of constraints $W$.  Then any
  null vector $u \in\Lambda_C$ of $M_C$
  satisfies  $u M_C  \phi = u M \phi = 0$
for any $\phi \in W^{\perp}$.
But then $u\in (W^{\perp})^{\perp}= W$,  so $u$ is a constraint
vector.  A similar proof follows when $M$ has nontrivial nullspace,
though $u$ can also have a component then in this nullspace.  }
While this situation does not arise in practice in any of the models
we analyze here, we do not have any way of strictly ruling it out,
particularly for models with one or more $\U(1)$ factors, so this
possibility must be kept in mind throughout the analysis.

\subsection{Quantization of fluxes and matter multiplicities}

\label{sec:quantization-1}

One thorny issue, which has not been fully resolved to our knowledge
anywhere in the literature, is the precise quantization condition on
the fluxes and the consequent constraints on the multiplicities of
chiral matter.  Even in the most well-understood $\SU(5)$ F-theory GUT
constructions, this question is left open in analyses of which we are
aware.
Note that this question arises whether or not there are
%is independent of
 issues related to the
shifted quantization condition \cref{eq:c2quant}.

We do not fully resolve this quantization issue here but we do shed
some light on the question and provide a set of sufficient conditions
for matter with certain multiplicities to exist.
In the basis for $\Lambda_S$ given by the surfaces $S_{IJ}$, the
coefficients $\phi^{IJ}$ are always
integers for purely vertical fluxes (or in some cases half-integers, when $c_2 (X)$ is not even)\footnote{For the most part we frame  the discussion in terms of cases where $c_2 (X)$ is an even class, so that the quantization issue of \cref{eq:c2quant} leaves $G$ as an integer cohomology class; it should be kept in mind however that when $c_2 (X)$ is not an even class, some of the flux background parameters must be half-integer, i.e. $\phi^{IJ} \in \Z + \tfrac{1}{2}$.  We consider explicit examples of this in \cref{sec:su5}.}, so that
%modulo homological equivalence
 the lattice vectors $\phi$ live in the lattice $H_{2,2}^{\text{vert}}(X,\Z) =
 \Lambda_S/ \sim$ obtained by quotienting out homologically trivial
 $\phi$.  However,
a basic observation is that the matrix $M_\text{red}$ that gives an inner
 product on this space (and which maps $\phi$ to some corresponding
 $\Theta$) does not in general have determinant equal to $\pm1$, so
 that the possible values of $\Theta$ that can be realized
 generically imply a nontrivial
 quantization on possible chiral matter multiplicities induced by
 vertical fluxes.  Furthermore, the symmetry constraints
 \labelcref{eq:Poincare,eq:gauge} impose further constraints on the
 allowed values of $\phi$ and hence the resulting nonzero
 $\Theta$
 and associated chiral multiplicities may be subject to additional quantization
constraints.

More explicitly, in many situations such as that of a purely
nonabelian gauge group,
the condition that certain $\Theta$s
must vanish, needed to preserve local Lorentz and gauge symmetry of the
4D theory, can be written schematically in the form
\begin{equation}
\left(\begin{array}{c}
0\\\Theta''
\end{array} \right)
= \left(\begin{array}{cc}
M' & Q\\
Q^T & M''
\end{array} \right)
\left(\begin{array}{c}
\phi'\\
\phi''
\end{array} \right) \,,
\label{eq:constraint-ordered}
\end{equation}
so we have
\begin{equation}
M' \phi' + Q \phi'' = 0\,.
\label{eq:mq}
\end{equation}
In the basis for $\Lambda_S$ given by the surfaces $S_{IJ}$, the
coefficients $\phi^{IJ}$ comprising $\phi', \phi''$ are always
integers.  When the matrix $M'$ has a non-unit determinant, we
can think of the image  of $M'$ acting on vectors $\phi'$ in $\Z^k$ as a
$k$-dimensional lattice $\Lambda'$.  We can solve the equation
(\ref{eq:mq}) for integer values of $\phi'$ if and only if $Q \phi''
\in \Lambda'$.  This gives a quantization condition on the flux
coefficients $\phi''$ that is both necessary and sufficient to have an
integer solution for $\phi'$.  Thus, we can determine a condition on
$\phi$, and hence on the nonzero $\Theta$s that parameterize the
chiral matter, which is sufficient to guarantee the existence of an
allowed flux background in $H_{2,2}^{\text{vert}}(X,\Z)$.  As we see in the more explicit analyses of
\cref{sec:exampleADE}, in cases of a simple gauge group like $\sfG = \SU(N)$,
this kind of analysis leads to a natural understanding of the
quantization condition on the fluxes from the appearance of the Cartan
matrix of $\sfG$ in the role of at least a block of the matrix $M'$.
The story is somewhat more complicated in the presence of $\U(1)$ factors.

\begin{figure}
\begin{center}
	\begin{tikzpicture}[scale=2.5]
		\node[] at (2.8,0) {\scalebox{1}{$H_{2,2}^{\text{vert}}(X)$}};
		\node[] at (0,2.6) {$H_{4}^{\text{hor}}(X) \oplus H_{2,2}^{\text{rem}}(X)$};
		\draw[dashed,thick,color=gray] (2*.5,-2*.1) -- (2*.5,2*1.1);
		\draw[dashed,thick,color=gray] (2*1,-2*.1) -- (2*1,2*1.1);
		\draw[dashed,thick,color=gray] (2*0,-2*.1) -- (2*0,2*1.1);
		\draw[dashed,thick,color=gray] (-2*.1,2*.5) -- (2*1.1,2*.5);
		\draw[dashed,thick,color=gray] (-2*.1,2*1) -- (2*1.1,2*1);
		\draw[dashed,thick,color=gray] (-2*.1,2*0) -- (2*1.1,2*0);
		\draw[big arrow,scale=2,very thick] (-2*.05,0) -- (1.2,0);
		\draw[big arrow,scale=2,very thick] (0,-2*.05) -- (0,1.2);
		\node[] at (2*.65,0.8) {$(\tfrac{1}{2},\tfrac{1}{2})$};
		\node[] at (2*1,-2*.15) {$(1,0)$};
		\node[] at (2*0,-2*.15) {$(0,0)$};
		%\node[] at (2*0,-2*.2) {$(0,0)$};
		%\node[] at (-2*.25,2*.5) {$(\tfrac{1}{2},\psi)$};
		%\node[] at (-2*.25,2*1) {$(\tfrac{1}{2},\psi)$};
		\node[draw,circle,fill=blue,color=blue,scale=.8] at (2,0) {};
		\node[draw,circle,fill=blue,color=blue,scale=.8] at (0,0) {};
		\node[draw,circle,fill=red,color=red,scale=.8] at (1,1) {};
	\end{tikzpicture}
\end{center}
\caption{Toy example of a possible realization of $ H_4(X,\Z)$ as an integral unimodular lattice (note that we implicitly include the shift by $c_2/2$), where we take the bilinear pairing to be $M_{\text{red}} \oplus M_{\text{red}}^{\perp} = \text{diag}(2,2)$. In this example we denote lattice vectors by $ (\phi,\psi)$, with $\phi \in H_{2,2}^{\text{vert}}(X), \psi \in H_{4}^{\text{hor}}(X) \oplus H_{2,2}^{\text{rem}}(X)$. As can be seen by requiring the inner product $2 \phi^2 + 2 \psi^2$ to take integer values, the restriction of $(\phi,\psi)$ to $(\phi,0)$ (represented by blue dots in the above graph) requires $\phi \in H_{2,2}^{\text{vert}}(X,\Z) \cong \Z$ to preserve the integrality of the lattice, i.e. $\phi \in \tfrac{1}{2} \Z$ is forbidden. However, there exist lattice vectors with $\psi \ne 0$ for which $\phi \in \tfrac{1}{2} \Z$, for instance the vector $(\tfrac{1}{2},\tfrac{1}{2})$ represented by the red dot in the above graph. This example shows that vertical flux backgrounds $\phi$ with rational coefficients $\phi^{IJ} \in \Q$, which preserve the integrality of both the lattice and chiral indices, could in principle exist.}
\label{fig:half}
\end{figure}
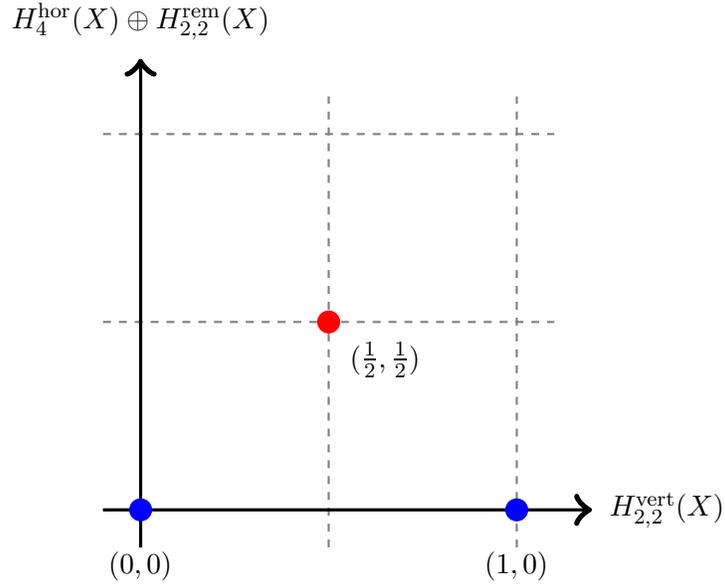

The analysis just summarized focuses only on vertical fluxes.  From
Poincar\'{e} duality of $H_{2,2}(X,\C)$, we know that there must exist a
flux so that $\int_SG = 1$ for any primitive element $S$ in $H_{2,2}
(X, \Z)$.  As mentioned above, however, the intersection form is not
in general unimodular on
$H^{2, 2}_\text{vert} (X,\Z)$.  Thus, a complete analysis of the set of
possible chiral matter multiplicities available
may require including flux backgrounds with components in
$H^{2, 2}_\text{rem} (X,\Z) \oplus
H^{4}_\text{hor} (X,\Z)$ and/or fractional coefficients in terms of the basis $\PD(S_{IJ})$ for $H^{2, 2}_\text{vert}(X)$---see \cref{fig:half} for an illustration of this point.
This is discussed in more detail in the
case of $\SU(5)$ in \cref{sec:su5}.
Flux backgrounds with such fractional coefficients have been analyzed previously
in, e.g., \cite{CveticEtAlQuadrillion}.
In that
context, in the notation of this paper, fractional values of $\phi$
are considered and the necessary constraints that $M \phi$
gives integer values (i.e. that $G$ integrated over any surface in
$H^\text{vert}_{2, 2}(X)$ is integral) are imposed.  However, not all such fractional values of
$\phi$ necessarily correspond to allowed fluxes.  As an example of this point, consider the self-dual lattice defined by the symmetric bilinear form
$\diag(2, 2)$. This lattice consists of all vectors $(x, y)$ with $x, y \in \Z/2,x
+ y \in\Z$.  The vector $(1/2, 0)$ has integer inner product with the
elements of a non-unimodular basis $(1, 0), (0, 1)$, but it is not an
element of the given self-dual lattice.  For similar reasons, 
the conditions that $M \phi$ is integral are not by themselves sufficient to
guarantee that $\phi$ is an integer homology class.
This question is further complicated by the lack of understanding of
the components of $\phi$ in
$H_{2, 2}^\text{rem} (X,\Z) \oplus
H_{4}^\text{hor} (X,\Z)$.  Thus,
it is
difficult to ascertain exactly which fractional values of $\phi$
correspond to vectors in the unimodular lattice $H_4 (X,\Z)$. 
We discuss this further in some specific examples in \cref{sec:su2}
and \cref{sec:su5}.
When $H_{2, 2}^\text{vert}(X,\Z)= H_4 (X,\Z) \cap H_{2, 2} (X,\R)$,
and $\phi$ is allowed to be a general element of $H_4 (X,\Z)$,
then the unimodularity of $H_4 (X,\Z)$ implies that the proper conditions for the vertical component $\phi_\text{vert}$ are that it should lie in the dual lattice to $H_{2, 2}^\text{vert}(X,\Z)$ and also in the constrained lattice $\Lambda_C$, but is not subject to any further apparent constraints.  This does not, however, imply that even in this case any chiral multiplicity is possible, without further information about whether the matter surface has components in $H^{2, 2}_\text{rem} (X,\Z) \oplus
H^{4}_\text{hor} (X,\Z)$; we leave a more detailed investigation of these integrality conditions to future work.

\subsection{Codimension-three $(4, 6)$ loci}
\label{sec:46}

Many F-theory geometries contain $(4, 6)$ (or
higher) singularities in the elliptic fibration over codimension-three loci in the base \cite{Candelas:2000nc,TaylorWangMC}; these are
often associated with non-flat fibers in the resolution
\cite{Lawrie:2012gg}.
In this paper, we focus on geometries without codimension-three $(4, 6)$
singular loci in the elliptic fibration. We note that we have also analyzed a
variety of situations, such as universal models with gauge groups $\sfG
= \SU(N\geq 7), \SO(N \geq 12), \gE_7$ and other cases that do
have codimension-three $(4, 6)$ loci, where we find that
there is an additional allowed flux background parameter and the rank of $M_\text{red}$
is larger than expected from the 4D anomaly cancellation
conditions.  
%A similar extra flux background parameter and increased rank arises
%in the single exceptional case $\SO(11)$ without codimension-three $(4,
%6)$ points where we find a mismatch between the available anomaly-free
%families of charged matter and the rank of $M_\text{red}$. 
A more detailed analysis of these
models is left for future work.

\subsection{Summary of new results}
\label{summary}
We summarize here the main results of the paper:

\begin{itemize}

%\item{} We compute what is to our knowledge the first known crepant resolution of the universal $\SM$ model constructed in \cite{Raghuram:2019efb}, realized as a hypersurface in an ambient $\bP^2$ bundle.

%\item{} As part of our methodology, we present a specific adaptation of\patrick{Replaced `generalize' with `adapt' and added a few more words to be more accurately describe our contribution to the literature.} the pushforward technology described in \cite{Esole:2017kyr} to compute intersection numbers and other relevant characteristic numbers for crepant resolutions of singular elliptic CY fourfolds in which the elliptic fiber is realized as a general cubic hypersurface in $\bP^2$.

\item{} We show, by way of example, that the pushforward technology of \cite{Esole:2017kyr} 
  can be used to easily compute the vertical fluxes of resolutions of
  singular Weierstrass models with any nonabelian gauge symmetry
  subgroup over an arbitrary smooth base. We also show that $\U(1)$ gauge
  factors can be incorporated into the analysis in a manner that
  depends explicitly on the intersections of the associated height
  pairing divisors with the curve classes of the base. We present an
  explicit expression for the vertical fluxes in terms of the
  pushforwards of the intersection numbers of the resolved elliptic CY
  fourfold to the base; in the special case of a purely nonabelian
  gauge group, these intersection numbers only depend on the
  intersections of the canonical class of the base and the classes of 
  the gauge divisors wrapped by seven-branes whose gauge bundles correspond to the simple factors of the F-theory gauge group.

\item{} We find that the reduced intersection pairing $M_\text{red}$ on the
  vertical middle cohomology $H^{2, 2}_\text{vert} (X,\Z)$ is
  independent of resolution (up to a change of
  basis)
in all cases
  we consider explicitly.
  %, although the full set of quadruple intersection numbers
%  are not generally resolution-independent.  
We furthermore show  that this
  resolution-independence holds
for all
  F-theory models with nonabelian gauge symmetry and generic matter, 
when the physically-relevant $M_\text{phys}$ is resolution invariant
and obeys certain compatibility conditions related to the weight lattice of
the gauge algebra.
%  Thus, the resolution-independence of $M_\text{red}$,
%  which as far as we know is not proven mathematically, seems
%  to follow from the physical condition of resolution-invariance of 4D
%  F-theory vacua associated with a given CY fourfold Weierstrass
%  model.
We conjecture that the resolution-independence of $M_\text{red}$
(and hence also of the full matrix $M$ including the nullspace built
from vertical cycles)
holds
more generally for F-theory models with arbitrary gauge groups,
including those with $\U(1)$ factors, and give some explicit examples supporting this conjecture.

%The pushforward technology dramatically simplifies (and may be necessary for) the task of computing fluxes for the universal $\SM$
 % model, whose resolution does not admit a holomorphic zero section.

%\item{} We combine these novel methods of resolution with existing
%techniques \cite{Grimm:2011fx}\wati{[pass 1] other references?} for relating fluxes to chiral matter indices in the reduced
%3D theory to give a complete formulation of matter multiplicities in
%various classes of 4D theories in a way that only depends upon (resolution-independent)
%characteristic data associated with the base.

\item{} Exploiting M-theory/F-theory duality, we match 3D Chern--Simons couplings with the vertical fluxes to obtain the chiral matter multiplicities associated to various examples of universal $\sfG$ models, some not previously studied in the literature. In particular we study low rank examples of models with simple, simply laced gauge group and generic matter (see \cref{tab:fluxtable}), as well as the universal $(\SU(2) \times \U(1)) / \Z_2$ model.

\item{} 
%With one exception ($\sfG =\SO(11)$), 
We find that in all cases we
  study,
%without codimension-three (4, 6) loci\wati{added this caveat that was
%missing}
  the
  number of independent vertical fluxes remaining after imposing
  constraints necessary to preserve local Lorentz and gauge symmetry
  in 4D---equivalently, the rank of the nondegenerate intersection
  pairing of the vertical cohomology of the resolved elliptic CY
  fourfold with integer coefficients, minus the number of
  independent symmetry
  constraints---is greater than or equal to the number of allowed
independent families of 4D
  chiral matter multiplicities plus the number of non-minimal
  codimension-three singularities in the F-theory base.
In all these cases, allowed fluxes produce matter
combinations that span the linear space of anomaly-free matter
representations.\footnote{The
    resolutions we study of models with codimension-three $(4, 6)$ loci
    are non-flat fibrations in which the fibers over the $(4, 6)$ loci
    contain a K\"ahler surface as an irreducible component. See the comments in \cref{sec:46} for further discussion.}
%This means that
%the number of independent allowed fluxes precisely matches the number
%of families of chiral matter allowed through anomaly cancellation,
%except for the special case $\sfG = \SO(11)$
%and theories with codimension-three $(4, 6)$ points.
%%, which we will treat in more detail in \cite{46}.
%The
%  $\SO(11)$ model, despite not admitting chiral $\SO(11)$
%  representations nonetheless has nontrivial vertical flux and thus
%  is an apparent exception to this \patrick{Replaced ``general characterization''.} pattern.
%%As
%  possible resolution to this puzzle we speculate that the physical
%  origin of the nonzero vertical flux in these models is due to
%  additional chiral degrees of freedom whose physical origin is at the
%  moment unclear
 %(See \cite{DelZotto:2018tcj} for a
%  discussion of additional puzzles related to the $\SO(11)$
 % F-theory model; see also the comments in \cref{sec:46}.)

% \item{} In all examples we study, we find that the nondegenerate intersection pairing on the vertical cohomology is independent of the choice of resolution up to an integer change of basis, and we conjecture this is generally true.

%\item{} We develop a systematic framework for addressing questions
%  about flux quantization and matter multiplicities, and give a
%  sufficient condition for the existence of solutions with
%  multiplicities.

\end{itemize}

%%%%%%%%%%%%%%%%%%%%%%%%%%%%%%%%%%%%%%%%%%%%%%%%%%%%%%%%%%%%%%%%%%%%%%%%%%%%%%
%%%%%%%%%%%%%%%%%%%%%%%%%%%%%%%%%%%%%%%%%%%%%%%%%%%%%%%%%%%%%%%%%%%%%%%%%%%%%%
%%%%%%%%%%%%%%%%%%%%%%%%%%%%%%%%%%%%%%%%%%%%%%%%%%%%%%%%%%%%%%%%%%%%%%%%%%%%%%

\section{Fluxes preserving 4D local Lorentz and gauge symmetry}
\label{sec:4dflux}

\Cref{sec:4dflux,sec:constraints-homology} present two different
perspectives on the relation between flux backgrounds $\phi^{IJ}$ and
fluxes $\Theta_{IJ}$ corresponding to the two paths from the upper
left to the lower right of the
commuting diagram in \cref{fig:commute}. In this section, we describe the
sublattice of flux backgrounds $\Lambda_C \subset \Lambda_S$, which is the
preimage of the lattice of fluxes $M \Lambda_S$ preserving 4D
local Lorentz and gauge symmetry. The matrix elements of the inner
product matrix $M_C$ on the constrained space depend on the
pushforwards $W_{IJKL}$ of quadruple intersection numbers of a smooth
elliptic CY fourfold $X$ resolving a Weierstrass model with gauge
symmetry $\sfG = (\sfG_\text{na} \times \U(1)) / \Gamma$ (cases with
additional $\U(1)$ factors are a straightforward generalization of the
results presented here.) Note that in this section and the next, we do not concern ourselves with the
shifted quantization condition \labelcref{eq:c2quant} but simply treat
$\Lambda_S$ as an integral lattice, with the understanding that
sometimes this quantization condition gives an overall half-integer
shift that must be incorporated in specific contexts.

In \cref{fluxesint} we review the  relationship between
vertical fluxes $\Theta_{IJ}$ and intersection numbers of the types of
smooth elliptic CY fourfolds $X$ with which we concern
ourselves. \cref{constraintsolutions} presents an explicit expression
for $M_C$ that is valid in most of the cases we consider. In \cref{sec:homologyrel} we discuss further the
relationship between the nullspace of $M_C$ and linear constraints on
$\Theta_{IJ}$, along with the relationship of these constraints to 4D
anomaly cancellation.

\subsection{Computing vertical fluxes with intersection theory}
\label{fluxesint}

Given a smooth CY fourfold $X$ and a basis of divisors $\hat D_I$
where $I = 0,1,\alpha, i_s$ (we take $I=a= 1$ to be the only index, if
there is one,
denoting a U(1) section---see the discussion immediately below
\labelcref{eq:Wazir} for more details about the index structure), we
may expand a vertical flux background $G \in
H^{2,2}_{\text{vert}}(X,\Z)$ in a basis of wedge products of
$(1,1)$-forms dual to divisors, $\PD(\hat D_I)$, where `PD' denotes the Poincar\'{e} dual.\footnote{Lefshetz's theorem on
  (1,1)-classes applied to projective varieties such as $X$ guarantees that given a basis of divisors $\hat D_I$ there always corresponds a Poincar\'e dual basis of harmonic $(1,1)$ forms $\PD(\hat D_I)$---see
  e.g. \cite{Bizet:2014uua} for a related discussion.}
In our analysis here we formally work in the Chow ring, which exhibits the intersection properties of elements of (co)homology that have a description in terms of
algebraic subvarieties.  The reason for this is that the pushforward technology
that we use, which is described in more detail in
 \cref{intersection}, is defined with respect to the Chow ring.  For
 the purposes of the analysis here, however, the only elements of the
Chow ring that concern us are the classes of divisors $\hat{D}_I$ and
their intersections
 $S_{IJ} = \hat D_I \cap \hat D_J \in \Lambda_S$, which can be
understood directly as elements of the homology groups $H_{3, 3}
(X,\Z)$ and $H_{2, 2} (X,\Z)$ respectively.

As described in \labelcref{eq:constraints-theta-m},
the integrals
of  a flux background $G$ over the cycles of vertical
surfaces can be
evaluated in terms of intersection products of $\hat D_I$
    \begin{equation}
        \Theta_{IJ} = \int_{S_{IJ}} G = \int_{X} G \wedge \PD(S_{IJ}) =\phi^{KL} S_{KL} \cdot S_{IJ}= \phi^{KL} \hat D_K \cdot \hat D_L \cdot \hat D_I \cdot \hat D_J\,,
    \end{equation}
and so we may parametrize a candidate
vertical flux $G$ in terms of its Poincar\'{e} dual class $\phi$ in the
Chow ring of $X$ as
$%  \begin{equation}
         \phi = \phi^{IJ} S_{IJ}
$,
%   \end{equation}
leading to the more succinct expression
    \begin{equation}
        \Theta_{IJ} =  \phi \cdot S_{IJ}\,.
    \end{equation}
%Above, $\hat D_I \cdot \hat D_J$ is the class of surface $S_{IJ}$ in
%the Chow ring of $X$, and `$\cdot$' indicates the intersection
%product.
This correspondence between integrals over cycles and
intersection products implies that the intersection pairing matrix $M
: \Lambda_S \times \Lambda_S \rightarrow \Z$ can be described as a
matrix with indices given by pairs $IJ$, where the matrix elements
are expressed in terms
of quadruple intersection numbers,
    \begin{equation}
        M_{(IJ)(KL)} =S_{IJ} \cdot S_{KL}\,.
    \end{equation}
Thus, essentially every computation relevant for determining the
multiplicities of chiral matter can be characterized in terms of
linear algebra and performed using intersection
theory.

In what follows, we assume that the smooth fourfold $X$ is a resolution of a singular Weierstrass model belonging to a family defined by the characteristic data $(K, \Sigma_s, W_{01})$, where $K$ is the canonical class of $B$, $\Sigma_s$ is the class of the gauge divisor in $B$ associated to the nonabelian gauge subalgebra $\mathfrak{g}_{s}$ and $W_{01}$ is the class of the pushforward $\pi_*(\hat D_0\cdot \hat D_1)$ of the intersection of the zero and generating sections. We moreover assume that the resolved elliptic CY fourfold $\pi : X \rightarrow B$ can be realized as a hypersurface inside an ambient fivefold that is the blowup of a $\bP^2$ bundle. These assumptions allow us to evaluate the quadruple intersection numbers explicitly by computing their pushforward to the Chow ring of the base $B$,
    \begin{equation}
        \pi_*( \hat D_I \cdot \hat D_J \cdot \hat D_K \cdot D_L ) =W_{IJKL}=  W_{IJKL}^{\alpha \beta \gamma} D_{\alpha} \cdot D_{ \beta} \cdot D_{ \gamma}\,,
    \end{equation}
where the right side of the above equation can be expressed as a
linear combination of triple intersection products of the classes of
the characteristic data $(K, \Sigma_s, W_{01})$. Furthermore, since
$X$ is an elliptic fibration, for certain multi-indices $IJKL$ the
pushforwards $W_{IJKL}$ have additional structure that remains
applicable for all known crepant resolutions. For convenience we
suppress the explicit pushforward map $\pi_*$ when the appropriate
ring is otherwise clear from the context. See \cref{intersection} for
additional mathematical details about the pushforward map $\pi_*$ and
the structure of the intersection numbers.

\subsection{Explicit solutions of the symmetry constraints}
\label{constraintsolutions}

The main result of this subsection is an explicit expression for the
matrix $M_C$ and the resulting possible
fluxes $\Theta_{IJ}$ that are the integrals of flux backgrounds
$\phi^{IJ}$ restricted to live in the sublattice $\Lambda_C$
satisfying the symmetry constraints \labelcref{eq:Poincare,eq:gauge}.

We now sketch the essential features of the computation; the details of this derivation can be found in
\cref{fluxderivation}. As we have seen, the symmetry constraints
\labelcref{eq:Poincare} and \labelcref{eq:gauge} imply $\Theta_{I\alpha} = 0$.
By ordering $\Theta_{IJ}$
so that those that $\Theta_{I\alpha}$ are listed first and likewise for $\phi^{I\alpha}$, as a matrix equation the symmetry constraints take the schematic form \labelcref{eq:constraint-ordered}, which we reproduce here for convenience:
\begin{equation}
\left(\begin{array}{c}
0\\\Theta''
\end{array} \right)
= \left(\begin{array}{cc}
M' & Q\\
Q^T & M''
\end{array} \right)
\left(\begin{array}{c}
\phi'\\
\phi''
\end{array} \right) \,.
%\label{eq:constraint-ordered}
\end{equation}
In solving the symmetry constraints \labelcref{eq:Poincare} and
\labelcref{eq:gauge} it is often convenient to eliminate (when possible)
the parameters $\phi^{I\alpha}$, whose indices match those of the
$\Theta_{I\alpha}$, which we require to vanish. We sometimes refer to
objects carrying indices $\hat I \hat J$ (i.e., pairs $IJ$ for which
$I, J \ne \alpha$) as ``distinctive'', and all others choices of
index $I\alpha$ as ``non-distinctive''. For example, in the above matrix
equation, $\phi''$ and $\Theta''$ have distinctive indices. However,
even when we cannot solve for all non-distinctive $\phi$ parameters
explicitly, we nevertheless sometimes denote by $\Theta''$ the
set of fluxes obeying the symmetry constraints.
Note also that even when we can solve for all the
non-distinctive $\phi$ parameters  it is sometimes useful to solve for
a different set of $\phi$s; see, e.g., \cref{eq:21-indirect}.

%We find in general for the systems we are considering that the
In cases for which the matrix block $M'$ is
nondegenerate
we can solve the equation \labelcref{eq:constraint-ordered} for the non-distinctive $\phi'$ parameters in terms
of the distinctive $\phi''$ parameters, giving
\begin{equation}
\phi' = -(M')^{-1} Q \phi'' \,.
\label{eq:solution-abstract}
\end{equation}
When $|\det M'| = 1$, there is an integer solution in $\phi'$ for
any $\phi''$.  When $|\det M'| > 1$, however,
the integrality condition imposed on $\phi''$ by requiring that the symmetry constraints be solved over $\Z$ is subtler, as discussed in
\cref{sec:quantization-1}.  
In some cases, such as imposing the constraints on $M_\text{red}$
after removing null vectors for a purely nonabelian group, the
corresponding matrix $M'$ in the non-distinctive directions is
non-degenerate and invertible, and this procedure of solving for the
$\phi'$ flux backgrounds can be performed
exactly as in \cref{eq:solution-abstract}.  In other cases, in
particular when we consider the constraints directly on $M$ and null
vectors still are included among the $\phi$s, the matrix $M'$ is
degenerate and cannot be inverted.  In many such cases we can impose
the constraints by simply using the pseudoinverse of $M'$ for
$(M')^{-1}$, which for a symmetric matrix basically means taking the
inverse on the orthocomplement of the null space and the zero matrix
on the null space.  This is equivalent to simply removing the null
space and then taking the inverse. Note that this works when the null vectors of $M'$ are also null vectors of $M$. 

We can use this more general sense of \cref{eq:solution-abstract} to write
an expression for the restriction of $M$ to $\Lambda_{C} \subset
\Lambda_S$ when $M'$ is nondegenerate and thus invertible on the
orthocomplement of the null space.
In the following we denote by $(M')^{-1}$ the pseudoinverse of $M'$.
In
particular, as outlined in \cref{sec:vertical-pairing} we can define
the $m \times (m - k)$ matrix
\begin{equation}
\label{eq:c}
C = \left(\begin{array}{c}
-(M')^{-1} Q\\
\Id_{m-k}
\end{array}\right) \,,
\end{equation}
which defines an embedding of  $\Z^{m-k}$ associated with the
$m-k$
distinctive directions into a rational extension of
the full lattice $\Lambda_S$, $C:\Z^{m-k} \rightarrow
\Lambda_C (\Q)
\subset \Lambda_S (\Q)$.  
Note that null directions in $M'$ are associated with constraints that
are automatically satisfied, so the corresponding combinations of
$\phi'$ vanish, in accord with the definition of the pseudoinverse.
As discussed in \cref{sec:quantization-1}, when  $\det M' = \pm 1$
(for the non-null part of $M'$)
this map gives a one-to-one correspondence between $\Z^{m-k}$ and
$\Lambda_C$; otherwise, the domain of $C$ must be taken to be the
subset
$\operatorname{dom} C = C^{-1} (\Lambda_C)$.
In general, given such a  mapping $C$, we can give an explicit description of the
 the inner product form
\begin{equation}
M_C =C^\transpose M C= M'' -Q^\transpose (M')^{-1} Q\,,
\label{eq:mc}
\end{equation}
which
 gives the intersection pairing of flux backgrounds in the constrained space
 $\Lambda_{C}$ as parameterized by $\phi''$, recalling that in some
 situations there may be additional discrete constraints on the
 $\phi''$ values allowed for a valid flux background.
We analyze these integrality conditions in more
detail for
 nonabelian gauge groups in \cref{integrality}, and for specific examples in \cref{sec:exampleADE,F6model}.
In much of the discussion, however, we elide this subtlety.

Carrying this description slightly further,
 we can also define an $m \times m$ matrix
\begin{equation}
P =\left( 0_{m\times k} \hspace*{0.1in} C \right) \,,
\end{equation}
which is idempotent ($P^2 = P$) and gives by right multiplication of
the matrix $M$
\begin{equation}
MP = P^\transpose MP =
\left(\begin{array}{cc}
0 & 0\\
0 & M_C
\end{array} \right).
\label{eq:MP}
\end{equation}
This extends the embedding map $C$ to be defined on all of
$\Lambda_S$, where the extra (non-distinctive) parameters are
essentially thrown out in the map, which becomes a projector; this
form of $M_C$ will be useful in some places.  In particular note that
the $\Theta$s that result from the action of $MP$ on a given set of
$\phi$s satisfying the constraint equations span the set of possible
vertical fluxes.  Recalling that $M_\text{phys}$ can be defined as the
inner product on $\Lambda_C\mathclose{}/\mathopen{}\sim$ after taking
the quotient by homologically trivial cycles, we have
\begin{equation}
\rk M_\text{phys}  = \rk M_C \,.
\end{equation}
This rank corresponds to the number of linearly independent
families of allowed fluxes.
%which as stated above we conjecture
%matches the number of linearly independent anomaly-allowed 4D chiral
%matter families.

%In the derivation sketched above, we assumed that $\sfG$ did not contain any $\U(1)$ factors. The generalization of these solutions to cases with $\U(1)$'s is slightly more complicated for the following reason: When $\U(1)$ factors are included, the analog of $M'$ (i.e., the matrix block $M'$ in \labelcref{eq:constraint-ordered} suitably generalized to act on directions $I=a$ associated to the $\U(1)$ sections) depends on the global topology of $B$, whereas by contrast for $\sfG$ purely non-abelian, $M'$ only depends on the local gauge divisors $\Sigma_s$. As a consequence, the matrix block $M'$ can sometimes be degenerate depending on how the height pairing divisors associated to the $\U(1)$'s intersect the lattice of curve classes in $B$. We elaborate on this complication in more detail below.

We present now
a formal expression for the matrix elements of $M_C$ in the case of a
gauge group $\sfG = (\sfG_\text{na} \times \U(1))/\Gamma$ for generic
characteristic data.
This set of expressions is valid whenever $M'$ is nondegenerate and
invertible (or pseudo-invertible, using null vectors of $M'$ that are also null vectors of $M$).
This condition always holds when $\sfG$ is purely
  nonabelian and the F-theory geometry admits a holomorphic zero section, and is true in most situations we have considered with
  simple bases and/or generic characteristic data
  when the gauge group contains $\U(1)$ factors.
As shown in the example in \cref{sec:21-exception}, however, there are some cases
where $M'$ is degenerate even after removing null vectors in the non-distinctive directions; in such situations we can still analyze the
spectrum by solving for a different set of $\phi$s, but the
formulae given here do not apply in this form.
  %In explicitly imposing the flux equations, we find it convenient to split the fluxes $\Theta_{IJ}$ into two pieces, where $\Theta_{IJ}^\text{c}$ only depends on ``characteristic'' parameters $\phi^{\hat K\hat L}$ (i.e., gravity and gauge indices $\hat K, \hat L = 0,i$) and the remainder $\Theta_{IJ}^\text{nc}=\Theta_{IJ} - \Theta_{IJ}^\text{c}$ has some dependence on ``non-characteristic'' parameters $\phi^{KL}$ (i.e., base indices $K=\alpha$ or $L=\alpha$). The advantage of sequestering the non-characteristic contributions is that the remaining characteristic contributions can be written solely in terms of intersection products in the Chow ring of $B$ involving the characteristic divisor classes $\Sigma_s, K$ used to define the CY fourfold described in \labelcref{chardata}.

The fluxes satisfying the symmetry constraints take the form\footnote{Hatted indices are of type
  $\hat I = 0,1,i_s$, i.e., a restriction of the usual indices to the
  case $I\ne \alpha$.}
    \begin{equation}
        \label{eq:thetaprime}
        \Theta_{ \hat I \hat J} = {M_{C}}_{(\hat  I \hat
                  J)(\hat K \hat L)} {\phi}^{ \hat K \hat L} \,.
    \end{equation}
In the above equation, the matrix elements of $M_C$ can be expressed as
    \begin{equation}
        \label{eq:geotheta1}
        {M_C}_{( \hat I \hat J) (\hat K \hat L) }  = { M_{C_\text{na}}}_{( \hat I \hat J) (\hat K  \hat L) } - { M_{C_\text{na}}}_{( \hat I \hat  J)(1\alpha)} M_{C_\text{na}}^{+(1\alpha)(1\beta)}  { M_{C_\text{na}}}_{(1\beta)( \hat K \hat L)}\,,
    \end{equation}
where $ { M_{C_\text{na}}}=  C_\text{na}^\transpose M  C_\text{na}$ is
the restriction of $M$ to the sublattice $\Lambda_{ C_\text{na}}$ of
backgrounds only satisfying the purely nonabelian constraints
$\Theta_{i_s \alpha} =0$. The components of $ M_{C_\text{na}}$ are
    \begin{equation}
        \begin{aligned}
            \label{eq:omegabar}
            {M_{C_\text{na}}}_{ ({I} {J}) (K L)} &= W_{  I  J K L}- W_{  I  J|i_{s}} \cdot W^{i_{s}| j_{s'}} W_{  K Lj_{s'}} - W_{0  I  J} \cdot W_{  K L} - W_{  I  J} \cdot  W_{0  K L} \\
            &\quad+ W_{00}\cdot W_{  I  J} \cdot W_{  K L}
        \end{aligned}
    \end{equation}
    where in particular
    \begin{align}
    \begin{split}
        \label{eq:omegabar2}
        { M_{C_\text{na}}}_{ (1 \alpha) (KL)} &= D_\alpha \cdot W_{\bar 1 KL}\\
         &=D_\alpha \cdot (-W_{1|k_{s''}}  W^{k_{s''}| i_s} W_{i_s  KL} + W_{1  IJ}-W_{0 KL}+(W_{00}-W_{01}) \cdot W_{ KL})
    \end{split} \\
    \begin{split}
        \label{eq:omegabar3}
        { M_{C_\text{na}}}_{(1\alpha)(1\beta)}&= D_\alpha \cdot D_\beta \cdot W_{\bar 1 \bar 1}\\
        &= D_\alpha \cdot D_\beta \cdot (- W_{1|k_{s''}} W^{k_{s''}|i_s} W_{1 i_s }+2(W_{00} - W_{01}) )
    \end{split}
    \end{align}
and  $M_{C_\text{na}}^{+(1\alpha)(1\beta)}$ is the inverse of ${
  M_{C_\text{na}}}_{(1\alpha)(1\beta)}$.

The structure of the various pushforwards $W_{IJ}$ is explained in more detail in
\cref{ellipticintersection}; for example in \cref{eq:omegabar3},
$W_{\bar 1 \bar 1}$ is equal to (minus) the height pairing
divisor associated to the $\U(1)$. For a purely nonabelian gauge group, there are no indices of the form
$(1 \alpha)$, the second term in \cref{eq:geotheta1} can be dropped, and
$M_C = M_{C_\text{na}}$ from \cref{eq:omegabar}.
The fact that the restriction of $M'$ (see \cref{eq:constraint-ordered})
to the nonabelian part of the theory (i.e., taking all
indices $I \alpha$ except $1 \alpha$) contains a non-trivial invertible submatrix for generic characteristic data over arbitrary $B$ can be
deduced from the explicit form of the components of $M'$, which are all
resolution-independent, as discussed in more detail in \cref{sec:nonabelian}.

The presence of a $\U(1)$ factor introduces additional complications,
as we now describe in more detail. The submatrix
$M_{C_\text{na}}^{+(1\alpha)(1\beta)}$ is generically the inverse of
the matrix ${ M_{C_\text{na}}}_{(1\alpha)(1\beta)} = [[W_{\bar 1
  \bar 1} \cdot D_\alpha \cdot D_\beta]]$.\footnote{Barred
  (``physical'') indices are of type $\bar{I} = \bar{0},
  \bar{1}, \alpha, i_s$. In the basis $\hat D_{\bar{I}}$,
  $\hat D_{\bar{0}}$ is the KK $\U(1)$ divisor and $\hat
  D_{\bar{1}}$ is the abelian $\U(1)$ divisor (i.e., the image of
  the generating section under the Shioda map), whereas in the basis
  $\hat D_{I}$, $\hat D_0$ is simply the zero section and $\hat D_1$
  is the generating section. The matrices $\sigma^{I}_{\hat I}$ in
  \labelcref{physbasis} and their inverses can be used to convert
  between these two bases.}
For bases with $h^{1,1}(B)$ not too large relative to $h^{1,1}(W_{\bar 1 \bar 1}$\footnote{Since $W_{\bar 1 \bar 1})$ is the class of a surface in $B$, whenever $h^{1,1}(W_{\bar 1 \bar 1})< h^{1,1}(B)$ the matrix $[[W_{\bar 1 \bar 1} \cdot D_\alpha \cdot D_\beta]]$ will be singular.} and generic characteristic data, this matrix is invertible.
When $M_{C_\text{na}(1\alpha)(1\beta)}$
is not invertible, however, the expression \labelcref{eq:geotheta1} is no longer valid; in
such cases a further analysis must be done, which often involves
 solving for a different set
of $\phi$ components, though the essentially the same procedure (i.e. solving the symmetry constraints $\Theta_{I\alpha} = 0$ by eliminating certain components of $\phi$) still works.
An explicit  example of this type of situation is illustrated in \cref{sec:21-exception}.
 We expect that while some null vectors of
$M_{C_\text{na}(1\alpha)(1\beta)}$ can be dealt with by solving some
of the constraints $\Theta_{1 \alpha} = 0$ for other $\phi$s, this
can be done for at most the total number of
parameters $\phi^{\hat{I}\hat{J}}$, $ \frac{1}{2}
(\rk\sfG + 2)(\rk\sfG + 1) $.  Null vectors of
$M_{C_\text{na}(1\alpha)(1\beta)}$ that cannot be treated in this manner, i.e., by solving
for other $\phi^{\hat I \hat J}$, should correspond to extra null vectors of
$M$.  It is also possible that even in the cases where null vectors of
$M_{C_\text{na}(1\alpha)(1\beta)}$ can be treated by solving for
parameters $\phi^{\hat I \hat J}$, this may increase
the number of null vectors of $M$ and decrease the number of
independent possible fluxes $\Theta$ (since $ \rk M = \dim M - \operatorname{nullity} M $ is equal to the number of independent flux backgrounds plus the number of independent constraints, so that
an increase in the number of null vectors corresponds either to a
decrease in the number of independent constraints or a decrease in the
number of independent fluxes); we have not encountered any
explicit examples where this behavior occurs, though we have not attempted to systematically construct such examples. We
explore the detailed structure of null vectors of $M$ in more detail
in \cref{sec:constraints-homology}.

Note that the analysis here can in general lose information about the
integer quantization on the $\phi$s, since in principle the inverse
matrices $W^{i_{s}| j_{s'}}$ and
$M_{C_\text{na}}^{+(1\alpha)(1\beta)}$ may be rational and not
integer valued.  We address these issues more explicitly in  \cref{sec:constraints-homology} in the context of the analysis where the nullspace
is removed first to give the reduced matrix $M_\text{red}$.

\subsection{Homology relations and anomaly cancellation}
\label{sec:homologyrel}

As discussed in \cref{sec:linear-anomaly}, the null vectors of $M_C$,
considered as elements of $\Lambda_C \subset\Lambda_S$, encode the full
set of F-theory constraints on the possible vertical fluxes
$\Theta_{IJ}$, which must include at least the anomaly cancellation
conditions but in principle may impose stronger constraints.  (See
\cite{Bies_2017} for a closely related discussion about anomalies in
F-theory.)  When we can explicitly solve for a subset of the $\phi$
variables and write an expression for $M_C$ in terms of the remaining
variables, such as is done in terms of the distinctive parameters
$\phi''$ in the preceding section, we can gain explicit information
that is relevant for understanding 4D chiral matter
multiplicities---in particular, the nullspace of such an $M_C$
contains complete information about the linear constraints satisfied
by the F-theory fluxes, as we now explain in more detail.  This
approach to understanding the number of independent families of chiral
matter available in universal F-theory models for a given $\sfG$
complements the related analysis of this question using $M_\text{red}$
as discussed in the following section.  In the remainder of this
discussion we assume that we have an explicit description of $M_C$ in
terms of a subset of the flux degrees of freedom, as realized
concretely in the preceding subsection in cases where $M'$ is
(pseudo-)invertible, so that in this subspace $\Theta'' = M_C
\phi''$ and the remaining $\Theta$s vanish.

Notice that since $M_C$ is symmetric, any null vector $\nu$ satisfying
$ M_C \nu = \nu^\transpose M_C =0$ must also satisfy $\nu^\transpose
\Theta'' = \nu^\transpose M_C \phi'' = 0$.  Thus, identifying the
nullspace of $M_C$ is equivalent to identifying the linear constraints
that must be satisfied by the fluxes $\Theta''$. This can be
accomplished in all purely nonabelian models admitting a resolution with a holomorphic zero section by
using the explicit expression for the nontrivial matrix
elements of $M_C$ given in \labelcref{eq:omegabar}.

The physical significance of the nullspace equations $ \nu^\transpose
\Theta'' =0$ is that they are the complete set of linear conditions
that must be obeyed by the symmetry constrained fluxes; provided it is possible to express the chiral
multiplicities as rational linear combinations of fluxes as in
\labelcref{eq:chiconvert}, this further implies that the nullspace
equations lead to the full set of linear constraints that must be
obeyed by the chiral matter multiplicities.  Since all allowed
F-theory models are by assumption consistent with 4D anomaly
cancellation, the nullspace equations include as a subset the linear
4D anomaly constraints.

This observation has immediate applications to the question of whether
or not F-theory geometry imposes additional linear constraints on
chiral matter multiplicities beyond those associated with 4D anomaly
cancellation, as the nullspace equations can easily be recovered from
\labelcref{eq:omegabar}. When $\sfG$ is purely nonabelian and the corresponding resolution admits a holomorphic zero section, the fact that
\labelcref{eq:omegabar} is true for arbitrary base $B$ implies that the
linear constraints on the chiral multiplicities can in principle be
read off for all $\sfG$ models in full generality, provided a
resolution $X$ can be identified such that the chiral indices can be
expressed in terms of the vertical fluxes $\Theta''$. In
\cref{sec:exampleADE} we make extensive use of this structure to
confirm that for all universal $\sfG$ models of this type that we study,
% (barring the exceptional case of the $\SO(11)$ model),
 F-theory geometry imposes no additional linear constraints on
the chiral multiplicities of matter charged under $\sfG$ beyond the 4D anomaly cancellation
constraints; we also find this to be true for all models we study with
$\U(1)$ gauge factors, as discussed in \cref{sec:abelianmodel}.

  For models with a $\U(1)$ gauge factor, some additional care
is needed since, as explained towards the end of
\cref{constraintsolutions},
%the matrix elements of $M_C$ are
%``base-dependent'' (i.e., require a complete specification of the
%triple intersection intersection numbers of $B$). This suggests it is
 it does
not seem  possible to easily compute a fully general form for
the nullspace of $M_C$ for a model
with $\U(1)$ gauge factors over an arbitrary base. Nevertheless, in
many circumstances it does appear possible to first solve for
$\Lambda_{C_\text{na}}$, then further restrict
$\Lambda_{C_\text{na}}$ to the sublattice $\Lambda_{C_\text{na}}
\cap \{\phi^{1\alpha} = 0\}$, for which the remaining symmetry
constraints $\Theta_{1\alpha} =0$ can be solved over arbitrary $B$
without modifying the nullspace equations $\nu^\transpose  \Theta''
=0$.
The basic idea here is that as long as there exists a linearly independent subset of null vectors of $M$ that span the $S_{1\alpha}$ directions,
setting $\phi^{1 \alpha} = 0$ for all $\alpha$ will not reduce the rank of the set of
$\Theta$s that are realized by acting with $M$ on
$\Lambda_{C}$, and hence will not change the nullspace
equations
 $\nu^\transpose  \Theta''
=0$ that encode linear constraints on the matter multiplicities.  We
 expect that generically the null vectors should have this property,
 and while we cannot prove that this is always the case we have not
 encountered any instances where this does not hold.  Thus, we can
often simplify the analysis of the linear constraints from null vectors by
 restricting to  background fluxes satisfying $\phi^{1\alpha} = 0$.
 (Note, however, that even though we do not expect this to modify the
 number of linear constraints, this strategy will not keep track of
 the precise lattice of allowed fluxes, for reasons similar to the
 analysis following \cref{eq:9664}.)
  With the restriction to $\Lambda_{C_\text{na}}
\cap \{\phi^{1\alpha} = 0\}$, the $\U(1)$ symmetry constraints take the
form
    \begin{align}
        M_{C_\text{na}(1\alpha)(\hat I \hat J) } \phi^{\hat I \hat J} = 0\,.
    \end{align}
In this case, the expressions for the symmetry constrained fluxes
induced by flux backgrounds restricted to the sublattice
$\phi^{1\alpha}=0$ only depend polynomially on triple intersections of the
characteristic data since setting $\phi^{1\alpha}=0$ eliminates
dependence on the matrix $W_{\bar 1 \bar 1 } \cdot D_\alpha
\cdot D_\beta$; therefore we can again compute the symmetry-preserving fluxes
in terms of the
characteristic data without committing to a specific choice of
$B$. Provided there are
null vectors with components spanning the $S_{1\alpha}$ directions as
described above, we can
then easily determine the linear constraints in this simpler setting
with the understanding that the same constraints apply to the
unrestricted fluxes as well, at least for generic characteristic
data. We do not attempt to specify the precise conditions under which
this is true; rather, we simply note that we have yet to identify any
counterexamples, i.e., any specific models with more restrictive
linear constraints among the fluxes (when $\phi^{1\alpha} \ne 0$) than
those implied by anomaly cancellation.  We give an explicit example of
this type of analysis in \cref{F6model}.

\section{Reduced intersection pairing}
\label{sec:constraints-homology}

%Mathematically, there are several ways in which we can think about the
%constraints and projection on the lattice $\Lambda_S$ with the inner
%product defined through $M$.  Operationally, it is sometimes
%convenient to first perform the projection by modding out by vectors
%in the nullspace of $M$.  In \cref{sec:nonabelian} we explicitly
%characterize the vectors in this nullspace when the gauge group $\sfG$ is
%purely nonabelian, and then consider in the subsequent subsections the
%explicit solution of the constraint equations for the resulting
%reduced matrix $M_\text{red}$.
%On the other hand, in some situations we can profit by first imposing
%the constraints.

In the previous section we explained how to restrict the lattice of
vertical flux backgrounds $\Lambda_S$ to the sublattice $\Lambda_{C}$
of vertical M-theory flux backgrounds that lift to consistent F-theory
flux backgrounds compatible with unbroken 4D local Lorentz and gauge
symmetry $\sfG$. Specifically, we showed how to compute the symmetric bilinear form matrix $M_C$ on $\Lambda_C$ so that
the symmetry constrained fluxes $\Theta''$ can be realized explicitly
as elements of the lattice $M_C \Lambda_{C}$.

In this section, we present a complementary approach, namely first
quotienting out the nullspace of $M$ to get the reduced inner product
matrix $M_\text{red}$,
and then imposing
$\Theta_{I\alpha} = 0$. The conceptual advantage of this approach
centers on the observation that $M_\text{red}$ (equivalently, the lattice
$H_{2,2}^{\text{vert}}(X,\Z)$, equipped with the intersection pairing $M_{\text{red}}$) appears to be independent of the choice
of resolution $X$.
It is also slightly easier in this approach to keep track of the
integer quantization on the fluxes.
 Furthermore, $M_\text{red}$ may be
used to consider F-theory models with flux backgrounds that break part
of the gauge symmetry, though we do not explore such configurations
here.

In \cref{sec:methodology}, we briefly describe how to obtain the vertical
cohomology as a lattice quotient, $H_{2,2}^{\text{vert}}(X,\Z) =
\Lambda_{S}\mathclose{}/\mathopen{}\sim$,  with some details of this analysis
relegated to \cref{sec:lattice-reduce}.
In
\cref{invariance}, we summarize the evidence suggesting that
$M_\text{red}$ is independent of the choice of resolution up to an
integral change of basis. Although we are unable to produce a
completely general expression for $M_\text{red}$, in
\cref{sec:nonabelian,abelianfactor} we describe the nullspace of the
intersection pairing $M$ in as much detail as we are able for various $\sfG$ models, and we defer specific examples to
\cref{sec:exampleADE,sec:abelianmodel}. \cref{rank} presents an
immediate physical application of the invariance of $M_\text{red}$.

\subsection{Nullspace quotient and integrality structure}
\label{sec:methodology}

Considered as an abstract lattice
quotient, the integrality
structure of $\Lambda_{\text{red}}: =\Lambda_S\mathclose{}/\mathopen{}\sim$ is automatically respected and the
quantization condition on flux backgrounds $\phi \in \Lambda_{\text{red}}$ is clear---it is simply
the condition that $\Lambda_{\text{red}}$ only contains integral elements. It is not
always completely straightforward, however, starting from a given
matrix $M$ and associated nullspace, to compute an integer
basis for $\Lambda_\text{red} = H_{2, 2}^\text{vert} (X,\Z)$ and the
associated symmetric bilinear form $M_\text{red}$ explicitly.  For
example, when the nullspace of an integer matrix describing the
bilinear form on a lattice is determined, any null vector that
contains a unit entry in some coordinate $IJ$ can be modded out by
simply removing that vector.  If there are no obvious unit entries,
however, the projection to integer homology is less transparent, and
typically one must identify an appropriate basis for the quotient
lattice.
A general methodology for performing this quotient and determining the
resulting inner product matrix $M_\text{red}$ is described in
\cref{sec:lattice-reduce}.
In general, this will require a choice of basis vectors for
$\Lambda_\text{red}$ that have multiple nonzero components in the
original basis for $\Lambda_S$.
In all the cases we have studied explicitly, it is possible to
identify a subset of the basis vectors of $\Lambda_S$ that form a good
basis for $\Lambda_\text{red}$; while we have not tried to prove that
this is always possible it simplifies the analysis in the cases where
this works.

\subsection{Resolution independence}
\label{invariance}

The quotient of the lattice $\Lambda_S$ by the nullspace of
$M$ gives the lattice of vertical classes
%sublattice
$H_{2,2}^{\text{vert}}(X,\Z)$.  The restriction of the intersection
pairing on $M$ gives a nondegenerate symmetric bilinear form $M_\text{red}$
that maps pairs of elements of $H_{2,2}^{\text{vert}}(X,\Z)$ to
$\Z$. An intriguing feature of $M_\text{red}$ is that in all examples
we study, $M_\text{red}$ appears to be independent of the choice of
resolution $X$ up to an integer change of basis. It is thus tempting
to conjecture that given any two resolutions $X, \tilde{X}$ of a singular
Weierstrass model with corresponding nondegenerate intersection
pairing matrices (resp.)  $M_\text{red}, \tilde{M}_\text{red}$, there exists
a matrix $U$ such that
    \begin{align}
        \tilde{M}_\text{red} = U^\transpose M_\text{red} U \,, \quad U \in
                \GL(h^{2,2}_{\text{vert}}(X), \Z)\,,
\label{eq:mu-conditions}
    \end{align}
where $h^{2,2}_{\text{vert}}(X) =
h^{2,2}_{\text{vert}}(\tilde{X})$.  Note that since $\GL(n, \Z)$ is a group, every $U$ must be invertible and therefore $\det U = \pm{}1$. While we have not checked the
resolution-independence of $M_\text{red}$ for every possible resolution
of every $\sfG$ model we study, nor for all choices of characteristic
data $(K,\Sigma_s, W_{01})$, for all cases in which we compute the
matrices $M_\text{red}, \tilde{M}_\text{red}$ explicitly, we find that there is indeed
an invertible integer matrix $U$ satisfying
\cref{eq:mu-conditions}. Note that it is somewhat easier to check that
the determinant, rank, and signature are equal for any pair
$M_\text{red}, \tilde{M}_\text{red}$. Since $M_\text{red}$ is symmetric,
Sylvester's law of inertia implies that any two intersection pairing
matrices with these common features are congruent to one another via
an invertible real (not necessarily integer) matrix $U$. However, this
is not enough to show that $U$ is an integral matrix, so to show that
$M_\text{red}$ and $\tilde{M}_\text{red}$ are equivalent it appears necessary to
explicitly compute a matrix $U$ satisfying $\tilde{M}_\text{red} = U^\transpose
M_\text{red} U$ and confirm that it is a unimodular matrix.
%We compute the form of such a matrix explicitly for nonabelian gauge
%groups in \cref{proof}.
%and to our knowledge the only way
%to accomplish this is to compute $U$ explicitly for a specific choice
%of characteristic data.

When   $M_\text{red}, \tilde{M}_\text{red}$ are related by an integer change of basis $U$, it furthermore follows that the associated degenerate
matrices $M, \tilde{M}$ are also related by an integer change of basis. This can be seen by first putting  each of
the
$M$ matrices in
the canonical form \labelcref{eq:pmp} with $M_\text{red}$ in the upper left
block, as described in
\cref{sec:lattice-reduce}, and then using a linear transformation with $U$ in the upper
left block and the identity in the remaining part of the matrix to
relate the two canonical forms of $M, \tilde{M}$.

In purely nonabelian cases $\sfG = \sfG_\text{na} = \prod_{s}
\sfG_s$, a general form for a
matrix $U$ relating  two different versions of $M_\text{red}$ can be
constructed explicitly provided that we  make the physically natural
%
% ``physical'' assumption that the matter spectrum and chiral indices
%are resolution invariant and further assume that for all resolutions
%$X$ under consideration the chiral indices can be expressed as linear
%combinations of the fluxes, $\chi_{\sfr} = x_{\sfr}^{i_s j_t}
%\Theta_{i_s j_t}$, and that there exists a subset of irreps such that
%the $x_{\sfr}^{i_s j_t}$ are $1$ for a particular $i_s j_t$ and $0$
%for all others.\andrew{We should think about this more and edit
%  appropriately} \andrew{Can we connect this to the basis used in the
%  quotient?} In such cases and
assumption that $M_\text{phys}$ is the same for both resolutions; we
carry out this analysis in \cref{proof}.
%With this, the structure
%of the equations defining the nullspace of $M_C$ is restrictive
%enough to fix the form of $U$, up to some undetermined
%signs.\patrick{Added this caveat, as the form of $U$ is not uniquely
%  fixed.[W]} 
As discussed in more detail in
 \cref{nonabwithchiral},
the resulting $U$ is only constrained to be rational and not integral
from these considerations, and a certain compatibility condition is
required for $U$ to be integral. In all cases we have considered,
however, we have found an integral $U$ of this form.
In more general models
with $\U(1)$ gauge factors and rational zero sections, we do not know
of such an explicit construction of $U$; nevertheless, a similar
 structure should hold in those cases, 
and for specific choices
of characteristic data it still appears to be true that
$M_\text{red}$ is independent of the specific choice of $X$ as we
illustrate in the context of the $(\SU(2) \times \U(1))/\Z_2$
model in \cref{sec:abelianmodel}.

If $M_\text{red}$ is indeed resolution independent, this further
suggests that the vertical cohomology $H^{2,2}_{\text{vert}}(X,\Z)$ of
any elliptic CY fourfold $X$ resolving a singular Weierstrass model
with gauge symmetry $\sfG$ is in some sense a mathematical invariant
characterizing properties of the singular locus of
$X_0$.

\subsection{Purely nonabelian gauge groups}
\label{sec:nonabelian}

Before discussing the more general case including a $\U(1)$ gauge factor,
we study some properties of the nullspace of the intersection pairing $M$ that hold
in the situation that the gauge group is purely nonabelian,
    \begin{equation}
        \sfG = \sfG_\text{na} = \prod_s \sfG_s\,,
    \end{equation}
and the zero section is holomorphic.

\subsubsection{Null space structure of $M$
with purely nonabelian gauge group}
\label{sec:null-nonabelian}

When the group $\sfG$
is purely nonabelian,
 the intersection pairing $M$ between pairs of vertical cycles $S_{00}$, $S_{0 \alpha}$, $S_{0 i_s}$,
$S_{\alpha \beta}$, $S_{\alpha i_s}$, $S_{i_s j_t}$ can be expressed as
\begin{equation}
  M=
\left(  \begin{array}{c c c c c c}
K^3 &  [K^2 \cdot D_\alpha] & 0 &  [K \cdot D_\alpha \cdot D_\beta] & 0 & 0\\
\hspace*{0.0in}
[K^2 \cdot D_{\alpha'}] & %
[[K \cdot D_\alpha \cdot D_{\alpha'}]] & 0 & %
    [[D_\alpha \cdot D_{\alpha'} \cdot D_{\beta}]] & 0 & 0\\
    0 & 0 & 0 & 0 & 0 & 0 \\
\hspace*{0.0in}
    [K \cdot D_{\alpha'}\cdot D_{\beta'}] & [[D_\alpha \cdot
        D_{\alpha'} \cdot D_{\beta'}]] & 0 & 0 & 0 & [[W_{\alpha' \beta' i_s j_{t}}]]\\
    0 & 0 & 0 & 0 & [[W_{\alpha' \alpha i'_{s'} i_{s}}]] & [[W_{\alpha' i'_{s'} i_s j_t}]\\
    0 & 0 & 0 & [[W_{\alpha \beta i'_{s'} j'_{t'}}]] & [[W_{\alpha i_s i'_{s'} j'_{t'}}]] & [[W_{i'_{s'} j'_{t'} i_s j_t}]]
\end{array} \right).
\label{fullmat}
\end{equation}
(Above, single brackets $[\cdot]$ denote a sub-vector and double
brackets $[[ \cdot ]]$ denote a submatrix; moreover, unprimed free indices correspond to rows while primed free indices correspond to columns.)
Note that, as described in more detail in
\cref{intersection}, the only intersection numbers in \cref{fullmat} that are
resolution-dependent are those that contain at least three indices of
type $I =i_s$; the values in the upper left are all included explicitly,
and we have
\begin{equation}
W_{\alpha \beta i_sj_t} = D_\alpha \cdot D_\beta \cdot W_{i_sj_t} =W_{i_s|j_t} D_\alpha \cdot D_\beta \cdot \Sigma_s=  -\delta_{st} D_\alpha \cdot D_\beta
\cdot \Sigma_s  \kappa_{ij}^{(s)}\,,
\label{eq:omegakappa}
\end{equation}
where
    \begin{align}
        \kappa_{ij}^{(s)} =-W_{i_s | j_s}
    \end{align}
is the inverse Killing metric
 of the gauge factor $s= t$ (which is equal to the
Cartan matrix for ADE groups)
and
$\Sigma_s$ is the divisor supporting that gauge factor.

%In general, $M$ is degenerate,
%and as we have discussed, the set of homology cycles $S_{IJ}$ is not
%a basis for
%$H_{2,2}^\text{vert}(X)$; thus one needs to eliminate the null
%directions of $M$ in order to determine the actual nondegenerate
%intersection pairing $M_\text{red}$.
The nullspace of $M$ is the
set of solutions to the equation
    \begin{align}
         M_{(IJ)(KL)} \nu^{KL} =0\,.
    \end{align}
Some elements of the nullspace correspond to linear combinations of
intersections
$S_{\alpha \beta}$ that are trivial in the base
homology.\footnote{For example, when $B = (\bP^1)^{\times 3}$ with
  classes $H_i, i = 1, 2, 3$ corresponding to points in the three
  factors crossed with $\F_0 \cong\bP^1\times\bP^1$ from the other two factors, the only nontrivial
  intersection is $H_1 \cdot H_2\cdot H_3 = 1$, and the curves $H_i \cap
  H_i$ are trivial in homology.}
From the conditions of Poincar\'{e} duality on the base and
nondegeneracy of the triple intersection product as discussed in
\cref{sec:nondegeneracy}, the number of independent homology classes
represented by $S_{\alpha \beta}$ and $S_{0 \alpha}$ are both equal to
$h^{2,2} (B) =h^{1, 1} (B)$;  null directions associated
with trivial homology classes in the linear space of $S_{\alpha
  \beta}$ can thus be removed, though for notational simplicity we
continue to use the same symbol $S_{\alpha \beta}$ for the reduced
basis.
Similarly, there are (linear combinations of) intersections $
S_{\alpha i_s}$ that correspond to trivial classes $D_\alpha \cdot
\Sigma_s$ in the base.  In general, the number of independent nontrivial 
classes $S_{\alpha i_s}$ is at most  $h^{1, 1}(\Sigma_s)$, but may be
smaller.
% Finally, further linear combinations of the intersections
% $S_{\alpha i_s}$ may  be in the nullspace when they are associated
% with curves in $\Sigma_s$ that are
% orthogonal
% to all curves of the form $D_\alpha \cdot \Sigma_s$.
All these null
vectors depend only on the geometry of the base.

After removing the nullspace elements associated with the base
geometry, which are independent of resolution, we can proceed further
by solving explicitly for more general nullspace elements; we find
that additional elements of the nullspace are generated by the
vectors
% [\textcolor{red}{be careful about rational numbers from
%    inverting $W_{i,j}$}]
%\wati{[pass 1] I'm now uncertain about integrality conditions coming from
%  nullspace beyond what comes from Cartan and constraints.  Discuss?}
    \begin{align}
    \begin{split}
    \label{linrel}
        \nu_{\langle 0 \rangle} &= (1 , - [K^\alpha], 0 ,0,0,0) \\
        \nu_{\langle i'_{s'}\rangle} &= (0,0,[\delta_{i'_{s'}}^{j_t}],0,0,0) \\
        \nu_{C\langle a \rangle}&=\nu_{C\langle a\rangle}^{i'_{s'} j'_{t'}} (0,[W_{i'_{s'} j'_{t'}}^\alpha ], 0,-[W_{i'_{s'} j'_{t'}}^\alpha K^\beta], -[W^{k_u|k'_{u'}} W_{i'_{s'} j'_{t'}|k'_{u'}}^\alpha] ,[\delta_{i'_{s'} j'_{t'}}^{k_u l_v}] )\,,
    \end{split}
    \end{align}
where the expression in parentheses in the third line above may be
viewed as the components of a basis of symmetry-constrained 4-cycles
$S_{C i_s j_t} = CS_{i_sj_t}\in \Lambda_C \subset\Lambda_S$ given by  
	\begin{equation}
	\label{eqn:SC}
		S_{Ci_sj_t} = W_{i_s j_t}^{\alpha} (S_{0\alpha} -
                K^\beta S_{\alpha \beta} )-W^{k_v|l_u} W^\alpha_{i_s
                  j_t |l_u} S_{\alpha k_v} + S_{i_s j_t}  \,,
	\end{equation} 
which satisfy
	\begin{equation}
		S_{C i_s j_t} \cdot S_{C k_u l_v} = S_{i_s j_t} \cdot
                S_{Ck_u l_v} = M_{C (i_s j_t)(k_u l_v)} \,,
	\end{equation} 
and $\nu^{i_s j_t}_{C\langle a \rangle}$ are the coefficients of null
vectors $M_C$, i.e.
    \begin{align}
    \label{4Cid}
        0 &= \nu_{C\langle a\rangle}^{i'_{s'} j'_{t'}} ( W_{k_u l_v }^\alpha K^\beta
        W_{i'_{s'}|j'_{t'}} - W^{m_w|k'_{u'}}
        W_{k_ul_v|k'_{u'}}^\alpha W_{m_w  i'_{s'}|j'_{t'}}^\beta+
        W_{k_u l_v i'_{s'}| j'_{t'}}^{\alpha \beta} )
    \end{align}
or equivalently
    \begin{align}
    \begin{split}
    \label{MCnull}
        0 %&= c^{k_u l_v} ( W_{k_u l_v }\cdot K W_{i'_{s'}|j'_{t'}} - W^{m_w|k'_{u'}} W_{k_ul_v|k'_{u'}} \cdot W_{m_w  i'_{s'}|j'_{t'}}+ W_{k_u l_v i'_{s'}| j'_{t'}})\\
            &=\nu_{C\langle a\rangle}^{i'_{s'} j'_{t'} } M_{C(k_u l_v ) (i'_{s'} j'_{t'})}\,.
    \end{split}
    \end{align}
In the above expression we have used the fact that the expression in parentheses
in \cref{4Cid} is $M_C$ in the special case of a purely nonabelian
gauge group and holomorphic zero section---see \labelcref{eq:omegabar}
and note here $W_{00} = K$.
The above computation shows the null vectors of $M_C$ are, in these cases, in
one-to-one correspondence with the null vectors of $M$ (note that in
this situation the subtlety of zero-norm null vectors of $M_C$
described in the second paragraph of  \cref{sec:linear-anomaly} does
not arise since in the notation of \cref{eq:c}, the sub-matrix $M'$
restricted to the non-distinctive parameters $\phi^{I \alpha}$ is invertible
after removing the null vectors that depend on the base
geometry, as well as those from the first two rows of \cref{linrel}).
In principle the appearance of the inverse matrix $W^{k |
  k'}$ in the third set of vectors (\ref{linrel}) may mean that
even when (\ref{4Cid}) is satisfied for all $\nu_{C\langle a\rangle}^{ij}$, these
null vectors are rational with integer $\nu_{C\langle a\rangle}^{ij}$, so that the $ij$ fluxes cannot simply be
projected out, as discussed in
\cref{sec:methodology}.  In all cases we have examined explicitly,
however, the entries are integer despite the presence of the inverse
matrix; we suspect that this occurs generally, though we have not
tried to prove this statement.

%On the one hand, explicit knowledge of the nullspace vectors allows us
%to directly construct the reduced matrix $M_\text{red}$.  From another
%point of view, these null vectors provide us with a set of further
%constraints on the fluxes.

As discussed in \cref{sec:homologyrel}, the structure of the nullspace
elements corresponds with constraints on the fluxes $\Theta_{IJ}$.  In particular,
the property $M=
M^\transpose$ implies that the above nullspace equations must also be
satisfied by the fluxes: 
    \begin{align}
    \label{homrel}
        S_{KL} \cdot S_{IJ} \nu^{IJ} =0~~\implies ~~\Theta_{IJ} \nu^{IJ} = \nu^{IJ} S_{IJ} \cdot S_{KL} \phi^{KL} =0\,.
    \end{align}
The   linear relations on fluxes coming from the first two classes of
null vectors in \cref{linrel}, namely
    \begin{align}
        \Theta_{IJ} \nu_{\langle 0\rangle}^{IJ} =\Theta_{00} - K^\alpha \Theta_{0\alpha} = 0,~~~~\Theta_{IJ} \nu_{\langle i'_{s'}\rangle}^{IJ}= \Theta_{0 i'_{s'}} =0
    \end{align}
are true in the special case of a holomorphic zero section; see, e.g.,
\cite{Grimm:2011sk}.
The possible coefficients $\nu_{C\langle a\rangle}^{i'_{s'} j'_{t'}}$ appearing in the third
linear condition
    \begin{align}
        \label{eq:thirdcondition}
        \Theta_{i'_{s'}j'_{t'}}  \nu_{C\langle a\rangle}^{i'_{s'} j'_{t'}} =\phi^{k_u l_v} M_{C(k_u l_v)(i'_{s'} j'_{t'})} \nu_{C\langle a\rangle}^{i'_{s'} j'_{t'}} = 0
    \end{align}
can be determined
in any given situation by explicitly identifying the nullspace vectors
of the form in the last line in \cref{linrel}.
In cases where there are no constraints on these coefficients, these
conditions force all fluxes $\Theta_{ij}$ to vanish and there is no
chiral matter.

While in principle for any base and characteristic data the nullspace
of the intersection matrix $M$ is straightforward to compute directly,
because of the relation \labelcref{MCnull}, the structure of the
constrained
matrix $M_C$ studied in the previous section can be used to analyze
the nullspace of $M$ in many cases.  In some cases, for the purposes of practical
computation, this analysis can be simplified
when rows/columns of $M_{C}$
vanish identically.
The subset of indices $i'_{s'}j'_{t'}$ for which
$M_{C(k_u l_v)(i'_{s'} j'_{t'})} = 0$ vanishes for all $k_u l_v$ are
indices for which the coefficients $\nu_{C\langle a\rangle}^{i'_{s'} j'_{t'}}$ can be set
equal to unity. In these cases, the appearance of the Kronecker delta
function in explicit coefficients of $S_{Ci'_{s'} j'_{t'}}$ given in
\cref{linrel} indicates that the basis elements $S_{i'_{s'}
  j'_{t'}}$ are redundant and may be removed from the generating
set. On the other hand, the subset of indices $i'_{s'} j'_{t'}$ for
which $M_{C(k_u l_v)(i'_{s'} j'_{t'})}$ does not vanish for all
$k_ul_v$ are those for which nontrivial elements can be found spanning
the nullspace of $M_C$ by taking appropriate linear combinations of
$S_{C i'_{s'} j'_{t'}}$. Removing these primitive directions from the
lattice $\Lambda_C$ completes the nullspace quotient and leaves behind
a basis of homologically nontrivial cycles spanning $\Lambda_C/
\sim$. Since $M_C$ can be used to indirectly define these remaining
elements via \cref{MCnull}, it follows that explicitly computing
$M_C$ automatically determines $M_\text{red}$; we elaborate on this
point in \cref{nonabwithchiral}.
% These homology relations geometrically encode, among other potential
%physical constraints, the 4D anomaly cancellation conditions.

 We next examine the structure of $M_\text{red}$ in the context of specific models with purely nonabelian gauge symmetry.

\subsubsection{Nonabelian groups without chiral matter}

Let us first assume that
all of the vectors $S_{C i'_{s'} j'_{t'}}$ appearing in
the third class of vectors listed in \cref{linrel} are null vectors; i.e.,
 the expression in parentheses in \labelcref{4Cid}
vanishes identically for all coefficients $\nu^{i'_{s'}j'_{t'}}_C$, so that $S_{Ci'_{s'}j'_{t'}} =0$ in homology.
We can then use the null vectors to eliminate the $ (\rk\mathsf
G_\text{na})^2 + \rk\sfG_\text{na} + 1$ redundant elements
$S_{00}, S_{0 i_s}, S_{i_sj_t}$ to form a basis consisting of the at most $h^{1,1}(B)(
h^{1,1}(B)+ \rk\sfG_\text{na} +1)$ classes $S_{0\alpha},
S_{\alpha \beta}, S_{\alpha i_s}$. (Since there may exist additional null vectors, in addition to $S_{0\alpha}$ we keep only homologically
nontrivial basis elements among $S_{\alpha \beta}, S_{\alpha i_s}$, and hence the total number basis elements may be less than $h^{1,1}(B)(
h^{1,1}(B)+ \rk\sfG_\text{na} +1)$.) This reduces the intersection matrix
$M$ to the intersection pairing
%\footnote{Note that $W_{i'_{s'}| i_s} \Sigma_s$ is shorthand for the component $ (\oplus_t h_t \Sigma_t)_{i'_{s'} i_s}$, with $h_t$ being minus the inverse Lie algebra metric and $\Sigma_t$ the irreducible component of the discriminant locus corresponding to 7-branes carrying the gauge symmetry factor $\sfG_{\text{na}, t} \subset \sfG_\text{na}$.}
    \begin{align}
    \label{resindependenthatM}
        M_\text{red} = [[D_\alpha \cdot D_{\alpha'}]] \otimes \begin{pmatrix} K & [D_\beta] & 0 \\ [D_{\beta' }] & 0 &0 \\
        0 & 0 &[[W_{i'_{s'}i_s}]] \end{pmatrix}\,,
    \end{align}
where the Kronecker product $\otimes$ in the above expression is understood to imply the intersection product $\cdot$ component-wise. Note that the integrality condition on flux backgrounds in
$H^{2,2}_\text{vert}(X,\Z)$ is preserved through the projection to this
basis when all the components in the vectors appearing in
\cref{linrel} are integer, which as discussed above occurs in all the
cases we have studied.
The pairing $M_\text{red}$ is
 manifestly independent of any choice of resolution $X$ with a holomorphic zero section, since in such cases $W_{i'_{s'} i_s}= \delta_{s'
  s}\kappa^{(s)}_{i' i} \Sigma_s$ and the characteristic data $K, \Sigma_s$ remains unchanged.

Since the intersection pairing on $B$ is
nondegenerate, as discussed in \cref{sec:nondegeneracy},
and we have explicitly removed null combinations of $S_{\alpha i_s}$,
$M_\text{red}$ is manifestly nondegenerate and resolution-independent.
It follows immediately that the
symmetry constraints \labelcref{eq:Poincare,eq:gauge} $\Theta_{0\alpha} = \Theta_{\alpha \beta} =
\Theta_{\alpha i_s} =0$ force all independent fluxes to vanish.
Stated different, the symmetry constraints together with
\cref{eq:thirdcondition}
imply $\Theta_{i_s j_t} = 0$.  Hence there are no nontrivial fluxes in
these cases, and consequently no chiral matter in the resulting 4D F-theory models, as assumed.
%   \begin{align}
%       \phi^{\alpha \beta} D_\alpha \cdot D_\beta \cdot D_{\alpha'} =0 \\
%       \Theta_{\alpha' \beta'} &=\phi^{0 \alpha}  D_\alpha \cdot D_{\alpha ' } \cdot D_{\beta '} =0\\
%       \Theta_{\alpha' i'_{s'} } &= \phi^{\alpha i_s} D_{\alpha} \cdot D_{\alpha'} \cdot W_{i'_{s'} i_s} =0
%   \end{align}
We give explicit examples of systems of this kind in
\cref{sec:exampleADE}, in particular for the groups $\sfG =
\SU(N<5)$.
%\footnote{Another way to verify that a given model does not contain chiral matter is to check that the rank of
%    $M_\text{red}$ is equal to the number of symmetry constraints
%    $\Theta_{I\alpha}=0$, provided there are no $(4, 6)$ singularities. There appears to be one puzzling exception to this pattern,
%    namely the case $\sfG = \SO(11)$; 
%%the physical interpretation of the vertical flux in these models is
%%not currently understood.
%we will address the $\SO(11)$ model along with the physics of $(4,6)$ points
%models elsewhere.}
% further in a separate publication \cite{46}.

\subsubsection{Nonabelian groups with chiral matter}
\label{nonabwithchiral}

Next we consider the case that the expression in parentheses
in \cref{4Cid} is non-vanishing for some non-empty subset of indices
$i'_{s'} j'_{t'}$. This implies that there are allowed nontrivial
flux backgrounds $\phi$ and corresponding fluxes $\Theta$; however,
not all of the nonzero fluxes $\Theta_{i_s j_t}$ are
independent in $M_C \Lambda_C$. We can remove all null vectors and project $\phi = \phi^{IJ}S_{IJ}$ onto a
basis
of surfaces $S_{0\alpha}, S_{\alpha \beta}, S_{\alpha i_s}, S_{j_t k_u}$ (again keeping only homologically non-trivial $S_{\alpha \beta}, S_{\alpha i_s}$), leading to the intersection pairing     \begin{align}
    \label{Mrednonabelian}
        M_\text{red} =  \begin{pmatrix}
            [[   D_{\alpha'} \cdot K \cdot D_\alpha ]] & [[D_{\alpha'}  \cdot D_{\alpha} \cdot D_\beta ]] & 0 & 0 \\
            [[  D_{\alpha' } \cdot D_{\beta'} \cdot D_{\alpha } ]] &0 &0 & [[ W_{\alpha' \beta' j_t k_u}]] \\
            0 & 0& [[ W_{\alpha'  i'_{s'}\alpha i_s} ]] & [[ W_{\alpha' i'_{s'} j_t k_u}]] \\
        0 & [[ W_{j'_{t'} k'_{u'}\alpha \beta} ]] &  [[W_{j'_{t'} k'_{u'}\alpha i_s}]] &[[ W_{j'_{t'} k'_{u'}j_t k_u }]]
        \end{pmatrix}\,,
%        \begin{pmatrix}
%           [[ K \cdot D_\alpha \cdot D_{\alpha'} ]] & [[ D_{\alpha} \cdot D_{\alpha'} \cdot D_\beta ]] & 0 & 0 \\
%           [[ D_{\alpha } \cdot D_{\alpha' } \cdot D_{\beta'}]] &0 &0 & [[ W_{\alpha' \beta' j_t k_u}]] \\
%           0 & 0& [[ W_{\alpha' \alpha i'_{s'} i_s} ]] & [[ W_{\alpha' i'_{s'} j_t k_u}]] \\
%       0 & [[ W_{\alpha \beta  j'_{t'} k'_{u'}} ]] &  [[W_{\alpha i_s  j'_{t'} k'_{u'}}]] &[[ W_{ j'_{t'} k'_{u'}  j_t k_u }]]
%       \end{pmatrix}\,.
    \end{align}
where we keep in mind that only a linearly independent subset of the
$(\rk\sfG_\text{na})(\rk\sfG_\text{na} +1)/2$ possible 4-cycles $S_{j_t k_u}$ is
represented in the above expression for $M_\text{red}$.
 In principle this choice of a subset of the
basis elements may not be compatible with the integral lattice
structure through the projection, but as mentioned above this kind of
issue does not occur for any of the cases we have considered
explicitly and we can always choose such a basis in these cases.  The
specific set of
 independent fluxes of the form $\Theta_{i_s j_t}$ (equivalently,
the set of independent 4-cycles of the form $S_{i_s j_t}$) depends
on the characteristic data of the resolution $X$, and hence we cannot
be more  precise at this point without specifying the characteristic
data of the elliptic fibration, although we expect that for every
F-theory model the number of independent fluxes is independent of
resolution. Nevertheless, we can see clearly that
imposing the symmetry conditions on the reduced
intersection pairing $M_\text{red}$ leaves behind a subset of
independent fluxes in the ``pure Cartan'' (i.e. $S_{i_sj_t}$) directions that
parametrize the combinations of 4D chiral indices realized by the F-theory compactification.
%Imposing the Poincar\'{e} and gauge symmetry conditions $\Theta_{\alpha \beta} = \Theta_{\alpha i_s} = 0$ at this stage leaves behind a subset of linearly independent fluxes of the form $\Theta_{i_s j_t}$ (i.e., depending only on pure Cartan indices). The interpretation of the M-theory background as a low-energy effective 3D theory then implies that the fluxes $\Theta_{i_s j_t}$ are independent linear combinations of chiral indices of the form \labelcref{chiinTheta} that satisfy the anomaly cancellation conditions implicitly encoded in the homology relations \labelcref{homrel}.
We give more explicit examples of systems of this kind in
\cref{sec:exampleADE}, see in particular \cref{tab:fluxtable}.

While it is not obvious that \labelcref{Mrednonabelian} is resolution
independent, 
%the additional assumptions described towards the end of
%\cref{invariance}, which essentially amount to the condition that the
%physics of an F-theory model are resolution-independent,\wati{added
%  last phrase}\patrick{I don't think this is quite accurate, because
%  the assumptions are stronger than the condition that the physics is
%  (i.e., the chiral indices are) simply
%  resolution-independent. Discuss?} are sufficient to enable us to
as shown in \cref{proof}, with some natural physical assumptions
(essentially that $M_\text{phys}$ is the same for the two resolutions)
we can
determine an explicit form for 
%the integer
a change of basis matrix $U$
%needed to convert
that converts between two different presentations of
$M_\text{red}$ associated to any pair of resolutions $X,\tilde X$
for which $M_\text{phys}, \tilde M_{\text{phys}}$  are related by an integral linear
transformation
$\tilde{M}_\text{phys}= U_p^\transpose M_\text{phys} U_p$.  The
transformation $U$ has the schematic form
%\begin{align}
%\label{Umatrix}
%        U%(\tilde i_s \tilde j_t )
%= \begin{pmatrix}
%            [[\delta_{0\alpha'0\alpha }]] & 0 & 0 &0 \\
%            0 & [[\delta_{\alpha' \beta' \alpha \beta}]] & 0 &0 \\
%            0& 0 &[[ \delta_{\alpha' i'_{s'}\alpha i_s }]] & 0 \\
%            [[U_{i'_{s'} j'_{t'} 0 \alpha} ]] & [[U_{i'_{s'} j'_{t'} \alpha \beta} ]] &[[U_{i'_{s'} j'_{t'} \alpha i_s}]]  &[[\delta_{i'_{s'} j'_{t'}i_sj_t} ]]
%        \end{pmatrix},
%    \end{align}
%where
% the argument of
% $U$ on the left hand side of the above equation
%is there to remind us that
\begin{align}
\label{Umatrix}
U   =
\begin{pmatrix}
1 & u\\
0 & U_p
\end{pmatrix} \,,
\end{align}
where $u$ may contain rational parts with a denominator of
$\det \kappa$.
%
%The details of the derivation of $U$ can be
%found in \cref{proof}.
As discussed in \cref{proof}, $U$ is an integral matrix when a certain
compatibility condition is satisfied between the off-diagonal blocks
on the lowest row and rightmost column of the two presenations of
$M_\text{red}$, for an allowed choice of equivalence $U_p$ (which has
an ambiguity up to automorphisms of $M_\text{phys}$).  In all cases we
have analyzed this compatibility condition is satisfied for some
$U_p$, and the resulting $U$ is an integer change of basis, but we do
not have a complete proof that this is generally the case.

In a related fashion,  there is a transformation of
the form \labelcref{Umatrix} that takes $M_\text{red}$ to a canonical product
form
     \begin{align}
    \label{Mredcanonical}
        M_\text{red}^\text{cp} =  
U^{\text t} M_\text{red} U =
\begin{pmatrix}
            [[   D_{\alpha'} \cdot K \cdot D_\alpha ]] & [[D_{\alpha'}  \cdot D_{\alpha} \cdot D_\beta ]] & 0 & 0 \\
            [[  D_{\alpha' } \cdot D_{\beta'} \cdot D_{\alpha } ]] &0
            &0 & 0 \\
            0 & 0& [[ W_{\alpha'  i'_{s'}\alpha i_s} ]] & 0 \\
        0 & 0 &  0 & M_\text{phys}/(\det \kappa)^2
        \end{pmatrix}\,,
    \end{align}
where we simply use the  upper right components of $U$ to transform away the
off-diagonal bottom row and right column of \cref{Mrednonabelian}.
This inner product matrix must be treated with respect to the lattice
$\Lambda^\text{cp} =U^{-1} \Z^n$, which is not in general an integer lattice in this case.  This form
is, however, useful since the symmetry constraints can be solved
trivially by setting all components except the last of the flux
background $\phi \in\Lambda_\text{cp}$ to vanish; an explicit example
of this is illustrated in \cref{sec:5-independence}.
The appearance of $\det \kappa$ in the bottom right component comes
from the fact that in general the off-diagonal values of $U$, associated with this transformation to the canonical product form in \cref{Mredcanonical}, are
rational with denominator $\det \kappa$, and $\Lambda_{\rm phys} =
((\det \kappa)\Z)^m$,
as discussed further in
\cref{proof}.  

\subsection{Gauge groups with a $\U(1)$ factor}
\label{abelianfactor}

For the more general case of Weierstrass models with gauge group $\sfG = ( \sfG_\text{na} \times \U(1) )/ \Gamma$, we find in practice it is typically easier to compute resolutions of physically-equivalent singular models in which the elliptic fiber is realized as a general cubic in $\bP^2$, see, e.g., \cite{KleversEtAlToric}. These models generically admit rational (as opposed to holomorphic) sections associated to $\U(1)_\text{KK}, \U(1)$ and consequently the structure of the pushforwards of quadruple intersection numbers involving the divisors $\hat D_0, \hat D_1$ are not known in full generality as is the case in models with a single holomorphic zero section. For example, in these cases
    \begin{align}
        W_{000 \gamma} \ne K^2 \cdot D_\gamma,~~~~ W_{0000} \ne K^3
    \end{align}
and so on. This unfortunately complicates the computation of
$M_\text{red}$ as our incomplete understanding of intersection
products involving $\hat D_0, \hat D_1$ makes the solutions to the nullspace
equations unclear, and thus at present we are unable to present even a
formal general expression for $M_\text{red}$ for models with $\U(1)$
factors over arbitrary $B$ with arbitrary characteristic
data. Nevertheless, we can follow the procedure to
construct $M_\text{red}$ outlined in
\cref{sec:methodology} for any specific $B$ and $\sfG$, and we find in all examples we have
considered that $M_\text{red}$ is also resolution independent for
models with a $\U(1)$ gauge factor---see \cref{sec:abelianmodel} for some
examples.  It is natural to conjecture that this is generally the case
although a more complete proof is clearly desirable.

\subsection{Dimension of $\Lambda_\text{phys}$}
\label{rank}

One immediate application of the conjectural resolution invariance of
$M_\text{red}$ is for understanding the number of independent F-theory
vertical flux backgrounds and fluxes that can arise in a given
model. After computing $M_\text{red}$, one can impose the symmetry
constraints in order to further restrict the lattice
$H_{2,2}^{\text{vert}}(X,\Z)$ to the sublattice $\Lambda_\text{phys}$
of independent F-theory flux backgrounds; the restriction of the
action of $M_\text{red}$ to $\Lambda_\text{phys}$ can be expressed as
a matrix $M_\text{phys}$.

The number of independent fluxes $\Theta$ subject to the symmetry constraints is equal to $\rk
M_\text{phys}$.
While in principle, as discussed in \cref{sec:linear-anomaly},
$M_\text{phys}$ can have null vectors associated with constraints
characterized by zero-norm non-null elements of $\Lambda_S$, we have
not encountered any situations where this occurs.  Indeed, this is
impossible in purely nonabelian theories since all non-distinctive
flux background parameters $\phi'$ are determined by the constraints as linear
functions of the $\phi''$ as in \cref{eq:solution-abstract}.  We
suspect, but have not proven, that this also does not happen in
theories with $\U(1)$ factors.  When there are no such null vectors of
$M_\text{phys}$, then  we have
    \begin{align}
        \rk M_\text{phys}
= \dim \Lambda_\text{phys}  &= 
\text{\# independent fluxes}\\
\nonumber & =
\rk M_\text{red} - 
\text{\# independent constraints} \,.
    \end{align}
The number of independent constraints is at most the number of basis
elements
$S_{0\alpha}, S_{\alpha \beta}, S_{\alpha
  i}$, i.e., $h^{1,1}(B) +  \frac{1}{2} h^{1,1}(B) (h^{1,1}(B) +1 +
2\rk\sfG)$, but in general can be smaller when there are homologically
trivial cycles $S_{\alpha \beta}, S_{\alpha i}$ as discussed in
\cref{sec:null-nonabelian}. This number must be resolution-independent
and can be identified directly from the structure of $M_\text{red}$
and the geometry of $B$.
%\wati{I'm not sure of the relevance of the subsequent statements and
%  the counting--discuss? I also think we want to disentangle the
%  relevance of M red and the statements about the number of
%  independent fluxes a little more, e.g., other footnote}
%However, we also find in all cases that
%$H_{2,2}^{\text{vert}}(X,\Z)$ at the very least includes the integer
%span of the basis elements $S_{0\alpha}, S_{\alpha \beta}, S_{\alpha
%  i}$. For example, in the case of a purely nonabelian gauge group,
%this point is evident from the rows of the matrix
%\labelcref{Mrednonabelian}. Therefore we may express the number of
%independent F-theory fluxes in a resolution-invariant manner as
%\wati{again, I'm not sure about counting, also shouldn't rk $G$ be
%  multiplied by number of $\alpha$'s with nontrivial intersection with
%$\Sigma$?}
%    \begin{align}
%        \rk M_\text{phys} = \rk M_\text{red} - \left(h^{1,1}(B) +  \frac{1}{2} h^{1,1}(B) (h^{1,1}(B) +1) + \rk\sfG \right) = \text{dim} \, \Lambda_\text{phys},
%    \end{align}
%assuming of course that $h^{1,1}(B), \rk\sfG$ are resolution-invariant
%quantities.

%Aside from the $\SO(11)$ model (which will be
%treated further elsewhere),% in \cite{46})
 All 4D F-theory models
 with generic matter that we have studied 
% in which codimension-three $(4, 6)$ points are absent
 have the property that the rank of $M_\text{phys}$ is greater than or equal to the number of independent realized families of
chiral matter multiplicities:
    \begin{equation}
    \label{eq:conjecture}
\rk M_\text{phys} \geq \text{\# of families of 4D chiral matter multiplets realized in F-theory.}
    \end{equation}
The fact that $\text{rk}\, M_{\text{phys}}$ is at least equal to the
number of independent chiral matter multiplicities seems to follow
from the assumption that all matter surfaces $S_{\sfr}$ have a
non-trivial vertical component. 
%On the other hand, the fact that $\text{rk} \, M_{\text{phys}}$ is no larger than the number of independent chiral matter multiplicities suggests that every independent vertical flux $\Theta_{ij}$ corresponds to some family of charged chiral matter. 
We have furthermore found in all of these cases that the number of independent families of chiral matter multiplicities realized in F-theory matches the number of independent families satisfying 4D anomaly cancellation. We
know of no natural geometric reason why this should always
be true; the observation that
in all cases considered this holds can be thought of as a statement
regarding the absence of swampland type models in which entire
families of anomaly-free 4D supergravity theories would lack an
F-theory realization.

If it is indeed true that $M_{\text{red}}$ is resolution-independent and moreover that the number of independent families of chiral matter is bounded above
by the rank of $M_\text{phys}$, computing $M_\text{red}$
may serve as an efficient strategy for scanning the F-theory landscape
for vacua that impose stronger constraints than 4D anomaly
cancellation without requiring the additional step of identifying the
matter surfaces $S_{\sfr}$.

\section{Computing chiral indices}
\label{3Dcompare}
In \cref{sec:4dflux} and \cref{sec:constraints-homology} we gave a
prescription for computing the lattice of vertical F-theory
flux backgrounds for $\sfG$ models with gauge group $\sfG = (
\sfG_\text{na} \times \U(1))/\Gamma$ and chiral matter transforming in
representation $\oplus \sfr^{\oplus n_{\sfr}}$. Here, we review a
method to compute the multiplicities
    \begin{align}
        \chi_{\sfr} =\int_{S_{\sfr}} G = n_{\sfr} - n_{\sfr^*}
    \end{align}
of the 4D chiral matter representations $\sfr$ in terms of the fluxes $\Theta''$, without explicit knowledge of the matter surface $S_{\sfr}$.

\cref{3DCS} reviews the relationship \cite{Grimm:2011fx} (see also \cite{Grimm:2011sk,Cvetic:2012xn}) between Chern--Simons couplings appearing in the low-energy effective 3D $\cN=2$ supergravity action describing M-theory compactified on a CY fourfold in a nontrivial flux background $G$, and the vertical fluxes $\int_{S_{IJ}} G =\Theta_{IJ}$. In \cref{sec:CS}, we explain how to compute the chiral indices by solving the linear system obtained by matching the vertical fluxes to one loop exact field theoretic expressions for CS couplings appearing in the 3D $\cN=2$ supergravity action, using a similar strategy to that used in \cite{Cveti__2014}.

\subsection{3D Chern--Simons terms and M-theory fluxes}
\label{3DCS}

The key step in our analysis that enables us to determine the chiral indices $\chi_{\sfr}$  in terms of vertical fluxes without explicit knowledge of the matter surfaces $S_{\sfr}$ is the identification \cite{Grimm:2011sk,Grimm:2011fx}
    \begin{align}
    \label{identification}
        \Theta_{\bar{I}\bar{J}} =  -\Theta_{\bar{I}\bar{J}}^{\text{3D}}\,, \quad \bar{I} = \bar{0}, i\,.
    \end{align}
On the right hand side of the above equation, $\Theta_{\bar{I}\bar{J}}^{\text{3D}}$ are Chern--Simons (CS) couplings that characterize the 3D effective action describing M-theory compactified on $X$ at low energies (recall that the index $\bar I =\bar{0}$ denotes the KK $U(1)$, see \cref{physbasis}).

The identification \labelcref{identification} holds for all M-theory
compactifications on CY fourfolds $X$ with nontrivial flux backgrounds
$G$, and follows from the dimensional reduction of 11D supergravity on
$X$.  In the special case that $X$ is a resolution of a
singular elliptic CY fourfold, M-theory/F-theory duality implies that
the low-energy effective 3D theory is a Kaluza--Klein (KK) theory equivalent
to a circle compactification of the 4D $\cN=1$ supergravity
theory describing a flux compactification of F-theory on the singular
fourfold. Because of this duality, the one-loop exact quantum dynamics
on the F-theory Coulomb branch gets mapped to the classical dynamics of
M-theory; in particular, this means that the contributions of massive
fermions on the F-theory Coulomb branch are captured by classical CS
couplings $\Theta_{\bar I \bar J}^{\text{3D}}$.

Concretely, given a collection of real Coulomb branch moduli $\varphi$ corresponding to the holonomies of Cartan $\U(1)$ gauge fields around the KK circle, the F-theory Coulomb branch is characterized by a collection of massive BPS hyperinos, with masses given by
    \begin{equation}
        \label{eq:BPSmass}
        m_\text{hyp} = n m_\text{KK} + \varphi \cdot w\,, \quad n \in \Z\,, \quad \varphi \cdot w = \varphi^i w_i\,, \quad i = 1, \dots, \rk \sfG\,,
    \end{equation}
where $w_i$ may be regarded as the Dynkin coefficients of a weight in a basis of fundamental weights, associated with the charges (under $\U(1)^{\text{rk} \,\mathsf G}$) of each hyperino on the Coulomb branch. In terms of the Cartan charges $(n, w_i)$ above, the one-loop exact CS couplings are given by \cite{Cvetic:2012xn}
    \begin{equation}
        \label{eq:qIJ1loop}
        \begin{aligned}
            \Theta_{i j}^\text{3D} &=   \sum_{w} ( \tfrac{1}{2} + \floor{|r_\text{KK}\varphi \cdot w| })  \,\sign(\varphi \cdot w)w_{i} w_{j}\,, \\
            \Theta_{ \bar{0}i}^\text{3D}  &=\sum_{w}  (  \tfrac{1}{12} +\tfrac{1}{2} \floor{|r_\text{KK} \varphi \cdot w|} ( \floor{|r_\text{KK}\varphi \cdot w| } + 1) ) w_{i}\,, \\
            \Theta_{ \bar{0} \bar  0}^\text{3D}  &=\sum_{w}  \tfrac{1}{6} \floor{|r_\text{KK}\varphi \cdot w| } (\floor{|r_\text{KK}\varphi \cdot w|} + 1) (2 \floor{|r_\text{KK}\varphi \cdot w|} + 1)\,,
        \end{aligned}
    \end{equation}
where $r_\text{KK} := 1/m_\text{KK}$ is the KK radius.

The sign and floor functions in the above expressions encode the dependence of the CS couplings on the phase of the vector multiplet moduli space parametrized by the Coulomb branch moduli $\varphi_i$ and KK modulus $m_\text{KK}$; we return to the issue of explicitly evaluating these functions shortly. For now, we point out that the CS couplings can be expressed as linear combinations of the chiral indices $\chi_{\sfr}$ by making the replacement $\Sigma_w \rightarrow \sum_{\sfr}\mult{\sfr} \sum_{w \in \sfr}$ (where $n_{\sfr}$ is the multiplicity of each type of representation $\sfr$ appearing in the 4D spectrum and we only sum over each distinct representation $\sfr$ once) and using the fact that the summands are odd under $\sfr \to \sfr^*$. Combining \cref{identification} and \cref{eq:qIJ1loop}, we may thus write
    \begin{equation}
        \Theta_{\bar{0} \bar{0}} = x_{\bar{0} \bar{0}}^{\sfr} \chi_{\sfr}\,, \quad \Theta_{\bar{0}i} = x_{\bar{0}i}^{\sfr} \chi_{\sfr}\,, \quad \Theta_{ij} = x_{ij}^{\sfr} \chi_{\sfr}\,,
    \end{equation}
and under our assumption that all matter surfaces have components in $S_{IJ}$\footnote{See \cref{sec:puzzle} for a possible counterexample to this assumption.} it is possible to invert the
coefficients $x^{\sfr}_{ij}$ so that
    \begin{equation}
        \chi_{\sfr} =x_{\sfr}^{ij} \Theta_{ij} =x_{\sfr}^{ij} S_{Cij} \cdot S_{kl} \phi^{kl} = S_{\sfr } \cdot \phi\,,
    \end{equation}
   where on the right hand side of the above equation we have used the fact that the matter surfaces are given by 
   	\begin{equation}
		S_{\sfr} = x_{\sfr}^{ij} S_{Cij}
	\end{equation}
and $S_{Cij}$ are defined in \cref{eqn:SC}.

\subsection{Computing 3D Chern--Simons terms using triple intersection numbers}
\label{sec:CS}
The explicit expressions for $\Theta_{\bar I \bar J}^{\text{3D}}$ given in the previous subsection depend on the values of the moduli-dependent functions $\text{sign}(\varphi \cdot w)$ and $\floor{|r_\text{KK} \varphi \cdot w |}$ as input. These functions partially  characterize the field theoretic regime (i.e., the ``phase'') of the F-theory Coulomb branch described by the 3D KK theory, or in geometric terms the regime of the K\"ahler moduli space to which the resolution $X$ corresponds. The Coulomb branch phase can in principle be computed geometrically by studying the fibers of $X$ over the codimension-two components of the discriminant locus in the base $B$, which carry local matter transforming in the representation $\sfr = \sfr \oplus \sfr^* $.

Unfortunately, this procedure is often delicate and sometimes
difficult to carry out systematically, so we instead use an
alternative approach that relies on the assumption that the
hypermultiplet representations characterizing the gauge sector of a 6D
supergravity theory can be recovered from a 5D KK theory, at least for
representations $\sfr$ that correspond to local matter in the
F-theory geometry. In particular, we exploit the fact that the matter
representations are encoded in codimension-two components of the
discriminant locus in the base $B^{(2)}$ of an elliptic CY threefold
to extract the phase of the Coulomb branch from the triple
intersections of Cartan divisors $\hat D_i$. Closely following the strategy described in \cite{Cveti__2014}, we now explain in detail
how to use this trick to compute the sign and floor functions
appearing in the field theoretic expressions for the 3D CS couplings
in the previous subsection.

Recall that in the case of M-theory compactified on an elliptic CY
threefold $X^{(3)}$, M-theory/F-theory duality (similar to the case of
a CY fourfold) identifies the triple intersection numbers with
one-loop quantum corrected CS couplings in 5D,
    \begin{align}
    \label{5Dmatch}
         \hat D_{\bar I} \cdot \hat D_{\bar J} \cdot \hat D_{\bar K} = k^{\text{5D}}_{\bar I \bar J \bar K},~~~~ \bar I =\bar{0}, i\,,
    \end{align}
where field theoretic expressions analogous to \cref{eq:qIJ1loop} have also been worked out for the 5D one-loop CS couplings \cite{Grimm:2013oga} (see also \cite{Witten:1996qb,Intriligator:1997pq,Bonetti:2011mw}):
\begin{align}
\begin{split}
\label{5DCS}
    k_{ijk}^\text{5D} &=  -\sum_{w} (  \floor{|r_\text{KK} \varphi \cdot w |} +\tfrac{1}{2}) \,  \sign(\varphi \cdot w) w_i w_j w_k \\
    k_{\bar{0}ij}^\text{5D} &=-\sum_{w} (  \tfrac{1}{12} +\tfrac{1}{2} \floor{|r_\text{KK} \varphi \cdot w|} ( \floor{|r\varphi \cdot w| } + 1) ) w_{i}w_j \\
    k_{\bar{0} \bar{0}i}^\text{5D}&=-\sum_{w}  \tfrac{1}{6} \floor{|r_\text{KK}\varphi \cdot w| } (\floor{|r_\text{KK}\varphi \cdot w|} + 1) (2 \floor{|r_\text{KK}\varphi \cdot w|} + 1) \,\sign(\varphi \cdot w) w_i\,.
\end{split}
\end{align}
    Importantly, the sign and floor functions appearing in \labelcref{5DCS} are the same as those in \labelcref{eq:qIJ1loop}, which means they can equally well be determined from the 5D CS couplings provided the 5D and 3D CS couplings correspond to the same Coulomb branch phase in an appropriate sense.

It turns out to be possible to determine the 5D CS terms from the types of (partial) resolutions $X$ we consider, as the sequences of blowups we use to obtain $X$ for a given $\mathsf G$ model defined over a threefold base $B$ can also be used to obtain resolutions $X^{(3)}$ of the same $\mathsf G$ model defined over a twofold base $B^{(2)}$.\footnote{Recall that in our case we only consider resolutions of singularities through codimension-two sub-loci of the discriminant locus in $B$. This is actually true for bases of arbitrary dimension, $B^{(d)}$, so long as the sequence of blowups used to obtain a (partial) resolution is formally identical through codimension two. In the case of a twofold (i.e. $d=2$), this implies that the resulting threefold $X^{(3)}$ is a genuine resolution.} Consequently, for a given $\mathsf G$ model and a common sequence of blowups resolving singularities through codimension two, the triple intersection numbers of $X^{(3)}$ are closely related to the quadruple
intersection numbers $W_{\bar I \bar J \bar K \alpha}$
of $X$. More precisely, the pushforwards $W_{\bar I \bar J \bar K}$ are
formally identical to the pushforwards of the triple intersection
numbers of $X^{(3)}$ to the base, with the key difference
that the pushforwards $W_{\bar I \bar J \bar K}$ are ``promoted'' from numbers to classes
of curves in the threefold base $B$. In the fourfold case, one then
computes quadruple intersection numbers involving three divisors
carrying nonabelian Cartan indices by computing the intersections of
these classes with other divisors in the base, i.e., $W_{\bar I
  \bar J \bar K} \cdot D_\alpha$. One can use this fact to
make the formal identification
    \begin{align}
    \label{5Didentification}
        k^{\text{5D}}_{\bar I \bar J \bar K} \rightarrow W_{\bar I \bar J \bar K}
    \end{align}
provided we replace specific coefficients in the sums \labelcref{5DCS}
with the intersection products of classes of certain curves in $B$.
If, as in the 4D case, we organize the expressions for the CS
couplings in \labelcref{5DCS} into sums over representations by making
the replacement $\sum_{w} \to \sum_{\sfR} n_{\sfR} \sum_{w
  \in \sfR}$ (where $n_{\sfR}$ is the multiplicity of
hypermultiplets in the 6D spectrum and we
only sum over each distinct quaternionic representation $\sfr$ once), then we simply need to promote
$n_{\sfr}$ to the classes of matter curves
$C_{\sfr}$ (matter curves are discussed in
\cref{sec:matter-multiplicities}; see also \cref{intersection} for an explicit description of how $C_{\sfr}$\footnote{Note that $C_{\sfr}$ are known for large classes of singular F-theory models \cite{Grassi:2011hq} and (in contrast to $S_{\sfr}$) are manifestly resolution-independent.} appear in the expressions for $W_{i_sj_tk_u}$.)\footnote{The fact that the pushforward technology used to
  evaluate the intersection numbers does not rely explicitly on the
  dimension of $B$ is a key part of what makes it such an efficient computational tool for this purpose.}

Alternatively, we could rephrase this discussion as indicating that
the formal expressions $W_{\bar I \bar J \bar K}$
should match the triple intersection numbers that arise when the threefold base
$B$ is instead ``demoted'' to a twofold $B^{(2)}$. Either way, the
upshot is that the sign and floor functions are captured by the terms
$W_{\bar I \bar J \bar K}$, as is made clear by the
matching \labelcref{5Didentification}. Since the linear system
\labelcref{5Didentification} does not involve any undetermined
parameters, the system can be solved explicitly for the values of
$\text{sign}(\varphi \cdot w)$ and $\floor{|r_\text{KK} \varphi \cdot
  w|}$. Thus, we find that matching triple intersections with the low
energy effective 5D physics of M-theory compactified on an elliptic CY
threefold $X^{(3)}$ allows us to circumvent the task of computing the
sign and floor functions directly from geometry, and we may
subsequently use these values as input for the 3D case.

We illustrate this procedure for the $\SU(2)$ model in \cref{sec:su2}.

\section{Models with simple gauge group}
\label{sec:exampleADE}

We apply our systematic approach for analyzing flux backgrounds
described in the previous sections in several examples of models with
simple nonabelian gauge groups, $\sfG = \sfG_\text{na}$. In
\cref{simpleADE}, we explain why the only simple $\sfG$ models with
generic matter admitting nontrivial chiral multiplicities are the
simply-laced groups $\sfG_\text{na} = \SU(N), \SO(4 k + 2), \gE_6$,
with $N \ge 5, k \ge 2$. \cref{simplefluxes} describes the common
features of the F-theory fluxes $\Theta''$ for these models; the full
set of results can be found in \cref{tab:fluxtable}. We turn our attention
to specific examples in \cref{sec:su2,sec:su5,su6,so10,e6}.

\subsection{Chiral matter for simply-laced gauge groups}
\label{simpleADE}

The groups $\SU(N), N \ge 5$, $\SO(4 k + 2), k
\ge 2$, and $\gE_6$ are precisely the compact simple Lie groups for
which we expect a one-dimensional family of anomaly-consistent chiral matter
spectra with generic matter in 4D; all other compact simple Lie groups have no
chiral solutions to the anomaly cancellation conditions with only generic
matter representations. For reference, generic matter in these models includes the following complex representations:
    \begin{itemize}
        \item{} $\SU(N)$: fundamental and two-index antisymmetric;
        \item{} $\SO(4k+2)$: spinor;
        \item{} $\text{E}_6$: fundamental.
    \end{itemize}

To see that these are the only gauge groups admitting chiral generic matter, note first that the set of generic matter
for a simple gauge group comprises three representations if the group has an
independent quartic Casimir and two representations otherwise (this can be
seen in the 6D context as coming from the fact that the anomaly cancellation
conditions depend on quadratic and quartic invariants of the gauge group). One
of these representations is always the adjoint, which is self-conjugate, and
thus there are at most two representations that can contribute chirally to the
spectrum in any case. For groups with an independent cubic Casimir, the number
of independent chiralities is further reduced by one by the 4D anomaly
cancellation equations. One can then carry out a case-by-case analysis of the
compact simple Lie groups to determine the number of independent chiralities
in each case.

For $\SU(N), N \ge 5$, there is an independent quartic Casimir,
giving two complex generic representations, and an independent cubic Casimir,
reducing the number of independent chiral families to one. For $\SU(2)$, every
representation is self-conjugate; for $\SU(3)$, there is an independent cubic
Casimir but no independent quartic Casimir; and for $\SU(4)$, there is an
independent cubic Casimir and the two-index antisymmetric representation is
self-conjugate; thus, in all these cases, there are no chiral solutions. The
group $\SO(N)$ only has complex representations for $N = 4 k + 2$, with only
the spinor being complex among generic matter representations, and has no
independent cubic Casimir, thus having a one-dimensional family of generic
chiral spectra for these $N$.	 None of the exceptional groups
has an independent quartic Casimir, giving only one generic representation
other than the adjoint, and of these, only $E_6$ has complex representations;
the $E_6$ fundamental is complex, and $E_6$ has no independent cubic Casimir,
leaving a one-dimensional family of generic chiral spectra.

Thus, we expect a
one-dimensional family of chiral solutions for the simple gauge groups
$\SU(N), N \ge 5$, $\SO(4 k + 2), k \ge 2$, and $\gE_6$, and no chiral
solutions for all other compact simple Lie groups.

\subsection{Summary of F-theory fluxes for simple nonabelian models}
\label{simplefluxes}

\subsubsection{Fluxes in universal (simple) $\sfG$ models}
\label{sec:fluxes-universal}

Universal $\sfG$ models with simple nonabelian gauge symmetry can be described in
F-theory using Tate models, i.e., Weierstrass models presented in Tate
form \cite{BershadskyEtAlSingularities}
    \begin{align}
        y^2 z + a_1 xyz + a_3 y z^2 - (x^3 + a_2 x^2 z + a_4 x z^2 + a_6 z^3 ) = 0
    \end{align}
with a choice of tuning
    \begin{align}
        a_n = a_{n,m_n} \sigma^{m_n}\,,
    \end{align}
which characterizes the sections $a_n$ of
the $n$th tensor power of the anticanonical bundle of
the base $B$ in the vicinity of the gauge divisor $\sigma = 0$. Note
that the divisor class of $\sigma = 0$ is $[\sigma] = \Sigma$ and
hence the divisor classes of the tuned sections are
    \begin{align}
        [a_{n,m_n}] = n(-K) -m_n \Sigma\,.
    \end{align}
In all nontrivial cases that we study without codimension-three $(4, 6)$ singularities, and excluding the case $\sfG = \SO(11)$, we use the methods of \cref{pushapp,sec:resolutions} to show there is a one-dimensional family of independent F-theory fluxes preserving the 4D gauge group $\sfG_\text{na}$ that take the form
    \begin{align}
        \Theta_\text{phys} = \phi \Sigma \cdot  [p(a_{n,s_{n}})] \cdot [p'(a_{n,s_{n}})], ~~~~ \phi \in \Z\,,
    \end{align}
where in the above formula the bracketed expressions are the classes of polynomials $p$ of the sections $a_{n,s_{n}}$ that depend on the choice of gauge group.
%   \begin{align}
%       D= [p(a_{n,s_{n}})], ~~~~ D' = [p'(a_{n,s_{n}})].
%   \end{align}
See \cref{tab:fluxtable} for results for various groups $\sfG_\text{na}$.

The physical significance and interpretation of the above results is perhaps more transparent in the lattice $M_C\Lambda_C$ of symmetry-constrained fluxes. For example, consistent with the
argument in the previous subsection, that each model admits at most a one-parameter family of chiral
multiplicities, we find in each case we study that the non-trivial fluxes $\Theta_{ij} \in M_C \Lambda_C$ can be expressed as
	\begin{equation}
        \label{eq:general-indices}
		\Theta_{ij}=M_{C(ij)(kl)}\phi^{kl}  \propto \frac{\ell(\phi^{kl})}{\det\kappa} \Sigma \cdot [p(a_{n,s_{n}})] \cdot [p'(a_{n,s_{n}})] =
                \chi_{\sfr}\,,
	\end{equation}
for some complex representation $\sfr$ and where
$\ell(\phi^{kl})$
is a linear combination of the parameters $\phi^{kl}$ whose precise
form depends on $\sfG$;
here, since $\sfG$ is simply-laced, $\kappa_{ij} = - W_{i|j}$ is the Cartan
matrix for $\sfG$. Moreover, since $\text{rk}\, M_C =1$ and $M_C^{\text{t}} = M_C$, the coefficients of the parameters $\phi^{ij}$ in the linear expressions $\ell(\phi^{ij})$ are identical to the proportionality constants relating
different nontrivial $\Theta_{ij}$, it follows that straightforwardly that
	\begin{equation}
	\label{selfdualG}
		\Theta_{kl} \phi^{kl} = \frac{\ell(\phi^{ij})^2}{\det\kappa} \Sigma \cdot [p(a_{n,s_{n}})] \cdot [p'(a_{n,s_{n}})]= \int_X G \wedge G\,,
	\end{equation}
which is non-negative provided $\Sigma \cdot [p] \cdot [p'] \geq 0$ 
%as required for the consistency of the Weierstrass model
\footnote{In order
  for Tate models to be consistent and not give rise to a larger gauge
  algebra, it is necessary that the line bundles to which the
  $a_{n,s_n}$ appearing in the table
are associated have non-empty spaces of sections. Mathematically, this
can be expressed as the requirement that the divisor classes
$[a_{n,s_n}]$ are effective; the resulting intersection products are
negative as long as these divisors are non-rigid.}; this is consistent with the assertion that $G$ is self-dual \cite{Gukov:1999ya}, which in turn implies $\int G \wedge G = \int G \wedge *G = \int |G|^2 \geq 0$; see e.g. \cref{SU5square} for a specific example of \cref{selfdualG} in the context of the $\SU(5)$ model. As we explain in \cref{integrality}, $\ell(\phi^{ij}) / \det\, \kappa$ is an integer and hence $\Theta_{ij}$ are manifestly integer-valued. 
%While it is challenging to fully specify the quantization conditions
%constraining the coefficients $\phi^{ij}$ in general, as noted in
%\cref{sec:quantization-1} we can combine the structure of the
%intersection pairing of surfaces in the fourfold with the symmetry
%constraints \labelcref{eq:Poincare}, \labelcref{eq:gauge} to obtain a
%necessary quantization condition for $\phi^{ij}$, assuming that the
%flux $G$ is an integer cohomology class (i.e., $\phi^{IJ} \in \Z$).

%With the exception of the $\SO(11)$ model, in models that do not admit nontrivial chiral matter multiplicities
%and that are also free of codimension-three $(4, 6)$ singularities we
%expect that $M_C = 0$, i.e., that $\Lambda_C$ is completely spanned by null
%vectors of $M$; this would imply that
%$\Theta_{ij} = 0$ for all $ij$. This expectation is validated in a
%number of examples: for instance, we have confirmed that the
%$\SU(N<5), \SU(6)^\circ, \SO(12), \text{Sp}(6), E_7, E_8, \text{F}_4$, and $\text{G}_2$ models do not admit any nontrivial F-theory fluxes
%  preserving the gauge symmetry. This property seems to be true
%  generally, consistent with the expectation that the sublattice
%  $\Lambda_\text{phys}$ should be
%  empty for 4D models not admitting chiral matter.
%As discussed in
%  \cref{summary} a striking exception that remains to be
%  satisfactorily understood is the $\SO(11)$ model.

\subsubsection{Integrality conditions for symmetry preserving flux backgrounds}
\label{integrality}

Before proceeding to examples, let us justify the integrality
condition $\ell(\phi^{ij})/ \det\kappa \in \Z$, which ensures
that our expression \labelcref{eq:omegabar} for $M_{C(IJ)(KL)}$ leads to
integer fluxes $\Theta_{ij}$ in \labelcref{eq:general-indices}. This
integrality condition
is of course guaranteed, provided that the symmetry constraints
$\Theta_{I\alpha} =0$ are solved over $\Z$ (assuming $\phi^{IJ} \in \Z$), since $M$
is an integer matrix, but nevertheless for the
sake of clarity we spell out explicitly how the integrality condition
propagates through to the final expressions in the case where the
constraints are imposed first and there is an additional quantization
condition on the domain of the mapping $C$ defined in \cref{eq:c}.
This integrality condition is an explicit example of the type of
quantization constraint discussed earlier in \cref{sec:quantization-1}
and \cref{constraintsolutions}.

To see how this works, we use the symmetry constraints to derive a condition on the combination of intersection numbers and parameters in the numerator of \labelcref{eq:general-indices}. First notice that the local Lorentz
symmetry constraint $\Theta_{\alpha \beta} =0$ implies (see
\labelcref{newconstraints1,newconstraints2})
$\phi^{0\gamma} =- \phi^{00} K^\gamma + \phi^{ij} \kappa_{ij}
\Sigma^\gamma$ and $\phi^{\beta \gamma} = - \phi^{ij} \kappa_{ij}
K^\beta\Sigma^\gamma $. Since these expressions are polynomial in the
distinctive parameters, it is evident that solving the
constraints does not impose any conditions on
$\phi^{ij}$. We thus turn our attention to the gauge symmetry
constraints $\Theta_{\alpha i} = 0$, which imply $\phi^{\beta j}
\kappa_{ij} \Sigma^\gamma = - \phi^{jk} \Delta^{\beta}_{\sfR}
\rho^{\sfR}_{ijk} \Sigma^\gamma$ and for nonzero
$\Sigma^\gamma$ further imply
    \begin{align}
        \phi^{\beta j}\kappa_{ij}= - \phi^{jk} \Delta_{\sfR}^\beta \rho_{ijk}^{\sfR}\,,
    \end{align}
where we note that $\rho_{ijk}^{\sfr}$ in the above equation is
defined by the intersection numbers $W_{ijk} \cdot D_\alpha =
\rho_{ijk}^{\sfr} C_{\sfr} \cdot D_\alpha$; see
\cref{3Cnumbers}. Since $\phi^{\beta j}$ is assumed to be an
integral lattice vector for every $\beta$, the right-hand side of the
above equation must lie in the root lattice of the group $\mathsf
G$. Comparing the above equation to the list of necessary
and sufficient conditions in \cref{tab:root-lattice} of \cref{sec:root-lattice} for a lattice vector to lie in the root lattice of a simple Lie group,
%For simple and simply-laced groups $\mathsf G = , a sufficient and necessary condition for a vector $v^i$ to lie in the root lattice can be expressed as a single condition $c^i v_i \equiv 0 \mod \det[[\kappa_{ij}]]$ for an appropriate choice of integers $c^i$.\andrew{This is not true, as $\SO(4 k + 2)$ requires two conditions, and is also invalid for multiply laced groups as the determinant should be of the Cartan matrix (equivalently, it is the order of the center of the group. [P]} This condition is clearly satisfied for the left-hand side of
we obtain for each universal (simple and simply-laced) $\mathsf G$ model an integrality condition of the form
    \begin{align}
        \frac{\Delta_{\sfR}^\beta \rho_{ijk}^{\sfR} c^i\phi^{jk} }{\det\kappa } \in \Z\,,
    \end{align}
where the choice of coefficients $c^i$ depends on $\sfG$. In all cases we study, we find that $\Delta_{\sfR}^\beta \rho_{ijk}^{\sfR} c^i\phi^{jk} \equiv D^\beta \ell(\phi^{ij}) \mod \det\kappa$, and hence for generic coefficients $\Delta_{\sfR}^{\beta} \rho^{\sfR}_{ijk}$ the above condition reduces to
    \begin{align}
%merge conflict location
%       \frac{\ell(\phi^{ij})}{\det[[\kappa_{ij}]]} \in
%                \Z \,.
        \frac{\ell(\phi^{ij})}{\det\kappa}  \in
      \,         \Z \,.
    \end{align}
The above integrality condition must be satisfied for any allowed
set of integer fluxes $\phi^{ij}$ that preserve 4D local Lorentz and gauge
symmetry, guaranteeing that all chiral matter indices
(\ref{eq:general-indices}) automatically take integer values. This is demonstrated explicitly in the case of $\sfG = \SU(5)$ in \cref{sec:su5}.

This analysis in general gives a sufficient condition, for each of the
$\mathsf G$ models studied here, for the chiral matter spectrum to have certain
multiplicities.  As discussed in \cref{sec:quantization-1}, however,
inclusion of fluxes in $H_4^{\text{hor}}(X,\Z) \oplus H_{2,2}^\text{rem} (X,\Z)$ may permit a
broader set of possible chiral multiplicities.

\begin{table}
    \centering
        \scalebox{.94}{$
            \begin{array}{|c|c|c|c|c|}
                \hline
                \substack{\text{Kodaira}\\\text{fiber}}& \sfG & \Delta^{(2)} &\Theta_\text{phys}\, (\textcolor{red}{\Theta_{(4,6)}}) & \substack{\text{geometric}\\\text{constraints}}\\\hline
                %\text{I}_2\,,\,\text{III}  & \mathfrak{su}(2) & & &\text{---}  &\text{---}\\\hline
                %& \mathfrak{su}(2) \\\hline
                %\text{I}_3^{\text{s}}\,,\,\text{IV}^{\text{s}}& \mathfrak{su}(3) &&&\text{---} & \text{---}\\\hline
                %\text{IV}^{\text{s}}& \mathfrak{su}(3) \\\hline
                %\text{I}_4^{\text{s}} & \mathfrak{su}(4) &&& \text{---} & \ text{---} \\\hline
                \textcolor{blue}{\text{I}_{5}^{\text{s}}}& \textcolor{blue}{\SU(5)} & \begin{array}{l}a_1^4(a_{6,5} a_1^2\\-a_{3,2} a_{4,3} a_1
                \\+a_{2,1} a_{3,2}^2)\end{array}&\phi \Sigma \cdot [a_1] \cdot [a_{6,5}]  & \chi_{\textbf{5}} + \chi_{\textbf{10}} =0  \\\hline
                \textcolor{blue}{\text{I}_{6}^{\text{s}}}& \textcolor{blue}{\SU(6)}& \begin{array}{l} a_1^4 (a_1^2 a_{6,6}\\-a_1 a_{3,3} a_{4,3}\\-a_{4,3}^2) \end{array} &\phi \Sigma \cdot [a_1] \cdot [a_{4,3}] & \chi_{\textbf{6}} + 2 \chi_{\textbf{15}} =0 \\\hline
                \textcolor{red}{\text{I}_{6}^{\text{s}}}& \textcolor{red}{\SU(6)^\circ} &\begin{array}{l} a_1^3 (a_{6,6} a_1^3\\-a_{3,2} a_{4,4} a_1^2\\+a_{2,2} a_{3,2}^2 a_1\\-a_{3,2}^3)\end{array}& (\textcolor{red}{\phi \Sigma \cdot [a_1] \cdot [a_{3,2}^3]}) & \chi_{\textbf{6}} =0\\\hline
                \textcolor{blue}{\text{I}_{7}^{\text{s}}}& \textcolor{blue}{\SU(7)} & \begin{array}{l} a_1^4 (a_1^2 a_{6,7}\\-a_1 a_{3,3} a_{4,4}\\+a_{2,1} a_{3,3}^2)\end{array}& \begin{array}{c}\phi \Sigma \cdot [a_1] \cdot [a_{6,7}] \\ (\textcolor{red}{\phi' \Sigma \cdot [a_1] \cdot [a_{2,1}]} )\end{array}& \chi_{\textbf{7}} + 3 \chi_{\textbf{21}} = 0\\\hline
                \text{I}_{6}^{\text{ns}} & \text{Sp}(6)
                &\begin{array}{l} a_2^2 (a_2 a_{3,3}^2\\-a_{4,3}^2\\+4
                   a_2 a_{6,6})\end{array} &
\textcolor{red}{(?)}&\text{---}\\\hline
            %   \text{I}^{\text{ns}} & \mathfrak{sp}(4)&&&\text{---}& \text{---}\\\hline
            %   \text{I}^{\text{ns}}& \mathfrak{sp}(5) &&&\text{---}& \text{---} \\\hline
            %   \text{I}_0^{*\text{ss}}& \mathfrak{so}(7)&&&\text{---} & \text{---} \\\hline
            %   \text{I}_0^{*\text{s}} & \mathfrak{so}(8) &&& \text{---} & \text{---}\\\hline
            %   \text{I}_1^{* \text{ns}} & \mathfrak{so}(9)&&&\text{---}&\text{---} \\\hline
                \textcolor{blue}{\text{I}_1^{* \text{s}}} &\textcolor{blue}{ \SO(10)}&a_{2,1}^3 a_{3,2}^2 & \phi \Sigma \cdot [a_{2,1}] \cdot [a_{6,5}]    &\text{any $\chi_{\textbf{16}}$} \\\hline
                \text{I}_2^{* \text{ns}} & \SO(11) &\begin{array}{l} a_{2,1}^2 (4 a_{2,1} a_{6,5}\\-a_{4,3}^2)\end{array} &\phi \Sigma \cdot [a_{2,1}] \cdot [a_{6,5}] &\text{---}\\\hline
                \textcolor{red}{\text{I}_2^{* \text{s}}} & \textcolor{red}{\SO(12)} &\begin{array}{l} a_{2,1}^2 (4 a_{2,1} a_{6,5}\\-a_{4,3}^2)\end{array}& (\textcolor{red}{\phi \Sigma \cdot [a_{2,1}] \cdot [a_{4,3}^2]}) &\text{---}\\\hline
                %\text{I}_3^{* \text{ns}} & \mathfrak{so}_{13} &a_{2,1}^3 \left(a_{3,3}^2+4 a_{6,6}\right)& \text{?} & \text{?}& a_{2,1}=0\\\hline
            %   \textcolor{blue}{\text{I}_3^{* \text{s}}} & \textcolor{blue}{\mathfrak{so}_{14}} &a_{2,1}^2 \left(4 a_{2,1} a_{6,7}-a_{4,4}^2\right)& ? &?&a_{2,1}=0 \\\hline
                \textcolor{blue}{\text{IV}^{* \text{s}}} & \textcolor{blue}{\text{E}_6} &a_{3,2}^4 &\phi \Sigma \cdot [a_{3,2} ] \cdot [a_{6,5}] &\text{any $\chi_{\textbf{27}}$}\\\hline
                \textcolor{red}{\text{III}^{*}} & \textcolor{red}{\text{E}_7} &a_{3,3}^4& (\textcolor{red}{\phi \Sigma \cdot [a_{4,3}] \cdot [a_{6,5}]}) &\text{---} \\\hline
                %\text{II}^{*} & \mathfrak{e}_8 &a_{6,5}^2&?&? &a_{6,5}=0\\\hline
                \text{IV}^{* \text{ns}} & \text{F}_4 &a_{6,4}^2&
\textcolor{red}{(?)}&\text{---} \\\hline
                \text{I}_{0}^{* \text{ns}} & \text{G}_2 &4 a_{4,2}^3+27 a_{6,3}^2&0 &\text{---}\\\hline
            \end{array}
        $}
    \caption{F-theory fluxes for universal $\sfG$ models with arbitrary
      characteristic data. (Partial) resolutions of these models are taken from \cite{Esole:2017kyr}. The
      final column matches the linear 4D anomaly conditions in all known
      examples. $\Delta^{(2)}$ is the codimension-two component of the
      discriminant locus restricted to the gauge divisor $\sigma
      =0$. Models that
      admit 4D chiral matter are indicated in blue and satisfy
      $\Theta_{\text{phys}} = \chi_{\sfr_*}$ (where
      $\chi_{\sfr_*}$ is the minimal chiral index), while
      models whose fluxes are proportional to some number of
      codimension-three $(4, 6)$ loci are indicated in
      red (note that the F$_4$ and Sp$(n)$ model fluxes appear to correspond to $(4,6)$ points that have not been resolved, and hence for which the explicit form of the flux is still unknown.) The $\SO$-type groups listed above
      range from those of smallest rank that admit nontrivial flux to
      those of largest rank for which the corresponding model does not
      have $(4, 6)$ loci in codimension two; the same is true of the
      $\Sp$-type groups, with the caveat that none have been
      identified that admit nontrivial flux (note that the bases of
      divisors and resolutions for the $\text{I}_{2n}^{\text{ns}}$ and
      $\text{I}_{2n+1}^\text{ns}$ models appear to be identical.) The
      $\text{E}_8$ model is suppressed because it has codimension-two
      $(4, 6)$ singularities for generic characteristic data. $\SU(N)$
      models with $N>6$ contain codimension-three $(4, 6)$ points;
    however for $\SU(6)^\circ, \SU(7)$ the geometric constraints are stated under the restriction that the characteristic data are chosen to ensure
  that  these points are absent, noting that under analogous conditions a
    similar pattern of fluxes may persist for $\SU(N>7)$. Note that
the $\SO(11)$ model flux does not correspond to chiral matter.  
}
\label{tab:fluxtable}
\end{table}

\subsection{$\SU(2)$ model}
 \label{sec:su2}

We now discuss explicit examples.
We begin with a very simple example that has been well studied in the
literature, but which nevertheless illustrates the issue of unimodularity of the
intersection pairing $M_\text{red}$, namely the universal $\SU(2)$
model. (For additional background about $\SU(N)$ models and their
resolutions, see \cref{sec:appendix-sun}.)
For $\mathsf G = \SU(2)$, we find that the reduced
intersection pairing $M_\text{red}$ is resolution-invariant and the
constraints that local Lorentz and $\SU(2)$ symmetry are unbroken in 4D forces
all the flux backgrounds $\phi$ to vanish, so there are no nontrivial
vertical fluxes and no chiral matter.  Identical conclusions follow for
$\mathsf G=\SU(3)$ and $\SU(4)$.

\subsubsection{Absence of chiral matter}

The SU(2) model is characterized by an I$_2$ Kodaira singularity over
the divisor $\Sigma$ and has matter in the representations $\textbf{2} ,
\textbf{3}$ with weights
    \begin{align}
    \begin{split}
        w^{\textbf{2}}_{+} &= (1),~~~~ w^{\textbf{2}}_- = (-1)\\
        w^{\textbf{3}}_{+}&= (2),~~~~w^{\textbf{3}}_0 =(0),~~~~w^{\textbf{3}}_{-}=(-2)\,.
    \end{split}
    \end{align}
The unique resolution $X_1 \rightarrow X_0$ admitting a holomorphic zero section consists of a single blowup. In this case, there is a single $\SU(2)$ Cartan divisor $\hat D_i$ whose nonzero quadruple intersection numbers are (see \cref{pushsu2} and above for details on how to evaluate the pushforwards explicitly)
    \begin{align}
        \begin{split}
        \label{SU2int}
            W_{\alpha \beta ii} &= W_{ii} \cdot D_{\alpha} \cdot D_{\beta} = -2 \Sigma \cdot D_{\alpha} \cdot D_{\beta} \\
            W_{\alpha iii} &=W_{iii} \cdot D_\alpha =  2  \Sigma \cdot (2 K-\Sigma) \cdot D_\alpha\\
            W_{iiii}&=W_{iiii} = 2 \Sigma \cdot ( -4 K^2 + 2 K \cdot \Sigma - \Sigma^2)\,,
        \end{split}
    \end{align}
with the remaining intersection numbers involving $\hat D_i$ vanishing.

The $\SU(2)$ model provides a simple illustration of the procedure,
discussed towards the end of \cref{3Dcompare}, for using low-energy
effective 5D physics as a shortcut to determine the values of the sign
and floor functions appearing in the field theoretic expressions for
the 3D CS couplings. In this case, the floor functions vanish because
the zero section is holomorphic. Furthermore, matching the
pushforwards of the above intersection numbers with the 5D CS
couplings (where the multiplicities $n_{\sfR}$ are replaced by
the matter curves $C_{\sfR}$) fixes the values of the sign
functions to be $\text{sign}(\varphi \cdot w_{\pm{}}^{\sfR}) =
\pm{}1$. In detail,
    \begin{align}
    \begin{split}
        k_{\bar{0}ii} &= - \frac{1}{12}  \sum_{w\in \sfr} C_{\sfr} \sum_{i=\pm{}} (w^{\sfr}_i)^2\\
        &=- \frac{1}{12} \left[\frac{1}{2} \Sigma \cdot (\Sigma + K ) \left( 2^2 + (-2)^2 \right) +   \Sigma \cdot (-8 K - 2 \Sigma ) \left( 1^2 +(-1)^2 \right)  \right]\\
        &=(-2 \Sigma) \cdot (-\frac{1}{2} K)\\
        &=W_{\bar{0}ii}
    \end{split}\\
    \begin{split}
        k_{iii} &=- \frac{1}{2}  \sum_{w\in \sfR} C_{\sfR} \sum_{i=\pm{}} (w^{\sfR}_i)^3 \text{sign}(\varphi \cdot w_i)   \\
        &= - \frac{1}{2}\left[ \frac{1}{2} \Sigma \cdot (\Sigma + K ) \left( 2^3 - (-2)^3 \right) +   \Sigma \cdot (-8 K - 2 \Sigma ) \left( 1^3 - (-1)^3 \right) \right]\\
        &=2  \Sigma \cdot (2 K-\Sigma)\\
        &=W_{iii}\,.
    \end{split}
    \end{align}

The matrix $M$ from this resolution takes
the form \cref{fullmat}, where the $W$ entries with two or more $i$
indices in the bottom right blocks are all even. Note also that the second Chern
class in the basis of \cref{fullmat} is given by $c_2 (X) =(27, [-39K^\alpha], 0, [(c_2
(B))^{\alpha \beta} + 11K^\alpha K^\beta], [7K^\alpha],0)$; since for any any
F-theory base the class $c_2 (B) + K^2$ is even \cite{Collinucci:2010gz}, it follows that the constrained fluxes $\Theta_{I \alpha}$ are
integral even when $c_2 (X)$ is not even.  Thus, we can always remove
the null vectors and impose the constraints by setting $\Theta_{I
  \alpha} = 0$ without worrying about half-integer shifts.

It is straightforward to verify that there is one null vector of the
form $\nu_{C \langle a \rangle}$, so $M_\text{red}$ takes the resolution-independent form
\cref{resindependenthatM} and $M_C = M_\text{phys} = 0$.  It follows
that 4D SU(2) models exist but do not admit chiral matter. This
conclusion is well known for F-theory models with a tuned SU(2) gauge
invariance and is reviewed in \cite{WeigandTASI}.  This result is
consistent with expectations from 4D anomaly cancellation since the
$\textbf{2}$ of $\SU(2)$ is a self-conjugate representation.

\subsubsection{Unimodularity and integrality of the intersection pairing}
\label{sec:2-unimodularity}

We pause briefly to illustrate issues of unimodularity and integrality of the reduced intersection pairing $M_\text{red}$ in the context of the $\SU(2)$ model. Although the lattice $H^{4}(X,\Z)$ is unimodular, the simple example of $\SU(2)$ illustrates that fact the intersection pairing acting on $H^{2,2}_{\text{vert}}(X_1,\Z)$ is generically not unimodular since
the absolute value of the determinant of $M_\text{red}$ is generically greater than one. As an example, for a gauge divisor $\Sigma = H \subset B = \bP^3$ where $H$ is the hyperplane class, we find
    \begin{align}
    \label{SU2P3mat}
    M_\text{red} = \begin{pmatrix} -4 & 1 &0 \\ 1 & 0 & 0 \\ 0 & 0 &-2 \end{pmatrix},~~~~\det M_\text{red}= 2\,.
    \end{align}
While one might imagine that we have
simply chosen the wrong basis for $H^{2, 2}_\text{vert}
(X_1,\Z)$, the story is slightly subtler.

To further analyze the situation, we briefly digress to
a related situation in the case of 6D F-theory compactifications,
focusing in particular on the parallel case where we have a tuned
$\SU(2)$ F-theory model over the base $B = \bP^2$.  In this case, the
triple intersection numbers of the Cartan divisors $\hat D_i$
have the related value $W_{iii}= 2$.  Thus, the intersection
number of the Cartan divisor $\hat D_i$ with the curve $C_{ii}= \hat D_i \cap
\hat D_i$ is 2.  Unlike in the 4D case, the dimension of $H^4
(X^{(3)})$ is equal to that of $H^{2}(X^{(3)})$ by Poincar\'{e}
duality, and again by Poincar\'{e} duality we know that there must be a curve $C$ satisfying $C
\cdot \hat D_i = 1, C \cdot \hat D_{I \ne i} =0$ corresponding to a
(possibly massive) state in the fundamental representation of
$\SU(2)$ (see
\cite{Morrison:2021wuv} for a related discussion).  Thus, in
this situation the curve $C_{ii}$ is not a primitive curve in $H_2
(X^{(3)},\Z)$, but rather $C= C_{ii}/2$ is such a primitive curve
and is Poincar\'{e} dual to $\hat D_i$.

This same story cannot hold, however, in the 4D
SU(2) model.  We do expect that there is a matter surface $S$
associated with matter in  the fundamental representation of
$\SU(2)$. This surface cannot simply be identified with $S_{ii}/2$,
however, since the intersection of that surface with itself under the
matrix \labelcref{SU2P3mat} is $(1/2)\times (-2) \times (1/2) = -1/2$.  Thus,
the Poincar\'{e} dual of the surface $S_{ii}\in H_{2, 2}^\text{vert}
(X_1,\Z)$ is not itself contained entirely in $H_{2, 2}^\text{vert} (X_1,\Z)$
and we see that the orthogonal decomposition \labelcref{ortho} of $H_{4}(X_1,\Z)$
with respect to the intersection pairing does not hold over $\Z$.

As
we discuss below in the context of $\SU(5)$ models with chiral matter, this point
indicates that the assumption $\phi^{IJ} \in \Z$ may be
too restrictive for our analysis to explore all possible chiralities.
%and reveals that we cannot access a broader range of chiralities
%simply by allowing the parameters $\phi^{ij} \in \Q$ as opposed
%to $\phi^{ij} \in \Z$. 
Rather, it appears to be necessary to extend
the analysis to account for contributions from the orthogonal
complement of $H_{2,2}^{\text{vert}}(X)$ in $H_{4}(X)$, which to our
knowledge has yet to be completely understood.

\subsection{$\SU(5)$ model}
\label{sec:su5}

The $\SU(5)$ model (see \cref{sec:appendix-sun}) is the simplest example of a universal SU($N$) model with chiral
matter. The full set of resolutions of the SU(5) model admitting a
holomorphic zero section were worked out in \cite{Esole:2014hya} (see
also \cite{Esole:2011sm,Hayashi:2013lra}) and the chiral indices were
computed for a subset of these resolutions in \cite{Grimm:2011fx} using similar methods to those described in this paper, as well as by other methods in e.g. \cite{Blumenhagen:2009yv,Grimm:2009yu}.

The SU(5) model describes chiral matter in the fundamental
($\textbf{5}$) and two-index antisymmetric (\textbf{10})
representations.  
The 4D chiral anomaly cancellation condition requires that
    \begin{align}
    \label{SU5anomaly}
         \chIndex{\bm{5}} +\chIndex{\bm{10}} = 0\,.
    \end{align}
We begin by analyzing the model with a specific resolution, $X_4
\rightarrow X_0$, which was
described as the toric `phase I' resolution in \cite{Grimm:2011fx} and as
the resolution `$\scB_{1,3}$' in \cite{Esole:2014hya}. The signs
associated to the central charges of BPS particles transforming in the (complex) representations $\textbf{5}, \textbf{10}$
can be found in \cref{SUNsigns}.
In \cref{sec:5-independence} we consider several other
resolutions and show explicitly that $M_\text{red}$ is the same up to
an integral choice of basis for each of these resolutions.

\subsubsection{Chiral matter multiplicities}

Plugging intersection numbers into \labelcref{eq:omegabar} for the
specific resolution just mentioned, we learn that there are four nontrivial
constrained fluxes $\Theta_{ij}$ (Cartan divisors take indices
$i=2,\dots,5$) that satisfy three linear relations, in agreement with
the solution described in \cite{Grimm:2011fx}:
    \begin{align}
    \label{SU5fluxrelations}
        \Theta_{33} = - \Theta_{35}= -\Theta_{44} =\Theta_{45}\,.
    \end{align}
% \footnote{The fact that
%  the coefficients of $\phi^{ij}$ appearing in
%  \labelcref{SU5fluxintegral} are the multiplicative inverses of the
%  coefficients in the linear relations \labelcref{SU5fluxrelations}
%  results from the symmetry of the the intersection pairing, $M_c$,
%  after it is constrained into the subspace of fluxes $\phi$ for which
%  the Poincar\'{e} and gauge symmetry conditions are satisfied---see
%  \labelcref{eq:omegabar}. Given a matrix $M_c$ that is both symmetric
%  and contains null directions, one can easily reproduce the observed
%  relationship between $\Theta_{ij}$ and $\phi^{ij}$.}
In particular (compare also to \cite{Marsano_2011}),
    \begin{align}
    \label{SU5fluxintegral}
        \Theta_{33} = \frac{\ell(\phi^{ij})}{5} \Sigma \cdot [a_1] \cdot [a_{6,5}]= \frac{1}{5}  (\phi^{33} - \phi^{35}-\phi^{44} + \phi^{45} ) \Sigma \cdot K \cdot (6 K + 5 \Sigma) \,.
    \end{align}
Note that the proportionalities in \cref{SU5fluxrelations} imply 
	\begin{align}
	\label{SU5square}
	 \phi^{33} \Theta_{33} + \phi^{35} \Theta_{35} + \phi^{44} \Theta_{44} + \phi^{45} \Theta_{45} = ( \phi^{33} - \phi^{35} - \phi^{44} + \phi^{45} )\Theta_{33} = \ell(\phi^{ij} ) \Theta_{33}\,.
	\end{align}
Comparing \cref{SU5fluxrelations} to the one-loop 3D CS couplings,
we learn that
\begin{equation}
\chIndex{\bm{5}} =-\Theta_{33}, \hspace*{0.1in} \chIndex{\bm{10}} = -\Theta_{44} \,,
\end{equation}
 and thus we recover the 4D anomaly cancellation equation
\labelcref{SU5anomaly}.
As explained in \cref{integrality}, solving the the gauge symmetry
constraints $\Theta_{i\alpha}=0$ over $\Z$ ensures that the flux \labelcref{SU5fluxintegral}
is integer-valued despite the factor of 5 in the denominator; we go through this analysis in some detail here to
illustrate this point. First note that a necessary and
sufficient condition for an
integral vector $v_i$ to lie in the $\mathfrak{su}(5)$ root lattice is (see \cref{sec:root-lattice}) $v_2 + 2 v_3 + 3 v_4 + 4 v_5 \equiv 0 \, (\text{mod}\, 5 )$, which is equivalent to the condition
\begin{equation}
-2
v_2+ v_3+4v_4+2v_5 \in 5 \Z\,.
\label{eq:5-condition}
\end{equation}
We can use this condition
and the logic following (\ref{eq:mq}) to determine a further constraint on the
parameters $\phi^{ij}$ by noting that from (\ref{fluxconstraints4}) and
(\ref{CartanW}), the gauge symmetry constraints imply
\begin{equation}
\Theta_{\alpha i}^{\text{d}} = \phi^{\beta j} (D_{\beta} \cdot D_{\alpha} \cdot
\Sigma) \kappa_{ij} \,,
\end{equation}
where the superscript `d' indicates that $\Theta^{\text{d}}_{\alpha
  i}$ is the part of $\Theta_{\alpha i}$ that only depends explicitly
on the distinctive parameters $\phi^{\hat I \hat J}$. It follows that
$\Theta^{\text{d}}_{\alpha i}$ lies in the tensor product of the
$\mathfrak{su}(5)$ root lattice (for the $i$ index) and the sublattice
of $H_{1, 1}$ spanned by $\Sigma \cap D_\alpha$ (for the $\alpha$
index).  Applying the linear combination from the condition
(\ref{eq:5-condition}) to the fluxes $\Theta_{\alpha i }^{\text{d}}$
we find
    \begin{align}
    \begin{split}
        -2\Theta^{\text{d}}_{\alpha 2} + \Theta^{\text{d}}_{\alpha 3} + 4 \Theta^{\text{d}}_{\alpha 4} + 2 \Theta^{\text{d}}_{\alpha 5}
%       &=\Sigma \cdot  D_{\alpha }\cdot \left(5 \phi^{12} D_{\alpha }-5 \phi^{14} D_{\alpha } +15 K \phi^{22}\right.\\
%       &~~~\left.-10 K \phi^{23} -5 K \phi^{25}+19 K \phi^{33}-10 K \phi^{34} \right.\\
%       &~~~\left.+K \phi^{35}+11 K \phi^{44} -K \phi^{45}+5 K \phi^{55} \right.\\
%       &~~~\left. +10 \Sigma  \phi^{22}-5 \Sigma  \phi^{23}+10 \Sigma  \phi^{33}-5 \Sigma  \phi^{34}\right)\\
        &=-(\Sigma \cdot D_\alpha) \cdot K  (\phi^{33} -\phi^{35} - \phi^{44}+ \phi^{45} ) \,.
    \end{split}
    \end{align}
and thus $K (\phi^{33} - \phi^{35} - \phi^{44} +\phi^{45})$
must lie in $5 H^{1, 1} (B,\Z)$.  This condition, which is necessary
to ensure that the full SU(5) gauge symmetry is preserved, is
sufficient to guarantee that the chiral matter multiplicities
determined by the flux \labelcref{SU5fluxintegral} are
integer-valued.  Note that for a generic base $B$, $K$ is not 5 times an
integral divisor, so the parameters must typically satisfy the condition
    \begin{align}
    \label{SU5quantcond}
        (\phi^{33} - \phi^{35} - \phi^{44} +\phi^{45}) \in 5 \Z\,.
    \end{align}

\subsubsection{Flux quantization}
\label{sec:flux-quantization-5}

We illustrate the non-unimodularity of $M_\text{red}$ and the
quantization of the parameters $\phi^{ij}$ in some further
detail with a concrete one-parameter family of $\SU(5)$ examples. Consider the case
$B = \bP^3, K =- 4 H, \Sigma = nH$ where $H$ is the hyperplane
class of $\bP^3$.
%After using the nullspace equations $\Theta \nu=0$ to
%eliminate redundant classes $S_{IJ}$ (while carefully maintaining
%integrality conditions)
%we arrive at a basis $S_{0 \alpha},
%S_{\alpha \alpha}, S_{\alpha i}, S_{33}$ (with the index $\alpha$ for the only base divisor)
Using the parametrization of the nullspace of $M_C$ given in \labelcref{linrel} along with the explicit results \labelcref{SU5fluxrelations}, we find that a suitable basis for $H_{2,2}^{\text{vert}}(X,\Z)$ is $S_{0 \alpha},
S_{\alpha \alpha}, S_{\alpha i}, S_{35}$ (with the index $\alpha$ for the only base divisor)
in terms of which the intersection pairing is
given by
\begin{equation}
M_\text{red} =  \left(
\begin{array}{ccccccc}
 -4 & 1 & 0 & 0 & 0 & 0 & 0 \\
 1 & 0 & 0 & 0 & 0 & 0 & 0 \\
 0 & 0 & -2 n & n & 0 & 0 & 4 n \\
 0 & 0 & n & -2 n & n & 0 & -4 n \\
 0 & 0 & 0 & n & -2 n & n & 4 n \\
 0 & 0 & 0 & 0 & n & -2 n & -4 n \\
 0 & 0 & 4 n & -4 n & 4 n & -4 n & -4 n^2 \\
\end{array}
\right)
% M_\text{red} =\left(
%\begin{array}{ccccccc}
% -4 & 1 & 0 & 0 & 0 & 0 & 0 \\
% 1 & 0 & 0 & 0 & 0 & 0 & -2 n \\
% 0 & 0 & -2 n & n & 0 & 0 & 8 n-n^2 \\
% 0 & 0 & n & -2 n & n & 0 & 12 n-4 n^2 \\
% 0 & 0 & 0 & n & -2 n & n & 3 n^2-16 n \\
% 0 & 0 & 0 & 0 & n & -2 n & -4 n \\
% 0 & -2 n & 8 n-n^2 & 12 n-4 n^2 & 3 n^2-16 n & -4 n & -14 n^3+116 n^2-256 n \\
%\end{array}
%\right).
% \left(\begin{array}{c c c c c c c}
%   - 4 & 1 & 0& 0 & 0 & 0 & 0\\
%   1 & 0 & 0& 0 & 0 & 0 & 0\\
%   0 & 0 & -2 & 1 & 0 & 0 & 4\\
%   0 & 0 &  1 & -2 &  1 & 0 & - 4\\
%0 &   0 & 0 &  1 & -2 &  1  & 4\\
%0 &0 &   0 & 0 &  1 & -2   & -4\\
%0 & 0 & 4 & -4 & 4 & -4 & -4
%\end{array}\right).
\label{eq:mr-2}
\end{equation}
In this example, $K$ is not 5 times  an integer divisor, and the
integrality condition \labelcref{SU5quantcond} follows from assuming
that $\phi^{IJ} \in \Z$. This can be seen explicitly as follows:
imposing the local Lorentz and gauge symmetry constraints, one finds that
the flux backgrounds $\phi^{\alpha i}$ can be solved for in terms of the flux backgrounds
$\phi^{ij}$
    \begin{align}
    \begin{split}
        %-\frac{\phi^{15}}{2} = \phi^{14} = - \phi^{13} = \frac{\phi^{12}}{2} = \frac{4}{5} \phi^{33}.
        % \phi^{\alpha 2} = - \frac{2}{5} (-16 + 5n) \phi^{33},~~\phi^{\alpha 3} = -\frac{3}{5} (-8 + 5n) \phi^{33} ,~~ \phi^{\alpha 4} = - \frac{44}{5} \phi^{33} ,~~ \phi^{\alpha 5} = - \frac{32}{5} \phi^{33}.
        \phi^{\alpha 2} = \frac{8}{5} \phi^{35},~~~~ \phi^{\alpha 3} = - \frac{4}{5} \phi^{35},~~~~ \phi^{\alpha 4} = \frac{4}{5} \phi^{35},~~~~ \phi^{\alpha 5} = - \frac{8}{5} \phi^{35}\,.
    \end{split}
    \end{align}
The second Chern class for this class of models over $\bP^3$ in the
reduced basis 
 for the resolution $\scB_{1,3}$ is
\begin{equation}
c_2(X_4) = 2 (n-22) S_{H3}+2 (3 n-34) S_{H4}+2 (n-20) S_{H5}+48 S_{0H}+182 S_{HH}-16 S_{H2}+S_{35}\,.
\end{equation}
Since all coefficients except that of $S_{35}$ are even integers, we
see that the proper shifted lattice for allowed values of $\phi$ keeps
all $\phi^{IJ}$ integral except $\phi^{35}$, which must take a
half-integral value.

The above relations imply that for any nontrivial integer solution to
these conditions we have\footnote{Note that the quantization condition is insensitive to the fact that we eliminated the redundant homology classes before imposing the Poincar\'{e} and gauge symmetry conditions. In particular, the homology relations can be used to show that $\phi^{35}, \phi^{44}, \phi^{45}$ are each proportional to $\phi^{33}$, and these proportionality factors can be used to convert \labelcref{SU5quantcond} into a relation of the form \labelcref{P3SU5quantcond}.}
    \begin{align}
    \label{P3SU5quantcond}
        \phi^{35} = \frac{5}{2} (2k + 1)  \in  \frac{5}{2} (2 \Z  + 1)\,.
    \end{align}
The single flux spanning $M \Lambda_\text{phys}$ is then given
by
    \begin{align}
        \Theta_{35} = \chIndex{\bm{5}} = \frac{4}{5} n (24-5n)
        \phi^{35}
= 2n (24 -5n) (2k + 1) \,,
\label{eq:33-n}
    \end{align}
so the chiral matter multiplicities are necessarily integral.
%Note that when  $n > 4$, the coefficient in \cref{eq:33-n} becomes
%    negative, which seems to conflict with the condition that $G$ must
%    be self-dual (and not anti-self-dual); thus, the self-duality
%    condition seems to impose further constraints on the allowed
%    fluxes.  We leave a further investigation of this kind of
%    situation in more general contexts for further investigation.
%where we note $\Delta_{\textbf{5}} \cdot \Delta_{\textbf{10}} \cdot \Sigma =  76$, in agreement with the general solution \labelcref{SU5fluxintegral}.
%$\Theta_{45}= 4 (x +2 y)$, where $x, y$ are
%integers satisfying $12 x +5 y = 0$. [Relate $\Theta_{45}= \chi_R$].
%It follows that the number of generations of chiral matter takes the
%form $\chi = 76 z,$ where $z = - x/5$ is an integer.\wati{[pass 1] stopped
%  here.  Integrality question is can we have a smaller value than 76
%  for the number of generations?}
%
%
%[Summarize different resolutions, write  general formula for number of generations in terms of intersection theory on the base.]
%
%(Quantization of flux?)
%
%[Generalizations?]

On the other hand,
clearly unimodularity of $H_{2,2}^{\text{vert}}(X_4,\Z)$ is not satisfied as for any $n \ge 1$
    \begin{align}
    \det M_\text{red} =n^5 (20 n-96) \ne \pm{}1\,.
    \end{align}
For example, for $n = 1$ the determinant is $-76$.  This means that
Poincar\'{e} duality guarantees that there are flux backgrounds that are not of
the simple form characterized by (half-)integer $\phi^{ij}$.
This is parallel to the situation discussed for $\SU(2)$ models above
in \cref{sec:2-unimodularity}.  Taking for example
the $n = 1$ case, as for SU(2) the elements of the dual lattice
to the
lattice $H_{2,2}^{\text{vert}}(X_4,\Z)$ with inner product
(\ref{eq:mr-2}) do not have integer norms.  Thus,
the Poincar\'{e}
dual to e.g.\ the surface $S_{33}$ must project to a fractional
vector in $H_{2,2}^{\text{vert}}(X_4)$  and therefore must also contain a component of
$H_{2,2}^\text{rem}(X_4) \oplus H_{4}^\text{hor}(X_4)$.

An interesting
question, which to our knowledge is not addressed anywhere in the
literature and to which the answer seems unknown, is whether or not including
such flux backgrounds can produce chiral matter multiplicities that are more
general than those given by, e.g., \cref{eq:33-n}.\footnote{As discussed in \cref{sec:quantization-1}, some necessary conditions on fractional coefficients for flux backgrounds in $H_{2,2}^{\text{vert}}(X,\Z)$ have been considered in \cite{CveticEtAlQuadrillion}, but not all flux backgrounds satisfying these conditions need be permissible.}
For example, for $n = 1$ the allowed chiral multiplicities from this
analysis should be $38, 114, \ldots$.
Naively it might
seem that Poincar\'{e} duality would suggest that arbitrary integer
matter multiplicities should be possible since there is always an
integral flux background in $H_{4}(X_4,\Z)$
that gives $\Theta_{33} = 1, \Theta_{\alpha i} = 0$.  
It may be,
however, that the components of $H_4(X_4,\Z)$ that must be
turned on for this flux background Poincar\'{e} dual to $S_{33}$ (recall the discussion of Poincar\'e duality and its relation to flux quantization at the of \cref{sec:quantization-1}) would break gauge invariance
(as discussed, e.g., in \cite{Braun:2014xka})
or some other necessary feature of the F-theory vacuum so that such
further chiral multiplicities would be ruled out.  Appealing to
a heterotic dual description also does not immediately clarify this
question, since (as demonstrated in e.g. \cite{Grimm:2011fx}) the chiral multiplicities achieved through the spectral cover construction
match the F-theory chiral multiplicities coming from purely vertical flux backgrounds, though
it is possible that additional chiral multiplicities could be realized through
more general bundle constructions. We leave further investigation of
these questions for future work.

\subsubsection{Resolution-independence of the reduced intersection pairing}
\label{sec:5-independence}

In this section we demonstrate the resolution independence of $M_\text{red}$ for the three resolutions $\scB_{1,3}, \scB_{1,2}, \scB_{2,1}$ described in \cite{Esole:2014hya}. In order to compute $M_\text{red}$ for these three cases, we first write down the symmetry constrained fluxes:
    \begin{align}
        \begin{split}
            \scB_{1,3} ~&:~ \Theta_{33} =- \Theta_{44} =- \Theta_{35} = \Theta_{45}\\
            \scB_{1,2} ~&:~ \Theta_{34} = -2 \Theta_{44} = -\Theta_{35} = \Theta_{45} \\
            \scB_{2,1} ~&: ~ \Theta_{23} = -2 \Theta_{33} = - \Theta_{24} = \Theta_{34}\,.
        \end{split}
    \end{align}
The indices $jk$ of the above fluxes determine the basis elements $S_{0\alpha}, S_{\alpha \beta}, S_{\alpha i}, S_{jk}$ spanning $H_{2,2}^{\text{vert}}(X,\Z)$ in each of these three resolutions. We illustrate this specifically in the case $B = \bP^3, \Sigma= nH$, where $H =D_\alpha$ is the hyperplane class of $\bP^3$. First, we compare the resolutions $\scB_{1,3}$ and $\scB_{1,2}$, for which a common basis is $S_{0\alpha}, S_{\alpha \beta}, S_{\alpha i}, S_{35}$. We find (see \labelcref{eq:mr-2})
    \begin{align}
    \begin{split}
    \label{Mred13}
        M_\text{red}(\scB_{1,3}) &= M_\text{red}(\scB_{1,2}) = \left(
\begin{array}{ccccccc}
 -4 & 1 & 0 & 0 & 0 & 0 & 0 \\
 1 & 0 & 0 & 0 & 0 & 0 & 0 \\
 0 & 0 & -2 n & n & 0 & 0 & 4 n \\
 0 & 0 & n & -2 n & n & 0 & -4 n \\
 0 & 0 & 0 & n & -2 n & n & 4 n \\
 0 & 0 & 0 & 0 & n & -2 n & -4 n \\
 0 & 0 & 4 n & -4 n & 4 n & -4 n & -4 n^2 \\
\end{array}
\right),    \end{split}
    \end{align}
i.e., the two intersection pairing matrices are identical for these two resolutions. On the other hand, to compare the two resolutions $\scB_{1,2}$ and $\scB_{2,1}$, we must use a different basis. A suitable basis in which to compare $M_{\text{red}}(\scB_{1,2}), M_{\text{red}}(\scB_{2,1})$ is $S_{0\alpha}, S_{\alpha \beta}, S_{\alpha i}, S_{34}$, for which we find
    \begin{align}
    \label{Mred21}
        M_\text{red}(\scB_{1,2}) =M_\text{red}(\scB_{2,1}) =\scalebox{.88}{$ \left(
\begin{array}{ccccccc}
 -4 & 1 & 0 & 0 & 0 & 0 & 0 \\
 1 & 0 & 0 & 0 & 0 & 0 & n \\
 0 & 0 & -2 n & n & 0 & 0 & 0 \\
 0 & 0 & n & -2 n & n & 0 & 16 n-2 n^2 \\
 0 & 0 & 0 & n & -2 n & n & 3 n^2-20 n \\
 0 & 0 & 0 & 0 & n & -2 n & 4 n \\
 0 & n & 0 & 16 n-2 n^2 & 3 n^2-20 n & 4 n & -6 n^3+72 n^2-224 n \\
\end{array}
\right)$}\,.
	\end{align}
Again, we find that for an appropriate choice of basis the
intersection pairing matrices are identical. This implies that were we
to identify $M_\text{red}$ for the resolutions $\scB_{1,3}$ and
$\scB_{2,1}$, we would be forced to identify a change of
basis, from \labelcref{Mred13} to \labelcref{Mred21}; the explicit
matrix $U$ presented in \labelcref{Umatrix} does the job for a particular choice of sign in $U_p = (\pm{} 1)$:
solving for the undetermined coefficients in $U$ we find that
they take integer values compatible with the congruence
	\begin{align}
		M_\text{red}(\scB_{2,1}) = U^\transpose  M_\text{red}(\scB_{1,3}) U\,.
	\end{align}
%First consider the case $B = \bP^3$. For $\scB_{1,2}$ we use the basis $S_{0\alpha}, S_{\alpha \alpha}, S_{\alpha i}, S_{45}$ and for $\scB_{2,1}$ we use the basis $S_{0\alpha}, S_{\alpha \alpha}, S_{\alpha i}$....
%   \begin{align}
%       \begin{split}
%       M_\text{red}(\scB_{1,2} ) &=  \left(
%\begin{array}{ccccccc}
% -4 & 1 & 0 & 0 & 0 & 0 & 0 \\
% 1 & 0 & 0 & 0 & 0 & 0 & n \\
% 0 & 0 & -2 n & n & 0 & 0 & 0 \\
% 0 & 0 & n & -2 n & n & 0 & 16 n-2 n^2 \\
% 0 & 0 & 0 & n & -2 n & n & 3 n^2-20 n \\
% 0 & 0 & 0 & 0 & n & -2 n & 4 n \\
% 0 & n & 0 & 16 n-2 n^2 & 3 n^2-20 n & 4 n & -6 n^3+72 n^2-224 n \\
%\end{array}
%\right)\\
%M_\text{red}(\scB_{2,1}) &= \left(
%\begin{array}{ccccccc}
% -4 & 1 & 0 & 0 & 0 & 0 & 0 \\
% 1 & 0 & 0 & 0 & 0 & 0 & n \\
% 0 & 0 & -2 n & n & 0 & 0 & 2 n^2-8 n \\
% 0 & 0 & n & -2 n & n & 0 & 4 n-n^2 \\
% 0 & 0 & 0 & n & -2 n & n & 4 n \\
% 0 & 0 & 0 & 0 & n & -2 n & 0 \\
% 0 & n & 2 n^2-8 n & 4 n-n^2 & 4 n & 0 & -2 n^3+16 n^2-32 n \\
%\end{array}
%\right).
%\end{split}
%   \end{align}

A related change of basis illuminates further
the question of flux quantization discussed in \cref{sec:flux-quantization-5}.
As discussed in general in \cref{nonabwithchiral} and
\cref{proof}, there is a (non-integral)
change of basis $U$ of the form \labelcref{Umatrix} from both of the forms
\labelcref{Mred13} and \labelcref{Mred21} to a canonical product form
\labelcref{Mredcanonical}, given here by
    \begin{align}
        M_\text{red}^\text{cp} 
=U^{\text{t}} M_\text{red} U =\scalebox{.88}{$ \left(
\begin{array}{ccccccc}
 -4 & 1 & 0 & 0 & 0 & 0 & 0 \\
 1 & 0 & 0 & 0 & 0 & 0 & 0 \\
 0 & 0 & -2 n & n & 0 & 0 & 0 \\
 0 & 0 & n & -2 n & n & 0 & 0 \\
 0 & 0 & 0 & n & -2 n & n & 0 \\
 0 & 0 & 0 & 0 & n & -2 n & 0 \\
 0 & 0 & 0 & 0 & 0 & 0 & n (96-20n)/5\\
\end{array}
\right)$}.
\label{eq:mcr5}
	\end{align}
Because $U$ is non-integral in these cases, the lattice $\Lambda^\text{cp}$ of flux
backgrounds on which \cref{eq:mcr5} acts is not $\Z^n$.  On the other
hand, the constraint equations simply set all but the last of the flux
background parameters to vanish.  The non-integer elements of $U$
involve terms of the form $m/5,  m \in\Z$ in the lowest row, and the
set of physical flux background parameters $(0, \ldots, 0, \phi)
\in\Lambda_\text{cp}$ are thus constrained so $\phi \in  5
(2\Z + 1)/2$, analogous
to the constraint \labelcref{SU5quantcond}. 

From this
analysis we see that the chiral multiplicity given by,
e.g., $\chIndex{\bm{5}}=\Theta_{35}$ is $\chIndex{\bm{5}}=2n (24 -5n)
(2k + 1)$, with $ k \in\Z$, in agreement with
\cref{eq:33-n}.  We can relate the canonical form \labelcref{eq:mcr5}
to the lattice $\Lambda_\text{phys} =\Z$, under which the intersection
form becomes $M_\text{phys} = (5n (96-20n))$.
We expect on physical grounds that any valid F-theory resolutions
should give rise to the same physics and
the same $M_\text{phys}$.  Note that while any two such
resolutions would admit  non-integer transformations $U, V$ taking
$M_\text{red}$ to the canonical form \cref{eq:mcr5}, this is not quite
sufficient to prove that $M_\text{red}$ are the same in those two
resolutions since there is no guarantee from what we have said here
that $UV^{-1}$ is an integer matrix, as discussed further in
 \cref{nonabwithchiral} and \cref{proof}.

 It is also worth noting that \cref{eq:mcr5} gives an example of the
 self-duality condition as discussed in more general terms in
\cref{sec:fluxes-universal}.  Namely, the $\SU(5)$ Weierstrass model
on $ \bP^3$ is only consistently defined without enhancement
when $n \leq 4$, in which case $M_\text{phys}$ has a positive matrix
entry and the flux background $\phi$ is self-dual.

\subsection{$\SU(6)$ model}
\label{su6}

The $\SU(6)$ model describes chiral matter in the fundamental
(\textbf{6}) and two-index antisymmetric (\textbf{15})
representations.
The 4D anomaly conditions give
    \begin{align}
    \chIndex{\bm{6}} + 2\chIndex{\bm{15}} =0.
\label{eq:su6-anomaly}
    \end{align}
The signs of the BPS central charges associated to the fundamental and two-index antisymmetric representations can be found in \cref{SUNsigns}. Plugging the intersection numbers into \labelcref{eq:omegabar}, we find that the only nonzero fluxes $\Theta_{ij}$ (Cartan indices are $i= 2,\dots, 6$) satisfy the linear relations
    \begin{align}
    \label{SU6fluxes}
        \Theta_{33} = - \Theta_{34} = \Theta_{35} = \Theta_{45} = - \frac{1}{3} \Theta_{55} = - \Theta_{36} = \Theta_{56}\,,
    \end{align}
where
    \begin{align}
    \begin{split}
    \label{SU6fluxintegral}
        \Theta_{33} &= \frac{\ell(\phi^{ij})}{6}  \Sigma \cdot [a_1] \cdot [a_{4,3}^2]\\
        &= \frac{1}{6} ( \phi^{33} - \phi^{34} + \phi^{35} - \phi^{36} + \phi^{45} - 3 \phi^{55} + \phi^{56} )\Sigma \cdot K \cdot (8K + 6\Sigma)\,.
    \end{split}
    \end{align}
Matching fluxes with the corresponding one-loop CS couplings implies $0=\Theta_{22} = - \chIndex{\bm{6}} - 2 \chIndex{\bm{15}}$, which, using \cref{SU6fluxes}, reproduces the 4D anomaly cancellation condition
\labelcref{eq:su6-anomaly}.
Inverting the linear system arising from matching the fluxes with 3D one-loop CS couplings, we find that the chiral indices can be expressed geometrically as
    \begin{align}
        -\frac{\chIndex{\bm{6}}}{2} = \chIndex{\bm{15}} =  \Theta_{33}\,.
    \end{align}
Note that the gauge symmetry condition \labelcref{eq:gauge} implies, in an analogous fashion to the $\sfG = \SU(5)$ case,
%As a check that the gauge symmetry conditions \labelcref{eq:gauge} ensure that $\Theta_{33}$ is integer-valued, note that an integral vector $v_i$ lies in the $\SU(6)$ root lattice iff $ v_2 + 2v_3 + 3 v_4 + 4v_5 - v_6 \in 6 \Z$. Applying this condition to the fluxes $\Theta_{\alpha i}$ one finds
%   \begin{align}
%   \begin{split}
%       &\Theta_{\alpha 2} + 2\Theta_{\alpha 3} + 3 \Theta_{\alpha 4} + 4\Theta_{\alpha 5} - \Theta_{\alpha 6} \\
%%      =~&\Sigma \cdot D_\alpha \cdot (-3 \phi ^{15} D_{\alpha }+3 \phi^{16} D_{\alpha }+2 K \phi ^{33}-2 K \phi ^{34}+2 K \phi ^{35}\\
%%      &-2 K \phi ^{36}+2 K \phi ^{45}-4 K \phi ^{56}-6 K \phi ^{66}+3 \Sigma  \phi ^{44}+3 \Sigma  \phi ^{45}\\
%%      &-3 \Sigma  \phi ^{55}\-3 \Sigma  \phi ^{56}+6 \Sigma  \phi ^{66}) \\
%       \equiv~&( \Sigma \cdot D_\alpha) \cdot 2K (\phi ^{33}-\phi ^{34}+\phi ^{35}-\phi ^{36}+\phi ^{45}- \phi ^{56}) \, \mod \, 6,
%   \end{split}
%   \end{align}
%which for generic $K$ in turn implies
    \begin{align}
         2K^\alpha (\phi ^{33}-\phi ^{34}+\phi ^{35}-\phi ^{36}+\phi ^{45}- \phi ^{56}) \in 6 \Z\,,
    \end{align}
ensuring that $\Theta_{33} \in \Z$.

For comparison, we also comment on the flux for an alternative Tate tuning \cite{HuangTaylorLargeHodge} (see also \cite{MorrisonTaylorMaS,AndersonGrayRaghuramTaylorMiT}) of the $\SU(6)$ model, denoted the $\SU(6)^\circ$ model, that has matter in the exotic three-index antisymmetric representation rather than the usual two-index antisymmetric representation. Because the three-index antisymmetric representation is self-conjugate, one would naively expect the space of vertical F-theory fluxes to be empty. However, it turns out the $\SU(6)^\circ$ model contains codimension-three $(4, 6)$ singularities, leading to a nontrivial flux presumably given by the integral of the flux background over the surface component of the non-flat fiber visible at the $(4, 6)$ point in the resolution $X_5$. See \cref{tab:fluxtable} for additional details.
%\patrick{This section should probably be taken out and left for future work on the $(4,6,12)$ points.} For comparison, we also study the exotic $\SU(6)^{\circ}$ tuning of \cite{HuangTaylorLargeHodge}, with matter in the representations $\textbf{6}, \textbf{20}$, and $\textbf{35}$. Note that the (exotic) three-index antisymmetric representation $\textbf{20}$ is self-conjugate and hence we expect to not find chiral matter in this case. Adopting the same notation for the basis of divisors as in the ordinary SU(6) model described above, we find
%   \begin{align}
%       \Theta_{33} = -\frac{1}{4} \Theta_{44} = \frac{1}{2} \Theta_{45} = - \Theta_{55} = - \Theta_{36} = \frac{1}{2} \Theta_{46} = - \Theta_{56}
%   \end{align}
%where
%   \begin{align}
%       \Theta_{33} = \frac{1}{6} ( 9 K + 6 \Sigma)\cdot K \cdot \Sigma (\phi^{33}- 4 \phi^{44} + 2 \phi^{45} - \phi^{55}-\phi^{36}  + 2 \phi^{46}  -\phi^{56} ).
%   \end{align}
%The matching between geometric fluxes and 3D CS couplings indicates
%   \begin{align}
%       \Theta_{33} = - \chIndex{\bm{6}}.
%   \end{align}
%However, the same matching of terms also implies, e.g.,
%   \begin{align}
%   \chIndex{\bm{6}}  =     \Theta_{66}^{\text{3D}} = \Theta_{66} \overset{\text{!}}{=} 0.
%   \end{align}
%The only way to make the above equation consistent is to set $\chIndex{\bm{6}} =0$, which is consistent with 4D anomaly cancellation.

\subsection{$\SO(10)$ model}
\label{so10}

The SO(10) Tate model is characterized by a I$_1^{* \text{split}}$
singularity over a gauge divisor $\Sigma$ and contains chiral matter
in the spinor representation ($\textbf{16}$); the multiplicity of
matter in this representation is unconstrained by anomalies. We label
Cartan divisors with indices $i =2,\dots, 6$. The signs of the BPS
central charges associated to the spinor can be found in
\cref{SO10signs}.

Using \labelcref{eq:omegabar}, we find that the F-theory fluxes satisfy the homology relations
    \begin{align}
        \Theta_{22} = -\Theta_{24} = \Theta_{25} = \Theta_{44} = - \Theta_{46}=- \Theta_{55}  =\frac{1}{2} \Theta_{66}
    \end{align}
where
    \begin{align}
    \begin{split}
    \label{SO10fluxintegral}
        \Theta_{22} &=\frac{\ell(\phi^{ij})}{4} \cdot [a_{2,1}] \cdot [a_{6,5}] \\
        &=\frac{1}{4} (\phi ^{22}-\phi ^{24}+\phi ^{25}+\phi ^{44}-\phi ^{46}-\phi ^{55}+2 \phi ^{66}) \Sigma \cdot (2 K+\Sigma )\cdot (6 K+5 \Sigma )\,.
    \end{split}
    \end{align}
%Applying the fact that an integral vector $v_i$ lies in the SO(10) root lattice iff $2 v _2+4 v _3+2 v _4+v _5-v _6 \in 4 \Z$ to the fluxes $\Theta_{\alpha i}$, we find
%   \begin{align}
%   \begin{split}
%       &2 \Theta _{\alpha 2}+4 \Theta _{\alpha 3}+2 \Theta _{\alpha 4}+\Theta _{\alpha 5}-\Theta _{\alpha 6} ,
%%      =\,&- D_\alpha \cdot \Sigma (4 \phi ^{13} D_{\alpha }-4 \phi ^{16} D_{\alpha }+6 K \phi ^{22}-4 K \phi ^{23}+2 K \phi ^{24}\\
%%      &-2 K \phi ^{25}+4 K \phi ^{33}-4 K \phi ^{34}+6 K \phi ^{44}-4 K \phi ^{45}-2 K \phi ^{46}\\
%%      &-2 K \phi ^{55}+12 K \phi ^{56}-8 K \phi ^{66}+7 \Sigma  \phi ^{22}-4 \Sigma  \phi ^{23}\\
%%      &+\Sigma  \phi ^{24}-\Sigma  \phi ^{25}+8 \Sigma  \phi ^{33}-4 \Sigma  \phi ^{34}\\
%%      &+7 \Sigma  \phi ^{44}-4 \Sigma  \phi ^{45}-3 \Sigma  \phi ^{46}+\Sigma  \phi ^{55}\\
%%      &+8 \Sigma  \phi ^{56}-6 \Sigma  \phi ^{66})\\
%%      =\,&  (D_\alpha \cdot \Sigma) \cdot  (2 K+\Sigma )\left(\phi ^{22}-\phi ^{24}+\phi ^{25}+\phi ^{44}-\phi ^{46}-\phi ^{55}+2  \phi ^{66} \right) \mod \, 4,
%   \end{split}
%   \end{align}
%which for generic $2 K + \Sigma$ implies
Matching with 3D one-loop CS terms, we find
    \begin{align}
        \chi_{\textbf{16}} = - \Theta_{22}\,.
    \end{align}
The gauge symmetry condition \labelcref{eq:gauge} implies
    \begin{align}
        (2 K + \Sigma)^\alpha (\phi ^{22}-\phi ^{24}+\phi ^{25}+\phi ^{44}-\phi ^{46}-\phi ^{55}+2  \phi ^{66}) \in 4 \Z\,,
    \end{align}
hence $\Theta_{22}$ is integer-valued. We again find no linear constraints on the $\SO(10)$ chiral spectrum other than those implied by anomaly cancellation.  
%and
%   \begin{align}
%       \begin{split}
%           2 \Theta _{\alpha 5}-2 \Theta _{\alpha 6} &= 2 D_\alpha \cdot \Sigma \cdot (-2 \phi _{1,5} D_{\alpha }+2 \phi _{1,6} D_{\alpha }+2 K \phi _{2,2}-2 K \phi _{2,4}\\
%           &~~~+2 K \phi _{2,5}+2 K \phi _{4,4}-2 K \phi _{4,6}+4 K \phi _{5,5}-6 K \phi _{5,6}\\
%           &~~~+10 K \phi _{6,6}+\Sigma  \phi _{2,2}-\Sigma  \phi _{2,4}+\Sigma  \phi _{2,5}\\
%           &~~~+\Sigma  \phi _{4,4}-\Sigma  \phi _{4,6}+\Sigma  \phi _{5,5}-4 \Sigma  \phi _{5,6}+8 \Sigma  \phi _{6,6}) \\
%           &=2 D_\alpha \cdot \Sigma^2 (\phi _{2,2}-\phi _{2,4}+\phi _{2,5}+\phi _{4,4}-\phi _{4,6}+\phi _{5,5}) \mod \, 4.
%       \end{split}
%   \end{align}
%Combining the above two conditions, we find that
%   \begin{align}
%       K \cdot \Sigma \cdot D_\alpha \Sigma  \left(-\phi _{2,2}+\phi _{2,4}-\phi _{2,5}- \phi _{4,4}+ \phi _{4,6}+\phi _{5,5}-2 \phi _{6,6}\right)
%   \end{align}

\subsection{$\text{E}_6$ model}
\label{e6}

Our final purely nonabelian example is $\mathsf G = \text{E}_6$, which is the only
exceptional group with complex representations and hence the only
exceptional group admitting chiral matter preserving the full gauge
symmetry. The E$_6$ Tate model is characterized by a IV$^{*
  \text{split}}$ singularity over gauge divisor $\Sigma$ and contains
chiral matter in the fundamental ($\textbf{27}$) representation, with
a multiplicity unconstrained by local anomalies. Additional details
about the resolution and corresponding signs of BPS central charges
can be found in \cref{E6appendix}.

Computing intersection numbers and substituting their values into the
expression in \labelcref{eq:omegabar}, we find that the nontrivial constrained fluxes satisfy the homology relations
    \begin{align}
        \frac{1}{2} \Theta_{22}  = -\Theta_{25} = - \Theta_{33}= \Theta_{35}=-\Theta_{55}=\Theta_{56}=- \frac{1}{2} \Theta_{66}
    \end{align}
where
    \begin{align}
    \begin{split}
    \label{E6fluxintegral}
        \Theta_{35}& =\frac{\ell(\phi^{ij})}{3} \Sigma \cdot [a_{3,2} ] \cdot [a_{6,5}] \\
         &=\frac{1}{3}  ( 2 \phi^{22} - \phi^{25} -\phi^{33} + \phi^{35} - \phi^{55} + \phi^{56} - 2 \phi^{66} )   \Sigma\cdot (3 K + 2\Sigma)  \cdot( 6 K + 5 \Sigma)\,.
    \end{split}
    \end{align}
Comparing with the corresponding 3D one-loop CS couplings, we find
    \begin{align}
        \chIndex{\bm{27}} = \Theta_{35}\,,
    \end{align}
in agreement with, e.g., Eq.~(4.12) in \cite{Kuntzler:2012bu}. Note that the gauge symmetry conditions \labelcref{eq:gauge} imply
    \begin{align}
        \Sigma^\alpha (2 \phi ^{22}-\phi ^{25}-\phi ^{33}+\phi ^{35}-\phi ^{55}+\phi ^{56}-2 \phi ^{66} ) \in 3 \Z\,,
    \end{align}
which ensures that $\Theta_{35}$ is integer-valued.\footnote{Curiously, in the special case that $\Sigma^\alpha \in 3 \Z$ the gauge symmetry condition does not appear to place any special conditions on the parameters $\phi^{ij}$.}

\section{Models with a $\U(1)$ gauge factor}
\label{sec:abelianmodel}

We now turn to the more general case of models with gauge symmetry
$\sfG = ( \sfG_\text{na} \times \U(1) )/ \Gamma$. As discussed in
previous sections, these models are complicated by the fact that the
fluxes do not simply depend on the mutual triple intersections of the
characteristic data $(K,\Sigma_s, W_{01})$, but rather also depend on
the intersection products of all divisors $D_\alpha \in B$ with the
height pairing divisor $W_{\bar 1 \bar 1}$ associated to the
$\U(1)$ factor---this is a reflection of the global geometric nature
of $\U(1)$ gauge factors in F-theory, in contrast to the local nature
of nonabelian gauge factors $\sfG_s \subset \sfG_\text{na}$. One notable
consequence is that the nullspace of $M_C$ is not obviously computable
for such models in a very general way without explicitly specifying $B$.
  A possible workaround to this complication, as
discussed at the end of \cref{sec:homologyrel}, is to further restrict
to the sublattice $\Lambda_S \cap \{ \phi^{1\alpha} =0\}$; we describe
an example of
this  analysis in \cref{sec:21-rcc}.  In the rest of
this section we focus attention on specific bases $B$, where we can
explicitly carry out the full flux analysis.

In \cref{F6model} we analyze the $(\SU(2) \times \U(1) )/\Z_2$ model
in detail.  In \cref{321model}, we briefly summarize the results of
the forthcoming paper \cite{Jefferson:2022yya} in which we use the methods of
this paper to analyze the universal $\SM$ model from
\cite{Raghuram:2019efb}.

\subsection{$(\SU(2) \times \U(1) )/\Z_2$ model}
\label{F6model}

Perhaps the simplest example of a model with gauge group $\sfG$
containing a $\U(1)$ gauge factor is the $F_{6}$ model studied in
\cite{KleversEtAlToric}, with $\sfG= (\SU(2) \times \U(1))/\Z_2$ and
matter transforming in the representations $\textbf{1}_{1},
\textbf{1}_2, \textbf{2}_{\frac{1}{2}},
\textbf{2}_{-\frac{3}{2}}$. A related F-theory model was recently analyzed in \cite{Esole:2019rzq}.

The 4D anomaly cancellation conditions impose the constraints that the
chiral matter multiplicities must correspond to an integer multiple of
the family
\begin{equation}
(\chi_{\textbf{1}_{1}},
\chi_{\textbf{1}_{2}},
\chi_{\textbf{2}_{\tfrac{1}{2}}},
\chi_{\textbf{2}_{-\tfrac{3}{2}}})
= (2, -1, -3, -1) \,.
\label{eq:21-family}
\end{equation}
The characteristic data of
%the singular limit $X_0$
this class of F-theory models consists of the canonical class $K$ and the two divisor classes $ S_7, S_9$,
in
terms of which the $\SU(2)$ gauge divisor is given by $S_8 = -K +S_9 -
S_7$. Explicitly, the singular $F_6$ model $X_0$ is realized
as a hypersurface in an ambient $\bP^2$ bundle over arbitrary smooth base
$B$, given by 
	\begin{equation}
		s_1 u^3 + s_2 u^2 v + s_3 u v^2 + s_4 v^3 + s_5 u^2 w + s_6 uvw + s_7 v^2 w + s_8 u w^2 =0
	\end{equation}
where $[u:v:w]$ are homogeneous coordinates of the ambient space fibers defined by the hyperplane classes
    \begin{equation}
    \label{eqn:F6fiberP2}
        [u] = \boldsymbol{H} +\boldsymbol{K} + \boldsymbol{S}_9,~~~~[ v ] = \boldsymbol{H} - \boldsymbol{S}_7 + \boldsymbol{S}_9,~~~~ [w  ] = \boldsymbol{H}
    \end{equation}
 (note $\boldsymbol{H}$ is the hyperplane class of the fibers and $\boldsymbol{K}, \boldsymbol{S}_7,\boldsymbol{S}_9$ are the pullbacks of the classes $K,S_7,S_9$ in the base to the Chow ring of the ambient space) and the divisor classes of the sections $s_m$ appearing the above hypersurface equation, namely $S_m=[s_m]$, are given by
    \begin{equation}
    	S_1 = -3 K - S_7 - S_9,~~~~ S_4 = 2 S_7 - S_9
    \end{equation}
along with
	\begin{equation}
		S_2 = \frac{1}{3} (2S_1 + S_4),~~~~ S_3 = \frac{1}{3} (S_1 + 2 S_4),~~~~ S_5 = \frac{1}{2} (S_1 + S_8) ,~~~~ S_6 = \frac{1}{2} (S_7 + S_8)\,.
	\end{equation}
For a good model with gauge group $\sfG= (\SU(2) \times \U(1))/\Z_2$,
the characteristic data is constrained so that the divisor classes
$S_1,S_4,S_7, S_8$ are effective.\footnote{These divisor classes are associated with the vertices of the dual polytope of the toric fiber defining the $F_6$ model in \cite{KleversEtAlToric}. %In
  %the notation of \cite{KleversEtAlToric}, these conditions can be
  %understood from the conditions that the parameters $S_7, S_8, S_1,
  %S_4$ associated with the vertices of the dual polytope associated
  %with the fiber must be effective.
  }

\subsubsection{Resolution, Chern--Simons terms, chiral index}
\label{sec:21-rcc}

The resolution $X_2 \to X_0$ described in \cite{KleversEtAlToric} entails a sequence of two
blowups acting on the $\bP^2$ fibers so that the resulting smooth
model $X_2$ may be viewed as a hypersurface in an ambient projective
bundle with fibers isomorphic to $\bP_{F_6}$, i.e., the blowup of
$\bP^2$ at two points. The hypersurface equation defining $X_2$ can be
computed systematically by exploiting the fact that to every two-dimensional toric variety $\bP_{F_i}$ is associated a canonically
defined genus one curve in $\bP_{F_i}$ that can be realized as a a
zero section of the anticanonical bundle.

In order to use the pushforward technology to compute intersection numbers and other relevant characteristic numbers associated to $X_2$, we regard the singular model $X_0$ as a hypersurface of the ambient projective bundle $Y_0 = \bP(\scV) \rightarrow B$, with $\bP^2$ fibers described by \cref{eqn:F6fiberP2}. Combining this data with the classes of the generators of the centers of the blowups comprising the resolution $X_2 \to X_0$, it is straightforward to explicitly compute the pushforwards of the intersection numbers in a suitable basis and evaluate the geometric expressions for the fluxes using \labelcref{eq:omegabar}--\labelcref{eq:omegabar3}.

The computation of the 3D Chern--Simons terms following the strategy of matching intersection numbers of the form $W_{\bar{I} \bar{J} \bar{K} \alpha}$ to 5D Chern--Simons terms $k^{\text{5D}}_{\bar{I}\bar{J}\bar{K}}$ (where $\bar{I}, \bar{J}, \bar{K} = \bar{0}, \bar{1}, i$) in this case is more involved due to the fact that the resolved model $X_2$ has a rational, as opposed to holomorphic, zero section. This is because one needs to determine, in addition to the signs of the BPS central charges, the ratio of their magnitudes to the KK modulus $m_\text{KK}$. Fortunately there is a simple geometric computation one can do to determine which particles (descending from M2 branes wrapping irreducible holomorphic curves in the M-theory background) have nontrivial KK charge. Notice that the pushforward of the intersection of the zero section and generating section is given by
    \begin{align}
        \pi_*(\hat D_0 \cdot \hat D_1) = W_{01} =  S_7\,.
    \end{align}
From the above expression we can infer that the primitive BPS particles in the representation $\sfR'$ carrying nontrivial KK charge must be associated to matter loci of the schematic form
    \begin{align}
        C_{\sfR'} =S_7 \cdot (\cdots)\,.
    \end{align}
Exploiting the fact that the spectrum of the $F_6$ model is known, we see that the classes of the relevant matter loci fitting this criterion are
    \begin{align}
        C_{\textbf{2}_{-\frac{3}{2}}} = S_7 \cdot ( -K - S_7 +S_9),~~~~ C_{\textbf{1}_2} = S_7 \cdot (2 S_7 -S_9)\,.
    \end{align}
%In fact, it was shown in \cite{KleversEtAlToric} that there is an exceptional curve $F_+$ wrapped by the zero section $\hat D_0$ with Cartan charges in the $\textbf{2}_{-\frac{3}{2}}$ of $\SU(2) \times \U(1)$. As a consistency check, we can use the pushforward techniques described in \cref{pushapp} to compute the Cartan charges of $F_+$ in Chow ring of the ambient bundle $Y \supset X$ of the resolution. Denoting by $\hat D_{I = 0,1,2}$ (resp.) the zero section, generating section, and $\SU(2)$ Cartan divisor, we find the Cartan charges of $F_+$ are\footnote{In this context $C_{\textbf{2}_{-\frac{3}{2}}}$ denotes the pullback of the class of the matter locus to the ambient space Chow ring, i.e., $C_{\textbf{2}_{-\frac{3}{2}}} =\boldsymbol{S}_7 \cdot ( -\boldsymbol{K} - \boldsymbol{S}_7 +\boldsymbol{S}_9)$.}
%   \begin{align}
%           ( \sigma_0^{I} \hat D_I , \sigma_1^I \hat D_I , \hat D_2 ) \cdot C_{\textbf{2}_{-\frac{3}{2}}} \cdot \boldsymbol{E}_2 \cdot \boldsymbol{D}_\alpha = (- 1, \frac{3}{2}, -1) \boldsymbol D_{\alpha},
%   \end{align}
%where $\boldsymbol{E}_2$ is the class of the exceptional divisor of the second blowup whose restriction to the class of the hypersurface is the zero section, i.e., $\hat D_0 = \boldsymbol{E}_2 \cdot \boldsymbol{X}$. The above Cartan charges indicate that $F_+$ is the highest weight of the $\textbf{2}_{-\frac{3}{2}}$.
The above analysis implies that the BPS particles transforming in the representations $\sfR' = \textbf{2}_{-\frac{3}{2}},\textbf{1}_2$  have nontrivial KK charge. It follows that these are the only particles for which the KK mass is not larger than the Coulomb branch mass; we may therefore set $\floor{| r_\text{KK} \varphi \cdot w_i^{\sfR} |} =0$ for all other representations $\sfR$. Utilizing this simplifying assumption, we find a perfect match between the 5D Chern--Simons terms and the triple intersection numbers involving precisely three Cartan divisors, provided we use the signs in \cref{F6signs} and set
    \begin{align}
        \floor{|r_\text{KK} \varphi \cdot w^{\textbf{2}_{-\frac{3}{2}}}_+ |}  = 1
    \end{align}
where $w^{\textbf{2}_{-\frac{3}{2}}}_+ = (-\tfrac{3}{2},1)$ is the highest weight.

While we have not found a way to determine a general form for the the nullspace of $M_C$ for arbitrary
characteristic data without specifying $B$,
as discussed in \cref{sec:homologyrel}
we can
 attempt to get a general picture of the
nullspace by restricting to the sublattice
 $\Lambda_S \cap \{ \phi^{1\alpha}
=0\}$.
Combining the pushforwards of the intersection numbers with
formulae
for the constrained fluxes in \labelcref{eq:omegabar,eq:omegabar2,eq:omegabar3}
reveals that after imposing the additional restriction
$\phi^{1 \alpha} = 0$, the fluxes satisfy (the $\SU(2)$ Cartan index is $i=2$)
    \begin{align}
    \label{F6null}
        2\Theta_{00} = -2 \Theta_{01} = 2\Theta_{11} =2 \Theta_{02} =  \Theta_{22}\,.
    \end{align}
Continuing to work in the restricted case $\phi^{1\alpha} =0$, comparing the fluxes with the corresponding 3D
Chern--Simons theory shows that the multiplicity of chiral matter
in $\textbf{2}_{-\tfrac{3}{2}}$ is
    \begin{align}
   	2  \chi_{\textbf{2}_{-\tfrac{3}{2}}} = \Theta_{22}\,.
    \end{align}
We thus expect in general that F-theory models can realize the
one-parameter family \cref{eq:21-family} of anomaly-free chiral matter fields for this
gauge group with generic matter.
Since in the unrestricted case $\phi^{1\alpha} \ne 0$ we cannot write down a completely general expression for the fluxes without specifying $B$, we next turn our attention to specific examples.
%As discussed in \cref{constraintsolutions}, the matrix $MP^{\text{na}}_{(1\alpha)
%  (1\beta)}$ in \labelcref{eq:omegabar3} is not in general invertible
%for all choices of the characteristic data $S_7, S_9$, which
%prevents us from writing down a
%completely general
%explicit expression for the fluxes associated to the
%resolved $F_6$ model defined over an arbitrary base.
%Nonetheless, the
%formulas for the constrained fluxes in
%\labelcref{eq:omegabar}--\labelcref{eq:omegabar3} can be used to easily
%obtain the fluxes for a specific choice of base, where we
%can explicitly solve the remaining constraint equations $\Theta_{1
%  \alpha} = 0$.

\subsubsection{Example: $B = \bP^3$}
As a first simple example, we study the
case $B = \bP^3, K = -4 H, S_7 = s_7 H, S_9 = s_9 H$.  We define
$n =s_9 + 4- s_7$ for convenience, and parameterize the results in
terms of $n, s_7$ (the parameter $n$ corresponds to the parameter
$s_8$ in the $F_6$ Weierstrass model of \cite{KleversEtAlToric}, which
as discussed above
is the degree of the divisor class $S_8 = n H$ associated with the $\SU(2)$ factor; the $\U(1)$
factor is associated with the height pairing parameter $h :=-2 W_{\bar 1 \bar 1}\cdot H^2 = 16 +
4s_7-n$).  Such a model is defined for integer values of $n, s_7$
satisfying the conditions $n, s_7, 16-n-2s_7 = 24 -3n/2-h/2, 4 + s_7-n
= (h-3n)/4 > 0$; from these conditions we see that the height pairing
also satisfies $h = 16 + 4s_7-n > 0$.  In this set of cases, the matrix $M_{C_\text{na}(1H)(1H)} =
    (n-4s_7-16)= -h/2$ would fail to have an inverse when $n =
    4s_7 + 16$, but this does not happen in the parameter range of
    interest as the height pairing divisor is always
    positive/effective, so for these models the matrix $M_{C_\text{na}(1\alpha)
      (1\beta)}$ is always invertible, making it possible in all cases to solve for $\phi^{1H}$ with this expression as a denominator. However, as we demonstrate below, in this case the formal rational expression for the flux (which must take integer values) is not an invariant property of the solution, but rather a feature of our choice of solution. As an alternative, we can solve the equation $\Theta_{1H} = 0$ by eliminating a different flux background parameter so as to produce a polynomial expression for $\Theta_{22}$ that is manifestly integer-valued.

As discussed in  \cref{sec:fluxes-algebra}, we can analyze this model by following one of two approaches: either we first impose the symmetry constraints and then study the nullspace of $M_C$ in order to determine linear constraints on the fluxes, or we first quotient out the nullspace of $M$ and then impose the symmetry constraints.
    Despite producing identical results, both approaches have their respective unique advantages. In the following discussion we briefly describe three
    versions of the calculation (two of which follow the former approach, with the third version following the latter approach) in order to illustrate the
    different aspects of the problem.
In particular, the full set of constraints determining the
quantization of the number of chiral matter fields is clearest in the
analysis beginning with $M_\text{red}$ in this class of examples,
though this may not be the case for other choices of $\sfG$ or $B$.

As stated more generally in \cref{sec:vertical-pairing}, throughout the analysis of this section we disregard the possible half-integer shift in $\phi$ that may be required when $c_2(X)$ is not even; this can easily be incorporated, as described for the $\SU(5)$ model in \cref{sec:su5}. \newline

\noindent \textbf{Rational solution.} Solving directly for $\phi^{1H}$ gives
    \begin{align}
    \begin{split}
        \Theta_{22}& =
\chi_{\textbf{2}_{-\tfrac{3}{2}}} = \frac{2 n s_7 (16-n-2s_7) (4 +s_7-n)}{h} \ell(\phi^{\hat I \hat J} )\\
         \ell(\phi^{\hat I \hat J} ) &= \phi^{00} - \phi^{01} + \phi^{02} + \phi^{11} +2 \phi^{22}\,,
    \end{split} \label{eq:21-direct}
    \end{align}
As in the discussion above, this is well-defined since $h >
  0$.

As for the purely nonabelian cases described previously, the
symmetry constraints and integer conditions on the flux backgrounds $\phi^{IJ}$
are sufficient to guarantee that this expression is integer- (and
even-)valued, although we have not identified as simple a structure
underlying this integrality as the group theoretic structure
underlying e.g. the combination of flux background parameters appearing in \cref{SU5quantcond}.
In this case, the symmetry constraints $\Theta_{H 2} =0$ imply
additional conditions, including
    \begin{align}
        2  s_7(4 + s_7-n) \ell(\phi^{\hat I \hat J} ) \in
                h\Z \,,
    \end{align}
which in turn implies that $ \Theta_{22}$ is integer-valued, and is
in fact an integer multiple of $n (16-n-2s_7)$.
The full set of constraints is seen most easily after imposing the
homology equivalence conditions on $S_{IJ}$, as discussed further below. \newline

\noindent \textbf{Polynomial solution.} Alternatively, we can solve $\Theta_{1H} = 0$
in this case for the flux
$\phi^{22}$, which gives us the result
    \begin{align}
    \begin{split}
        \Theta_{22}& =
        \chi_{\textbf{2}_{-\tfrac{3}{2}}}
= \frac{2}{3}  (16-n-2s_7) (4-n + s_7)
\ell'(\phi^{ I  J} )\\
         \ell'(\phi^{ I J} ) &= \phi^{1H} + (s_7-4)
                 \phi^{11} +n \phi^{12}\,.
\label{eq:21-indirect}
    \end{split}
    \end{align}
In this analysis, the condition $\Theta_{1H} = 0$ implies that
$(16-n + 4s_7)\ell'(\phi^{I  J} ) = 3n s_7 k$ where $k$ is an
integer combination of fluxes, so generically we expect $\Theta_{22}$
to be  an integer multiple of $2n s_7(16-n - 2s_7) (4-n+s_7)$.

The two presentations of the chiral multiplicity (\ref{eq:21-direct})
and (\ref{eq:21-indirect}) must give equivalent answers after all
quantization conditions are properly taken account of.  The second
expression is simpler since only a factor of 3, and not the $\U(1)$
height pairing $h$, appears in the denominator.  On the other hand,
the expression for $\ell$ is simpler than that of $\ell'$ as it does
not depend on the characteristic data.
For the full analysis of the quantization conditions we now turn to
the analysis using $M_\text{red}$. \newline

\noindent \textbf{Vertical homology and flux quantization.} An important difference in resolutions only admitting a rational zero
section from those with a holomorphic zero section such as those
associated with the purely nonabelian groups studied in previous
sections is that solving the symmetry conditions
\labelcref{eq:Poincare,eq:gauge} over the integers generically imposes
additional constraints on the parameters beyond those necessary
to ensure integrality of the solutions.

To complete the discussion of this simple example we describe the
complete quantization condition following from the integrality of the
fluxes $\phi^{IJ}$.  We can first explicitly remove the nullspace of
$M$ by
dropping the fluxes $\phi^{01}, \phi^{02}, \phi^{11}, \phi^{12},
\phi^{22}$, which each appear with a coefficient of 1 in a nullspace
vector with all other entries integer.  With this simplification,
the matrix $M_\text{red}$ in the basis $S_{0H}, S_{HH},
S_{H2}, S_{1H}, S_{00}$ becomes
    \begin{align}
    \label{21mr}
    M_\text{red} = \begin{pmatrix} -4 & 1 &0 & s_7 &  16-ns_7
          \\ 1 & 0 & 0 &   1 & -4 \\ 0 & 0 &-2 n &n & -ns_7\\
s_7 & 1 & n & -4 & s_7 (n-4)\\
16-ns_7 & -4 & -ns_7 & s_7 (n-4) & -64 + 12ns_7
-n^2 s_7-ns_7^2 \end{pmatrix}\,,
    \end{align}
and
\begin{equation}
\det M_\text{red}= -n^2 s_7 (16-n - 2s_7) (4-n+s_7)\,.
\end{equation}
Note that the top left 4 $\times$ 4 block is
resolution-invariant
and
 the top left 3 $\times$ 3 block corresponds to the generalization of \labelcref{SU2P3mat} to
arbitrary $n$.  Given this form of the reduced matrix,
we can directly solve the constraint equations
$\Theta_{0H},\Theta_{HH}, \Theta_{H2} = 0$
 for $\phi^{0H}, \phi^{HH}, \phi^{H2}$.  The first two of
these each can be described as
an integer linear combination of remaining fluxes, and the third can
be solved as an integer whenever the flux combination
$\phi^{1H} -s_7 \phi^{00}$ is even.  The remaining
equation $\Theta_{1H} = 0$ becomes
\begin{equation}
3ns_7 \phi^{00} = h \phi^{1H} \,.
\end{equation}
These are therefore the only nontrivial constraints on these fluxes.
With this simplification for removing the nullspace, the parameters
$\ell, \ell'$ become $\phi^{00}, \phi^{1H}$ respectively.

From these constraints for any fixed values of $n, s_7$ we can
explicitly determine the quantization of the chiral multiplicity
encoded by $\Theta_{22}$.  For example, when $n$ is odd, $h$ is odd as
well and the even parity constraint on $\phi^{1H}-s_7 \phi^{00}$ is
automatically satisfied, so
when $h$
and $3ns_7$ furthermore
have no common divisors, it follows that $\Theta_{22}$ can
be an arbitrary integer multiple of $2ns_7 (16-n - 2s_7) (4-n+s_7)$,
up to bounds determined by the tadpole condition (and where, as noted above, to simplify the discussion we have ignored possible half-integer shifts for non-even $c_2(X)$).

As explicit examples, if $n = s_7 = 1\ (h = 19)$, the chiral index will be a
multiple of $2 \times 13 \times 4 = 104$, and if $n = 1, s_7 = 4\ (h =
43)$, the chiral index will be a multiple of 392. If, however, e.g., $n
= 3, s_7 = 2\ (h = 21)$, then $h$ has a common factor with $3ns_7$, in
particular, $4-n + s_7 = 3, 16-n-2s_7 = 9,$ and $\Theta_{22}$ can be
any multiple of $81$ (instead of 243), up to tadpole
constraints.
And when $n = 4, s_7 = 1\ (h = 16)$, the even parity constraint imposes the
additional condition that $\phi^{1H}$ must be even, so $\Theta_{22}$
is a multiple of $4ns_7 (16-n-2s_7) (4-n + s_7) = 160$.

\begin{table}
\begin{center}
$
    \begin{array}{|cc|}
        \hline
            \textbf{1} & \textbf{2} \\\hline
                \left(\begin{array}{c|cc} \frac{\varphi \cdot w}{|\varphi \cdot w |}  & w_1 & w_2 \\\hline
                + & 1 & 0  \\\hline
                    +& 2 & 0 \end{array} \right) &  \left( \begin{array}{c|cc}  \frac{\varphi \cdot w}{|\varphi \cdot w |}  & w_1 & w_2 \\\hline +& \frac{1}{2} & 1 \\ -& \frac{1}{2} & -1 \\\hline + & - \frac{3}{2} & 1 \\ + & -\frac{3}{2} & -1 \end{array}\right) \\\hline
    \end{array}
    $
    \caption{Signs and Cartan charges associated to the BPS spectrum of the resolved $F_6$ model with gauge group $\sfG = (\SU(2) \times \U(1))/\Z_2$ analyzed in \cite{KleversEtAlToric}. The Cartan charges are the Dynkin coefficients of the weights in the representations $\textbf{1}_{1}, \textbf{1}_2, \textbf{2}_{\frac{1}{2}}, \textbf{2}_{-\frac{3}{2}}$ of $\SU(2)$ and the signs correspond to the signs of the BPS central charges $\varphi \cdot w$ for a given choice of Coulomb branch moduli $\varphi^i$.}
    \label{F6signs}
\end{center}
\end{table}

\subsubsection{Example: $B = \tilde{\F}_n$}
\label{sec:21-exception}

We next consider a one-parameter family of examples where the F-theory base is taken to be a Hirzebruch threefold, $ B= \tilde{\F}_n$, in order to illustrate how the submatrix $M_{C_\text{na}(1\alpha)(1\beta)}$ can fail to be invertible for certain choices of characteristic data $K, S_7, S_9$. A Hirzebruch threefold is a generalization of a Hirzebruch surface $\F_n$ (i.e., a $\bP^1$ fibration over a $\bP^1$ base) in which the base of the $\bP^1$ fibration is taken to be $\bP^2$ instead of $\bP^1$. For our purposes, we simply need to know the intersection theory of $\tilde{\F}_n$. To make an analogy, note that Hirzebruch $\F_n$ has two independent classes $F, E$, where $F$ is the class of the $\bP^1$ fiber (meaning that $F$ is the divisor class of a point in the $\bP^1$ base) and $E$ is the class of the $\bP^1$ base (meaning that $E$ is the divisor class of a point in a $\bP^1$ fiber). These two classes have the following intersection properties:
	\begin{equation}
		F^2 =0,~~~~ F \cdot E = 1,~~~~ E^2 = -n, ~~~~ n \in \Z_{\geq 0}\,.
	\end{equation}
The threefold $\tilde{\F}_n$ similarly has two independent
divisor classes $D_2:= F, D_1:= E$ satisfying
	\begin{equation}
	\label{eqn:genFn}
		D_2^3 =0 ,~~~~ D_2^2 \cdot D_1 = 1,~~~~ D_2\cdot D_1^2 = -n,~~~~ D_1^3 = n^2,~~~~ n \in \Z_{\geq 0}\,.
	\end{equation}
Since $\tilde{\F}_n$ is a toric variety, the canonical class $K$ of $\tilde{\F}_n$ is as usual given by minus the sum of all divisors corresponding to one-dimensional cones of the toric fan:
	\begin{equation}
		K = - \sum D_{\alpha} =  - ( D_2 + D_2 + D_2 + D_1 + ( D_1 + n D_2) ) = - (3 +n) D_2 - 2 D_1\,.
	\end{equation}
(Note that the above results can easily be derived by adapting the pushforward technology described in \cref{pushapp} to the projectivization $\bP(\scV) \rightarrow B^{(2)}$ of a rank one vector bundle $\scV = \scL \oplus \scO_{B^{(2)}}$.) We expand the divisors $S_m$ in the basis $D_\alpha$, $S_m = s_{m\alpha} D_\alpha$. In terms of this basis of divisors, the constraints on the characteristic data for a good
$(\SU(2) \times \U(1))/ \Z_2$ model are then
\begin{align}
\label{eqn:Fnconstraints}
(s_{71}, s_{72}) & > (0, 0) \\
(s_{81}, s_{82}) & > (0, 0)  \nonumber\\
(8-2s_{71} -s_{81}, 12 + 4n-2s_{72} -s_{82}) & > (0, 0)  \nonumber\\
(2+s_{71} -s_{81},  3 + n+s_{72} -s_{82})  & > (0, 0) \,,  \nonumber
\end{align}
where a (Weil) divisor $S_m$ is effective if $s_{m\alpha} \geq 0$ and either
  $s_{m1} > 0$ or $s_{m2} > 0$.
Note that when $n > 3$ there is a non-Higgsable gauge factor on the
  divisor $D_2$, which may lead to an enhancement of the gauge
  symmetry in the class of universal  $(\SU(2) \times \U(1))/ \Z_2$
models.

In this family of examples, (minus) the height pairing divisor is then given by
	\begin{equation}
		W_{\bar 1 \bar 1} = \frac{1}{2} S_8 + 2 (K - S_7) = \left( -4 -2 s_{71} + \frac{s_{81}}{2} \right) D_1 + \left( -2 (3 + n) - 2 s_{72}+ \frac{s_{82}}{2} \right) D_2. 
	\end{equation}
Combining the above expression for $W_{\bar 1 \bar 1}$ with the $\tilde{\F}_n$ intersection numbers in \cref{eqn:genFn} we find 
	\begin{equation}
		[[M_{C_{\text{na}} (1\alpha)(1\beta)} ]]  = \scalebox{.8}{$\left(
\begin{array}{cc}
 n^2 (4 s_{71}-s_{81}+4)+n (-4 s_{72}+s_{82}-12) & n (-4 s_{71}+s_{81}-4)+4 s_{72}-s_{82}+12 \\
 n (-4 s_{71}+s_{81}-4)+4 s_{72}-s_{82}+12 & 4 s_{71}-s_{81}+8 \\
\end{array}
\right)$}
	\end{equation}
from which it follows 
\begin{equation}
		\det[[M_{C_\text{na} (1\alpha)(1\beta)} ]] =-\frac{1}{2} (4 (n+s_{72}+3)-s_{82}) (n (4 s_{71}-s_{81}+4)-4 (s_{72}+3)+s_{82})\,.
		\end{equation} 
Hence we see that if we choose the characteristic data such that 
	\begin{equation}
		s_{82} = \begin{cases} 12 + 4n + 4 s_{72}\\
		12 - 4n - 4n s_{71} + 4 s_{72} + n s_{81}
		\end{cases}
\label{eq:special-cases}
	\end{equation}
the matrix $M_{C_\text{na} (1\alpha)(1\beta)} $ will be singular.

We now turn our attention to some specific choices of $n \leq 3$ and
look in particular for flux compactifications on $(\SU(2) \times
\U(1))/ \Z_2$ models with characteristic data satisfying the
special conditions \cref{eq:special-cases} that lead to
non-invertibility of $M_{C_\text{na}(1\alpha)(1\beta) }$.

\paragraph{Example: $B = \tilde{\F}_0 \cong \bP^2 \times \bP^1$.} As a specific example, consider the case $n=0$. The matrix of intersection pairings with $W_{\bar 1 \bar 1}$ takes the form
\begin{equation}
        [[M_{C_\text{na}(1\alpha)(1\beta) }]] = -\frac{1}{2} \left(
\begin{array}{cc}
 0 &4 s_{72}-s_{82}+12 \\
 4 s_{72}-s_{82}+12 & 4 s_{71}-s_{81}+8 \\
\end{array}
\right)
   \end{equation}
with $\det[[M_{C_\text{na} (1\alpha)(1\beta)}]] =-(4
s_{72}-s_{82}+12)^2/4$.
This matrix is always invertible since $12 + 4s_{72}-s_{82} >
12-2s_{72}-s_{82} > 0$.
%While for generic characteristic data this matrix is invertible, we
%see that for characteristic data satisfying the fine-tuning
%	\begin{equation}
%		s_{82} = 12 + 4 s_{72}
%	\end{equation}
%the two off diagonal components vanish, leading to a singular matrix
%and hence we cannot solve the $\U(1)$ symmetry constraints by only
%eliminating flux background parameters of the form
%$\phi^{1\alpha}$. Nevertheless, 
%since the intersection matrix $M$ does
%not develop another null vector as a result of the above fine tuning
%it is in principle
%While we can thus always solve the constraint equations for $\phi_{1
%  \alpha}$, this in general leads to a rational solution; it is also
%possible to obtain a valid solution to the $\U(1)$ symmetry
%constraints by eliminating a distinctive parameter, giving a
%polynomial solution. One possible such solution is as follows:
%    \begin{align}
%        \phi^{1H_2} = (3 - s_{72}) \phi^{11} - (12 + 4 s_{72}) \phi^{12},~~~~ \phi^{22} = \frac{1}{2} ( -\phi^{00} + \phi^{01} - \phi^{02} - \phi^{11}),
%    \end{align}
%leading to the chiral multiplicity
%    \begin{align}
%        \Theta_{22} =\chi_{\textbf{2}_{-\tfrac{3}{2}}}= 12 s_{72} ( s_{72} + 3 ) ( \phi^{1H_1} + (s_{71} - 2) \phi^{11} + s_{81} \phi^{12}).
%    \end{align}
%\patrick{Need to explain that the above choice of fine tuning is
%  forbidden by constraints on the characteristic data.}\wati{included
%  this, rewrote.}
%
\paragraph{Example: $B = \tilde{\F}_3$.}
As another specific example consider the case $n=3$, for which 
	\begin{equation}
		  [[M_{C_\text{na}(1\alpha)(1\beta) }]] =-\frac{1}{2} \left(
\begin{array}{cc}
 3 (12 s_{71}-4 s_{72}-3 s_{81}+s_{82}) & -12 s_{71}+4 s_{72}+3 s_{81}-s_{82} \\
 -12 s_{71}+4 s_{72}+3 s_{81}-s_{82} & 4 s_{71}-s_{81}+8 \\
\end{array}
\right)\,.
\end{equation}
Generically the determinant of this matrix is non-vanishing, but there
 is a family of allowed choices of characteristic data for which the
 determinant vanishes.  For example, making the choices
		\begin{equation}
			S_7  =S_8 = -K~~\Leftrightarrow ~~
		(s_{71},s_{72}) = (s_{81},s_{82}) = (2,6)
	\end{equation}
leads to a singular matrix.  $M$ does not develop any additional null
vectors as a result of the above specialization,
 so it is possible to fully solve the $\U(1)$ symmetry conditions by
 eliminating distinctive parameters. In contrast to the previous
 specific example $B = \tilde{\F}_0$, this choice  for the
 characteristic data is not  forbidden by the constraints described at
 the beginning of this section and hence it appears that such a choice
 of parameters describes a consistent F-theory flux vacuum in which the
 $\U(1)$ gauge symmetry can be preserved in 4D in spite of $M_{C_\text{na}(1\alpha)(1\beta) }$ being singular; therefore, an explicit solution must include at least one nontrivial flux background other than $\phi^{1\beta}$ as in e.g. the polynomial solution \labelcref{eq:21-indirect}.
This provides an explicit example of the kind of situation mentioned
at the end of \cref{constraintsolutions}.

\subsubsection{Resolution independence of $M_\text{red}$}

We collect some evidence supporting the conjecture that $M_\text{red}$ (and hence $H_{2,2}^{\text{vert}}(X,\Z)$) is also resolution independent in the the more general setting of models with $\U(1)$ gauge factors. Here, we compare the resolution of the $(\SU(2) \times \U(1))/\Z_2$ model studied in the previous subsections, which we denote by $X_2$, and an alternative resolution $X'_3$ defined by the sequence of blowups
    \begin{align}
         X_3' \overset{(e_2,s_8|e_3)}{\longrightarrow} X_2'  \overset{(u,v|e_2)}{\longrightarrow} X_1' \overset{(u,s_4v + s_7w|e_1)}{\longrightarrow}  X_0
    \end{align}
where we follow the notation of \cite{KleversEtAlToric}.

For simplicity, let us specialize again to the case $B= \bP^3$, where we again denote the $\SU(2)$ gauge divisor by $S_8 = n H$. In a common basis $S_{0H},S_{HH},S_{H2},S_{1H},S_{11}$ we find
    \begin{align}
    \begin{split}
        M_\text{red}(X_2)&=\scalebox{.85}{$\left(
\begin{array}{ccccc}
 -4 & 1 & 0 & s_7 & s_7 (s_7-n) \\
 1 & 0 & 0 & 1 & -4 \\
 0 & 0 & -2 n & n & -4 n \\
 s_7 & 1 & n & -4 & n s_7-s_7^2-4 s_7+16 \\
 s_7 (s_7-n) & -4 & -4 n & n s_7-s_7^2-4 s_7+16 & -n^2 s_7+3 n s_7^2-4 n s_7-2 s_7^3+32 s_7-64 \\
\end{array}
\right)$}\\
        M_\text{red}(X_3') &=\scalebox{.85}{$ \left(
\begin{array}{ccccc}
 -4 & 1 & 0 & s_7 & s_7^2 \\
 1 & 0 & 0 & 1 & -4 \\
 0 & 0 & -2 n & n & -4 n \\
 s_7 & 1 & n & -4 & -n s_7-s_7^2-4 s_7+16 \\
 s_7^2 & -4 & -4 n & -n s_7-s_7^2-4 s_7+16 & -n^2 s_7-3 n s_7^2+12 n s_7-2 s_7^3+32 s_7-64 \\
\end{array}
\right)$}\,.
    \end{split}
    \end{align}
These two matrices are related by a change of basis
    \begin{align}
        M_\text{red}(X_2) = U^\transpose  M_\text{red}(X_3') U,~~~~ U = \left(
\begin{array}{ccccc}
 1 & 0 & 0 & 0 & 0 \\
 0 & 1 & 0 & 0 & 0 \\
 0 & 0 & 1 & 0 & 0 \\
 0 & 0 & 0 & 1 & 0 \\
 2 s_7 & n s_7 & -s_7 & 2 (4-s_7) & 1 \\
\end{array}
\right)\,.
    \end{align}
An analogous change of basis holds for other choices of base we have checked.

\subsection{$\SM$ model}
\label{321model}

One of the initial motivations of this paper was to analyze the 4D
massless chiral spectrum of the universal $\SM$ model of
\cite{Raghuram:2019efb}. This model is believed to be the most general
F-theory model with tuned
$\SM$ gauge symmetry and generic matter spectrum,
consisting of the representations appearing in the MSSM as well as
three additional ``exotic'' matter representations. The gauge sector
of the 4D $\cN=1$ supergravity describing this theory at low
energies admits three linearly independent families of anomaly free
combinations of chiral matter representations, so a flux
compactification of the $\SM$ F-theory model can be expected to yield
at most three independent combinations of chiral indices.

While the universal $\SM$ model can be defined by means of a
Weierstrass model, due to the presence of a $\U(1)$ gauge factor (much
like the $( \SU(2) \times \U(1))/\Z_2$ model), for the purpose
of computing a resolution it proves to be more convenient to start
with a construction of the singular F-theory background as a
hypersurface $X_0$ of an ambient $\bP^2$ bundle where the
elliptic fiber of $X_0$ is realized as a general cubic in the $\bP^2$ fibers of the ambient space. The hypersurface equation for $X_0$
can be obtained by unHiggsing the $\U(1)$ model with charge $q=4$
matter constructed in \cite{Raghuram34}. The characteristic data of
this model consists of the classes $K, \Sigma_2, \Sigma_3, Y$ where
$\Sigma_m$ is the gauge divisor class of the nonabelian factor
$\SU(m)$ and $Y =: W_{01}$ pulls back to the intersection of the
(rational) zero and generating sections of a resolution of $X_0$.  One
special subclass of these models are those with $Y = 0$, which have
only MSSM-type matter and have been studied using the toric $F_{11}$
fiber \cite{KleversEtAlToric, CveticEtAlQuadrillion}.

In a forthcoming publication \cite{Jefferson:2022yya}, following the approach
of this paper we present a complete analysis of the lattice of 4D
symmetry-preserving vertical fluxes and associated 4D chiral
multiplicities of the universal $\SM$ model over an arbitrary
threefold base $B$. Consistent with the emerging picture of the
landscape of F-theory vertical flux vacua described in this paper, one
of the main results of \cite{Jefferson:2022yya} is that all three families of
chiral matter representations can be realized in F-theory---further
evidence suggesting that the linear constraints on 4D chiral matter
multiplicities imposed by F-theory geometry coincide with the linear
constraints implied by 4D anomaly cancellation.

These results are found for arbitrary bases using the simplified analysis
associated with the restricted class of flux backgrounds $\phi^{1
  \alpha} = 0$, as well as for specific bases using the full analysis
of $M_{\text{red}}$ and keeping quantization conditions intact.

%%%%%%%%%%%%%%%%%%%%%%%%%%%%%%%%%%%%%%%%%%%%%%%%%%%%%%%%%%%%%%%%%%%%%%%%%%%%%%
%%%%%%%%%%%%%%%%%%%%%%%%%%%%%%%%%%%%%%%%%%%%%%%%%%%%%%%%%%%%%%%%%%%%%%%%%%%%%%
%%%%%%%%%%%%%%%%%%%%%%%%%%%%%%%%%%%%%%%%%%%%%%%%%%%%%%%%%%%%%%%%%%%%%%%%%%%%%%
\section{Conclusions and future directions}
\label{sec:conclusions}

\subsection{Summary of results}

We have described a novel and coherent approach for analyzing 4D
vertical flux compactifications in F-theory (that is, flux backgrounds
belonging to $H_{2,2}^{\text{vert}}(X,\Z)$) that preserve 4D local Lorentz and gauge symmetry. Our approach both offers
unique computational advantages, and sheds light on the
geometric nature of some of the resolution-invariant physics encoded
in the singularities of the F-theory background related to the 4D
massless chiral spectrum that has so far proven difficult to analyze
directly in the type IIB duality frame.

One of the key elements of our analysis is the integral lattice of
vertical 4-cycles, with symmetric bilinear form given by the symmetric matrix
of quadruple intersection numbers of the smooth CY fourfold $X$
interpreted as an intersection pairing on surfaces corresponding to the pairwise intersections of divisors.
 By Poincar\'e duality the nondegenerate part of
 this lattice is equivalent to the lattice
$\httv(X,\Z)$  of vertical flux backgrounds.
We conjecture that  this lattice, along with its nondegenerate inner
product given by the matrix $M_\text{red}$, is a resolution-invariant
structure for any singular elliptic CY fourfold encoding an F-theory
compactification.  This conjecture seems natural from the point of
view of type IIB string geometry, and is satisfied by a wide range of
explicit examples that we have considered in this paper. The
resolution-independence of $M_\text{red}$ also implies that the symmetric bilinear form $M$ on the formal space of intersection surfaces $S_{IJ}$,
which contains a nullspace corresponding to homologically trivial
surfaces, is resolution-independent. This further implies the existence of nontrivial
relations among the set of quadruple intersection numbers of the resolved
CY fourfold, even though these quadruple intersection numbers are not in general
resolution-independent (i.e., equivalent under an integral linear change of basis of the
divisors.) Understanding the geometry of this conjecture better and
its ramifications for the intersection structure of singular CY
fourfolds is an interesting problem for further investigation.

The resolution-independence of $M$ and $M_\text{red}$ is a sufficient
condition for the chiral matter content of a given class of F-theory
flux compactifications to be resolution-invariant, but as far as we
can tell is not directly provable from this geometric condition.
  The structure of $M_\text{red}$ we have studied here could be used to
  further study F-theory flux compactifications both in situations
  where the geometric gauge group remains unbroken in 4D by the fluxes, which is
  the primary focus here, as well for cases where the geometric gauge
  group is broken by vertical fluxes, which seems like an interesting
  direction for further research.
In cases where the flux does not break the gauge group, additional
constraints are placed on the fluxes.
Conceptually,  the approach we have taken here to studying such vacua
involves the interplay between two operations
applied to  the formal intersection pairing matrix $M$.
The first of
these two operations entails restricting to a sublattice of flux
backgrounds satisfying the constraints necessary and sufficient to
preserve 4D local Lorentz and gauge symmetry. This operation
is central to the standard approach used in much of the previous literature to analyze vertical flux
backgrounds in F-theory; our methods for computing intersection
numbers combined with the rigid structure of the elliptic fibration
enable us to write a formal expression for the elements of this
sublattice.
In contrast, the second of these operations involves taking the
quotient of the
lattice of vertical 4-cycles by homologically trivial cycles, the
result of which is the lattice of vertical homology classes
$H_{2,2}^{\text{vert}}(X,\Z)$ with inner product given by $M_\text{red}$.
While these two operations commute, studying the interplay between the
two different orders of these operations gives insight into the
structure of the connection between chiral matter and F-theory fluxes.

Computationally, the approach presented here is a synthesis of various techniques that have
appeared in the literature. Notably, we apply
recently developed algebrogeometric techniques for computing
intersection numbers of divisors in smooth elliptically fibered CY
varieties to classes of resolutions that can more easily be obtained
from geometric constructions of singular F-theory backgrounds in which
the elliptic fiber is realized as a general cubic in $\bP^2$---this procedure therefore provides a means to analyze a
broader class of F-theory constructions than is encompassed by the
usual Weierstrass model construction. Moreover, since these techniques
(like those used in \cite{Cveti__2014, LinWeigandG4})
express the intersection numbers of divisors in terms of triple
intersections of certain divisors in the base of the
elliptic fibration (i.e., the characteristic data), this approach can be used to conveniently organize
the landscape of F-theory vertical flux compactifications into
families of vacua with fixed gauge symmetry and matter representations
over an arbitrary base.

We have demonstrated the utility of this approach by analyzing
vertical flux backgrounds in numerous examples with simple gauge
symmetry group and generic matter.  We have also analyzed several examples of models with
a $\U(1)$ gauge factor, to illustrate the straightforward
generalization of these methods to models with
$\U(1)$ gauge factors; in principle a similar analysis is possible for
models with an arbitrary number of $\U(1)$ factors.
Of particular note among models with $\U(1)$
gauge factors is the universal $\SM$ model \cite{Raghuram:2019efb}
whose 4D massless chiral spectrum we analyze in a forthcoming
publication \cite{Jefferson:2022yya} using the methods described in this
paper. We find in all examples that the linear constraints on the
chiral matter multiplicities imposed by F-theory geometry exactly
match the 4D anomaly cancellation conditions, which suggests that it
may be possible to realize all anomaly-free combinations of 4D chiral
matter in F-theory, at least at the level of allowed linearly
independent families of the generic matter types for a given gauge
group.

\subsection{Future directions}

The existence of a resolution independent structure such as
$H_{2,2}^{\text{vert}}(X,\Z)$ is consistent with the expectation that the kinematics
of F-theory vacua are captured entirely in the singular elliptic CY
geometry encoded by the  axiodilaton over a general base in type IIB
string theory. To our
knowledge the conjecture that $H_{2,2}^{\text{vert}}(X,\Z)$ is resolution independent
has not previously been explored in the literature and would be useful
to prove rigorously, as this points to several potential future
avenues of investigation related to the physics of F-theory flux
compactifications:

\begin{itemize}
\item{} 
One of the outstanding challenges of F-theory is to give a
  complete and mathematically precise definition of this formulation
  of string theory.  While this is often done by taking a limit of
  M-theory (see, e.g., \cite{DenefLesHouches,Grimm:2010ks}), a more
  intrinsic definition may be possible from the point of view of type
  IIB string theory.  The progress made here in understanding the
  resolution-independent aspects of the singular elliptic
fourfold geometry $X_0$  may help in better understanding how matter
surfaces and chiral matter may be formulated and computed directly
from the type IIB point of view.
\item{} From a mathematical point of view, the
  resolution-independence of $M_\text{red}$ indicates that there is
  some intrinsic meaning to the lattice $\httv(X_0,\Z)$ of integral
  vertical surfaces and their intersection form on the singular
fourfold  geometry $X_0$.  This is particularly
  intriguing as the surfaces $S_{ij}$ most relevant for chiral matter
  in
F-theory flux
  vacua  project to trivial surfaces in the base and thus are hidden
  in the singular elliptic CY given by the F-theory Weierstrass
  model.  Developing a clear mathematical picture of this aspect of
  intersection theory of singular complex fourfolds poses an
  interesting challenge  on the mathematical side.
\item{} More concretely, the resolution-independence of $\httv(X,\Z)$
  suggests that there should be some way of  directly computing the
  intersection matrix $M_\text{red}$ without explicitly performing any
  blowups at all.  While many of the intersection numbers that form
  this matrix are resolution-independent, others are not, so
  identifying an organizing principle that would make possible a
  resolution-independent statement of the form of this matrix would be
  a significant step forward for the intrinsic understanding of
  singular F-theory flux vacua.
 
 \item{} We have focused in this paper on the intersection structure of CY fourfolds, which is relevant for 4D F-theory vacua. We may speculate, however, that the analogous homology group $H_{2,2}^{\text{vert}}(X^{(3)},\Z)$ for a CY threefold $X^{(3)}$ is also resolution-invariant. It may be possible to prove this resolution-invariance in a more direct and explicit way, and this may further shed light on the structure of $H_{2,2}^{\text{vert}}(X,\Z)$ for a CY fourfold. 
 
\item{} While in this paper we have focused on fluxes that preserve
  the geometric gauge group, so that the gauge invariance constraints
  $\Theta_{ I\alpha}= 0$ are all satisfied, it would be interesting to
  study flux vacua in which this condition is weakened.  In
  particular, as discussed in e.g.  \cite{Raghuram:2019efb}, while
  direct tuning of the Standard Model gauge group in F-theory is one
  way to get semi-realistic physics models, the bulk of the moduli
  space of CY fourfolds, and apparently the vast majority of the flux
  vacua, are dominated by bases that force large numbers
  of non-Higgsable gauge factors such as $E_6, E_7, E_8$ (see
  e.g. \cite{HalversonLongSungAlg, TaylorWangLandscape, TaylorWangVacua}); for these
  bases it is difficult or impossible to tune the Standard Model gauge
  group, but the group $\SM$ may be realized by turning on fluxes that
  break the gauge symmetry.  Some preliminary work in this direction
  for $E_8$ breaking was done in \cite{TianWangEString}, but the
  methods developed here may provide a very useful tool in more
  systematically pursuing this kind of analysis for flux breaking of
  non-Higgsable groups like $E_6$ and $E_7$.
\item{} Intriguingly, in all models we study we find that the
  symmetry-preserving fluxes appear to depend on resolution-invariant
  linear combinations of triple intersections of characteristic
  divisor classes in the base of the elliptic fibration, so that the
  minimum magnitude of the fluxes appears to be controlled by certain
  numbers of special points lying in the discriminant locus. Since the
  chiral indices themselves can be expressed as linear combinations of
  the fluxes, this suggests that the chiral indices in some sense ``count''
  special points in the F-theory base. One very clear illustration of
  this idea is given by $(4,6)$ points, as the symmetry-preserving
  fluxes in the models we have studied receive contributions
  proportional to the numbers of $(4,6)$ points in the
  base.% \cite{46}.

More generally, in many cases the multiplicity of chiral matter in
fixed representations is proportional to the number of points in the
base in the intersection of the associated matter curve and another
characteristic divisor, suggesting some explicit direct connection
between chiral matter fields and base geometry.
While these observations are not
  necessarily unique to our analysis, our computational methods have
  enabled us to survey a large enough number of examples to reveal
  patterns among different families of models, see, e.g., the
  expressions for the fluxes in \cref{tab:fluxtable}. Along these
  lines, it would be interesting and quite useful to understand how to
  associate the chiral indices to certain types of singularities
  visible directly from the F-theory limit, and it is possible that the
  resolution independence of $H_{2,2}^{\text{vert}}(X,\Z)$ will prove useful in this capacity.
\item{} Another direction in which this work could naturally be
  extended  involves the question of whether or not all families of
  anomaly-free matter can be realized in F-theory.  In all the
  examples we have studied, of simple gauge groups and groups with a
  single $\U(1)$ factor, we have found by considering both generic and
  specific choices of base that F-theory imposes no linear
  constraints on the chiral matter multiplicities beyond those
  expected by anomaly cancellation. This has implications for the
  analysis of the ``swampland,'' suggesting that at the level of
  linear families of matter F-theory naturally realizes the full set of
  possibilities that are consistent with low-energy constraints.  It
  would be good to check whether this continues to hold for more
  complicated models with more abelian factors, or even to find some
  general principle based on the resolution-invariance of
  $M_\text{red}$ that can match the rank of this intersection form
  with the number of expected families of chiral matter.
\item{} At finer level of detail, there are questions related to
  the quantization and multiplicities of chiral matter that could be
  explored further both mathematically and through more concrete
  physics models.
As we have discussed here (see in particular \cref{sec:quantization-1}), the quantization conditions on matter from
purely vertical fluxes may be weakened when the other components of
middle homology are incorporated and/or fractional vertical flux
coefficients are included, since by Poincar\'{e} duality there
should be in principle cycles with a single unit of flux through any
primitive matter surface, even though in general the determinant 
of $M_\text{red}$ has magnitude greater than 1.  Further analysis of the geometry and
associated physics of these kinds of questions could help elucidate
more detailed swampland type questions regarding which precise
multiplicities of matter can arise in given 4D supergravity models
realized from F-theory.
%
%\item{} \wati{added this and comment in 6.4.2}
%As noted in  a simple $\SU(5)$ example in
%\cref{eq:flux-quantization-5}, the condition that $G$ is self-dual
%seems in some cases to place upper limits on the  surfaces that can
%support certain gauge factors with nontrivial chiral matter content.
%This would be interesting to investigate and understand
%further in more general situations.
%

\item{} While in this paper we have focused on chiral matter in 4D
  theories, a full understanding of the low-energy physics of a given
  F-theory compactification also requires understanding the
  vector-like matter.  Though vector-like matter multiplicities are
  subtler than chiral matter, some recent progress has been made
  in this direction \cite{Bies:2014sra,Bies:2017fam,Bies:2020gvf,Bies:2021nje,Bies:2021xfh}.  It would be interesting to investigate
  whether there is resolution-independent structure, analogous to that
  studied here, that can be used
  to describe such vector-like multiplicities.

\item{} Finally, we note that for a pair of CY fourfolds related by
  mirror symmetry, their respective vertical and horizontal
  cohomologies are isomorphic \cite{Greene:1993vm,Braun:2014xka}. In this paper we
  restricted our focus to vertical flux backgrounds and did not
  attempt to explore the space of horizontal fluxes associated to a
  given 4D F-theory model. However, a more complete analysis of
  F-theory flux compactifications generically requires horizontal
  fluxes to be included in the picture. 
If it turns out that the
  vertical homology of a given CY fourfold is indeed resolution
  invariant, this would suggest that the corresponding horizontal
  homology of the mirror CY fourfold is also an invariant structure
  across certain regions of moduli space and may provide a strategy
  for studying horizontal fluxes, which have received comparatively
  less attention in the literature, and which may also give insight
  into the quantization issues mentioned above.
The intersection form on the horizontal part of
$H^4 (X,\Z)$ also plays an important role in recent work  that uses asymptotic
Hodge theory to describe string vacua in large field limits
\cite{Grimm:2019bey,Grimm:2020ouv}, and it would be
interesting to understand if similar resolution-independent
structure is relevant there.
	\end{itemize}

We thank Mirjam Cveti\v{c}, Mboyo Esole, Thomas Grimm, Jonathan
Heckman, David Morrison, Ling Lin, Shing Yan (Kobe) Li, Sakura Schafer-Nameki, and Timo Weigand for discussions and for
comments on earlier versions of this manuscript. This work was
supported by DOE grant DE-SC00012567. AT was also supported in part by
the Tushar Shah and Sara Zion fellowship and DOE (HEP) Award
DE-SC0013528. WT would like to thank the Aspen Center for Physics
(ACP) for hospitality during part of this work. The authors would all
like to thank the Witwatersrand (Wits) rural facility and the MIT
International Science and Technology Initiatives (MISTI)
MIT--Africa--Imperial College seed fund program for hospitality and
support during some stages of this project.

\appendix

\section{4D anomaly cancellation}

\label{4Danomalyreview}

We review anomaly cancellation in four dimensions, following primarily \cite{Cvetic:2012xn}.

Consider a 4D $\cN = 1$ theory with gauge algebra of the form
    \begin{equation}
        \ag = \bigoplus_{s} \ag_s \oplus \bigoplus_{\bar a }^{} \au(1)_{\bar a}\,,
    \end{equation}
with the $\ag_s$ being simple nonabelian gauge algebra factors, indexed by $s$, and with $\mathfrak{u}(1)_{\bar a}$ being abelian gauge factors, indexed by $\bar a$. As we are only considering local gauge anomalies, we need not specify the global structure of the gauge group $\mathsf G$ here.
% \textcolor{red}{We use lowercase Greek letters to index the nonabelian simple gauge factors and lowercase Roman letters to index the abelian factors.}
Matter in chiral multiplets transforms in irreducible representations of the form
    \begin{equation}
        \sfr = \bigotimes_{s}^{} \sfr_s \otimes \bigotimes_{\bar a}^{} q_{\sfr, \bar a} =: (\sfr_1, \sfr_2, \dots)_{(q_{\sfr,1},q_{\sfr,2},\dots)}\,.
    \end{equation}
% We use $\multDim{\sfr_s}$, $\multDim{\sfr_s, q_a}$, $\multDim{q_a}$, $\multDim{q_a, q_b, q_c}$ to respectively denote the multiplicity with which a left-chiral Weyl fermion transforming in representation $\sfr_s$, $\sfr_s \otimes q_a$, $q_a$, $q_a \otimes q_b \otimes q_c$ of gauge factor $\ag_s$, $\ag_s \times \au(1)_a$, $\au(1)_a$, $\au(1)_a \times \au(1)_b \times \au(1)_c$ occurs in the 4D spectrum. These multiplicities are defined to account for the dimensions of the representations of other gauge factors in all representations $\tilde{\sfr}$ of the full gauge algebra containing the relevant representation. Thus, for example,
%     \begin{equation}
%         \multDim{\sfr_s} = \sum_{\sfr \supset \sfr_s} \mult{\sfr} \Big(\prod_{\lambda \ne s} \dim \sfr_\lambda\Big)\,.
%     \end{equation}

In four dimensions, the gauge and gauge--gravitational mixed anomalies have contributions from chiral Weyl fermions via the familiar triangle diagrams, and additionally from Green--Schwarz tree-level diagrams exchanging $\U(1)$-gauged scalar axion fields $\rho_\alpha$. The conditions for the anomaly to cancel reduce to\footnote{The symbol `$\text{tr}_{\mathsf{f}}$' denotes a trace taken over the field strength $F_s$ transforming in the fundamental (i.e. defining) representation $\mathsf{f}$ of the gauge factor $\mathfrak{g}_s$ (see \cref{SGS}), and similarly for `$\text{tr}_{\sfr_s}$'. Note also that the traces over products of field strengths can be expressed as, e.g., $\text{tr}_{\sfr_s} \, F_s^p = \sum_{w \in \sfr_s} \left(\sum_{i_s} \varphi^{i_s} w_{i_s}\right)^p$.}\footnote{Note that in F-theory flux compactifications with abelian gauge factors associated to rational sections $\hat D_{\bar a}$, we must impose $\Theta_{\alpha \bar a} =0$ in order to ensure that the associated abelian gauge symmetry is not rendered massive in the low energy effective 4D theory by the St\"uckelberg mechanism.}
    \begin{equation}
        \label{eq:ACeqs}
        \begin{aligned}
            % 0 &= \sum_{\sfr_s} \multDim{\sfr_s} E_{\sfr_s}\,, \\
            % b_s^\alpha \Theta_{\alpha a} &= 2 \lambda_s \sum_{\sfr_s, q_a} \multDim{\sfr_s, q_a} A_{\sfr_s} q_a\,, \\
            % a^\alpha \Theta_{\alpha a} &= -\frac{1}{4} \sum_{q_a} \multDim{q_a} q_a\,, \\
            % b_{a b}^\alpha \Theta_{\alpha c} + b_{a c}^\alpha \Theta_{\alpha b} + b_{b c}^\alpha \Theta_{\alpha a} &= 2 \sum_{q_a, q_b, q_c} \multDim{q_a, q_b, q_c} q_a q_b q_c\,,
            0 &= \sum_{\sfr_s} \Big(\prod_{s' \ne s} \dim \sfr_{s'}\Big) \frac{\text{tr}_{\sfr_s}\, F_{s}^3}{\text{tr}_{\mathsf f}\, F_s^3}\,, \\
            b_s^\alpha \Theta_{\alpha \bar a} &= 2 \lambda_s \sum_{\sfr_s} \Big(\prod_{s' \ne s} \dim \sfr_{s'}\Big) w^{\sfr_s}_{\bar a}\, \frac{\text{tr}_{\sfr_s} \, F_s^2}{\text{tr}_{\mathsf{f}}\, F_s^2}\,, \\
            a^\alpha \Theta_{\alpha \bar a} &= -\frac{1}{4} \sum_{\sfr} (\dim \sfr) w^{\sfr}_{\bar a}\,, \\
            b_{\bar a \bar b}^\alpha \Theta_{\alpha \bar c} + b_{\bar a \bar c}^\alpha \Theta_{\alpha \bar b} + b_{\bar b \bar c}^\alpha \Theta_{\alpha \bar a} &= 2 \sum_{\sfr} (\dim \sfr) w^{\sfr}_{\bar a} w^{\sfr}_{\bar b} w^{\sfr}_{\bar c}\,,
        \end{aligned}
    \end{equation}
where $\Theta_{\alpha \bar a}$ is the gauging of the axion $\rho_\alpha$ under the abelian vector $A^a$ associated with the gauge factor $\au(1)_{\bar a}$,
    \begin{equation}
        \bigdiff\rho_\alpha = \diff\rho_\alpha + \Theta_{\alpha \bar a} A^{\bar a}\,,
    \end{equation}
while the $a^\alpha, b_s^\alpha, b_{\bar a \bar b}^\alpha$ are anomaly coefficients that specify the Green--Schwarz couplings of $\rho_\alpha$,
    \begin{equation}
    \label{SGS}
        S_\text{GS} = -\frac{1}{8} \int \frac{2}{\lambda_s} b_s^\alpha \rho_\alpha \trace_{\mathsf f}(F_s \wedge F_s) + 2 b_{\bar a \bar b}^\alpha \rho_\alpha F_{\bar a} \wedge F_{\bar b} - \frac{1}{2} a^\alpha \rho_\alpha \trace(R \wedge R)\,.
    \end{equation}
Here, $\lambda_s = 2 c^\vee_s / A_{\Adj_s}$, with $c^\vee_s$ the dual Coxeter number of $\ag_s$.
% , and the group theory coefficients $A_{\sfr}$ and $E_{\sfr}$ are defined for a given representation $\sfr$ by
%     \begin{equation}
%         \trace_{\sfr} F^2 = A_{\sfr} \trace F^2\,, \quad \trace_{\sfr} F^3 = E_{\sfr} \trace F^3\,.
%     \end{equation}
% In these expressions, $\trace_{\sfr}$ denotes a trace in the representation $\sfr$, while $\trace$ indicates a trace in the fundamental representation.

\section{Tensor structures in intersection products of divisors}

\label{intersection}

\label{ellipticintersection}

This appendix is an overview of various tensor structures
characterizing the pushforwards of intersection numbers of divisors in
a resolution $X$ of a singular elliptically fibered CY variety
defining an F-theory model with gauge group $\sfG$. Although the
structures we describe in this appendix are to our knowledge not
rigorously proven, we expect them to apply for the full class of CY
fourfolds we describe in this paper (see below for a precise statement
of our assumptions about the type of CY manifold for which these
structures apply). Furthermore, these structures have been verified in
a vast number of examples of intersection products for resolutions of
F-theory models studied in the literature. We
include relevant references where appropriate; however, since much of this
structure has been described in numerous places in the literature, we do not
attempt to be exhaustive. Note also that the explicit computations of intersection numbers that we carry out in
this paper using the techniques of \cref{pushapp} match with these
general tensor structures in all cases we have computed where the
structure is known, and appear to extend to other cases (e.g., 4-Cartan index
intersection numbers) where the general structure is not understood.

Let $X$ be a resolution of a singular elliptic CY $n$-fold $X_0$ with canonical projection
\begin{align}
\label{canproj}
\pi \colon X \rightarrow B\,.
\end{align}
For simplicity we assume that $X_0$ has a rational zero section and an additional rational section, although we stress that much of the structure described here is generalizable to cases in which there are an arbitrary number of rational sections. Following the Shioda--Tate--Wazir formula \cite{shioda1972}, we use the basis of divisors $\hat{D}_I$, where $I =0, a,\alpha,i_s$ respectively labels a choice of rational zero section, additional rational sections, pullbacks of divisors in the base, and Cartan divisors (i.e., the irreducible components of the pullback to $X$ of the irreducible components $\Sigma_s$ of the discriminant locus $\{\Delta = 0\} \subset B$). For simplicity, we assume that the intersections $\Sigma_s \cap \Sigma_{s'}$ are pairwise transverse. We refer to the divisor classes
    \begin{align}
    \label{chardata}
        \Sigma_s,~~\Sigma_{s'}, ~~ \dots~~K = -c_1(B),~~ W_{0a} := \pi_* (\hat D_0 \cdot \hat D_a)
    \end{align}
as the characteristic data of $X$. Note that $\hat D$ denotes a divisor class in the Chow ring of $X$, whereas $D$ denotes a divisor class in the Chow ring of $B$ (and similarly for higher codimension). Unless the distinction is otherwise unclear from the context, we generically use the same symbol for both a divisor and its class in the Chow ring, and moreover we typically do not explicitly indicate pullback maps. Repeated indices are summed over when one index is raised and the other is lowered.

A convenient method for evaluating intersection numbers in the Chow ring of a smooth elliptic variety $ X$ is to compute the pushforward $\pi_*$ of the intersection product to the Chow ring of $B$, where $\pi$ is defined in \cref{canproj}. This method is particularly useful because the projection formula (see, e.g., \cite{fulton})
        \begin{align}
        \label{projform}
            f_* ( f^*(C) \cdot \hat D ) = C \cdot f_*(\hat D)
        \end{align}
for classes\footnote{Note that $\hat D$ is a divisor class in the Chow ring of the space in the preimage of the map $f$.} $C,\hat D$ and $f$ an appropriate map implies that intersection products involving divisors $\hat D_\alpha$ that are the pullbacks of divisors $D_\alpha$ in the base inherit the intersection structure of $D_\alpha \subset B$. Hence, we can anticipate the pushforwards of intersection products to exhibit the general structure (for concreteness assume that $X$ is a CY fourfold, i.e., $n=4$)
    \begin{align}
    \begin{split}
    \label{Wint}
        \hat D_I \cdot \hat D_J \cdot \hat D_K \cdot \hat D_L &=W_{IJKL} := W_{IJKL}^{\alpha \beta \gamma} D_\alpha  \cdot D_\beta \cdot D_\gamma\\
        \hat D_\alpha \cdot \hat D_J \cdot \hat D_K \cdot \hat D_L&= W_{JKL} \cdot D_\alpha :=W_{JKL}^{\beta \gamma} D_\alpha \cdot D_\beta \cdot D_\gamma \\
    \hat D_\alpha \cdot \hat D_\beta \cdot \hat D_K \cdot \hat D_L&= W_{KL} \cdot D_\alpha \cdot D_\beta :=W_{KL}^{ \gamma} D_\alpha  \cdot D_\beta \cdot D_\gamma \\
        \hat D_\alpha \cdot \hat D_\beta \cdot \hat D_\gamma \cdot \hat D_L &= W_{L}  \cdot D_\alpha \cdot D_\beta \cdot D_\gamma:=W_{L} D_\alpha  \cdot D_\beta \cdot D_\gamma\,,
    \end{split}
    \end{align}
where $W_{IJKL},W_{JKL},W_{KL},W_{L}$ can be expressed as intersection products in the Chow ring of $B$---note that these intersection products only involve the characteristic data $K, \Sigma_s, W_{01}$.

To make contact with low-energy effective field theoretic descriptions of the low-energy effective field theory describing M-theory compactified on $X$, we can change our basis of divisors to the ``gauge'' basis $\hat D_{\bar I} = \sigma_{\bar I}^J \hat D_J$ defined by \cite{Grimm:2011sk,Bonetti:2011mw}
    \begin{align}
    \begin{split}
    \label{physbasis}
        \sigma_{ \bar 0}^{ I} &=(1 ,0,- \frac{1}{2} W_{00}^{\alpha},0)^{ I} \\
         \sigma_{\bar 1}^{ I} &=(-1,1,W_{00}^\alpha - W_{01}^{\alpha}, -W_{1|j_{s'}} W^{j_{s'}| i_s} )^{ I}\\
         \sigma_{\bar J}^{I}&=\delta_{\bar J}^I~~~~\text{for $\bar J \ne 0,1$}\,.
    \end{split}
    \end{align}
By linearity, the fluxes in the gauge basis are then linear combinations of the fluxes in the standard geometric basis. For example, $\hat D_{\bar 1} = \sigma_{\bar 1}^I \hat D_I$ is the image of the Shioda map described in \cite{MorrisonParkU1}. Using
\begin{align}
    \begin{split}
        (\sigma^{-1})_0^{\bar I} &= ( 1,0, \frac{1}{2} W_{00}^\alpha , 0 )^{\bar I} \\
        (\sigma^{-1})_{1}^{\bar I } &= (1,1, -\frac{1}{2}W_{00}^{\alpha} + W_{01}^\alpha,  W_{1|j_{s'}} W^{j_{s'}| i_s})^{\bar I}\\
        (\sigma^{-1})_{J}^{\bar I} &= \delta_{J}^{\bar I} ~~~~\text{for $J \ne 0,1$}\,.
    \end{split}
    \end{align}
one can invert the above linear transformation:
    \begin{align}
        (\sigma^{-1})_{J}^{\bar I} \sigma_{\bar I}^{K} = \delta_{J}^{K},~~~~ \sigma^{J}_{\bar K} (\sigma^{-1})_{J}^{\bar I} = \delta_{\bar K}^{\bar I}\,.
    \end{align}
We sometimes make the abuse of notation
    \begin{align}
    \label{allU(1)s}
        	\bar I = i = (\bar a,i_s)
    \end{align}
to collectively denote all abelian gauge indices as opposed to distinguishing between pure U(1) and nonabelian Cartan indices (note $\hat D_{\bar a} := \sigma_{\bar a}^{I} \hat D_I$, where for our purposes we need only consider a single generating section, i.e., $a=1$.)

Throughout the paper we make extensive use of the fact that intersection numbers of the form $W_{IJKL}$ where $I,J,K,L \ne 0,a,\alpha$ exhibit special tensor structures. (Note that various aspects of the structure of the pushforwards $W_{IJKL}$ have been pointed out and used extensively in the string theory literature, see, e.g., Section 3 of \cite{Cvetic:2012xn} and references therein.) In particular, it is useful to grade the intersection numbers $W_{IJKL}$ by their number of nonabelian Cartan indices $I=i_s$, which corresponds to the number of nonabelian Cartan divisors $\hat D_{i_s}$ appearing in the expression $W_{IJKL} = \hat D_I \cdot \hat D_J \cdot \hat D_K \cdot \hat D_L$. We now summarize some features of these tensor structures:

    \begin{itemize}

        \item \underline{Four nonabelian Cartan indices}. Without
          introducing specific assumptions about the chiral matter
          content of the 4D theory engineered by $X$, a general
          characterization of the four Cartan index tensor structure
          is to our knowledge presently unknown. It may be
          possible to combine assumptions about 4D anomaly cancellation
          and the existence of particular matter surfaces $S_{\sfr}$
          to predict a subset of the $W_{ijkl}$.
%For a specific resolution of the singularities associated with a fixed
%gauge group, these indices can in many situations
%be computed using the methods of \cref{pushapp}.

\item \underline{Three nonabelian Cartan indices}. Intersection
  products involving three nonabelian Cartan divisors in elliptically
  fibered CY manifolds have been the subject of a great deal of string
  theory literature. For example, over a twofold base $B^{(2)}$, these
  intersection products are intersection numbers of a CY threefold $X^{(3)}$, and they
  encode various aspects of the kinematics of 5D M-theory
  compactifications on $X^{(3)}$ dual to the Coulomb branch of 6D F-theory
  compactifications on $X^{(3)} \times S^1$. In our case (i.e., CY
  fourfolds $X$), intersection products involving three Cartan divisors
  are divisor classes and hence push forward to divisor classes in $B$, but nonetheless carry much of the same tensor structure as in
  the CY threefold case, as we now demonstrate.

Combining various results on the relationship between these
intersection numbers and the corresponding low-energy effective
physics (see e.g. \cite{Grimm:2013oga,Bonetti:2011mw,Grimm:2015zea})
with known results on the geometry of matter curves
$C_{\sfr}$ for local matter arising from transverse
intersections of divisors in $B$ \cite{Grassi:2011hq}, we can
infer\footnote{For instance, the most direct method to derive
  \cref{3Cnumbers} is to match 5D 1-loop CS terms to intersection
  products involving three nonabelian Cartan divisors as in
  \cref{5Dmatch} and below, and then to identify the ``coefficients''
  $C_{\sfR}$ as the classes of matter curves described in
  \cite{Grassi:2011hq}.} the following structure:
            \begin{align}
            \begin{split}
            \label{3Cnumbers}
                    W_{i_sj_{s'}k_{s''}} &= \rho_{i_s j_{s'} k_{s''}}^{\sfR_{s s'}} C_{\sfR_{ss'}}\\
                    C_{\sfR_{ss'}} &=\Sigma_s  \cdot \Delta_{\sfR_{ss'}} =\Sigma_s \cdot ( a_{\sfR_{ss'}} K + b_{\sfR_{ss'}} \Sigma_{s} + c_{\sfR_{ss'}} \Sigma_{s'})\,,
            \end{split}
            \end{align}
where $a_{\sfR_{ss'}},b_{\sfR_{ss'}},c_{\sfR_{ss'}}
\in \tfrac{1}{2}\Z$ and $\sfR_{ss'}$ denotes a hypermultiplet
representation transforming under the product gauge group $\sfG_s
\times \sfG_{s'}$ (by convention $\sfR_{ss}$ only transforms
under $\sfG_s$).
For a fixed gauge group and any of the generic matter types, the
coefficients
$a_{\sfR_{ss'}},b_{\sfR_{ss'}},c_{\sfR_{ss'}}$ can
be computed directly from the associated universal Weierstrass
model.

In the above expressions we sum over 5D $\cN=1$ hypermultiplet\footnote{In this notation, $\sfR = \sfr \oplus \sfr^*$ is a quaternionic representation, and hence sums over representations do not distinguish between a complex representation $\sfr$ and its conjugate $\sfr^*$.} representations $\sfR_{ss'}$, and $\rho^{\sfR_{ss'}}_{i_sj_{s'}k_{s''}}$ are triple intersection numbers that can be extracted from the pure Cartan expression for the prepotential $\cF$ of a 5D M-theory compactification (see \cite{Intriligator:1997pq}):
\begin{align}
\begin{split}
\label{IMSF}
        6 \cF_\text{Cartan}&= \varphi^{i_s} \varphi^{j_{s'}} \varphi^{k_{s''}} W_{i_sj_{s'}k_{s''}}  = -\frac{1}{2}   \sum_{\sfR_{ss'} } C_{\sfR_{ss'}}  \sum_{w \in \sfR_{ss'}} \sign(\varphi \cdot w) (w_{i_{s''}} \varphi^{i_{s''}})^3 \,.
\end{split}
\end{align}
For example, $\Delta_{\textbf{adj}_s} =  (\Sigma_s +K) /2$ and
$C_{\sfR_{ss'}} = \Sigma_{s} \cdot \Sigma_{s'}$. In the above
expression $w_l$ are the components of the weight $w$ in the basis of
fundamental weights, i.e., the basis canonically dual to simple
coroots $\alpha_i^{\vee}$ satisfying $\alpha_i^{\vee} \cdot w = w_i$. Note that $W_{i_s j_{s'} k_{s''}} \propto \partial_{\phi^{i_s}} \partial_{\phi^{j_{s'}}} \partial_{\phi^{k_{s''}}} \cF_\text{Cartan}$ is manifestly resolution-dependent, since the right hand side of \cref{IMSF} depends explicitly on $\text{sign}(\varphi \cdot w)$, which in turn depends on the specific phase of the Coulomb branch to which the intersection numbers correspond.

%The above formula implies
%   \begin{align}
%       \rho^{\sfR}_{ijk} = -\frac{3}{  \delta_{ijk} }  \sum_{w \in \sfR} \text{sgn}(w) w_i w_j w_k ,~~~~ \delta_{ijk} \equiv \begin{cases} 1 ~&\text{if $i \ne j \ne k$}\\
%       2 ~&\text{if $i = j \lor j= k \lor k = i$}\\
%       3 ~ &\text{if $i = j = k$}
%       \end{cases}.
%   \end{align}

    \item \underline{Two nonabelian Cartan indices}. For intersection
      numbers of the form $W_{i_sj_{s'}\alpha \beta} = W_{i_s j_{s'}}
      \cdot D_\alpha \cdot D_\beta$, we have the
      resolution-independent structure \cite{Cvetic:2012xn}
        \begin{align}
        \begin{split}
        \label{CartanW}
            W_{i_s j_{s'}} = W_{i_s| j_{s'}} \Sigma_s=-\delta_{ss'} \kappa_{ij}^{(s)} \Sigma_s \,,
        \end{split}
        \end{align}
    where $W_{i_s|j_{s'}}$ is (minus) the inverse Killing form\footnote{The tensor $\kappa_{ij}^{(s)} $ is the inverse of the metric tensor of the simple Lie algebra $\mathfrak{g}_{s} \subset \mathfrak{g}_\text{na} = \oplus_s \mathfrak{g}_{s}$ and appears, e.g., in the 5D scalar kinetic term $\int \kappa_{ij}^{(s)} \diff\varphi^{i_s} \wedge * \diff\varphi^{j_s} $.}  associated to $\sfG_\text{na}$ and satisfies the relation
        \begin{align}
            W_{i_s |j_{s'}} W^{j_{s'}| k_{s''}} = \delta_{i_s}^{k_{s''}}\,.
        \end{align}

    \item \underline{One or fewer nonabelian Cartan indices}. For intersection numbers of the form $W_{i_s\alpha \beta \gamma} = W_{i_s} D_\alpha \cdot D_\beta \cdot D_\gamma$ the tensor structure trivializes:
        \begin{align}
        \begin{split}
            W_{i_s} = 0\,.
        \end{split}
        \end{align}
    \end{itemize}
Understanding these tensor structures turns out to be crucial for characterizing the general form of the constrained fluxes for F-theory models with gauge group $\sfG = (\U(1) \times \sfG_\text{na})/\Gamma$. In particular, the fact that $W_{i_s|j_t}$ can be inverted is crucial to the calculation in \cref{fluxderivation}.

Note that we also make liberal use of the formal identity
    \begin{align}
        W_{i_s JKL} = \Sigma_{s} \cdot W_{JKL|i_s}\,,
    \end{align}
of which the definition $W_{i_sj_{s'}} = \Sigma_s W_{i_s|j_{s'}}$ in \cref{CartanW} is a special case. Finally, when the zero section $\hat D_0$ (or any section, for that matter) is holomorphic, we have
    \begin{align}
    \begin{split}
        W_{0000} &= K^3\,, \\
        W_{000\alpha} &= K^2 \cdot D_\alpha\,, \\
        W_{00\alpha \beta} &=K \cdot D_\alpha \cdot D_\beta\,, \\
        W_{0 \alpha \beta \gamma} &= D_\alpha \cdot D_\beta \cdot D_\gamma\,.
    \end{split}
    \end{align}

\section{Solution to the symmetry constraints}
\label{fluxderivation}

In this appendix we show that, given a resolution $X$ of a singular fourfold $X_0$ corresponding to a 4D F-theory compactification with gauge symmetry $\sfG = (\U(1) \times \sfG_\text{na})/\Gamma$, flux backgrounds $G$ that preserve the full Poincar\'{e} and gauge symmetry in the F-theory limit can typically be parametrized entirely by distinctive parameters $\phi^{\hat I \hat J}$, where $\hat I, \hat J = 0,1, i_s$.

Our starting point is the unconstrained expression for a flux $\Theta_{IJ}$, which can be split into terms that depend separately on distinctive and non-distinctive parameters as follows:
    \begin{align}
        \label{eq:gentheta}
        \Theta_{ I  J} &=\Theta_{ I  J}^\text{d} +\phi^{0\beta} W_{0 \beta  I  J} +\phi^{1 \beta} W_{1 \beta  I  J} + \phi^{\beta \gamma} W_{\beta \gamma  I  J}  + \phi^{\beta k_{s''}} W_{\beta k_{s''}  I  J}\,.
    \end{align}
The term $\Theta^{\text{d}}_{IJ}$ in the above expression only depends explicitly on distinctive parameters.

Our goal is to explicitly constrain the above expression to lie a subspace in which the symmetry constraints \labelcref{eq:Poincare,eq:gauge} are satisfied, by solving for the non-distinctive parameters in terms of the distinctive parameters. To see how this works,
we separate the local Lorentz and gauge symmetry constraints\footnote{For generic characteristic data, the set of fluxes $\Theta_{I\alpha}$ are linearly independent. However, for some special choices of characteristic data it is possible for certain linear combinations of the fluxes to vanish, say $\nu^{\alpha I} \Theta_{I\alpha} =0$, indicating the existence of additional null vectors for the intersection matrix $M$. In such cases, some of the constraints become redundant; in practice we drop these redundant constraints so that we only solve a linearly independent subset $\Theta_{I\alpha} =0$.}  into distinctive and non-distinctive contributions, leading to the following linear system:\footnote{Note that $W_{\alpha \beta \gamma \delta} = 0$ by definition. Moreover, we assume the (unproven) property $W_{\alpha \beta \gamma i_s} = W_{ \alpha \beta 0 i_s} =0$ for the smooth fourfolds $X$ we consider.}
\begin{align}
\label{fluxconstraints1}
    0 &= \Theta_{0 \alpha}= \Theta_{0 \alpha}^\text{d} + \phi^{0\beta} W_{0 \beta 0 \alpha} +\phi^{1 \beta} W_{1 \beta 0 \alpha} + \phi^{\beta \gamma} W_{\beta \gamma 0\alpha }\,,   \\
    \label{fluxconstraints2}
    0&= \Theta_{1\alpha} = \Theta_{1\alpha}^{\text{d}} +\phi^{0\beta} W_{0 \beta 1 \alpha} +\phi^{1 \beta} W_{1 \beta 1\alpha} + \phi^{\beta \gamma} W_{\beta \gamma 1\alpha }  + \phi^{\beta j_{s'}} W_{\beta j_{s'} 1 \alpha}\,, \\
    \label{fluxconstraints3}
    0 &= \Theta_{\alpha \beta} = \Theta_{\alpha \beta}^{\text{d}}+\phi^{0\gamma} W_{0 \gamma  \alpha \beta} +\phi^{1 \gamma} W_{1 \gamma  \alpha \beta}\,,  \\
    \label{fluxconstraints4}
    0&=\Theta_{\alpha i_s} = \Theta_{\alpha i_s}^{\text{d}} +\phi^{1 \beta} W_{1 \beta \alpha i_s}  + \phi^{\beta j_{s'}} W_{\beta j_{s'}  \alpha i_s}\,.
\end{align}
There are $3+ 2 h^{1,1}(B) + h^{1,1}(B)^2 + 2  \rk\sfG_\text{na} + h^{1,1}(B)  \rk\sfG_\text{na}+ ( \rk\sfG_\text{na})^2$ parameters $\phi^{IJ}$ in total. Since the number of independent constraints is $2h^{1,1}(B) + h^{1,1}(B)^2 + \rk\sfG_\text{na}  h^{1,1}(B)$, subtracting this number from the number of independent parameters $\phi^{IJ}$ generically leaves behind $3  + 2 \rk\mathsf  G_\text{na} + (\rk\sfG_\text{na} )^2$ independent parameters, precisely equal to the number of distinctive parameters $\phi^{\hat I \hat J}$. Thus our task can be reduced to solving the above linear system in such a way that the non-distinctive terms in \cref{eq:gentheta} are replaced by linear combinations of the terms $\Theta_{0\alpha}^{\text{d}},\Theta_{1\alpha}^{\text{d}},\Theta_{\alpha \beta}^{\text{d}},\Theta_{\alpha i_s}^{\text{d}}$.

We now derive an explicit algebraic expression for the symmetry-constrained fluxes. First, observe that $ \phi^{\beta j_{s'''}}\Sigma_{s'''} \cdot D_\alpha \cdot D_\beta$ $=$ $W^{i_{s'}| j_{s'''}} \phi^{\beta k_{s''}} W_{\beta k_{s''} \alpha i_{s'}}$, so that the final set of terms in \cref{fluxconstraints2} are given by $\phi^{\beta j_{s'}} W_{\beta j_{s'} 1 \alpha}$ $=$ $W_{1|j_{s'}} ( \phi^{\beta j_{s'}}  \Sigma_{s'} \cdot D_\alpha \cdot D_\beta)$ $=$ $W_{1|j_{s'}}W^{i_{s'''}| j_{s'}} \phi^{\beta k_{s''}} W_{\beta k_{s''} \alpha i_{s'''}}$. This allows us to replace the final set of terms in \cref{fluxconstraints2} with the first two sets of terms on the right-hand side of \cref{fluxconstraints4}. Next, by observing that $W_0 = W_1 = 1$ and $W_{00} = W_{11} = K$, as well as $W_{01\beta \gamma} = W_{01}^{\alpha} W_{0\alpha \beta \gamma}$, we able to use the constraint  \labelcref{fluxconstraints3} to eliminate the first two sets of non-distinctive terms from the linear combination $(\Theta_{0\alpha} + \Theta_{1\alpha})/2$. Finally, the linear combination $(\Theta_{0\alpha} - \Theta_{1\alpha} )/2$ can be used to simplify the resulting expressions as well as constrain the parameters $\phi^{1\alpha}$. In summary we find that the symmetry constraints can be re-expressed as
%   \begin{align}
%       G^{\beta j_{s'}} W_{\beta j_{s'} \alpha i_s} &= - [ \Theta_{\alpha i_s}^{\text c} + G^{1\beta} W_{1 \beta \alpha i_s} ] \\
%           G^{\beta \gamma} W_{0 \alpha \beta \gamma} &= - \frac{1}{2}[ \Theta_{0\alpha}^{\text c} + \Theta_{1\alpha}^{\text c}-  (K^{\beta} + W_{01}^{\beta} ) \Theta_{\alpha \beta}^{\text c}+ \pi_{1j_{s'}} h^{i_{s} j_{s'}} G^{\beta k_{s''}} W_{\beta k_{s''} \alpha i_{s}} ]\\
%           G^{0 \gamma} W_{0 \alpha \beta \gamma} &= - [ \Theta_{\alpha \beta}^{\text c} + G^{1 \gamma} W_{1 \alpha \beta \gamma} ]\\
%           \pi_{1j_{s'}} h^{i_{s} j_{s'}} G^{\beta k_{s''}} W_{\beta k_{s''} \alpha i_{s}} &= - [ \Theta_{1\alpha}^\text{c} - \Theta_{0\alpha}^{\text c} +(W_{01}^\beta - K^\beta) (-\Theta_{\alpha \beta}^{\text c}- 2G^{1\gamma}  W_{0 \alpha \beta \gamma}) ].
%   \end{align}
%
        \begin{align}
        \begin{split}
        \label{newconstraints1}
        \phi^{0 \gamma} W_{0 \alpha \beta \gamma} &= -  \Theta_{\alpha \beta}^{\text{d}} - \phi^{1 \gamma} W_{0\alpha \beta \gamma} \end{split} \\
        \begin{split}
        \label{newconstraints2}
            \phi^{\beta \gamma} W_{0 \alpha \beta \gamma} &= - \Theta_{0\alpha}^{\text{d}} + W_{00}^\beta \Theta_{\alpha \beta}^{\text{d}} +  (W_{00}^\beta - W_{01}^\beta)\phi^{1 \gamma}  W_{0 \alpha \beta \gamma} \end{split}\\
            \begin{split}
            \label{newconstraints3}
            \phi^{\beta j_{s'}} W_{\beta j_{s'} \alpha i_s} &= - \Theta_{\alpha i_s}^{\text{d}} - \phi^{1\beta} W_{1 \beta \alpha i_s} \end{split}  \\
            \begin{split}
            \label{newconstraints4}
            W_{1|j_{s'}} W^{ j_{s'}| i_s} \phi^{\beta k_{s''}} W_{\beta k_{s''} \alpha i_{s}} &= -  \Theta_{1\alpha}^\text{d} + \Theta_{0\alpha}^{\text{d}} +(W_{01}^\beta - W_{00}^\beta) \Theta_{\alpha \beta}^{\text{d}}\\
            &~~~~+2(W_{01}^\beta - W_{00}^\beta)\phi^{1\gamma}  W_{0 \alpha \beta \gamma} .\end{split}
    \end{align}
The equations \labelcref{newconstraints3} and \labelcref{newconstraints4} can be combined to recover the U(1) gauge symmetry constraint equations,
    \begin{align}
    \begin{split}
    \label{G1eq}
        %\phi^{1\beta} D_\beta \cdot D^{\bullet}  \cdot D_\alpha &= \phi^{\hat K \hat L} C^{\bullet}_{\hat K \hat L} \cdot D_\alpha.
    %\phi^{IJ} C_{IJ}^{\bullet} \cdot D_\alpha= (\phi^{1\beta} C^{\bullet}_{1\beta}   + \phi^{\hat K \hat L} C^{\bullet}_{\hat K \hat L}) \cdot D_\alpha=0
    \phi^{IJ} W_{\bar 1 IJ} \cdot D_\alpha= (\phi^{1\beta} W_{\bar 1 1 \beta}   + \phi^{\hat K \hat L} W_{\bar 1 \hat K \hat L}) \cdot D_\alpha=\Theta_{\bar 1 \alpha } =\sigma_{\bar 1}^J \Theta_{J \alpha} = 0
        \end{split}
    \end{align}
where in the above equation (compare to \cref{physbasis})
    \begin{align}
    \begin{split}
    \label{ncfluxconstraint}
        %D^{\bullet} &= W_{1i_s|k_{s''}} W^{k_{s''}| i_s}  + 2 (W_{01} - W_{00})\\
        %C^{\bullet} D_\alpha &= - W_{1,k_{s''}} h^{k_{s''} i_s} \Theta_{\alpha i_s}^{\text c} + \Theta_{1 \alpha}^{\text c} - \Theta_{0 \alpha}^{\text c}  +(W_{00}^\beta - W_{01}^\beta) \Theta_{\alpha \beta}^{\text c}\\
        W_{\bar 1 K L}&= -W_{1|k_{s''}}  W^{k_{s''}| i_s} W_{i_s  K  L} + W_{1  K  L}-W_{0 K  L}+(W_{00}-W_{01}) \cdot W_{ K L}\,.
    \end{split}
    \end{align}
%The constraint \labelcref{G1eq} implies the family of fluxes parametrized by characteristic parameters must be chosen in such a way that the above identity is always satisfied. Since $\phi^{1\beta} D_\beta$ is specified by $h^{1,1}(B)$ free parameters and there are $3 + 2 \rk\sfG_\text{na} + (\rk\sfG_\text{na})^2$ characteristic parameters, there are not enough free parameters $\phi^{1\beta}$ to compensate for a change in $C^{\bullet}$ and hence the coefficients $\phi^{1\beta}$ clearly vanish if $h^{1,1}(B)$ is not suitably large:
%   \begin{align}
%   \label{G1vanishes}
%       \phi^{1\beta}  =0 ~~\implies ~~C^{\bullet} = 0~~~~\text{if $h^{1,1}(B) < 3 + 2 \rk\sfG_\text{na} + (\rk\sfG_\text{na})^2$}.
%   \end{align}
%Notice that
%   \begin{align}
%       W_{\bar 1 \bar 1} =  \sigma^{I}_{\bar 1} W_{\bar 1 I}= W_{\bar 1 1} = (2(W_{00} - W_{01})- W_{1|k_{s''}} W^{k_{s''}|i_s} W_{1 i_s } )
%   \end{align}
%is (minus) the height pairing divisor associated to the U(1).
Using
\cref{newconstraints1,newconstraints2,newconstraints3,newconstraints4}
to eliminate all dependence on non-distinctive parameters, we find
that the symmetry-constrained fluxes $\Theta_{ \hat I \hat J} = M_{C(\hat I \hat J)(KL)}\phi^{KL}$ are defined by
\begin{align}
    \label{appeq:geotheta1}
        {M_C}_{( \hat I \hat J) (\hat K \hat L) }  = { M_{C_\text{na}}}_{( \hat I \hat J) (\hat K  \hat L) } - { M_{C_\text{na}}}_{( \hat I \hat  J)(1\alpha)} M_{C_\text{na}}^{+(1\alpha)(1\beta)}  { M_{C_\text{na}}}_{(1\beta)( \hat K \hat L)}\,,
    \end{align}
where $ { M_{C_\text{na}}}=  C_\text{na}^\transpose M  C_\text{na}$ is the restriction of $M$ to the sublattice $\Lambda_{ C_\text{na}}$ of backgrounds only satisfying the purely nonabelian constraints $\Theta_{i_s \alpha} =0$. The components of $ M_{C_\text{na}}$ are
    \begin{align}
    \begin{split}
    \label{appeq:omegabar}
         { M_{C_\text{na}}}_{ ({I} {J}) (K L)}&= W_{  I  J K L}- W_{  I  J|i_{s}} \cdot W^{i_{s}| j_{s'}} W_{  K Lj_{s'}} - W_{0  I  J} \cdot W_{  K L} - W_{  I  J} \cdot  W_{0  K L}\\
        &~~~ + W_{00}\cdot W_{  I  J} \cdot W_{  K L}
    \end{split}
   \end{align}
   where in particular
   \begin{align}
    \begin{split}
    \label{appeq:omegabar2}
         { M_{C_\text{na}}}_{ (1 \alpha) (KL)} &= D_\alpha \cdot W_{\bar 1 KL}\\
         &=D_\alpha \cdot (-W_{1|k_{s''}}  W^{k_{s''}| i_s} W_{i_s  KL} + W_{1  IJ}-W_{0 KL}+(W_{00}-W_{01}) \cdot W_{ KL})
    \end{split}\\
    \begin{split}
    \label{appeq:omegabar3}
        { M_{C_\text{na}}}_{(1\alpha)(1\beta)}&= D_\alpha \cdot D_\beta \cdot W_{\bar 1 \bar 1}\\
        &= D_\alpha \cdot D_\beta \cdot (- W_{1|k_{s''}} W^{k_{s''}|i_s} W_{1 i_s }+2(W_{00} - W_{01}) )\,.
    \end{split}
    \end{align}
%\begin{align}
%   \Theta_{ I  J} = MP_{( I  J) ( K  L) }  \phi^{KL},
%\end{align}
%where
%\begin{align}
%   \label{appeq:geotheta1}
%       MP_{( I  J) ( K  L) }  = MP^{\text{na}}_{( I  J) ( K  L) } - MP^{\text{na}}_{( I  J)(1\alpha)}MP_+^{\text{na}(1\alpha)(1\beta)} MP^{\text{na}}_{(1\beta)( K  L)},
%   \end{align}
%and where
%   \begin{align}
%   \begin{split}
%   \label{appeq:omegabar}
%        MP^{\text{na}}_{ ({I} {J}) (K L)}&= W_{  I  J K L}- W_{  I  J|i_{s}} \cdot W^{i_{s}| j_{s'}} W_{  K Lj_{s'}} - W_{0  I  J} \cdot W_{  K L} - W_{  I  J} \cdot  W_{0  K L}\\
%       &~~~ + W_{00}\cdot W_{  I  J} \cdot W_{  K L}
%   \end{split}\\
%   \begin{split}
%   \label{appeq:omegabar2}
%        M  P^{\text{na}}_{ (1 \alpha) (KL)} &= D_\alpha \cdot W_{\bar 1 KL}\\
%        &=D_\alpha \cdot (-W_{1|k_{s''}}  W^{k_{s''}| i_s} W_{i_s  KL} + W_{1  IJ}-W_{0 KL}+(W_{00}-W_{01}) \cdot W_{ KL})
%   \end{split}\\
%   \begin{split}
%   \label{appeq:omegabar3}
%       MP^{\text{na}}_{(1\alpha)(1\beta)}&= D_\alpha \cdot D_\beta \cdot W_{\bar 1 \bar 1}\\
%       &= D_\alpha \cdot D_\beta \cdot (- W_{1|k_{s''}} W^{k_{s''}|i_s} W_{1 i_s }+2(W_{00} - W_{01}) ).
%   \end{split}
%   \end{align}
Note that $W_{\bar 1 \bar 1}$ is equal to (minus) the height pairing divisor associated to the factor $\U(1) \subset \sfG$, and that $M_{C_\text{na}}^{+(1\alpha)(1\beta)}$ is the inverse (when it exists) of $M_{C_\text{na}(1\alpha)(1\beta)}$.

\section{Resolution-independence of $M_\text{red}$ in nonabelian models}
\label{proof}

In this appendix we show that the matrices $M_\text{red}$ associated
with different resolutions can be related through a basis change $U$
when a physically-motivated condition is satisfied.  This does not
completely prove that $M_\text{red}$ is resolution-independent; in
particular, $U$ may in general be rational, although in all cases we
have considered the change of basis is integral, and we suspect but
have not proven that
this is always the case.

Given a singular F-theory model characterized by a nonabelian group $\sfG
= \sfG_\text{na} = \prod_{s} \sfG_s$ and matter spectrum $\oplus
\sfr^{\oplus n_{\sfr}}$, we consider the %subset
set of possible F-theory
%resolution 
resolutions for which the chiral indices can be expressed
as linear combinations of the fluxes, $\chi_{\sfr} =
x_{\sfr}^{i_sj_t} \Theta_{i_sj_t}$, i.e., those resolutions where
the matter surfaces for all chiral matter representations contain a
vertical component.  (Thus for this analysis we ignore the potential existence of unusual resolutions, such as those described in \cref{sec:puzzle}, for which a subset of the matter surfaces do not contain a vertical component).
%We further assume that for each
%resolution there exists a subset of irreps such that the
%$c^{i_sj_t}_{\sfr} \in M_C \Lambda_C$ are primitive and without loss
%of generality may be taken to be unit vectors.\andrew{Can we add words
%  here clarifying this assumption to remove the potential cases where
%  the choices of $i, j$ don't correspond to the same $\chi$s and thus
%  we can't say the $\Theta$s on each side are pairwise equal?}
We  make the assumption that, for each pair of resolutions
of the same singular geometry, the resulting $M_\text{phys}$ is the
same up to a choice of integral basis.  While we do not have a general
proof that this must always be the case it is not much stronger than
the statement that the set of allowed flux backgrounds and chiral
multiplicities are the same for both resolutions, which we expect
on physical grounds
since the physics of any F-theory model should be resolution-invariant.\footnote{This
  does not rule out the possibility of situations where, despite the fact that the fluxes for two resolutions are the same, the intersection pairings on their respective lattices of flux backgrounds differ (e.g. $M_\text{phys}= (4),  (1)$ and $
  \chi_{\sfr} = 1, 4$.) However, we have not
  encountered such situations in any of the F-theory models we have
  studied.}  

Consider a pair of resolutions $X,\tilde X$
satisfying this criterion.  
For simplicity we assume that the gauge group has a single nonabelian
factor $G$, though a very similar analysis can be carried out for groups with
multiple nonabelian factors.
%This implies that for two subsets of
%indices $\{ i_sj_t \}, \{ \tilde i_s\tilde j_t\} \subset \{
%\hat I \hat J\}$ the fluxes $\Theta_{i_s j_t}, \tilde
%\Theta_{\tilde i_s\tilde j_t}$ are pairwise equal. Recalling
%that the expression for symmetry-constrained fluxes is $\Theta_{\hat I
%  \hat J} = {M_{C}}_{(\hat I \hat J)(\hat K \hat L)} \phi^{\hat K \hat
%  L} $ where
%    \begin{align}
%        {M_C}_{(\hat I \hat J)(\hat K \hat L)} &= W_{\hat I \hat J
%          \hat K \hat L} + W_{00} \cdot W_{\hat I \hat J} \cdot
%        W_{\hat K \hat L} - W^{i_s | j_{t}} W_{\hat I \hat J |i_s}T
%        \cdot \Sigma_s \cdot W_{\hat K \hat L | j_{t}}, 
%    \end{align}
%and using the fact that $M_C$ is guaranteed to be symmetric for
%nonabelian gauge groups, this implies that
For each of these resolutions, we consider
again the general form of $M_\text{red}$  \labelcref{Mrednonabelian}, namely
\begin{align}
\label{Mredcanon}
    M_\text{red} = \begin{pmatrix}
            [[   D_{\alpha'} \cdot K \cdot D_\alpha ]] & [[D_{\alpha'}  \cdot D_{\alpha} \cdot D_\beta ]] & 0 & 0 \\
            [[  D_{\alpha' } \cdot D_{\beta'} \cdot D_{\alpha } ]] &0 &0 & [[ W_{\alpha' \beta' j_t k_u}]] \\
            0 & 0& [[ W_{\alpha'  i'_{s'}\alpha i_s} ]] & [[ W_{\alpha' i'_{s'} j_t k_u}]] \\
        0 & [[ W_{j'_{t'} k'_{u'}\alpha \beta} ]] &  [[W_{j'_{t'} k'_{u'}\alpha i_s}]] &[[ W_{j'_{t'} k'_{u'}j_t k_u }]]
        \end{pmatrix}\,,
\end{align}
where we recall that unprimed indices denote columns and primed indices
denote rows. 
This matrix generally has the schematic structure
\begin{equation}
    M_\text{red}=
\left(\begin{array}{cc}
M' & Q\\
Q^\transpose & M''
\end{array} \right) \,,
\end{equation}
where $M'$ takes the form \labelcref{resindependenthatM} and
is non-degenerate and thus invertible.  We can then carry
out a change of basis using the matrix
\begin{equation}
U_1 = \begin{pmatrix}
\Id & u_1\\
0 &\Id
\end{pmatrix},
\label{eq:basis-change-1}
\end{equation}
namely
\begin{equation}
U_1^\transpose    M_\text{red} U_1 =
\begin{pmatrix}
M' & 0\\
0 & M''_1
\end{pmatrix} \,,
\label{eq:m1}
\end{equation}
where explicitly $u_1 = -(M')^{-1} Q$ and $M_1'' = M''
-u_1^\transpose M' u_1$.
Note that $u_1$, and hence $U_1$ are generically rational since
$(M')^{-1}$ contains a factor of $\det \kappa$ in the
denominator as discussed abstractly in \cref{sec:quantization-1}
and more explicitly in \cref{integrality}.
Because $M'$ is non-degenerate, the symmetry constraints are imposed
by simply setting the first set of coordinates to vanish.  Thus,
$M''_1$ is essentially  $M_\text{phys}$. The subtlety here is that since $(M')^{-1}$ has a denominator of $\det \kappa$, in most
cases, like the example described explicitly in
\cref{sec:5-independence}, we have 
\begin{equation}
M''_1 = M_\text{phys}/(\det
\kappa)^2\,.
\label{eq:1-physical}
\end{equation}
In such a situation, the condition that the change of basis
\cref{eq:basis-change-1} gives an integer vector means that $M''_1$
only acts on the sublattice of vectors $\phi''$ such that $u_1 \phi''$
is integer-valued in all components.

From this analysis we can now immediately see that the equivalence of
$M_\text{phys}$ between the two resolutions allows us to relate the
forms of $M_\text{red}$.  Assuming that \cref{eq:1-physical} holds for
both the resolutions $X, \tilde{X}$, and that the resulting
$M_\text{phys}$ are
related through an integral linear transformation
\begin{equation}
\tilde{M}_\text{phys}= U_p^\transpose M_\text{phys} U_p \,,
\label{eq:physical-transform}
\end{equation}
we have
\begin{equation}
\tilde{M}_\text{red}= U^\transpose M_\text{red} U \,,
\label{eq:red-transform}
\end{equation}
where
\begin{equation}
U = U_1 \begin{pmatrix}
\Id & 0\\
0 & U_p
\end{pmatrix}
\tilde{U}_1^{-1}  =
\begin{pmatrix}
1 & u_1 U_p-\tilde{u}_1\\
0 & U_p
\end{pmatrix}
\,.
\label{eq:red-u}
\end{equation}
Note that the matrix $U_p$ used in \cref{eq:physical-transform} is not
uniquely defined, and has an ambiguity up to the set of automorphisms
of the lattice $M_\text{phys}$.  Thus, we have a number of candidate
transformations $U$ of the form \cref{eq:red-u}.  As an explicit class
of examples note that even if $M_\text{phys}= \tilde{M}_\text{phys}$
and both matrices are diagonal with distinct eigenvalues, the matrix
$U_p$ may be any diagonal matrix composed of elements $\pm 1$.

This argument is almost enough to prove that the two forms of $M_\text{red}$
are equivalent under an integral linear change of basis, whenever the
$M_\text{phys}$ forms are equivalent.
The possible obstructions to this result of resolution-independence of
$M_\text{red}$ are related to the rational form of $u_1,
\tilde{u}_1$.  In general, we expect both $u_1$ and $\tilde{u}_1$ to
have rational terms with a denominator of $\det\kappa$.  The
fractional parts need to cancel for \cref{eq:red-u} to be an integer
transformation.  This implies a certain compatibility condition, whereby the elements of $M_\text{red}$ lie in the weight lattice
but not the root lattice of $\sfG$.
Because there is some ambiguity in the choice of $U_p$, as mentioned above,
for there to be an integral $U$ satisfying \cref{eq:red-u}, only one
of the possible $U_p$ choices needs to satisfy this necessary
compatibility condition.
In all
cases we have considered this condition is satisfied for at least one
choice of $U_p$, and the resulting $U$ is an
integral change of basis that explicitly demonstrates the
resolution-invariance of $M_\text{red}$ for the resolutions we have studied. We do not
have a general proof that this must always occur, but note that even if
this compatibility condition is not satisfied, \cref{eq:red-transform}
 still holds for the matrix $U$ defined in \cref{eq:red-u}. 
  Finally, note that there could also be a subtlety if in
 one of the resolutions the denominator in \cref{eq:1-physical} is
 cancelled but not in the other resolution, although it is difficult
 to imagine a circumstance where the resulting $M_\text{phys}$ would
 still match. We have not encountered any such situations.
%\vspace*{0.1in}

The analysis presented here assumes a single nonabelian gauge factor.
This argument can easily be generalized to multiple nonabelian gauge
factors.  For theories with an abelian factor, the story is less clear
as we do not have as general a way of understanding $M_\text{red}$,
and there are potential issues with the invertibility of the part of
the matrix analogous to $M'$.  Nonetheless, we suspect that a similar
approach will shed light on the resolution-independence of theories
with more general gauge groups.

Note that when the matrices $M_\text{red}$ associated with two
different resolutions are related by an integral change of basis, the
same is also true for the general intersection  matrices $M$
associated with the two resolutions.  This follows since, as discussed
in \cref{sec:lattice-reduce}, in each case there is an integral
transformation putting $M$ in the form \cref{eq:pmp}, so composing
these transformations with the appropriate integral $U$ on the subspace
containing $M_\text{red}$ gives an integral transformation relating
the two versions of $M$.

\section{Pushforward formulae}
\label{pushapp}

In this Appendix we describe some details of the computational approach we use to evaluate intersection products of divisor classes in resolutions of singular elliptically fibered projective varieties over a smooth base $B$, $X\rightarrow B$. Since it is in practice rather cumbersome to compute intersection products of divisors in blowups of elliptically fibered spaces like $X$ directly (i.e. in the Chow ring of $X$), we circumvent this difficulty by pushing these intersection products down to the Chow ring of $B$ (see \cite{fulton}, Remark 3.2.4, p. 55), where the intersection form is by assumption known explicitly. 

The computational methods we describe here are essentially a simple and straightforward adaptation of the formulae presented in \cite{Esole:2017kyr} (see also \cite{Fullwood:2011zb} for related discussions of pushforward formulas, as well as \cite{Esole:2018tuz,Esole:2018bmf} for more recent work that uses pushforward formulas to compute various characteristic numbers of elliptic fibrations). Many of the foundational results in intersection theory, algebraic and complex geometry upon which these formulae are based can be found in classic texts such as \cite{fulton,Griffiths:433962}. 

\subsection{Pushforward maps for resolutions of singular elliptic fibrations}

The pushforward maps we describe in this appendix can be realized explicitly as a composition of pushforward maps associated to two types of projection maps: \newline

\noindent \underline{Canonical projection of the elliptic fibration}. The first type of projection map is the canonical projection of the singular elliptic fibration, $X_0 \rightarrow B$. We can determine the pushforward $\varpi_*$ associated to this projection by exploiting the fact that the singular elliptic varieties $X_0$ we consider this paper are all realized as hypersurfaces inside an ambient projective bundle, $Y_0$, which can be viewed as the projectivization of a direct sum of line bundles $\scL_a \rightarrow B$,
    \begin{align}
     \label{linebundlesum}
        Y_0 = \bP(\scV) \overset{\varpi}{\rightarrow} B,~~~~        \scV = \oplus_{a=1}^3 \scL_a\,.
    \end{align}
Since $\scV$ is a direct sum of (complex) line bundles, standard results in complex geometry imply that the total Chern class is given by
	\begin{equation}
		c(\scV) = \prod_a (1 + \bm{L}_a)~~\implies ~~ \bm{L}_a = c_1(\scL_a)\,.
	\end{equation}
The above characterization of $Y_0$ as the projectivization of a direct sum of line bundles provides sufficient information for us to specify all types of divisors of $Y_0$ in which we are interested. The different types of divisors are as follows: One type of divisor we wish to consider is the pullback $\boldsymbol{D}_\alpha$ of a divisor $D_\alpha$ (which lives in the Chow ring of $B$) to the Chow ring of $Y_0$. Of particular importance is a special subset of the divisor classes, namely the first Chern classes $c_1(\scL_a) = \bm{L}_a = f(\bm{D}_\alpha)$ (here, $f$ is an unspecified linear function of certain divisors we refer to characteristic data of the elliptic CY $X_0$, which can themselves be expressed as linear combinations of the pullbacks $\bm{D}_\alpha$.)  The second type of divisor class we consider is $\bm{H} := c_1(\scO_{Y_0}(1))$ where $\scO_{Y_0}(1)$ is the twisting sheaf (i.e. the dual of the tautological line bundle) associated to the projectivization of $\scV$. It turns out that all characteristic classes of $X_0$ can be associated to formal power series of the classes $\bm{H}, \bm{D}_\alpha$, hence our first goal is to explain how to compute intersection numbers of these divisor classes by pushing them forward to the Chow ring of $B$ via the map $\varpi_*$; since the divisors $\bm{D}_\alpha$ are pullbacks of divisor classes living in the Chow ring of $B$, this task reduces to computing pushforwards of intersection products of the class $\bm{H}$, as we now describe.

Given a formal power series $\tilde Q(t) = \tilde Q_n t^n$, one can derive an explicit expression for the pushforward of $\tilde Q(\bm{H})$ to the Chow ring of $B$ by exploiting well known properties of the total Segre class $s(\scV) = c(\scV)^{-1}$. In particular, we use the fact that (see Chapter 3 of \cite{fulton})
	\begin{equation}
		\varpi_* \frac{1}{1-\bm{H}} =s(\scV) = \prod_a \frac{1}{1 + \bm{L}_a}\,,
	\end{equation}
along with the degree of the map $\varpi$, to obtain the formula\footnote{In more detail, we expand both sides of the pushforward identity 
	\begin{equation*}
		\varpi_* \frac{1}{1-\bm{H}} = \frac{1}{\prod_{a=1}^3 (1 + \bm{L}_a)}	
	\end{equation*} 
as formal power series, and then (using the fact that $\varpi$ is a degree two map, hence $\varpi_* 1 = \varpi_* \bm{H} =0$) match terms of equal degree to obtain a general relation of the form $\varpi_* \bm{H}^p = f_p(\bm{L}_a)$ that can be applied to any formal power series term-by-term in the expansion $\tilde Q(\bm{H}) = \sum_{p=0}^\infty \varpi^* Q_p \bm{H}^p$. Explicitly, we make the substitutions $\bm{H} \rightarrow \epsilon \bm{H}, \bm{L}_a \rightarrow \epsilon t_a$ so that
	\begin{align*}
		 \sum_{n=0}^\infty 	\varpi_*(\bm{H}^n) \epsilon^n&=\frac{1}{\prod_{a=1}^3 (1+ \epsilon t_a) } = \sum_{n= 0}^\infty s_n(-t_1,-t_2,-t_3) \epsilon^n= \sum_{n=0}^\infty \left( \sum_{a=1}^3 \frac{(-t_a)^{n+2}}{\prod_{b\ne a} (t_a - t_b) } \right) \epsilon^{n}
	\end{align*}
where $s_n(-t_1,-t_2,-t_3)$ are totally symmetric polynomials of degree $n$ in the three variables $-t_a$. Note that the right hand side of the above equation makes use of the identity $s_n(t_1,\dots,t_d) = \sum_{a=1}^d t_a^{n+d-1} \prod_{\substack{b\ne a}} \frac{1}{t_a-t_b}$, see Lemma 1.10 of \cite{Esole:2017kyr}, although for fixed $d$ the result above can easily be obtained by computing a partial fraction decomposition of the expression $\frac{1}{\prod_{a} (1+\epsilon t_a)}$. We then match the $\cO(\epsilon)$ terms and re-sum the resulting formal power series to obtain the succinct expression inside the limit on the right hand side of \cref{X0push}.
}
\begin{align}
\label{X0push}
    \varpi_* \tilde Q(\bm{H}) =\lim_{t_c \rightarrow \bm{L}_c} \sum_{a=1}^{3} \frac{Q(-t_a)}{\prod_{b\ne a} (t_a - t_b) }\,.
\end{align}
In the above equation $t_a$ are distinct formal variables and (setting $\tilde Q_n = \varpi^* Q_n$) the power series $\tilde Q(t) = \varpi^{*} Q_i t^i$ implicitly defines $Q(t) = Q_i t^i$ via the projection formula \cref{projform}. Observe that \cref{X0push} is a simple adaptation of the derivation in Theorem 1.11 of \cite{Esole:2017kyr} to the case where the line bundles $\scL_a$ are all distinct. Note that we present the right hand side of \cref{X0push} as a limit to accommodate special cases where $\bm{L}_a = \bm{L}_b$ for some subset of the first Chern classes $\bm{L}_{a}$.

Although in the above discussion we have assumed that the line bundles $\scL_a$ are generically all distinct, it is important to keep in mind that the scaling symmetry of the projectivization of the bundle $\scV$ implies that $\bP(\scV) \cong \bP(\scV \otimes \scI)$ where $\scI$ is any invertible line bundle. In particular, without loss of generality we may set $\scI =\scL_c^{-1}$ and reduce to the standard case
	\begin{align}
		\bP(\scV)\, \cong \, \bP(\scL \oplus \scL' \oplus \scO)\,,
	\end{align} 
which in turn brings the formula \cref{X0push} into contact with equivalent formulae that have appeared in related literature on elliptic fibrations---see for example (7.3) in Lemma 2.8 of \cite{Esole:2014dea}. The above standard form is as specific as we can be about the choice of line bundles $\scL, \scL'$ while still accommodating the full scope of $\sfG$ models we wish to describe. For example, models that include $\U(1)$ gauge factors such as the $(\SU(2) \times \U(1))/\Z_2$ model of \cref{F6model} typically have $\scL \ne \scL'$. Nonetheless in some cases we can specialize further in order to obtain more succinct formulae that are in practice easier to implement. One possibility entails specializing further to the case $\bm{L} = p \bm{L}'', \bm{L}' = p' \bm{L}''$ where $p, p'$ are non-negative integers, for which it is possible to obtain even simpler expressions for \cref{X0push}. A notable set of examples of this specialization are the Tate models, for which $\bm{L} = -2 \boldsymbol{K}, \bm{L}' = -3 \boldsymbol{K}$; in these cases \cref{X0push} reduces to the formula presented at the beginning of Theorem 1.11 in \cite{Esole:2017kyr}. \newline

\noindent \underline{Blowdowns.} We now turn to the second type of projection map, namely the projection associated to a blowup (i.e. the blowdown). The elliptic fibrations $X_0$ we study are typically singular and in practice require a resolution $X \rightarrow X_0$ implemented by sequence of blowups in order for the intersection products of divisors to be well-defined and calculable. We specifically consider blowups of the form
    \begin{align}
    \label{blowupnotation}
        X_{i+1} \overset{(g_{i+1,1}, \dots, g_{i+1,n_{i+1}}|e_{i+1})}{\longrightarrow} X_{i}
    \end{align}
    where the notation $(g_{i+1,1}, \dots, g_{i+1,n_{i+1}}|e_{i+1})$ is shorthand for the blowup $Y_{i+1} \rightarrow Y_{i}$ of the ambient space $Y_i$ along the center $\{ g_{i+1,1} = \cdots = g_{i+1,n_{i+1}} =0\}\subset  Y_i$ with exceptional divisor $e_{i+1} = 0$, and $X_{i+1} \subset Y_{i+1}$ is the proper transform of $X_i$ under the blowup. Importantly, note that we must restrict to blowups where the centers $g_i$ are (at most) linear polynomials in the homogeneous coordinates of the ambient space of the fiber. We abuse notation and make the replacements
    \begin{align}
        g_{i,j}  \to e_i g_{i,j}
    \end{align}
to implement the $i$th blowup. Each blowup (chosen appropriately) introduces a new divisor class
        \begin{align}
            \bm{E}_i = [e_i  ]
        \end{align}
and thus it is desirable to be able to compute pushforwards of formal multivariate power series depending on the classes $\bm{E}_i$ (again, as was the case with the first type of pushforward map $\varpi_*$ described above, the projection formula \cref{projform} implies that we are free to ignore divisor classes that are pullbacks and simply focus on the action of the pushforward map $f_i$ on classes $\bm{E}_i$). Fortunately, there is a similar formula to \cref{X0push}, derived by an analogous procedure, that can be used to compute pushforwards $f_{i+1*}$---we refer the interested reader to Section 3.1 of \cite{Esole:2017kyr} for details of the derivation. Given a blowup $Y_{i+1} \overset{f_{i+1}}{\rightarrow} Y_{i}$ along the center $g_{i+1,1} = \cdots = g_{i+1,n_{i+1}} =0 $ and a formal power series $\tilde Q(\bm{E}_{i+1})$ in the Chow ring of $Y_{i+1}$, the pushforward of $\tilde Q$ to the Chow ring of $Y_i$ is given by  (see \cite{Esole:2017kyr}, Theorem 1.8)
    \begin{align}
    \label{blowuppush}
        f_{i+1*} \tilde Q(\bm{E}_{i+1}) =  \sum_{k=1}^{n_{i+1}} Q(\bm{g}_{i+1,k}) M_{k} ,~~~~ M_{k} = \prod_{\substack{ m=1 \\m \ne k  }}^{n_{i+1}} \frac{\bm{g}_{i+1,m}}{ \bm{g}_{i+1,m}- \bm{g}_{i+1,k}}
    \end{align}
where in the above formula $n_{i+1}$ is the number of generators of the center of the $(i+1)$th blowup and we assume $\bm{g}_{i+1,k} = [ g_{i+1,k}]$ are all distinct. 

\cref{X0push,blowuppush} can be composed to compute the pushforward of any intersection product in the Chow ring of $X$ to the Chow ring of $B$. We briefly sketch how this computation works in practice before presenting some explicit examples. Recall that the resolutions we study in this paper are hypersurfaces $X \subset Y$ inside ambient projective (in fact, toric) bundles $Y$ equipped with the projection $ \pi : Y \rightarrow B$. In such cases, the divisors of $X$ can be realized concretely as the restriction of divisors in the Chow ring of $Y$ to the hypersurface $X$, namely
    \begin{align}
        \hat D_I = \hat{\boldsymbol{D}}_{I}\cap \bm{X},~~~~ \hat{\boldsymbol{D}}_I =  \ell_I(\bm{H},\bm{E}_i) 
    \end{align}
where $\ell_I$ is a linear polynomial and $\bm{X}$ is the divisor class of the resolved hypersurface $X \subset Y$. For example, a quadruple intersection product takes the form
	\begin{equation}
		\hat D_I \cdot \hat D_J \cdot \hat D_K \cdot \hat D_L =  \hat{\boldsymbol{D}}_I  \cdot  \hat{\boldsymbol{D}}_I  \cdot  \hat{\boldsymbol{D}}_I  \cdot  \hat{\boldsymbol{D}}_I  \cdot  \hat{\boldsymbol{X}} =: \tilde Q(\bm{E}_i, \bm{H}, \bm{D}_\alpha)\,,
	\end{equation} 
where $\cdot$ in the middle expression above (i.e. on the right hand side of the first equality) should be understood as the intersection product in the Chow ring of $Y$, and $\tilde Q$ on the far right hand side should be viewed as a formal power series in the Chow ring of $Y$. Assuming that the resolution $X \rightarrow X_0$ is obtained by means of a sequence of, say $r$ blowups, the pushforward $\pi_*$ of the projection $\pi : X \rightarrow B$ can be viewed as shorthand for a composition of pushforwards,
    \begin{align}
        \pi_* = \varpi_* \circ f_{1*} \circ \cdots \circ f_{(r-1)*} \circ f_{r*}\,,
    \end{align}
where the first pushforward $f_{*r}$ maps the expression from the Chow ring of $Y_r$ to the Chow ring of $Y_{r-1}$, the second pushforward maps the resulting expression to the Chow ring of $Y_{r-2}$, and so on. 

Ultimately, all of the characteristic classes of $X$ in which we are interested can be expressed as formal power series in the Chow ring of $Y$. For example, the total Chern class $c(X)$ is given by
    \begin{align}
    \label{totalChern}
        c(X) = \left( \prod_{i} (1 + \bm{E}_i ) \cdot \prod_{j=1}^{n_i} \frac{  (1 + \bm{g}_{i,j} - \bm{E}_i)}{(1 + \bm{g}_{i,j}) }  \right) \cdot \frac{c(Y_0)}{1 + \bm{X}} \cap  \bm{X}\,.
    \end{align}
where in the above expression
    \begin{align}
        c(Y_0) =\prod_k (1 + \bm{H} +\bm{L}_k) \cdot c(B)
    \end{align}
and we recall that $\bm{L}_k$ is defined in \cref{linebundlesum}. The Chern polynomial $c_t(X) = 1 + c_1(X) t + c_2(X) t^2 + \cdots$ can be used to extract terms of different degree from the above expression, either in the Chow ring of $Y$ or (after computing the pushforward) in the Chow ring of $B$.

\subsection{Example: $\SU(2)$ model}

We illustrate the pushforward technology by way of an example, namely the $\SU(2)$ model (this is the $N=2$ case of the $\SU(N)$ Tate models described in \cref{sec:appendix-sun}). Although this example has already been worked out explicitly in \cite{Esole:2017kyr}, we reproduce some of the details here in order to clarify our particular choice of notation. 

The Weierstrass equation defining the singular $\SU(2)$ model $X_0$ is
    \begin{align}
        y^2 z + a_1 x y z + a_{3,1} \sigma yz = x^3 + a_{2,1} \sigma x^2 + a_{4,1} \sigma xz^2 + a_{6,2} \sigma^2 z^3=0\,.
    \end{align}
We resolve this model by means of the blowup
    \begin{align}
        X_1 \overset{(x,y,\sigma|e_1)}{\longrightarrow} X_0\,,
    \end{align}
meaning that we make the replacements
    \begin{align}
        x \to e_1 x,~~~~ y \to e_1 y,~~~~ \sigma \to e_1 \sigma\,.
    \end{align}
Factoring out two powers of $e_1$ from Weierstrass equation of the total transform (i.e., subtracting two copies of the exceptional divisor), we see that the proper transform $X_1 \subset Y_1$ is described by
    \begin{align}
         y^2 z +  a_1 x y z + e_1 a_{3,1} \sigma yz = e_1 x^3 + e_1 a_{2,1} \sigma x^2 + a_{4,1} \sigma xz^2 +  a_{6,2} \sigma^2 z^3=0\,.
    \end{align}
The divisor class of the proper transform $X_1$ in the Chow ring of $Y_1$ is
    \begin{align}
        \boldsymbol X_1 = 3 \boldsymbol H - 6 \boldsymbol K - \boldsymbol E_1\,.
    \end{align}
The sole Cartan divisor of $X_1$ is
    \begin{align}
        \hat D_i = \boldsymbol{E}_1\cap \boldsymbol X_1\,.
    \end{align}

We use the pushforward technology described at the beginning of this section to compute the quadruple intersection number $W_{iiii}$. The first step is to evaluate the pushforward of $W_{iiii}$ to the Chow ring of $X_0$, which we denote explicitly by $f_{1*}$ (here we indicate the pushforward and pullback maps explicitly, keeping in mind that $\varpi_{*}$ is the pullback of the projection $\varpi: X_0 \rightarrow B$):
    \begin{align}
    \begin{split}
    \label{pushcompex}
        f_{1*}  (\hat D_i^4) &=f_{1*} (\boldsymbol E_1^4 \cdot \boldsymbol X_1)\\
        & = f_{1*} (f_1^*  (3 \boldsymbol H - 6 \varpi^*  K) \cdot \boldsymbol E_1^4 - \boldsymbol E_1^5)\\
        &=(3 \boldsymbol H - 6 \varpi^* K) \cdot f_{1*}(\boldsymbol E_1)^4 - f_{1*}(\boldsymbol E_1)^5\\
        &=    (3 \boldsymbol H - 6 \varpi^*  K) \cdot \sum_{k=1}^{3} \boldsymbol g_{1,k}^4 \cdot   \prod_{\substack{m=1\\m\ne k}}^{3} \frac{\boldsymbol g_{1,m}}{\boldsymbol g_{1,m} - \boldsymbol g_{1,k}}-  \sum_{k=1}^{3} \boldsymbol g_{1,k}^5 \cdot \prod_{\substack{m=1\\m\ne k}}^{3} \frac{\boldsymbol g_{1,m}}{\boldsymbol g_{1,m} - \boldsymbol g_{1,k}}\\
        &=: \tilde Q(\boldsymbol H)\,.
    \end{split}
    \end{align}
In the above expression $\boldsymbol g_{1,k}$ are the classes of the generators of the blowup center:
    \begin{align}
    \begin{split}
        \boldsymbol g_{1,1}& = [e_1x] = \boldsymbol H - 2 \varpi^*  K\\
        \boldsymbol g_{1,2} &= [e_1 y] = \boldsymbol H - 3\varpi^*  K\\
        \boldsymbol g_{1,3} &= [ e_1 \sigma  ] = \varpi^* \Sigma\,.
    \end{split}
    \end{align}
Thus far, we have computed the pushforward of the quadruple intersection $W_{iiii}$ to the Chow ring of $X_0$. In order to compute the pushforward of $W_{iiii}$ to the base, we now expand $\tilde Q(\boldsymbol H)$ as a formal power series in the variable $\boldsymbol H$ (with coefficients consisting of polynomials in the classes $\varpi^* K, \varpi^* \Sigma$) and evaluate the pushforward of each power of $\boldsymbol H$ to the Chow ring of $B$ using the formula \cref{X0push}. We do not include details of this computation here as it is completely analogous to the pushforward computation illustrated above in \cref{pushcompex}. In the end, we obtain
    \begin{align}
    \label{pushsu2}
        W_{iiii} =2 \Sigma \cdot ( - 4 K^2 + 2 K \cdot \Sigma - \Sigma^2)\,.
    \end{align}

\section{Resolutions of some Tate models}
\label{sec:resolutions}

The Tate form of the Weierstrass model $X_0$ is defined by the hypersurface equation
    \begin{align}
    \label{Tateexample}
        y^2 z + a_1 xyz + a_3 y z^2 - (x^3 + a_2 x^2 z + a_4 x z^2 + a_6 z^3 ) = 0
    \end{align}
in the ambient projective bundle $Y_0 = \bP(\scV) \rightarrow B$ with $\bP^2$ fibers parametrized by homogeneous coordinates $[x:y:z]$. Tate tunings of simple nonabelian gauge groups $\sfG = \SU(N), \SO(4k+2), \text{E}_6$ are characterized by Kodaira singularities of (resp.) types $\text{I}_N^{\text{split}}, \text{I}_{2k-3}^{* \text{split}}, \text{IV}^{* \text{split}}$ over the codimension-one locus $\Sigma \subset B$. Since the main feature of such models is the existence of a particular type of Kodaira singularity in codimension-one, the divisor class $\Sigma$ together with the canonical class $K$ are sufficient to characterize the features of the elliptic fibration $X$ in which we are primarily interested, and most other relevant mathematical quantities can be defined in terms of $K $ and $ \Sigma$ (or their dual line bundles). In particular, the divisor classes
    \begin{align}
        [a_n  ] = -n \boldsymbol{K}
    \end{align}
and the classes of the divisors $x,y,z = 0$ in the ambient space $Y_0$ are given by\footnote{We use bold symbols to denote divisor classes in the Chow ring of the ambient space.}
    \begin{align}
    \begin{split}
        [ x  ] &=  \bm{H} - 2 \bm{K} \\
        [y  ] &= \bm{H} - 3 \bm{K} \\
        [z ] &= \bm{H}
    \end{split}
    \end{align}
where $\bm{H} = c_1(\scO_{\bP(\scV) } (1))$ denotes the hyperplane class of the fibers of $Y_0$ and $\bm{K}$ is the pullback of the canonical class $K$. This implies that the divisor class of the zero locus of the Weierstrass equation is
    \begin{align}
         \boldsymbol X_0  =     3 \boldsymbol H - 6 \boldsymbol K\,.
    \end{align}
Note that $X_0$ is equipped with a holomorphic zero section $x = z =0$.

Our aim is to compute intersection numbers of divisors in resolutions $X \rightarrow X_0$ of the singular model defined by \labelcref{Tateexample}. There is a vast literature on crepant resolutions of CY singularities in the context of F-theory compactifications analyzed from various perspectives; see, e.g., \cite{Lawrie:2012gg,Hayashi:2014kca,Esole:2015sta,Esole:2016npg} for more comprehensive explorations of the networks of possible resolutions associated to the F-theory Coulomb branch. As noted in \cref{partial}, our resolutions are in general only partial resolutions in that we do not attempt to resolve all singular fibers that appear over codimension-three loci universally in certain Tate models, nor do we consider cases where additional tunings leading to singular fibers over loci of codimension two (or higher) in the base are forced by the specific choice of singular elliptic fibration. The resolutions we study are composed of a sequence of blowups of the ambient space $Y_0$ of the form \labelcref{blowupnotation}; these blowups restrict to blowups of various loci on the CY hypersurface $X_0$. To carry out the computation of intersection numbers, we select both a basis of divisors for the resolved space, $\hat D_{I = 0, \alpha, i} \subset X$, and a basis of divisors $\bm{H}, \boldsymbol{D}_\alpha, \bm{E}_{i} \subset Y$ for the proper transform $Y$ of the ambient space under the sequence of blowups, where $\bm{E}_i$ is the class of the exceptional divisor associated to the $i$th blowup $Y_i \rightarrow Y_{i-1}$. We now present some specific examples.

\subsection{SU($N$) model}
\label{sec:appendix-sun}

F-theory models with SU($N$) gauge group have been the subject of much attention in the literature, especially low rank examples
\cite{Lawrie:2012gg,Hayashi:2013lra,Esole:2011sm,Esole:2014bka,Esole:2014hya,Esole:2015xfa}. Apart
from special exceptions such as SU(6) with three-index antisymmetric
matter, SU($N$) models in F-theory are characterized by a I$_N^\text{split}$ singularity over $\Sigma = \{ \sigma =0 \} \subset B$ and can be realized explicitly using a Weierstrass model
    \begin{align}
    \label{SUTateform}
         y^2 z + a_1 xyz + a_3 y z^2 - (x^3 + a_2 x^2 z + a_4 x z^2 + a_6 z^3 ) = 0
    \end{align}
together with the Tate tuning
    \begin{align}
    \label{SUNTate}
        \text{I}_{N}^\text{split} ~:~ a_1 = a_1,~~~~ a_2 = a_{2,1} \sigma,~~ ~~a_{3,\floor{\frac{N}{2}}} \sigma^{\floor{\frac{N}{2}}},~~~~ a_{4, \ceil{\frac{N}{2}}} \sigma^{\ceil{\frac{N}{2}}},~~~~ a_6 = a_{6,N} \sigma^N\,.
    \end{align}
These models have a holomorphic zero section $x =z =0$, and contain matter in the adjoint, fundamental ($\boldsymbol N$), and two-index antisymmetric (\textbf{$N(N-1)/2$}) representations. The discriminant locus takes the form
    \begin{align}
        \Delta = \sigma^N ( \Delta^{(2)} + \cO(\sigma) ),~~~~ \Delta^{(2)} = - a_1^4 p_N\,,
    \end{align}
where the polynomial $p_N$ is defined to be
\begin{align}
 p_N = \begin{cases}
             - a_{4,\floor{\frac{N}{2}}}^2 + \cO(a_1) ~~&\text{$N$ even} \\
            a_{2,1} a_{3,\floor{\frac{N}{2}}}^2 + \cO(a_1)~~&\text{$N$ odd}.
        \end{cases}
    \end{align}
Thus, the fundamental and antisymmetric matter multiplets are localized on (resp.) the codimension-two loci $\sigma = p_N = 0$, $\sigma = a_1 = 0$, whose divisor classes in the Chow ring of $B$ are
    \begin{align}
    \begin{split}
    \label{SUNmattercurves}
        C_{\textbf{$N$}}  =\begin{cases} \Sigma \cdot (-8K - N \Sigma), &N \ne 3 \\
        \Sigma \cdot (-9 K - 3 \Sigma), &N= 3\end{cases}\,, \quad
        C_{\textbf{$\frac{1}{2}N(N-1)$}} =\Sigma \cdot (-K)\,.
    \end{split}
    \end{align}
We primarily study the family of resolutions $X_{N-1} \rightarrow X_0$ realized in \cite{Esole:2015xfa} by the sequence of blowups
    \begin{align}
    \label{SUNres}
        X_{N-1} \overset{(*,e_{N-2}|e_{N-1})}{\longrightarrow} \cdots  \overset{(x,e_2|e_3)}{\longrightarrow} X_2 \overset{(y,e_1|e_2)}{\longrightarrow} X_1\overset{(x,y,\sigma|e_1)}{\longrightarrow} X_0
        ,~~~~ * = \begin{cases} x~~ \text{($N$ even)} \\
        y~~ \text{($N$ odd)}
        \end{cases}\,.
    \end{align}
See \labelcref{blowupnotation} and the discussion immediately below for an explanation of the notation used to indicate blowup maps in the above equation. The class of the holomorphic zero section is
    \begin{align}
        \hat D_0 = \frac{\bm{H}}{3} \cap \bm{X}_{N-1}\,.
    \end{align}
Let $\hat D_{i}$ denote the Cartan divisors of $X_{N-1}$. Using the fact that $\bm{E}_i = \{ e_i = 0 \}$ is the class of the exceptional divisor of the $i$th blowup in the ambient space $Y_i$, we may write \cite{Esole:2015xfa}
    \begin{align}
    \label{SUNdiv}
        \hat D_i = \begin{cases} (\bm{E}_{2i -1} - \bm{E}_{2i}) \cap \bm{X}_{N-1} ~~~~&\text{$i < \ceil{\frac{N}{2}}$}\\
            \bm{E}_{N-1} \cap \bm{X}_{N-1} ~~~~ &\text{$i = \ceil{\frac{ N}{2}}$}\\
            (\bm{E}_{2N-2i} - \bm{E}_{2N- 2i + 1}) \cap \bm{X}_{N-1} ~~~~&\text{$i > \ceil{\frac{N}{2}}$}.
            \end{cases}
    \end{align}
The Cartan divisors correspond to simple coroots of $\SU(N)$, in a basis where the nonzero pushforwards $\pi_*(\hat D_i \cdot \hat D_j) = W_{i|j} \Sigma$ are given by
    \begin{align}
        [[W_{ij}]] = [[ \pi_*(\hat D_i \cdot \hat D_j ) ]]=-
         \begin{pmatrix} 2 & 1 &  & & && \\
         -1 & 2 & -1 & & & &  \\
         & -1 &2 &-1 &  &&   \\
         &&\ddots&\ddots&\ddots&& \\
         &&&-1&2&-1&
         \\&&&&-1&2&-1
         \\&&&&&-1&2 \end{pmatrix}\Sigma
    \end{align}
corresponding to the usual presentation of the $\SU(N)$ Cartan matrix.

The signs and Dynkin coefficients of the weights in the fundamental and antisymmetric representations of $\SU(5)$ and $\SU(6)$, needed to compute the field theoretic expressions for the 3D Chern--Simons terms associated to these models, are given in \cref{SUNsigns}.\footnote{Since a pair of irreps $\sfr, \sfr^*$ related by conjugation are characterized by the same partial ordering, a labeling convention assigning indices to the weights of $\sfr$ automatically determines a identical convention for the conjugate irrep $\sfr^*$.}
\begin{table}
    \begin{center}
    $
    \begin{array}{|c|cc|}\hline \sfG & \boldsymbol{N} & \boldsymbol{N(N-1)/2} \\\hline \SU(5) &\left(
\begin{array}{c|cccc}
\frac{\varphi \cdot w}{|\varphi \cdot w|} & w_2 & w_3 & w_4 & w_5 \\\hline
 +&1 & 0 & 0 & 0 \\
 +&-1 & 1 & 0 & 0 \\
 +&0 & -1 & 1 & 0 \\
 -&0 & 0 & -1 & 1 \\
 -&0 & 0 & 0 & -1 \\
\end{array}
\right)&\left(
\begin{array}{c|cccc}
\frac{\varphi \cdot w}{|\varphi \cdot w|} & w_2 & w_3 & w_4 & w_5 \\\hline
 +&0 & 1 & 0 & 0 \\
 +&1 & -1 & 1 & 0 \\
 +&-1 & 0 & 1 & 0 \\
 +&1 & 0 & -1 & 1 \\
 +&-1 & 1 & -1 & 1 \\
 -&1 & 0 & 0 & -1 \\
 -&-1 & 1 & 0 & -1 \\
 -&0 & -1 & 0 & 1 \\
 -&0 & -1 & 1 & -1 \\
 -&0 & 0 & -1 & 0 \\
\end{array}
\right) \\
\SU(6) &\left(
\begin{array}{c|ccccc}\frac{\varphi \cdot w}{|\varphi \cdot w|} & w_2 & w_3 & w_4 & w_5 & w_6 \\\hline
 +&1 & 0 & 0 & 0 & 0 \\
 +&-1 & 1 & 0 & 0 & 0 \\
 +&0 & -1 & 1 & 0 & 0 \\
 -&0 & 0 & -1 & 1 & 0 \\
 -&0 & 0 & 0 & -1 & 1 \\
 -&0 & 0 & 0 & 0 & -1 \\
\end{array}
\right) & \left(
\begin{array}{c|ccccc}
\frac{\varphi \cdot w}{|\varphi \cdot w|} & w_2 & w_3 & w_4 & w_5 &w_6\\\hline
 +&0 & 1 & 0 & 0 & 0 \\
 +&1 & -1 & 1 & 0 & 0 \\
 +&-1 & 0 & 1 & 0 & 0 \\
 +&1 & 0 & -1 & 1 & 0 \\
 +&-1 & 1 & -1 & 1 & 0 \\
 +&1 & 0 & 0 & -1 & 1 \\
 +&-1 & 1 & 0 & -1 & 1 \\
 +&0 & -1 & 0 & 1 & 0 \\
 -&1 & 0 & 0 & 0 & -1 \\
 -&-1 & 1 & 0 & 0 & -1 \\
 -&0 & -1 & 1 & -1 & 1 \\
 -&0 & -1 & 1 & 0 & -1 \\
 -&0 & 0 & -1 & 0 & 1 \\
 -&0 & 0 & -1 & 1 & -1 \\
 -&0 & 0 & 0 & -1 & 0 \\
\end{array}
\right)\\\hline
\end{array}
$
\end{center}
\caption{Signs and Cartan charges of the BPS particles associated to the representations $\boldsymbol{N}$ and $\boldsymbol{N(N-1)/2}$ in the $\SU(N)$ model resolutions \labelcref{SUNres} for $N=5,6$. The charges are the Dynkin $w_i$ coefficients of the weights $w$ and the signs correspond to the signs of the BPS central charges $\varphi \cdot w$ for a given choice of Coulomb branch moduli $\varphi^i$. The indices $i$ of the weights are chosen to match the indices labeling the Cartan divisors $\hat D_i$ in \labelcref{SUNdiv}, which are associated to the simple coroots of $\mathfrak{su}(N)$}
\label{SUNsigns}
\end{table}

\subsection{$\SU(6)^{\circ}$ model}
The exotic $\SU(6)$ Tate tuning is
    \begin{align}
        a_1 = a_1\,, \quad a_2 = a_{2,2} \sigma^2\,, \quad a_{3} = a_{3,\frac{N}{2} -1} \sigma^{\frac{N}{2} -1}\,, \quad a_{4} = a_{4,\frac{N}{2} +1} \sigma^{\frac{N}{2} +1}\,, \quad a_{6} = a_{6,N} \sigma^N
    \end{align}
for $N$ even and where $\sigma = 0$ is the codimension-one locus in the base over which there is a I$_{N}^\text{split}$ singularity. We study the resolution $X_5 \rightarrow X_0$ obtained by the following sequence of blowups:
    \begin{align}
    \label{exoticSU6res}
        X_5 \overset{(y,e_4|e_5)}{\longrightarrow} X_4 \overset{(y,e_3|e_4)}{\longrightarrow} X_3  \overset{(x,e_2|e_3)}{\longrightarrow} X_2  \overset{(y,e_1|e_2)}{\longrightarrow} X_1  \overset{(x,y,\sigma|e_1)}{\longrightarrow} X_0\,.
    \end{align}
The signs to which the above resolution corresponds are given in \cref{SU6signs}. The classes of the Cartan divisors $\hat D_{i=2, \dots, 6}$ are 
    \begin{align}
        \begin{split}
        \label{SU6div}
            \hat D_2& = (\bm{E}_1-\bm{E}_2) \cap \bm{X}_5\,, \\
            \hat D_3&= (\bm{E}_3 - \bm{E}_4)  \cap \bm{X}_5\,, \\
            \hat D_4&= (\bm{E}_4 - \bm{E}_5)  \cap \bm{X}_5\,, \\
            \hat D_5&= \bm{E}_5   \cap \bm{X}_5\,, \\
            \hat D_6&= (\bm{E}_2 - \bm{E}_3)  \cap \bm{X}_5\,.
        \end{split}
    \end{align}

\begin{table}
\begin{center}
    $
\left(
\begin{array}{c|ccccc}
\frac{\varphi \cdot w}{|\varphi \cdot w|} & w_2 & w_3 &w_4 & w_5 & w_6 \\\hline
+& 1 & 0 & 0 & 0 & 0 \\
+& -1 & 1 & 0 & 0 & 0 \\
 +&0 & -1 & 1 & 0 & 0 \\
 +&0 & 0 & -1 & 1 & 0 \\
 -&0 & 0 & 0 & -1 & 1 \\
 -&0 & 0 & 0 & 0 & -1 \\
\end{array}
\right)
$
\end{center}
\caption{Signs and Cartan charges of the BPS particles associated to the weights of the $\textbf{6}$ in the exotic $\SU(6)$ model resolution \labelcref{exoticSU6res}. The charges are the Dynkin coefficients of the weights of the $\textbf{6}$ and the signs correspond to the signs of the BPS central charges $\varphi \cdot w$ for a given choice of Coulomb branch moduli $\varphi^i$.}
\label{SU6signs}
\end{table}

\subsection{$\SO(4k+2)$ model}

The SO($4k+2$) Tate model is characterized by a I$^{*\text{split}}_{2k-3}$ Kodaira singularity singularity over a gauge divisor $\Sigma = \{ \sigma = 0\}\subset B$, and can be realized explicitly using a Weierstrass model
    \begin{align}
    \label{SOTateform}
         y^2 z + a_1 xyz + a_3 y z^2 - (x^3 + a_2 x^2 z + a_4 x z^2 + a_6 z^3 ) = 0
    \end{align}
together with the Tate tuning
    \begin{align}
    \label{SO4k+2Tate}
        \text{I}_{2k-3}^{*\text{split}} ~:~ a_1 = a_{1,1} \sigma,~~~~ a_2 = a_{2,1} \sigma,~~ ~~a_{3,k} \sigma^{k},~~~~ a_{4, k+1} \sigma^{k+1},~~~~ a_6 = a_{6,2k+1} \sigma^{2k+1}\,.
    \end{align}
These models have matter in the adjoint, fundamental ($\boldsymbol{4k+2}$), and spinor ($\boldsymbol{4^{k}}$). The discriminant locus takes the form
    \begin{align}
        \Delta = \sigma^{2k+3} (\Delta^{(2)}+ \cO(\sigma) ),~~~~ \Delta^{(2)} = 16 a_2^3 a_3^2 \,.
    \end{align}
The classes of the matter curves of the fundamental and spinor representations are
    \begin{align}
        C_{\boldsymbol{4k+2}} = \Sigma \cdot (-3K -k \Sigma) ,~~~~ C_{\boldsymbol{4^k}} = \Sigma \cdot (-2K- \Sigma )\,.
    \end{align}
We study the family of resolutions $X_{2k+1} \rightarrow X_0$ realized in \cite{Bhardwaj:2018yhy} by the sequence of blowups
    \begin{align}
    \begin{split}
    \label{SO4k+2res}
        X_{2k+1} &\overset{(e_{2k-2},e_{2k-1}|e_{2k+1})}{\longrightarrow}X_{2k} \overset{(y,e_{2k-1}|e_{2k})}{\longrightarrow} X_{2k-1} \overset{(x,e_{2k-2}|e_{2k-1})}{\longrightarrow}  \cdots \\
        & ~~~~~~~~\cdots \overset{(y,e_1|e_2)}{\longrightarrow} X_1\overset{(x,y,\sigma|e_1)}{\longrightarrow} X_0\,.
    \end{split}
    \end{align}
The class of the holomorphic zero section is
    \begin{align}
        \hat D_0 = \frac{\bm{H}}{3} \cap \bm{X}_{2k+1}\,.
    \end{align}
Specifically, we restrict our attention to the specific case $\SO(10)$ (i.e., $k=2$). For the $\SO(10)$ model we choose the basis of Cartan divisors ($i = 2, \dots,6$),
    \begin{align}
        \begin{split}
        \label{SO10div}
            \hat D_2 &= (-\boldsymbol E_1 + 2 \boldsymbol E_2 - \boldsymbol E_3 - \boldsymbol E_5) \cap \boldsymbol X_{5} \\
            \hat D_3 &= (\boldsymbol E_1 - \boldsymbol E_2 ) \cap \boldsymbol X_5 \\
            \hat D_4 &= \boldsymbol E_5 \cap \boldsymbol X_5 \\
            \hat D_5 &= \boldsymbol E_4 \cap \boldsymbol X_5 \\
            \hat D_6 &=(\boldsymbol E_3 - \boldsymbol E_4 - \boldsymbol E_5) \cap \boldsymbol X_5
        \end{split}
    \end{align}
in which the Cartan matrix is represented as
    \begin{align}
        [[W_{ij}]] = [[ \pi_*(\hat D_i \cdot \hat D_j) ]] =- \left(
\begin{array}{ccccc}
 2 & -1 & 0 & 0 & 0 \\
 -1 & 2 & -1 & 0 & 0 \\
 0 & -1 & 2 & -1 & -1 \\
 0 & 0 & -1 & 2 & 0 \\
 0 & 0 & -1 & 0 & 2 \\
\end{array}
\right) \Sigma\,.
    \end{align}
\begin{table}
    \begin{center}
        $
            \left(
\begin{array}{c|ccccc}
\frac{\varphi \cdot w}{| \varphi \cdot w|} & w_2 & w_3 & w_4 &w_5 & w_6  \\\hline
 + & 0 & 0 & 0 & 0 & 1 \\
 + & 0 & 0 & 1 & 0 & -1 \\
 + & 0 & 1 & -1 & 1 & 0 \\
 + & 0 & 1 & 0 & -1 & 0 \\
 + & 1 & -1 & 0 & 1 & 0 \\
 + & -1 & 0 & 0 & 1 & 0 \\
 + & 1 & -1 & 1 & -1 & 0 \\
 - & -1 & 0 & 1 & -1 & 0 \\
 - & 1 & 0 & -1 & 0 & 1 \\
 - & -1 & 1 & -1 & 0 & 1 \\
 - & 1 & 0 & 0 & 0 & -1 \\
 - & -1 & 1 & 0 & 0 & -1 \\
 - & 0 & -1 & 0 & 0 & 1 \\
 - & 0 & -1 & 1 & 0 & -1 \\
 - & 0 & 0 & -1 & 1 & 0 \\
 - & 0 & 0 & 0 & -1 & 0 \\
\end{array}
\right)
        $
    \end{center}
\caption{Signs and Cartan charges associated to BPS particles associated to the weights of the $\textbf{16}$ in the SO(10) model resolution \labelcref{SO4k+2res}. The charges are the Dynkin coefficients of the weights of the $\textbf{16}$ and the signs correspond to the signs of the BPS central charges $\varphi \cdot w$ for given Coulomb branch moduli $\varphi^i$. The indices $i$ of the weights are chosen to match the indices labeling the Cartan divisors $\hat D_i$ in \labelcref{SO10div}, which are associated to the simple coroots of $\mathfrak{so}(10)$.}
\label{SO10signs}
\end{table}
%For the $\SO(14)$ model we choose the basis of Cartan divisors ($i=2,\dots,8$)
%   \begin{align}
%       \begin{split}
%           \hat D_2 &= ( - \boldsymbol E_1 +2 \boldsymbol E_2 -2\boldsymbol E_3 + 2 \boldsymbol E_4-\boldsymbol E_5-\boldsymbol E_7) \cap \boldsymbol X_{7}\\
%           \hat D_3&=( \boldsymbol E_1- \boldsymbol E_2)\cap \boldsymbol X_{7}\\
%           \hat D_4 &=( \boldsymbol E_3-2 \boldsymbol E_4+ \boldsymbol E_5+ \boldsymbol E_7)\cap \boldsymbol X_{7}\\
%           \hat D_5 &=( \boldsymbol E_4 - \boldsymbol E_5 - \boldsymbol E_7)\cap \boldsymbol X_{7}\\
%           \hat D_6&= \boldsymbol E_7 \cap \boldsymbol X_7\\
%           \hat D_7 &=( \boldsymbol E_5- \boldsymbol E_6- \boldsymbol E_7)\cap \boldsymbol X_{7}\\
%           \hat D_8 &= \boldsymbol E_6\cap \boldsymbol X_{7}.\\
%       \end{split}
%   \end{align}
%In this basis, the Cartan matrix is represented as
%   \begin{align}
%       [[W_{ij} ]] = [[ \pi_* \hat D_i \cdot \hat D_j ]] = -\left(
%\begin{array}{ccccccc}
% -2 & 1 & 0 & 0 & 0 & 0 & 0 \\
% 1 & -2 & 1 & 0 & 0 & 0 & 0 \\
% 0 & 1 & -2 & 1 & 0 & 0 & 0 \\
% 0 & 0 & 1 & -2 & 1 & 0 & 0 \\
% 0 & 0 & 0 & 1 & -2 & 1 & 1 \\
% 0 & 0 & 0 & 0 & 1 & -2 & 0 \\
% 0 & 0 & 0 & 0 & 1 & 0 & -2 \\
%\end{array}
%\right)\Sigma.
%   \end{align}

\subsection{E$_6$ model}
\label{E6appendix}
The E$_6$ Tate model $X_0$ is characterized by a IV$^{* \text{split}}$ singularity over divisor $\Sigma = \{ \sigma  =0\} \subset B$ and can be defined by the following Weierstrass equation
    \begin{align}
    y^2z + a_{3,2} \sigma^2 y z^2- ( x^3 + a_{4,3} \sigma^2 xz^2 + a_{6,5} \sigma^5z^3 ) = 0\,.
    \end{align}
This model has a holomorphic zero section $x= z=0$ and matter spectrum $\textbf{27} \oplus \textbf{78}$. We study the resolution \cite{Esole:2017kyr}
    \begin{align}
    \label{E6res}
        X_6 \overset{(y,e_4|e_6)}{\longrightarrow}  X_5\overset{(y,e_3|e_5)}{\longrightarrow}  X_4  \overset{(e_2,e_3|e_4)}{\longrightarrow}  X_3 \overset{(x,e_2|e_3)}{\longrightarrow} X_2  \overset{(y,e_1|e_2)}{\longrightarrow} X_1 \overset{(x,y,\sigma|e_1)}{\longrightarrow}  X_0\,.
    \end{align}
The signs to which the above resolution corresponds are given in \cref{E6signs}.
    \begin{table}
    \begin{center}
    $
\left(
\begin{array}{c|cccccc}
\frac{\varphi \cdot w}{| \varphi \cdot w|} & w_2 & w_3 & w_4 &w_5 & w_6 & w_7 \\\hline
 +&1 & 0 & 0 & 0 & 0 & 0 \\
 +&-1 & 1 & 0 & 0 & 0 & 0 \\
 +&0 & -1 & 1 & 0 & 0 & 0 \\
 +&0 & 0 & -1 & 1 & 0 & 1 \\
 +&0 & 0 & 0 & -1 & 1 & 1 \\
 +&0 & 0 & 0 & 1 & 0 & -1 \\
 +&0 & 0 & 0 & 0 & -1 & 1 \\
 +&0 & 0 & 1 & -1 & 1 & -1 \\
 +&0 & 0 & 1 & 0 & -1 & -1 \\
 +&0 & 1 & -1 & 0 & 1 & 0 \\
 +&0 & 1 & -1 & 1 & -1 & 0 \\
 +&1 & -1 & 0 & 0 & 1 & 0 \\
-& -1 & 0 & 0 & 0 & 1 & 0 \\
 +&0 & 1 & 0 & -1 & 0 & 0 \\
 +&1 & -1 & 0 & 1 & -1 & 0 \\
-& -1 & 0 & 0 & 1 & -1 & 0 \\
-& 1 & -1 & 1 & -1 & 0 & 0 \\
 -&-1 & 0 & 1 & -1 & 0 & 0 \\
-& 1 & 0 & -1 & 0 & 0 & 1 \\
 -&-1 & 1 & -1 & 0 & 0 & 1 \\
 -&1 & 0 & 0 & 0 & 0 & -1 \\
 -&-1 & 1 & 0 & 0 & 0 & -1 \\
 -&0 & -1 & 0 & 0 & 0 & 1 \\
 -&0 & -1 & 1 & 0 & 0 & -1 \\
 -&0 & 0 & -1 & 1 & 0 & 0 \\
 -&0 & 0 & 0 & -1 & 1 & 0 \\
 -&0 & 0 & 0 & 0 & -1 & 0 \\
\end{array}
\right)
$
\end{center}
\caption{Signs and Cartan charges associated to BPS particles associated to the weights of the $\textbf{27}$ in the E$_6$ model resolution \labelcref{E6res}. The charges are the Dynkin coefficients of the weights of the $\textbf{27}$ and the signs correspond to the signs of the BPS central charges $\varphi \cdot w$ for given Coulomb branch moduli $\varphi^i$. The indices $i$ of the weights are chosen to be match the indices of the Cartan divisors $\hat D_i$ in \cref{E6div}, which are associated to the simple coroots of $\mathfrak{e}_6$.}
\label{E6signs}
\end{table}
Again, the notation for the blowup maps in the above equation is explained in \labelcref{blowupnotation} and the discussion immediately below. The class of the holomorphic zero section is
    \begin{align}
        \hat D_0 = \frac{\bm{H}}{3} \cap \bm{X}_{N-1}\,.
    \end{align}
The classes of the Cartan divisors $\hat D_{i=2, \dots, 7}$ are 
    \begin{align}
        \begin{split}
        \label{E6div}
            \hat D_2& = \bm{E}_5 \cap \bm{X}_6\\
            \hat D_3&= \bm{E}_6  \cap \bm{X}_6\\
            \hat D_4&= (- \bm{E}_1+2  \bm{E}_2 - \bm{E}_3 - \bm{E}_4) \cap \bm{X}_6\\
            \hat D_5&=(  \bm{E}_1- 2  \bm{E}_2 +  \bm{E}_3 + 2  \bm{E}_4 -  \bm{E}_6) \cap  \bm{X}_6\\
            \hat D_6&= ( \bm{E}_3 -  \bm{E}_4 - \bm{E}_5) \cap  \bm{X}_6\\
            \hat D_7&= (\bm{E}_1 - \bm{E}_2) \cap  \bm{X}_6\,.
        \end{split}
    \end{align}
The above Cartan divisors are labeled such that
    \begin{align}
        [[W_{ij}]] = [[ \pi_*(\hat D_i \cdot \hat D_j)]] =- \left(
\begin{array}{cccccc}
 2 & -1 & 0 & 0 & 0 & 0 \\
 -1 & 2 & -1 & 0 & 0 & 0 \\
 0 & -1 & 2 & -1 & 0 & -1 \\
 0 & 0 & -1 & 2 & -1 & 0 \\
 0 & 0 & 0 & -1 & 2 & 0 \\
 0 & 0 & -1 & 0 & 0 & 2 \\
\end{array}
\right)\Sigma\,.
\end{align}

\section{A puzzle: $\SU(5)$ model resolutions}
\label{sec:puzzle}

In this paper we restrict our attention to vertical M-theory flux backgrounds $G \in H^{2,2}_{\text{vert}}(X,\Z)$ where $X$ is an arbitrary smooth CY fourfold. In order for such backgrounds to lift to consistent F-theory backgrounds, we require that $X$ is a resolution of a singular CY fourfold $X_0$ and that $G$ satisfies certain conditions both necessary and sufficient to lift to an F-theory flux background.

Our procedure for computing the chiral indices $\chi_{\sfr}$ is predicated on the assumption that given an M-theory flux background $X, G$ satisfying the above conditions, the full set of chiral multiplicities $\chi_{\sfr}$ can be extracted in the M-theory duality frame by identifying an appropriate collection of matter surfaces $S_{\sfr}$ and computing integrals of the form $\chi_{\sfr} = \int_{S_{\sfr}} G$. Although the equation $\chi_{\sfr} = \int_{S_{\sfr}} G$ is expected to hold true for general $X$ and $G \in H^{2,2}(X,\R) \cap H^4(X,\Z)$ lifting to consistent F-theory flux vacua, in our case the restriction $G \in H^{2,2}_{\text{vert}}(X,\Z)$ necessitates further assumptions about the matter surfaces, namely that (recalling the orthogonal decomposition \cref{ortho}) since a complete basis of vertical fluxes is given by
	\begin{align}
		\Theta_{ij} = \int_{S_{ij}} G,	
	\end{align}
there exists a non-trivial choice of coefficients $x^{ij}_{\sfr}$ satisfying
	\begin{align}
		\chi_{\sfr} = x^{ij}_{\sfr} \Theta_{ij} = \int_{x^{ij}_{\sfr} S_{Cij}} G
	\end{align}
for all $\sfr$ in the 4D spectrum. Equivalently, \cref{ortho} suggests that our procedure only yields a non-trivial answer for the chiral index $\chi_{\sfr}$ provided the corresponding matter surface $S_{\sfr}$ has components in the vertical homology $H_{2,2}^{\text{vert}}(X,\Z)$:
	\begin{align}
		S_{\sfr} =x^{ij}_{\sfr} S_{Cij} =  x^{ij}_{\sfr} S_{ij} + \cdots,~~~~x^{ij}_{\sfr} S_{ij}  \ne 0\,,
	\end{align} 
where $\cdots$ indicates other components with indices $IJ \ne ij$. If no $x^{ij}_{\sfr}$ exist such that the above equation is satisfied, this suggests $S_{\sfr}$ does not have any components in $H_{2,2}^{\text{vert}}(X,\Z)$. 

A useful test of these assumptions is to compute the full set of gauge symmetry-preserving vertical fluxes all available resolutions of a given $\sfG$ model; if the assumptions are valid, then for any such resolution $X$ it will always be possible to find such a choice of $x^{ij}_{\sfr}$, so that the chiral indices may be expressed as a sum over $\Theta_{ij}$. One valuable model for which the full set of resolutions (up to codimension-three singularities in the base $B$) has been computed is the universal $\SU(5)$ model with generic matter \cite{Esole:2014hya}.

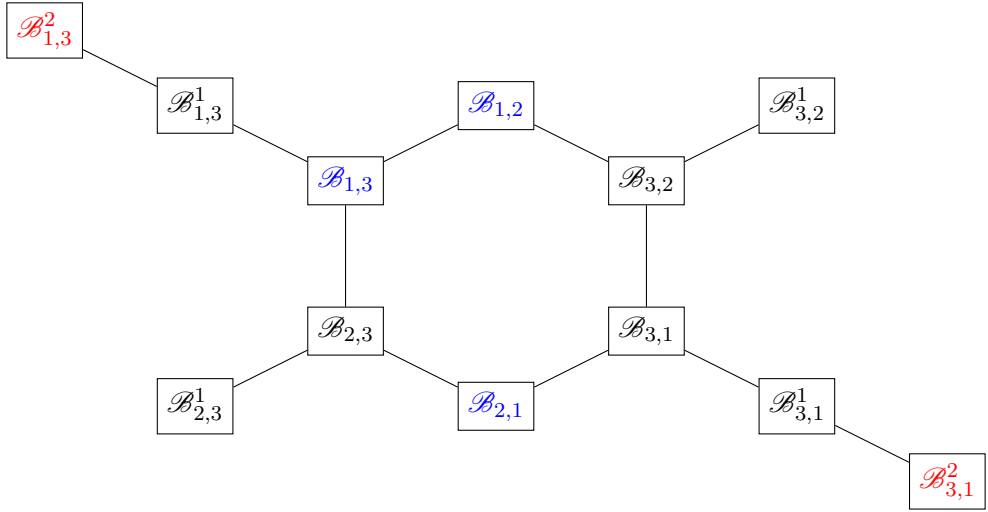
\begin{figure}
	\begin{center}
		\begin{tikzpicture}[yscale=2,xscale=4]
			\node[draw,rectangle](B13) at (-.5,.5) {$\textcolor{blue}{\scB_{1,3}}$};
			\node[draw,rectangle](B12) at (0,1) {$\textcolor{blue}{\scB_{1,2}}$};
			\node[draw,rectangle](B32) at (.5,.5) {$\scB_{3,2}$};
			\node[draw,rectangle](B23) at (-.5,-.5) {$\scB_{2,3}$};
			\node[draw,rectangle](B21) at (0,-1) {$\textcolor{blue}{\scB_{2,1}}$};
			\node[draw,rectangle](B31) at (.5,-.5) {$\scB_{3,1}$};
			\node[draw,rectangle](B321) at (1,1) {$\scB_{3,2}^1$};
			\node[draw,rectangle](B311) at (1,-1) {$\scB_{3,1}^1$};
			\node[draw,rectangle](B312) at (1.5,-1.5) {\textcolor{red}{$\scB_{3,1}^2$}};
			\node[draw,rectangle](B231) at (-1,-1) {$\scB_{2,3}^1$};
			\node[draw,rectangle](B131) at (-1,1) {$\scB_{1,3}^1$};
			\node[draw,rectangle](B132) at (-1.5,1.5) {\textcolor{red}{$\scB_{1,3}^2$}};
			\draw (B13)-- (B12) -- (B32) -- (B31) -- (B21) -- (B23) -- (B13);
			\draw (B132) -- (B131) -- (B13);
			\draw (B231) -- (B23);
			\draw (B31) -- (B311) -- (B312);
			\draw (B32) -- (B321);
		\end{tikzpicture}
	\end{center}
	\label{SU5flops}
	\caption{Network of resolutions for the $\SU(5)$ Tate model, as presented in Figure 5 of \cite{Esole:2014hya} (see also Figure 6). Each boxed node represents a particular resolution of the singular $\SU(5)$ model and each edge connecting a pair of nodes indicates a flop transition between two resolutions. A flop transition between two resolutions $\scB, \scB'$ is characterized by all curves belonging to a homology class $C$ of $\scB$ collapsing to zero volume and a new curve $C'$ being blown up whose volume in $\scB'$ is (formally) minus the volume of $C$ in $\scB$. In the context of the $\SU(5)$ model the curve classes of interest $C_w$ correspond to weights $w$ transforming in some representation of $\SU(5)$; in particular, the volume of such a curve is $\varphi^i w_i$, where $w_i$ are the Dynkin coefficients of the weight $w$. The resolutions in red are those for which the 3D CS terms $\Theta^{\text{3D}}_{ij} = 0$.}
\end{figure}	

Unfortunately, the pushforward technology used to compute quadruple intersection numbers as described in this paper cannot be applied to all of these resolutions, so we do not have a direct geometric computation of $\Theta_{ij}$ for the full network of resolutions of the $\SU(5)$ model. However, the analysis of \cite{Esole:2014hya} not only includes explicit descriptions of the resolutions, but also the Dynkin coefficients $\hat D_i \cdot C_{w} = w_i$ of the matter curves $C_{w}$ whose volumes $\text{vol}(C_w) = \varphi \cdot w$ shrink to zero as a result of the flop transitions connecting pairs of resolutions. Since we know for certain resolutions the field theoretic expressions \cref{eq:qIJ1loop} for the fluxes $\Theta_{ij} = - \Theta_{ij}^{\text{3D}}$ in terms of the Dynkin coefficients $w_i$ and the signs of the BPS central charges $\varphi \cdot w$ (see e.g. \cref{SUNsigns}), by starting with the known collection of signs we can use the fact that the sign of a single central charge $\varphi \cdot w$ flips as we move to an adjacent resolution, to determine the signs of the full set of central charges in the adjacent resolution---in other words, we use the fact that the signs of the central charges of a pair of resolutions related by a flop transition differ by a single sign flip. In this manner, by flipping the signs of appropriate central charges as we move around the graph in \cref{SU5flops}, we can determine the signs of all the $\SU(5)$ resolutions, which in turn permits us to compute the field theoretic expressions for $\Theta_{ij}$ for all $\SU(5)$ model resolutions, at least in principle.

At face value, simply knowing the field theoretic expressions $\Theta^{\text{3D}}_{ij}$ for all resolutions of a given $\sfG$ model does not appear to be particularly illuminating, because it tells us nothing about the corresponding geometric expressions for $\Theta_{ij}$ in the stringy UV completion. Nevertheless, one reasonable assumption we can make is that 4D anomaly cancellation is satisfied and hence the chiral indices $\chi_{\sfr}$ appearing in the field theoretic expressions
	\begin{align}
		\Theta_{ij}^{\text{3D}} = x_{ij}^{\sfr} \chi_{\sfr}
	\end{align} 
may freely be constrained to obey 4D cancellation while remaining consistent with the geometric expressions $\Theta_{ij} = \int_{S_{ij}} G$. This observation potentially leads to a puzzle: Suppose there exists a resolution (equivalently, a collection of signs of central charges) such that the coefficients $x_{ij}^{\sfr}$ are proportional to the coefficients of the pure gauge anomaly condition for all $ij$. In such cases, for anomaly cancellation to be satisfied we must have $\Theta_{ij} = 0$ for all $ij$, and hence it appears our assumption about the matter surfaces having non-trivial vertical components somehow fails.

In fact, this seems to be precisely the case for the two resolutions $\scB_{1,3}^2, \scB_{3,1}^2$ of the $\SU(5)$ model. We can see this in the case of $\scB_{1,3}^2$ by following the procedure described above. First note the signs appearing in \cref{SUNsigns} correspond to the resolution $\scB_{1,3}$, for which we find
	\begin{align}
	\begin{split}
 \Theta^{\text{3D}}_{22} &= -\chi_{\boldsymbol{5}}-\chi_{\boldsymbol{10}} \\
 \Theta^{\text{3D}}_{23} &= \frac{1}{2} \left(\chi_{\boldsymbol{5}}+\chi_{\boldsymbol{10}}\right) \\
 \Theta^{\text{3D}}_{33} &= -\chi_{\boldsymbol{5}} \\
 \Theta^{\text{3D}}_{24} &= 0 \\
 \Theta^{\text{3D}}_{34} &= \frac{1}{2} \left(\chi_{\boldsymbol{5}}+\chi_{\boldsymbol{10}}\right) \\
 \Theta^{\text{3D}}_{44} &= -\chi_{\boldsymbol{10}} \\
 \Theta^{\text{3D}}_{25} &= 0 \\
 \Theta^{\text{3D}}_{35} &= -\chi_{\boldsymbol{10}} \\
 \Theta^{\text{3D}}_{45} &= \frac{1}{2} \left(\chi_{\boldsymbol{10}}-\chi_{\boldsymbol{5}}\right) \\
 \Theta^{\text{3D}}_{55} &= \chi_{\boldsymbol{5}}+\chi_{\boldsymbol{10}} 
	\end{split}
	\end{align}
To go from the resolution $\scB_{1,3}$ to $\scB_{1,3}^1$, we must flip the sign of the central charge corresponding to the weight $w^{\boldsymbol{10}} = (0,-1,0,1)$, which according to \cref{SUNsigns} is negative and hence must become positive. Then, to go from the resolution $\scB_{1,3}^1$ to the resolution $\scB_{1,3}^2$, we must flip the sign of the central charge corresponding to the weight $w^{\boldsymbol{5}} = (0,0,-1,1)$ from negative to positive. Computing $\Theta^{\text{3D}}_{ij}$, we find 	\begin{align}
		\begin{split}
 \Theta^{\text{3D}}_{22} &= -(\chi_{\boldsymbol{5}}+\chi_{\boldsymbol{10}}) \\
 \Theta^{\text{3D}}_{23} &= \frac{1}{2} \left(\chi_{\boldsymbol{5}}+\chi_{\boldsymbol{10}}\right) \\
 \Theta^{\text{3D}}_{33} &= -(\chi_{\boldsymbol{5}}+\chi_{\boldsymbol{10}}) \\
 \Theta^{\text{3D}}_{24} &= 0 \\
 \Theta^{\text{3D}}_{34} &= \frac{1}{2} \left(\chi_{\boldsymbol{5}}+\chi_{\boldsymbol{10}}\right) \\
 \Theta^{\text{3D}}_{44} &= -(\chi_{\boldsymbol{5}}+\chi_{\boldsymbol{10}}) \\
 \Theta^{\text{3D}}_{25} &= 0 \\
 \Theta^{\text{3D}}_{35} &= 0 \\
 \Theta^{\text{3D}}_{45} &= \frac{1}{2} \left(\chi_{\boldsymbol{5}}+\chi_{\boldsymbol{10}}\right) \\
 \Theta^{\text{3D}}_{55} &= 0. \\
\end{split}
	\end{align}	
Comparing the above expressions to the anomaly cancellation condition $\chi_{\boldsymbol 5} + \chi_{\boldsymbol{10}} = 0$, we therefore expect $\Theta_{ij} = -\Theta_{ij}^{\text{3D}} = 0$ for all $ij$ in the resolution $\scB_{1,3}^{2}$. Analogous results hold for $\scB_{3,1}^2$.

One possible interpretation of the above computation is that $\Theta_{ij}^{\text{3D}} = - \Theta_{ij} = 0$ and hence according to our above reasoning the matter surfaces $S_{\sfr}$ do not contain any vertical components. This would in turn suggest that in general only a proper subset of possible resolutions can be used to access information about the 4D massless chiral spectrum, given purely vertical flux backgrounds. If this conclusion is true, the resolutions $\scB_{1,3}^2, \scB_{3,1}^2$ are counterexamples to the conjecture that $H_{2,2}^{\text{vert}}(X,\Z)$ is a resolution-independent structure and they furthermore contradict the assumption that all resolutions can be used to compute the 4D massless chiral spectrum (at least for strictly vertical flux backgrounds). On the other hand, since we do not currently have the means to compute lattice of vertical fluxes for $\scB_{1,3}^2, \scB_{3,1}^2$, it is possible that these resolutions exhibit some unusual features that might explain why $\Theta^{\text{3D}}_{ij} = 0$. We leave a satisfactory explanation of this puzzle to future work, although we stress that it is important to understand this aspect of the computation since the explanation could significantly affect key assumptions underlying our analysis.  

\section{Reducing the intersection pairing matrix}
\label{sec:lattice-reduce}
We now present the systematic approach to finding $M_\text{red}$ from the
degenerate intersection pairing matrix $M$. We seek to find a basis for the
integral lattice defined by $M$ such that $M_\text{red}$ is embedded as a
submatrix and all other entries are zero, i.e., we wish to find a unimodular
matrix $P \in \GL(\dim M, \Z)$ such that
    \begin{equation}
        P^\transpose M P = \begin{pmatrix}M_\text{red} & 0 \\ 0 &
          0\end{pmatrix}\,.
\label{eq:pmp}
    \end{equation}
If we have a (degenerate) basis matrix for $M$, i.e., a matrix $B$ satisfying
$M = B^\transpose B$ (which can be found using the Cholesky algorithm
generalized to positive semidefinite symmetric matrices), then this can be carried out
using standard lattice reduction algorithms such as the LLL algorithm
\cite{LLL,Bremner}. However, although $M$ defines an integral lattice, there
may be no integral basis $B$ for this lattice, and so determining $B$ can
require extracting square roots. In practice, one may wish to avoid this due
to issues with floating point arithmetic. 

An alternate approach is to begin by finding the LDLT decomposition of $M$,
yielding a lower unitriangular matrix $L$ and a diagonal matrix $D$ such that
$M = L D L^\transpose$. (The basis matrix $B$ could then be found as $B =
\sqrt{D} L^\transpose$, which can clearly introduce square roots.) A basis
satisfying our desired properties can be found by putting $B^\transpose$ in
(row-style) Hermite normal form $H = U B^\transpose$.
% , i.e., finding a unimodular matrix $U$ such that the
% matrix $H = U B^\transpose$ satisfies the following restrictions:
%   \begin{itemize}
%       \item
%           $H$ is upper triangular with any zero rows at the bottom,
%       \item
%           the leftmost nonzero entry (the \emph{leading coefficient}) in each row is positive and strictly to the right of the leading coefficient of the row above it,
%       \item
%           the entries below each leading coefficient are zero and the entries above each leading coefficient are nonnegative and strictly smaller than that leading coefficient.
%   \end{itemize}
Two basis matrices $B$ and $B'$ describe the same integral lattice if and only
if their transposes have the same Hermite normal form $H$, and thus the
transpose of the Hermite normal form $H^\transpose$ itself serves as an
appropriate choice of basis matrix. From the definition of the basis matrix,
we see then that $P = U^\transpose$ provides us with the congruence we were
seeking. While we can put the potentially real-valued matrix $B$
into Hermite normal form using integer Gaussian elimination or a modification
of the LLL algorithm \cite{HavasMajewskyMatthews}, the benefit of this approach is that using LDLT decomposition
allows us to find the appropriate $U$ without needing to extract the square
roots in $B$. Specifically, the $U$ that puts $B^\transpose$ into Hermite
normal form is also the matrix that puts $L$ into Hermite normal form,
$\tilde{H} = U L$. Thus, the desired unimodular congruence matrix $P =
U^\transpose$ can be found from the LDLT decomposition using only rational
matrices.

To summarize, the desired unimodular congruence $P$ can be found efficiently
either by using the LLL algorithm directly on the potentially real-valued
basis matrix $B$ (which may be found using the Cholesky decomposition), or by
finding the LDLT decomposition $M = L D L^\transpose$ and then using the LLL
algorithm to find the matrix $U$ putting $L$ into Hermite normal form and
setting $P = U^\transpose$; the latter approach avoids ever introducing
real-valued matrices.\footnote{It is worth noting that while this approach does provide a systematic method to find $M_\text{red}$, which amounts to finding a lattice basis for the null space of $M$, it does not provide a general method for checking if two integral lattices are congruent to one another, as there is ambiguity in the determination of the basis matrix $B$ (or equivalently in using the LDLT decomposition). Specifically, there are in general multiple valid choices of basis matrix $B$ that span different integral lattices in $\R^n$ but nevertheless reproduce the same Gram matrix $M$. Relating two lattices $M, M'$ via the approach outlined here is thus sufficient but not necessary to prove congruence over the integers.} In Mathematica, we can implement the latter approach to
find $U$ simply as
\verb+Transpose@First@HermiteDecomposition[First@LDLT[mat]]]+, where
\verb+mat+ is the input matrix and \verb+LDLT+ is a user-defined function
computing the LDLT decomposition of \verb+mat+ and returning \verb+{L, D}+.

As an example of this procedure, consider
    \begin{equation}
        \label{eq:MredExample}
        M =
        \begin{pmatrix}
            9   & 0   & -3 & -21 \\
            0   & 9   & -6 & -42 \\
            -3  & -6  & 5  & 35 \\
            -21 & -42 & 35 & 245
        \end{pmatrix}\,,
    \end{equation}
which has two independent null vectors, $(7, 14, 0, 3)$ and $(1, 2, 3, 0)$. The LDLT decomposition of this matrix is
    \begin{equation}
        M = L D L^\transpose\,, \quad
        L =
        \begin{pmatrix}
            1            & 0             & 0 & 0 \\
            0            & 1             & 0 & 0 \\
            -\frac{1}{3} & -\frac{2}{3}  & 1 & 0 \\
            -\frac{7}{3} & -\frac{14}{3} & 0 & 1
        \end{pmatrix}\,, \quad
        D = \diag(9, 9, 0, 0)\,.
    \end{equation}
The matrix $L$ can be put into Hermite normal form via a unimodular matrix $U$ as
    \begin{equation}
        H = U L\,, \quad
        H =
        \begin{pmatrix}
            \frac{1}{3} & \frac{2}{3} & 0 & 2 \\
            0           & 1           & 0 & 0 \\
            0           & 0           & 1 & 2 \\
            0           & 0           & 0 & 3
        \end{pmatrix}\,, \quad
        U =
        \begin{pmatrix}
            5 & 10 & 0 & 2 \\
            0 & 1  & 0 & 0 \\
            5 & 10 & 1 & 2 \\
            7 & 14 & 0 & 3
        \end{pmatrix}\,.
    \end{equation}
We find then that
    \begin{equation}
        U M U^\transpose =
        \begin{pmatrix}
            5 & 10 & 0 & 2 \\
            0 & 1  & 0 & 0 \\
            5 & 10 & 1 & 2 \\
            7 & 14 & 0 & 3
        \end{pmatrix}
        \begin{pmatrix}
            9   & 0   & -3 & -21 \\
            0   & 9   & -6 & -42 \\
            -3  & -6  & 5  & 35 \\
            -21 & -42 & 35 & 245
        \end{pmatrix}
        \begin{pmatrix}
            5  & 0 & 5  & 7 \\
            10 & 1 & 10 & 14 \\
            0  & 0 & 1  & 0 \\
            2  & 0 & 2  & 3
        \end{pmatrix} =
        \begin{pmatrix}
            5 & 6 & 0 & 0 \\
            6 & 9 & 0 & 0 \\
            0 & 0 & 0 & 0 \\
            0 & 0 & 0 & 0
        \end{pmatrix}\,,
    \end{equation}
as desired.

Although the method above provides a systematic approach to find $M_\text{red}$, for the cases we consider in this paper the result can be easily read off by inspection. We now briefly discuss two situations that one may encounter when carrying out this process: in some cases, $M_\text{red}$ cannot be found as a submatrix in $M$ and so a nontrivial basis change is required, and in some cases there may be rational (as opposed to integral) components of the lattice coordinates at intermediate stages of the process, with the final result nevertheless involving only integer values for the lattice coordinates.
As an example of the kind of issue that arises in the former case, consider
the integral lattice $\Gamma = \Z^2$ with the symmetric bilinear form
\begin{equation}
\begin{pmatrix}
9 & 6 \\
6 &  4
\end{pmatrix}
\label{eq:9664}
\end{equation}
and null vector $(2, -3)$.  The quotient lattice $\Gamma'
=\Gamma\mathclose{}/\mathopen{}\sim$ with $(x, y) \sim (x + 2 n, y - 3 n)$ can be described as a
lattice $\Z$ with inner product $(1)$, but this cannot be realized by
simply dropping one of the coordinates $x$ or $y$; rather, the
generator of the quotient lattice must be a vector of the form $(1, -1) + k (2, -3)$ (or its
negative). As mentioned in the main text, in all the cases we have
considered, we find that it is possible to choose a proper basis as a
subset of the original basis vectors when the full nullspace is
considered. At intermediate steps to reach such a basis, however, we
may find it useful to, e.g., project out a null vector by dropping a
coordinate in which the null vector has a non-unit value, which
naively would suggest fractional values for the coordinates in the
reduced lattice, but this generally is compensated by further nullspace
vector removal. This is the second issue raised above. As an example of this kind of procedure, consider a
4D lattice with null vectors $(7, 14, 0, 3)$ and $(1, 2, 3, 0)$ (as is the case for the example \labelcref{eq:MredExample}).  If
we first project out the first of these vectors by dropping the fourth
coordinate, we are left with a lattice of points $(x, y, z)$ where $x,
y$ may have non-integer parts $m / 3, 2 m / 3$.  We can then, however, use
the second null vector to drop the first coordinate, subtracting a
multiple $m / 3$ of this vector from the lattice vectors with
non-integer parts, and this automatically removes the non-integer
components from the second variable as well, so that $(y, z)$ are good
coordinates for the quotient lattice.
\footnote{In this simple case, this of course can also be seen easily
  by first projecting out the second null vector and then recomputing
  the null vectors in the reduced space, so that the first null vector
  in $(y, z, w)$ coordinates becomes $(0, -21, 3)$ indicating that
  $(0, -7, 1)$ is also in that second partially reduced lattice.  We
  have found it simplest, however, in specific cases of interest to
  simply start with the list of null vectors of the original matrix
  and go through intermediate non-integral lattice bases as described
  above.}  An example of a situation where the more complicated kind
of intermediate fractional lattice arises is described explicitly in
the case of the $\SM$ gauge group in a followup paper that applies the
methods developed here to universal Weierstrass models with that gauge
group. This kind of explicit computational approach for finding the
reduced basis is not essential in any way to our results but it makes
the explicit analysis of various cases easier. In general, the basis
of the quotient lattice and the resulting $M_\text{red}$ can always be
determined efficiently via the method described at the start of this section.

\section{Condition to lie in the root lattice}
\label{sec:root-lattice}

Here, we briefly discuss the conditions for an integer vector $v_i$ to lie in the root lattice of a simple group $\sfG_\text{na}$ in the basis of fundamental weights. The constraints can be simply summarized as
    \begin{equation}
        \label{eq:root-lattice-condition}
        (C^{-\transpose})_{i j} v_j \in \Z^n\,,
    \end{equation}
with $C_{i j}$ the Cartan matrix and $n$ the rank of $\sfG_\text{na}$. This follows because the rows of $C_{i j}$ are precisely the roots of $\sfG_\text{na}$ expressed in the basis of fundamental weights, and so a vector $v_i$ lies in the root lattice if and only if $v_i = (C^\transpose)_{i j} u_j$ for some integer vector $u_j$. Due to the appearance of the inverse Cartan matrix, the conditions following from \cref{eq:root-lattice-condition} are all modular congruence conditions mod $\det C$; note that $\det C$ is the order of the center of $\sfG_\text{na}$. Because $v_i$ must be an integer vector, these conditions can be reduced to a single condition for all cases except $\sfG_\text{na} = \SO(4 k + 2)$, for which there are two independent conditions mod $\Z_2$ (this is related to the fact that the center of $\SO(4 k + 2)$ is $\Z_2 \times \Z_2$, rather than $\Z_4$). The conditions are summarized in \cref{tab:root-lattice}.

\begin{table}
    \centering

    \begin{tabular}{*{2}{c}} \toprule
        Group         & Conditions \\ \midrule
        $A_n$         & $\sum_{j = 1}^n j v_j \in (n + 1) \Z$ \\
        $B_n$         & $\sum_{j = 1}^n j v_j \in 2 \Z$ \\
        $C_n$         & $v_n \in 2 \Z$ \\
        $D_{2 k}$     & $\sum_{j = 1}^{2 k} j v_j \in 2 \Z, \quad v_{2 k - 1} + v_{2 k} \in 2 \Z$ \\
        $D_{2 k + 1}$ & $(2 k - 1) v_{2 k} + (2 k + 1) v_{2 k + 1} \in 4 \Z$ \\
        $E_6$         & $\sum_{j = 1}^6 j v_j \in 3 \Z$ \\
        $E_7$         & $v_4 + v_6 + v_7 \in 2 \Z$ \\
        $E_8$         & --- \\
        $F_4$         & --- \\
        $G_2$         & --- \\ \bottomrule
    \end{tabular}

    \caption{List of modular congruence conditions that must be satisfied for an integer vector $v_i$ to lie in the root lattice (in the basis of fundamental weights) for each of the compact simple Lie groups. We use the conventions of \cite{Slansky} for the ordering of simple roots.}
    \label{tab:root-lattice}
\end{table}

\section{Notation}
Below is a list of notation commonly used throughout this document:
\label{sec:notation}
    \begin{itemize}
        \item{} $\boxed{X_0}$: Singular elliptic Calabi-Yau (CY) fourfold (i.e., complex dimension four) defining the compactification space of a 4D F-theory model. For the more general case of a $n$-fold where $n \ne 4$, we write $X_0^{(n)}$.
        \item{} $\boxed{Y_0}$: Ambient fivefold projective bundle (i.e., a bundle over the base $B$ in which the fibers are projective spaces) in which the singular CY fourfold is realized as a hypersurface, $X_0 \subset Y_0$.
        \item{} $\boxed{B}$: Threefold base of the singular elliptic CY fourfold, $X_0 \rightarrow B$. More generally, when $B$ is an $(n-1)$-fold with $n \ne 4$, we write $B^{(n-1)}$.
        \item{} $\boxed{D_\alpha}$: Basis of primitive divisors of $B$. We use the same symbol to denote a divisor and its class in the Chow ring.
        \item{}  $\boxed{D \cdot D'}$: Intersection product of pair divisors $D, D'$.
        \item{} $\boxed{[x]}$: Class of the divisor $x = 0$ in the appropriate Chow ring. For products $xy=0$, we have $[xy] = [x] + [y]$.
        \item{} $\boxed{K}$: Canonical divisor of $B$, $K= K^\alpha D_\alpha$.
        \item{} $\boxed{\Delta}$: Discriminant of the Weierstrass equation. The locus $\Delta =0$ in $B$ is the discriminant locus, over which the elliptic fibers of $X_0$ develop singularities.
        \item{} $\boxed{\Sigma_s}$: The divisor class of the codimension-one locus $\sigma_s = 0$ in the base supporting the simple gauge algebra $\mathfrak{g}_{s}$, i.e., $\Sigma_s = [ \sigma_{s}] = \Sigma_s^\alpha D_\alpha$.
        \item{} $\boxed{a_n}$: Sections of the anticanonical class, i.e., $[a_n] = n(-K)$. When the sections are tuned to vanish over a gauge divisor $\sigma = 0$, we write $a_n = a_{n,m_n} \sigma^{m}$ with $[a_{n,m_n}] = n(-K) - m_n \Sigma$.
        \item{} $\boxed{\Delta^{(2)}}$: Residual codimension-two components of the discriminant when restricted to a particular codimension-one component. For example, for Tate models with a simple gauge algebra over the codimension-one locus $\sigma = 0$ the discriminant can be written $\Delta = \sigma^m (\Delta^{(2)} + \cO(\sigma))$.
        \item{} $\boxed{X}$: Smooth elliptic CY fourfold $X \rightarrow X_0$ resolving the singular fourfold $X_0$. When the smooth fourfold is the result of an explicit finite sequence of blowups, we write $X_i$ to denote the proper transform of $X_0$ under the $i$th blowup.
        \item{} $\boxed{\pi}$: Canonical projection map from the smooth fourfold to the base, $\pi : X \rightarrow B$.
        \item{} $\boxed{\hat D_I}$: Standard geometric basis of primitive divisors in $X$, where the ``hat'' decoration distinguishes divisors in $X$ (more generally $X^{(n)}$) from divisors in $B$ (more generally $B^{(n-1)}$). For elliptic CY fourfolds, the indices $I = 0,a,\alpha,i_s$ label the zero section $\hat D_0$, generating sections $\hat D_a$, the pullbacks of divisors $\hat D_\alpha = \pi^* D_\alpha$ in the base $B$, and Cartan divisors $\hat D_{i_s}$.
        \item{} $\boxed{\hat D_{\hat I}}$: Distinctive divisors, which exclude the pullbacks of base divisors, i.e., $\hat I = 0,a,i_s$.
        \item{} $\boxed{\hat D_{\bar I}}$: Divisors in the ``gauge'' basis, $\hat D_{\bar I} = \sigma_{\bar I}^I \hat D_I$, where $\bar I = \bar{0}$ is associated to $\U(1)_\text{KK}$ and $\bar I =\bar 1$ is associated to $\U(1) \subset \sfG$. In particular, $\hat D_{\bar 1}$ is the image of $\hat D_1$ under the Shioda map.
        \item{} $\boxed{\sigma_{\bar I}^I}$: Change of basis matrix, from the standard geometric basis $\hat D_I$ to the ``gauge field'' basis $\hat D_{\bar I}$. The inverse matrix is $(\sigma^{-1})^{\bar I}_I$. See \labelcref{physbasis} and below.
        \item{} $\boxed{W_{IJKL}}$: Pushforward with respect to the projection $\pi$ of the quadruple intersection product, i.e., $ W_{IJKL} = \pi_*(\hat D_I \cdot \hat D_J \cdot \hat D_K \cdot \hat D_L)$. Since the pushforward is evaluated in the Chow ring of the base, we may write $W_{IJKL} = W_{IJKL}^{\alpha \beta \gamma} D_\alpha \cdot D_\beta \cdot D_\gamma$. See \cref{ellipticintersection}.
        \item{} $\boxed{W_{JKL|i_s}}$: Factor in the intersection product $W_{i_sJKL} = W_{JKL|i_s} \cdot \Sigma_s$. Note that $W_{JKL|i_s}=W_{JKL|i_s}^{\alpha \beta} D_\alpha \cdot D_\beta$.
        \item{} $\boxed{k_{\bar I \bar J \bar K}^{\text{5D}}}$: 5D one-loop Chern--Simons coupling.
        \item{} $\boxed{\Theta_{\bar I \bar J}^{\text{3D}}}$: 3D one-loop Chern--Simons coupling.
        \item{} $\boxed{W_{i_s|j_s}}$: (Minus the) elements of the inverse Killing form of the simple nonabelian subalgebra $\mathfrak{g}_{s}$, i.e., $W_{i_s|j_s} = - \kappa^{(s)}_{ij}$.
        \item{} $\boxed{\kappa_{ij}}$: Matrix elements of the inverse Killing form $\kappa$.
        \item{} $\boxed{W_{\bar 1 \bar 1}}$: Minus the height pairing divisor in $B$ associated to the $\U(1)$ gauge factor.
        \item{} $\boxed{[A_{a}]}$: A vector $A$ whose components are $A_a$. Not to be confused with the class of a divisor, as should hopefully be clear from the context.
        \item{} $\boxed{[[A_{ab}]]}$: A matrix $A$ whose elements are $A_{ab}$.
        \item{} $\boxed{Y}$: Ambient fivefold bundle in which the resolution $X$ is realized as a hypersurface, $X \subset Y$. In practice, $Y = Y_i$ is the total transform of the ambient projective bundle $Y_0$ under a composition of blowups, $f_i : Y_{i-1} \rightarrow Y_i$.
        \item{} $\boxed{\boldsymbol{D}}$: Divisor in the ambient fivefold $Y$ whose restriction to the hypersurface $X$ is a divisor in $X$, i.e., $D = \boldsymbol{D} \cap \boldsymbol{X}$.
        \item{} $\boxed{e_i}$: Local coordinate whose zero locus in $Y$ is (the proper transform of) the exceptional divisor $\boldsymbol{E}_i$.
        \item{} $\boxed{\varpi}$: Canonical projection map from the ambient fivefold to the base, $\varpi : Y \rightarrow B$.
        \item{} $\boxed{\sfG}$: F-theory gauge symmetry group encoded in the singularities of $X_0$.
        \item{} $\boxed{\sfG_\text{na}}$: Nonabelian subgroup of $\sfG$. We abuse notation and write $\sfG_\text{na}= \prod_s \sfG_{s}$ where the index $s$ labels the simple subgroups of $\sfG_\text{na}$.
        \item{} $\boxed{\mathfrak{g}}$: Lie algebra of the gauge group, $\mathfrak{g} = \text{Lie}(\sfG)$.
        \item{} $\boxed{\mathfrak{g}_\text{na}}$: Nonabelian subalgebra of $\mathfrak{g}$. Analogously, we write $\mathfrak{g}_\text{na} = \oplus_s \mathfrak{g}_{s}$.
        \item{} $\boxed{\sfr}$: Irreducible complex representation of the gauge symmetry group $\sfG$.
        \item{} $\boxed{n_{\sfr}}$: Multiplicity of irreps $\sfr$ appearing in a representation, i.e., $\sfr \oplus \cdots \oplus \sfr = \sfr^{\oplus \sfr}$.
        \item{} $\boxed{\chi_{\sfr}}$: Chiral multiplicity of matter representations $\sfr$, i.e., $\chi_{\sfr} = n_{\sfr} - n_{\sfr^*} = - \chi_{\sfr^*}$.
        \item{} $\boxed{\sfR}$: Quaternionic representation, $\sfR =\sfr \oplus \sfr^*$.
        \item{} $\boxed{C_{\sfR}}$: Class of the codimension-two locus in $B$ over which local matter transforming in the representation $\sfr$ or $\sfr^*$ is supported. Given a gauge divisor $\Sigma_{s}$, $C_{\sfR_s} = \Sigma_{s} \cdot (a_{\sfR_s} K + b_{\sfR_s} \Sigma_s)$ for some coefficients $a_{\sfR_s},b_{\sfR_s} \in \Q$.
        \item{} $\boxed{w^{\sfr}_i}$: Dynkin coefficients (i.e., coefficients in a basis of fundamental weights, which are canonically dual to the simple coroots) of the weight $w^{\sfr}$ of the representation $\sfr$.
        \item{} $\boxed{\varphi^i}$: Real Coulomb branch moduli. Equivalently, these are the coefficients of a scalar $\varphi$ expanded in a basis of simple coroots. Since the simple coroots of an algebra are canonically dual to the fundamental weights, given a weight $w^{\sfr}$, we have $\varphi \cdot w^{\sfr} =\varphi^i w_i^{\sfr}$.
        \item{} $\boxed{H^{2,2}_{\text{vert}}(X,\Z)}$: Vertical
          cohomology subgroup of the orthogonal decomposition
          $H^{2,2}(X,\C) = H^{2,2}_{\text{vert}}(X,\C)
          \oplus H^{4}_{\text{hor}}(X,\C) \oplus
          H^{2,2}_{\text{rem}}(X,\C)$. Note that
          $H^{2,2}_{\text{vert}}(X,\Z)$ is the linear span with
          integer coefficients of wedge products of the elements of
          $H^{1,1}(X,\Z)$. Given a basis of divisors $\hat
          D_I$, we write $\PD(S_{IJ}) = \PD(\hat D_I) \wedge \PD(\hat
          D_J)$.

Equivalently, $H_{2,2}^{\text{vert}}(X,\Z)$ is
spanned by $S_{IJ}$ modulo homological equivalence, $\phi \sim \psi \Leftrightarrow M (\phi -\psi) = 0$. As a lattice, $H_{2,2}^{\text{vert}}(X,\Z) = \Lambda_{S}\mathclose{}/\mathopen{}\sim$.

        \item{} $\boxed{\PD(\hat D)}$: Poincar\'{e} dual of the divisor $\hat D$.
        \item{} $\boxed{\Lambda_S}$: Lattice of 4-cycles $S_{IJ} = \hat D_I \cap \hat D_J$ equipped with the bilinear form $M$.
        \item{} $\boxed{\phi}$: Poincar\'{e} dual vertical flux background, $\phi = \PD(G) \in \Lambda_S$. We frequently abuse terminology and refer to $\phi$ as a flux background, rather than the Poincar\'{e} dual of a flux background $G$.
        \item{} $\boxed{\phi'}$: ``Non-distinctive'' flux backgrounds, i.e., backgrounds spanning the directions $IJ$ where $I = \alpha$ or $J= \alpha$.
        \item{} $\boxed{\phi''}$: ``Distinctive'' flux backgrounds, i.e., backgrounds spanning the directions $IJ$ where $I, J \ne \alpha$.
            \item{} $\boxed{M}$: Intersection pairing $M : \Lambda_S \times \Lambda_S \rightarrow Z$. As a matrix, we write $M_{(IJ)(KL)} = (\hat D_I \cdot \hat D_J) \cdot (\hat D_K \cdot \hat D_L)$.
        \item{} $\boxed{\Theta_{IJ}}$: Integral of a vertical flux background over the cycle $S_{IJ}$, i.e., $\Theta_{IJ} = \int_{S_{IJ}} G = \phi \cdot (\hat D_I \cdot \hat D_J) = M_{(IJ)(KL)} \phi^{KL}$.
        \item{} $\boxed{\Theta_{IJ}^{\text{d}}}$: The terms in the expansion of $\Theta_{IJ}$ that only depend on distinctive flux backgrounds, i.e., $\Theta_{IJ}^{\text{d}} = M_{(IJ)(\hat K \hat L)} \phi^{\hat K \hat L}$.
        \item{} $\boxed{\Lambda_C}$: The sublattice $\Lambda_C \subset
          \Lambda_S$ of flux backgrounds $\phi$ satisfying the
          symmetry constraints $\Theta_{I\alpha} =0$. See
          \labelcref{eq:Poincare,eq:gauge}. We sometimes write
          $\Lambda_C \cong P \Lambda_{S}$, where $P$ is an idempotent
          matrix, when we can solve the symmetry constraints
for all $\phi'$ in terms of $\phi''$.
%       \item{} $\boxed{MP^{\text{na}}}$: Restriction of the intersection pairing $M$ to the sublattice $\Lambda_C^{\text{na}}$ of fluxes satisfying all symmetry constraints except the $\U(1)$ constraints $\Theta_{1\alpha}=0$, via right action of the idempotent matrix $P^{\text{na}}$.
        \item{} $\boxed{M_C}$: Restriction of $M$ to the sublattice
          $\Lambda_C \subset \Lambda_S$.
        \item{} $\boxed{MP}$: Restriction of the intersection pairing
          $M$ to the sublattice $\Lambda_C$ via right action of the
          idempotent matrix $P$, which is defined in certain
          circumstances, and simplifies some computations.
In these cases, $M_C$ can
          be written explicitly as a submatrix in $MP
          = \begin{pmatrix} 0 & \\ & M_C\end{pmatrix}$ that only acts
            on distinctive flux backgrounds $\phi'' \subset
            \Lambda_{S}$, given the embedding $\phi =\begin{pmatrix}
            \phi' \\ \phi'' \end{pmatrix}$
Note the isomorphism $M_C \Lambda_C \cong M P \Lambda_S$, and furthermore that $(MP)^{\text{t}}= P^{\text{t}} M = MP$.
%        \item{} $\boxed{H_{2,2}^{\text{vert}}(X,\Z)}$: Vertical homology subgroup, spanned by $S_{IJ}$ modulo homological equivalence, $\phi \sim \psi \Leftrightarrow M (\phi -\psi) = 0$. As a lattice, $H_{2,2}^{\text{vert}}(X,\Z) = \Lambda_{S}\mathclose{}/\mathopen{}\sim$.
        \item{} $\boxed{\nu}$: Null vectors of $M$, i.e., $M \nu = 0$. Equivalently, vectors satisfying $\Theta_{IJ} \nu^{IJ} =0$ or $S_{IJ} \nu^{IJ} =0$. Each independent $\nu$ represents a homological equivalence relation.
        \item{} $\boxed{M_\text{red}}$: ``Reduced intersection
          pairing'', i.e., the restriction of the intersection pairing
          $M$ to $H_{2,2}^{\text{vert}}(X,\Z)
=          \Lambda_S\mathclose{}/\mathopen{}\sim$.
        \item{} $\boxed{\Lambda_\text{phys}}$: Sublattice $\Lambda_\text{phys} \subset
H_{2,2}^{\text{vert}}(X,\Z)=
          \Lambda_S\mathclose{}/\mathopen{}\sim$ of flux backgrounds both satisfying the symmetry constraints and quotiented by homological equivalence. We sometimes write $\Lambda_\text{phys} = \Lambda_C\mathclose{}/\mathopen{}\sim$.
        \item{} $\boxed{M_\text{phys}}$: Restriction of the intersection pairing $M$ to $\Lambda_\text{phys}$. Schematically, $M_\text{phys} = M_C\mathclose{}/\mathopen{}\sim$.
    \end{itemize}

\bibliographystyle{JHEP}
\bibliography{references}

\end{document}